\documentclass[english,pointlessnumbers, abstracton, headsepline,twoside,openright]{scrreprt}
\usepackage[LGR,T1]{fontenc}
\usepackage[latin9]{inputenc}
\usepackage{color}
\usepackage{babel}
\usepackage{array}
\usepackage{verbatim}
\usepackage{varioref}
\usepackage{float}
\usepackage{calc}
\usepackage{textcomp}
\usepackage{url}
\usepackage{multirow}
\usepackage{amsmath}
\usepackage{stackrel}
\usepackage{graphicx}
\usepackage{setspace}
\usepackage[authoryear]{natbib}
\makeatletter

\DeclareRobustCommand{\greektext}{%
  \fontencoding{LGR}\selectfont\def\encodingdefault{LGR}}
\DeclareRobustCommand{\textgreek}[1]{\leavevmode{\greektext #1}}
\ProvideTextCommand{\~}{LGR}[1]{\char126#1}

\providecommand{\tabularnewline}{\\}

\usepackage{babel}
\usepackage{lettrine}
\usepackage{siunitx}
\usepackage{ 
            ellipsis, fixltx2e, mparhack,   
            booktabs, longtable             
}  
\usepackage{textcomp}
\usepackage[automark]{scrpage2}

\usepackage{amssymb}
\usepackage{amsmath}
\usepackage{amsfonts}

\usepackage{float}
\usepackage{multirow}
\usepackage{caption}
\usepackage{gensymb} 
\usepackage{times}
\usepackage[dvipsnames]{xcolor}
\usepackage{tikz}

\usepackage{empty page}
\usepackage{zref-abspage}

\title{CAVAthesis}

\newcommand\myworries[1]{}

\clearscrheadfoot
\ohead{\headmark}
\ofoot[\pagemark]{\pagemark}

\renewenvironment{abstract}{
    \@beginparpenalty\@lowpenalty
      \begin{center}
\normalfont\sectfont\abstractname
        \@endparpenalty\@M
      \end{center}
}{
    \par
}

\usepackage{microtype}

\usepackage{ifpdf} 
\ifpdf 

 \IfFileExists{lmodern.sty}{\usepackage{lmodern}}
  {\usepackage[scaled=0.92]{helvet}
    \usepackage{mathptmx}
    \usepackage{courier} }
\fi

\addto\captionsngerman{ 
\renewcommand{\abstractname}{Abstract}
}

 \usepackage[colorlinks=true, bookmarks, bookmarksnumbered, bookmarksopen, bookmarksopenlevel=1,
  linkcolor=black, citecolor=black, urlcolor=blue, filecolor=blue,
  pdfpagelayout=OneColumn, pdfnewwindow=true,
  pdfstartview=XYZ, plainpages=false, pdfpagelabels,
  pdfauthor={LyX Team}, pdftex,
  pdftitle={LyX's Figure, Table, Floats, Notes, and Boxes manual},
  pdfsubject={LyX-documentation about figures, tables, floats, notes, and boxes},
  pdfkeywords={LyX, Tables, Figures, Floats, Boxes, Notes}]{hyperref}

\newcommand{\@ldtable}{}
\let\@ldtable\table
\renewcommand{\table}{ %
                 \setlength{\@tempdima}{\abovecaptionskip} %
                 \setlength{\abovecaptionskip}{\belowcaptionskip} %
                 \setlength{\belowcaptionskip}{\@tempdima} %
                 \@ldtable}

\usepackage{xcolor} 
 
\colorlet{chapter}{black!100} 
 
\addtokomafont{chapter}{\color{chapter}} 
 
\makeatletter
 \renewcommand*{\chapterformat}{%
   \begingroup
     \setlength{\unitlength}{1mm}%
     \begin{picture}(20,40)(0,5) 
       \setlength{\fboxsep}{0pt} 
      
       \put(22,15){\line(1,0){\dimexpr 
           \textwidth-20\unitlength\relax\@gobble}}%
       \put(0,0){\makebox(20,20)[r]{%
           \fontsize{28\unitlength}{28\unitlength}\selectfont\thechapter 
           \kern-.05em
         }}%
       \put(22,15){\makebox(\dimexpr 
           \textwidth-20\unitlength\relax\@gobble,\ht\strutbox\@gobble)[l]{%
             \ \normalsize\color{black}\chapapp~\thechapter\autodot 
           }}%
     \end{picture} 
 \let\cleardoublepage\relax
\let\clearpage\relax
   \endgroup 
 
} 
 
\let\cleardoublepage\relax
\usepackage{pdfpages}

\usepackage{shadethm}  

\usepackage{listings}
    \newcommand*\styleC{\fontsize{9}{10pt}\usefont{T1}{ptm}{m}{n}\selectfont }
    \newcommand*\styleD{\fontsize{9}{10pt}\usefont{OT1}{pag}{m}{n}\selectfont }

\usepackage[pdfpagelabels]{hyperref}
\hypersetup{colorlinks=true,linkcolor=black,citecolor=Violet,urlcolor=Blue,pdfstartview= FitH} 
\hypersetup{pdftitle =PhD Thesis,pdfdisplaydoctitle=true,plainpages=false}
\hypersetup{pdfpagelayout=TwoPageRight}

    \makeatletter
    \edef\Motscle{emph={\lst@keywords}}
    \expandafter\lstset\expandafter{%
      \Motscle}
    \makeatother

    \definecolor{Ggris}{rgb}{0.45,0.48,0.45}

\makeatletter
\def\cleardoublepage{\clearpage\if@twoside \ifodd\c@page\else\hbox{}
\vspace*{\fill}\vspace{\fill}\thispagestyle{empty}
\newpage\if@twocolumn\hbox{}
\newpage\fi\fi\fi}
\let\cleardoublepage\relax
\makeatother

{\textbf{Definition}\begin{itshape}}
{\end{itshape}}

{\textbf{Algorithm (sNM3F)}\begin{itshape}}
{\end{itshape}}

{\textbf{Proof of convergence}}
{}

\AtBeginDocument{
  
}

\makeatother

\begin{document}
\thispagestyle{empty} 
\setcounter{page}{1}
\begin{center}
\vspace{2cm}
\begin{table}[ht]
\centering 
\begin{tabular}{lp{3cm}r}
{\includegraphics[height=4cm]{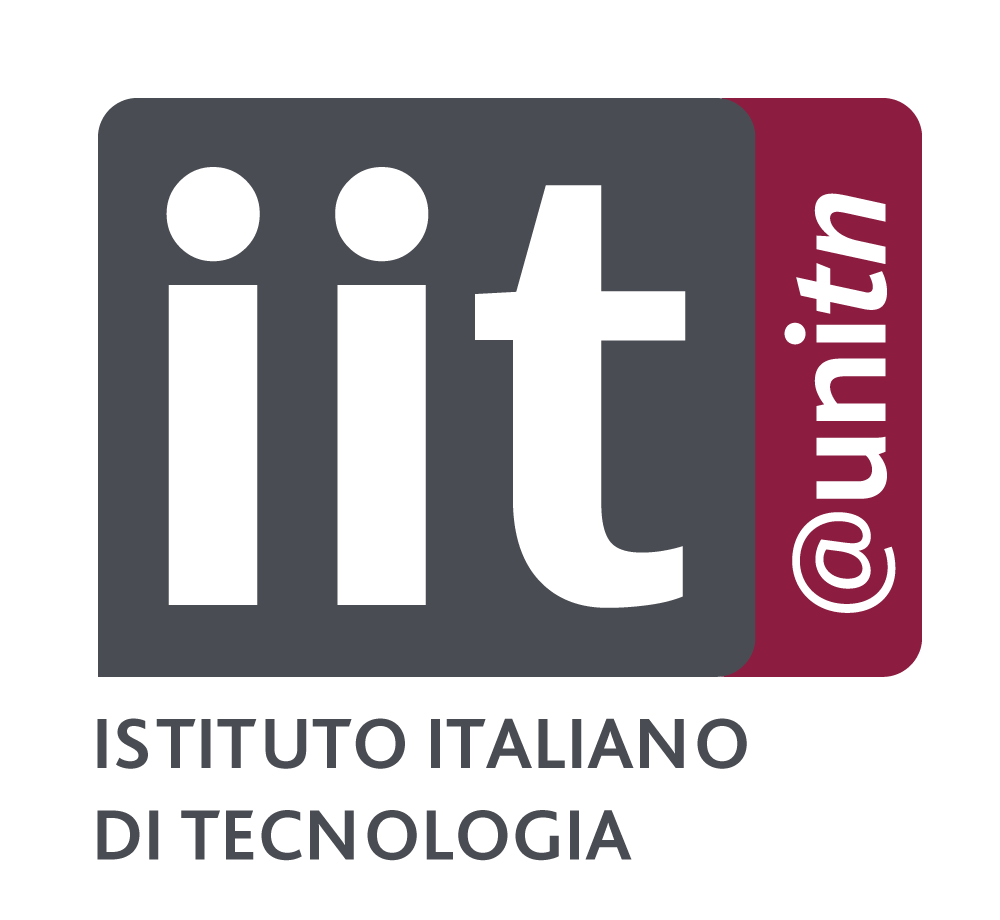}}&&
{\includegraphics[height=3.6cm]{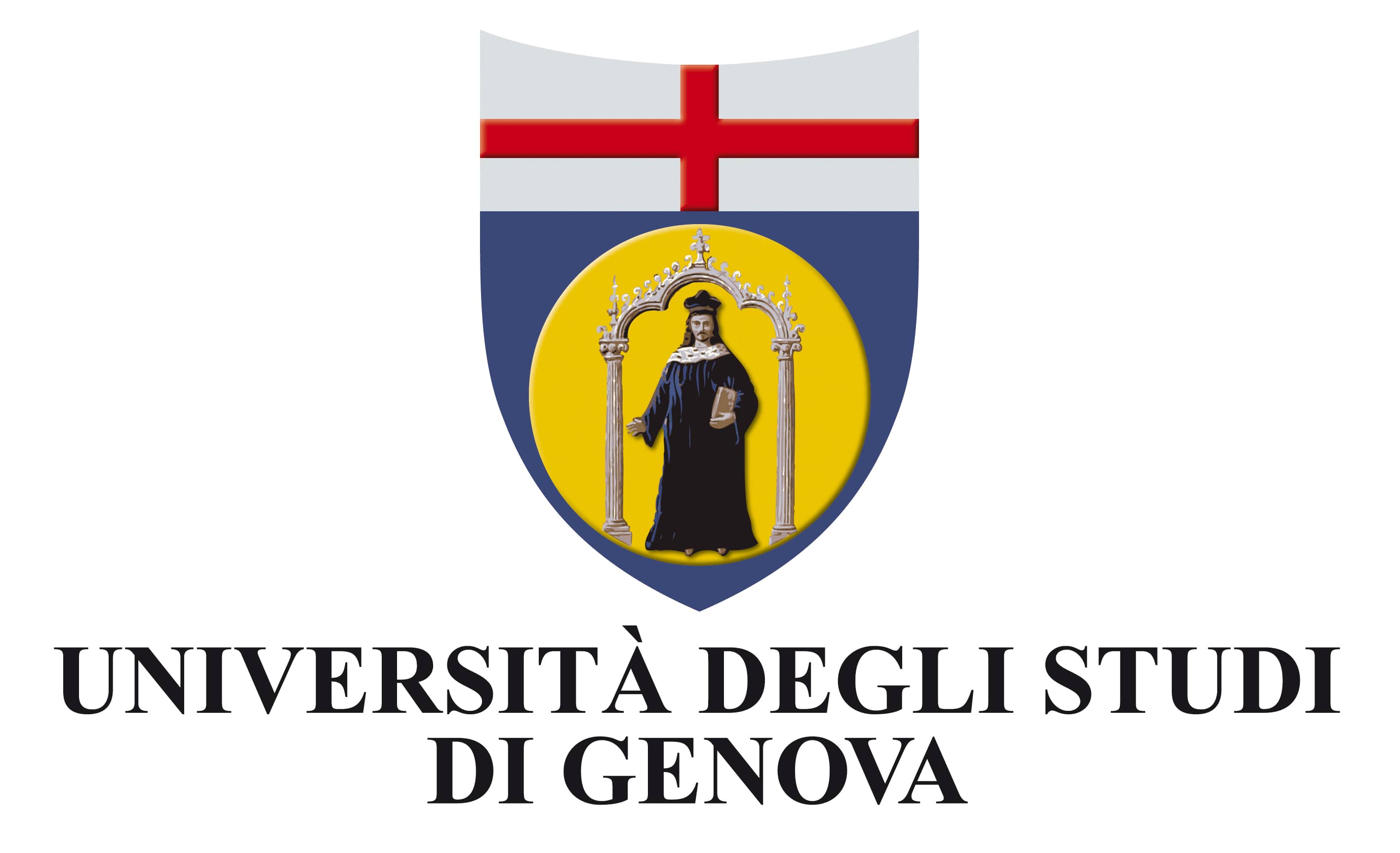}}\\
\end{tabular}
\end{table}
\vspace{1cm} 
\LARGE {\bfseries {A computational investigation of the relationships\\ 
between single-neuron and network dynamics\\
in the cerebral cortex}}\\[1cm] 
\large \bfseries Stefano Cavallari\\[2cm] 
\end{center} 
\begin{center}
\large \bfseries Advisor: Prof. Stefano Panzeri\\
\large \bfseries Co-advisor: Dr. Alberto Mazzoni\\[1.5cm]
\end{center} 
\begin{center}
A dissertation submitted in partial fulfillment\\
of the requirements for the degree of\\
Doctor of Philosophy (Ph.D.)\\[1cm] 
\large Doctoral School on ``Life and Humanoid Technologies''\\
\large Doctoral Course on ``Robotics, Cognition and Interaction Technologies''\\[1cm] 
April 2015\\ 
XXVII Cycle
\end{center} 
\setcounter{footnote}{0} 


\setlength{\evensidemargin}{0.5cm} 
\setlength{\oddsidemargin}{0.5cm} 

\newpage{}

~

\newpage{}

\begin{spacing}{0.5}
\begin{center}
\textit{\small{}}%
\begin{minipage}[c]{1\columnwidth}%
\begin{singlespace}
\begin{flushright}
\textit{\small{}\textquotedbl{}The brain is wider than the sky,}\\
\textit{\small{}For, put them side by side,}\\
\textit{\small{}The one the other will include}\\
\textit{\small{}With ease, and you beside.}
\par\end{flushright}{\small \par}
\begin{flushright}
\textit{\small{}The brain is deeper than the sea,}\\
\textit{\small{}For, hold them, blue to blue,}\\
\textit{\small{}The one the other will absorb,}\\
\textit{\small{}As sponges, buckets do.}
\par\end{flushright}{\small \par}
\begin{flushright}
\textit{\small{}The brain is just the weight of God,}\\
\textit{\small{}For, lift them, pound for pound,}\\
\textit{\small{}And they will differ, if they do,}\\
\textit{\small{}As syllable from sound.\textquotedbl{}\medskip{}
}
\par\end{flushright}{\small \par}
\begin{flushright}
\textit{\small{}Emily Dickinson, 1862}
\par\end{flushright}{\small \par}
\end{singlespace}
\end{minipage}
\par\end{center}{\small \par}
\end{spacing}

\newpage{}

~
\begin{quotation}
\pagebreak{}
\end{quotation}
\begin{abstract}
The brain in complex animals is organized in different areas, such
as visual areas, motor areas etc., which are believed to be responsible
for specific functions. Each area can have in turn its own organization,
but for all areas it is believed that their functions rely on the
dynamics of networks of neurons rather than on single neurons. On
the other hand, the network dynamics reflect and arise from the integration
and coordination of the activity of populations of single neurons.
Understanding how single-neurons and neural-circuits dynamics complement
each other to produce brain functions is thus of paramount importance. 

LFPs and EEGs are good indicators of the dynamics of mesoscopic and
macroscopic populations of neurons, while microscopic-level activities
can be documented by measuring the membrane potential, the synaptic
currents or the spiking activity of individual neurons. How can we
combine the information coming from microscopic, mesoscopic and macroscopic
levels to get better insights about the relationships between single-neuron
activity and the state of neural circuits? How can we model the relationship
between these two levels? How can we optimally analyze concurrent
recordings at multiple scales to gain insights on the relationship
between macroscopic/mesoscopic and microscopic brain dynamics?

In this thesis we develop mathematical modelling and mathematical
analysis tools that can help the interpretation of joint measures
of neural activity at microscopic and mesoscopic or macroscopic scales.
In particular, we develop network models of recurrent cortical circuits
that can clarify the impact of several aspects of single-neuron dynamics
(that depend on the details of synaptic dynamics) on the activity
of the whole neural population (i.e., at the mesoscopic level). We
then develop statistical tools to characterize the relationship between
the action potential firing of single neurons and mass signals. We
apply these latter analysis techniques to joint recordings of the
firing activity of individual cell-type identified neurons and mesoscopic
(i.e., LFP) and macroscopic (i.e., EEG) signals in the mouse neocortex.
We identified several general aspects of the relationship between
cell-specific neural firing and mass circuit activity, providing for
example general and robust mathematical rules which infer single-neuron
firing activity from mass measures such as the LFP and the EEG. 
\end{abstract}
\newpage{}

\cleardoublepage 
\phantomsection 
\addcontentsline{toc}{chapter}{Contents} 
\tableofcontents

\cleardoublepage 
\phantomsection 
\addcontentsline{toc}{chapter}{List of Figures} 
\listoffigures	
\cleardoublepage 
\phantomsection 
\addcontentsline{toc}{chapter}{List of Tables} 
\listoftables  
\cleardoublepage

\newpage{}

\ohead[]{} \begin{center} 
\textsf{\textbf{\large Abbreviations}}
\end{center}

\begin{tabular}{c|c}
BMI & Brain Machine Interface\tabularnewline
COBN & Conductance-Based Network\tabularnewline
CUBN & Current-Based Network\tabularnewline
EEG & Electroencephalography\tabularnewline
fMRI & Functional Magnetic Resonance Imaging\tabularnewline
FR & Firing Rate\tabularnewline
GLM & General Linear Model\tabularnewline
LIF & Leaky Integrate-and-Fire\tabularnewline
LFP & Local Field Potential\tabularnewline
MP & Membrane Potential\tabularnewline
MUA & Multi-Unit Activity\tabularnewline
NMSD & Normalized Mean Squared Distance \tabularnewline
OU & Ornstein-Uhlenbeck\tabularnewline
PSP & Post Synaptic Potential\tabularnewline
PV-pos & Parvalbumin-positive interneuron\tabularnewline
SOM-pos & Somatostatin-positive interneuron\tabularnewline
spk & Spike times\tabularnewline
SUA & Single-Unit Activity\tabularnewline
\end{tabular}

\newpage{}

~

\newpage{}

\chapter{Introduction}

\ohead{\headmark} 
\pagestyle{scrheadings}    

\lettrine[lines=2]{I}{n the} first hundred years of neuroscience,
many influential studies that shaped our current understanding of
brain function were performed by recording and considering the response
properties of single neurons in isolation. Examples of this progress
are for the classic work of Vernon Mountcastle or of David Huebel
and Thisten Wiesel on the receptive field and stimulus tuning properties
of neurons in different sensory modalities \citep{mountcastle1957modality,Talbot1968sense,mountcastle1978organizing,hubel1959receptive,hubel1962receptive,wiesel1963single,hubel1968receptive}
or the work of Michael Brecht and colleagues on the behavioral effect
of stimulating individual neurons \citep{houweling2008behavioural}.
However, considering each neuron in isolation can only lead us so
far. No neuron is an island. In recent studies, the idea of point
neurons performing a function is gradually being replaced by the conceptual
framework of considering neurons as part of the circuits they belong
to. 

Neurons belong to microcircuits, not all elements of which are necessarily
selective to sensory stimulus, and not all activations within the
microcircuits imply a direct control of the neuron by the sensory
stimulus, as it would be the case in feedforward processing. Feedforward
inputs to cortical microcircuits are typically weak, and there is
strong recurrent excitation and inhibition whose balance - which is
obviously critical for circuit dynamics - may depend on the external
input as well as on neuromodulation \citep{Logothetis08}. Thus the
activity of a neuron can only be understood in the context of the
state of the microcircuit, mesoscopic and macroscopic circuits it
belongs to \citep{Panzeri_Macke_Gross_Kayser_2015}. Given that most
brain functions likely arise from the concerted operation of many
microscopic and macroscopic circuits, even very dense recordings from
a single structure can only get us so far. 

The alternative to single-neuron recording, which is the dominating
one for studying neural mechanisms of cognition in humans, is to use
tools such as fMRI or EEG/MEG to collect macroscopic measures of massed
neural action over large regions (potentially, the whole brain). This
has the clear advantage of being able to capture concerted relationships
between macroscopic structures. However, the linkage between microscopic
neuronal activity and the measured massed action at each site is immensely
complex. Take for example EEG. The typical integration area of an
EEG electrode contains several million neurons, a few tens of billion
synapses, tens of km of dendrites and hundreds of km of axons. These
numbers by themselves suggest the difficulty of drawing analogies
between mass measures and microscopic neural activity. In addition,
the organization of microconnectivity within the microcircuit (or
EEG integration area) leads (as discussed above) to state-dependent
dynamics \citep{douglas1989canonical,douglas2004neuronal}. This means
that mesoscopic and macroscopic massed neural dynamics can in principle
arise from a large number of states or different circuit operations.
This in turn implies that the neural interpretation of massed noninvasive
methodologies may not be possible without concurrent electrical measurements
of activity of single neurons or small populations thereof \citep{Panzeri_Macke_Gross_Kayser_2015}.

Because of the above reasons, recent experimental efforts have been
aimed at the simultaneous recording of microscopic and mesoscopic-macroscopic
brain activity. Several recent studies reported new observations of
the relationships between the spiking dynamics of a few neurons and
of mesoscopic and macroscopic circuits they belong to \citep{schwartz2006spike,whittingstall2009frequency,Rasch08,rasch2009neurons,nauhaus2009stimulus,okun2010subthreshold,zanos2012relationships,waldert2013influence,hall2014real},
revealing for example that the information carried by single neurons
is state-dependent \citep{harris2011cortical} and it can only be
read out when single-neuron spikes are referred to indicators of microcircuit
state such as the phase of LFPs \citep{Montemurro08,Kayser09,Mazzoni2011,Panzeri_Macke_Gross_Kayser_2015}.
In addition, other studies have begun to use simultaneous recordings
of the local activity of small neural populations together with large-scale
measures of mass activity in multiple regions, and have given already
important insights into the relationship between local and global
brain dynamics. For example, one study \citep{canolty2010oscillatory}
used simultaneous recordings of spiking activity and LFPs from multiple
brain regions to reveal the important role of phase coordination among
oscillations in different regions for the selective recruitment of
cell assemblies. Another study \citep{logothetis2012hippocampal}
used concurrent electrophysiological measures and whole-brain fMRI
to reveal the patterns of whole-brain activity that happen in correspondence
to the firing of high-frequency sharp wave ripple events in the hippocampus. 

These new simultaneous recordings of signals at different scales hold
the key to relate microcircuit dynamics to massed neural activity
and to the interrelationships among macroscopic networks. However,
we still lack the appropriate mathematical tools to properly analyze
and interpret these recordings. In particular, we need better modeling
tools to relate single-neuron synaptic and spiking dynamics to the
dynamics of the whole circuit. We also need better analytical tools
to describe the empirical relationships between the microscopic activity
of specific cell types in cortex and the macroscopic or mesoscopic
circuit activation. In this thesis, we make progress along both these
directions.

The work presented in this thesis is organized as follows. 

In the next chapter, we begin by summarizing the main concepts of
computational neuroscience that we used to develop this work (i.e.,
LIF network models, mutual information and Wiener kernel methods)
to give the opportunity to readers without specialist background on
mathematical neuroscience to acquire the basic tools that will help
the understanding of the research presented in successive chapters. 

In chapter \ref{chapter_frontiers}, we describe a new neural network
modelling framework that permits the understanding of the impact of
specific details of synaptic dynamics onto mesoscopic-level circuit
activity. First we develop a new algorithm for comparing quantitatively
and fairly how different assumptions (i.e., mathematical expressions)
about synaptic dynamics affect circuit-level activity. Previous studies
have compared the effects of different choices of synaptic models
mainly at the single-neuron level or choosing parameters for network-dynamics
comparison in a rather arbitrary way, whereas we introduce a rigorous
framework to perform such comparison. By applying this formalism to
the study of the dynamics of recurrent networks, we find that the
network-scale first order statistics (population FR and spectrum of
the network oscillations) are robust to the changes in the single-neuron
synaptic properties, while both the correlation properties of neural
population interactions and the modulation of network oscillations
by external inputs strongly depend on the choice of the synaptic model
\citep{Cavallari_Panzeri_Mazzoni_2014}.

In chapter \ref{chapter_Fellin}, we first develop - and then apply
to data - new mathematical tools for the analysis of the relationship
between single-neuron spiking activity and mass signals (EEG or LFP).
We apply these tools to simultaneous recordings of LFPs and EEGs together
with the firing activity of identified classes of neurons during slow
wave oscillations. We find that the linear component of the relationship
between single-neuron activity and mass signals is remarkably stable
across cells and animals, allowing a blind estimation of the mass
signals from the spiking activity (both of excitatory and inhibitory
neurons) and vice versa. We also observe that the single-unit activity
tends to prevent changes in both the LFP and EEG signals. 

Finally, in chapter \ref{chap:Conclusions}, we summarized the results
presented in the previous chapters and discussed their implications,
as well as further questions arising from this work. We concluded
by describing very interesting and promising directions for further
investigations. 

\newpage{}

\chapter{Theoretical framework\label{chap_Theoretical-framework}}

\ohead{\headmark} 
\pagestyle{scrheadings}    

\lettrine[lines=2]{T}{he} brain is likely the most complex system
in the universe, and we are still far away from a general theory able
to explain from first principles the way it works. In absence of a
first-principles general theory, it has been however possible to make
progress by identifying mechanisms at multiple spatial and temporal
scales that likely influence brain dynamics, and then building quantitative
mathematical models of these mechanisms. These models can in turn
be compared to data and help to validate quantitatively and refine
initial hypothesis on the relationship between the brain's biophysics
and the brain's function. The effort on mathematical modelling of
neurons and neural networks has been vast (for recent books, see \citet{Dayan,quiroga2013principles,izhikevich2007dynamical,gerstner2014neuronal})
and cannot be reviewed in a single thesis.

In this section, however, we introduce a small set of elements of
the conceptual and mathematical formulation of neural function \citep{Bear2007}
and of the main features of the (single-neuron and network) models
\citep{Dayan} that we will use to investigate the relationship between
network dynamics and single-neuron activity (see chapter \ref{chapter_frontiers}).
We hope that this short introduction will help readers without a specialist
background to navigate through our original research, which we will
present in the next chapters. 

\section{Modelling neurons and neural circuits}

\subsection{Introduction}

The brains of all species composed primarily of two broad classes
of cells: neurons and glial cells. Glial cells come in several types,
and perform a number of critical functions, including structural support,
metabolic support, insulation, and guidance of development, neurons,
however, are usually considered the most important cells in the brain
\citep{Kandel} and in this work we will focus only on the neuronal
activity. 

The brain (like any biological systems) can be investigated with many
different instruments (e.g. microelectrodes, calcium-imaging, fMRI,
etc.) and each of them take pictures of the brain activity from a
different prospective (and on a different scale). It is similar to
what happen in Physics with the description of a physical system in
different reference frames, but in neuroscience we are still faraway
from knowing the transformation laws to move from a reference system
to another. Our point of view to investigate the brain activity (our
\textquotedbl{}reference system\textquotedbl{}) it is represented
by the electrical properties/activity of the neurons and of networks
of neurons measured through microelectrodes or glass pipette. Therefore
we are going to introduce a simple way to model the electrical properties
of the neurons. In particular, we will use the electrical circuits
theory to model the neurons and their networks (with some ad hoc assumptions
to model the action potential), so, in the end, we will describe the
neural networks as specific electrical circuits. Indeed the ingredients
we will adopt to build the model are electric charges, electric potentials,
capacitors, resistance, electric currents, etc... This is also the
most natural choice because
\begin{enumerate}
\item The neurons exchange action potentials, which are electric potential
\item The experimental data we will use to test our models come from recordings
of electric potentials
\end{enumerate}
The neurons communicate each other through electrical signals that
travel across the neuronal structures. The electrical signals are
i.e., positive ons (like Ca\textsuperscript{2+}, K\textsuperscript{+},
Na\textsuperscript{+}, Cl\textsuperscript{-}) that move driven by
an electric potential generated by an inhomogeneous spatial displacement
of positive and negative charges (ions). The neuronal structure that
allows this separation of positive and negative charge (needed to
generate the electrical potential) is the cell membrane.

The cell membrane is a lipid bilayer 3 to 4 nm thick that is essentially
impermeable to most charged molecules. This insulating feature causes
the cell membrane to act as a capacitor where the two electrical plates
are given by the internal and external surfaces of the membrane. The
(membrane) potential is precisely the difference between the electrical
potentials measured on these two surfaces of the membrane. Most of
the time, there is an excess concentration of negative charge inside
a neuron. By convention, the potential of the extracellular fluid
outside a neuron is defined to be zero, therefore, when a neuron is
at rest (that is the net flow of current across the cell membrane
is zero) the potential is negative with a value around -65 mV.

\begin{figure}
\begin{centering}
\includegraphics[scale=0.2]{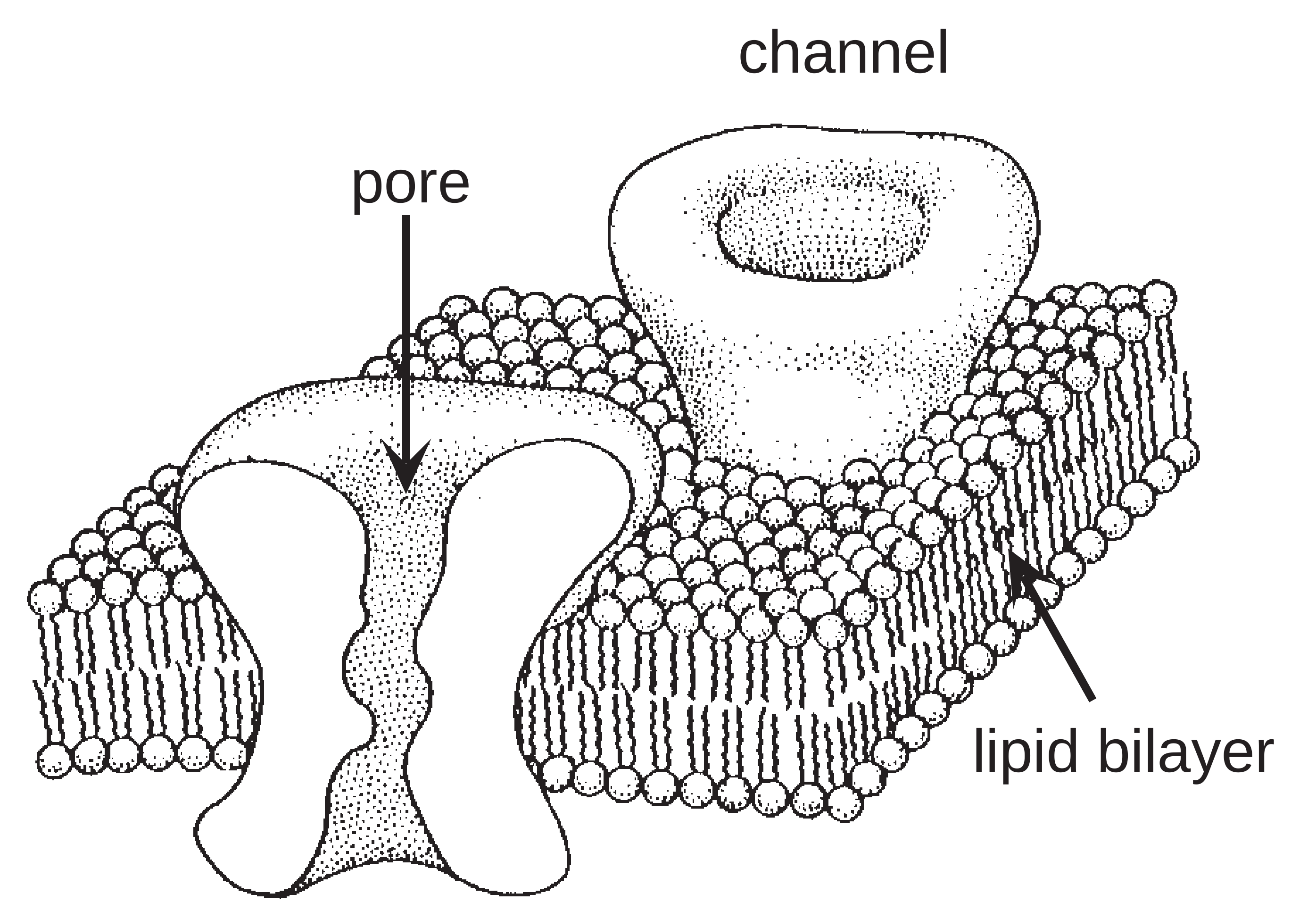}
\par\end{centering}
\centering{}\caption[Lipid bilayer]{\textbf{Schematic diagram of two ion channels embedded in a section
of lipid bilayer}. The ion channels are about 10 nm long. (Source:
\citealp{Dayan})\label{fig_lipid_bilayer}}
\end{figure}

The membrane is embedded with many ion-conducting channels (usually
highly selective, see figure \ref{fig_lipid_bilayer}) which affect
in a dynamic-dependent way the ionic permeability of the membrane
(that overall increases of about 10,000 times with respect to a pure
lipid bilayer). Indeed the movement of ionic charges and the generation
and transmission of action potentials in the neurons are ruled by
a lot of complex biological mechanisms which ultimately determine
the opening and the closing of the ionic channels: the cell membrane
actively shapes the flow of the ionic currents. Therefore the electric
signals spreading across neural networks is different both in the
way it travels and in composition (it is a ionic current, not a current
of free electrons) with respect to a current of free electrons flowing
in a conductive materials. Furthermore the membrane contains pumps
for selected ions whose role is to expend energy to maintain the difference
in the ion concentrations between the inner and the outer part of
the cell. 

In summary, the conductive properties of the channels can be affected
by many factors \label{conductance_changes}such as:
\begin{itemize}
\item the membrane potential
\item the intracellular concentration of various intracellular messengers
(e.g. Ca\textsuperscript{2+}-dependent channels)
\item the extracellular concentration of neurotransmitters
\item the presence of the ionic pumps
\end{itemize}

\subsection{Single-compartment models}

The membrane potential measured at different places within a neuron
can take different values. The single-compartment models are single-neuron
models where the entire neuron is described with a single membrane
potential, $V$. Therefore this approximation assumes that the neuron
has a (relatively) uniform membrane potential across their surfaces.
In general this may looks a rough approximation, and a way to evaluate
how good it is at the single-neuron level is to compute the electrotonic
distance \citep{koch2004biophysics}, nevertheless, depending on the
dynamics and the scale we are modeling, there are many situations
where the spatial variations in the membrane potential inside a neuron
is not thought to play an important function.

\begin{figure}
\begin{centering}
\includegraphics[scale=0.4]{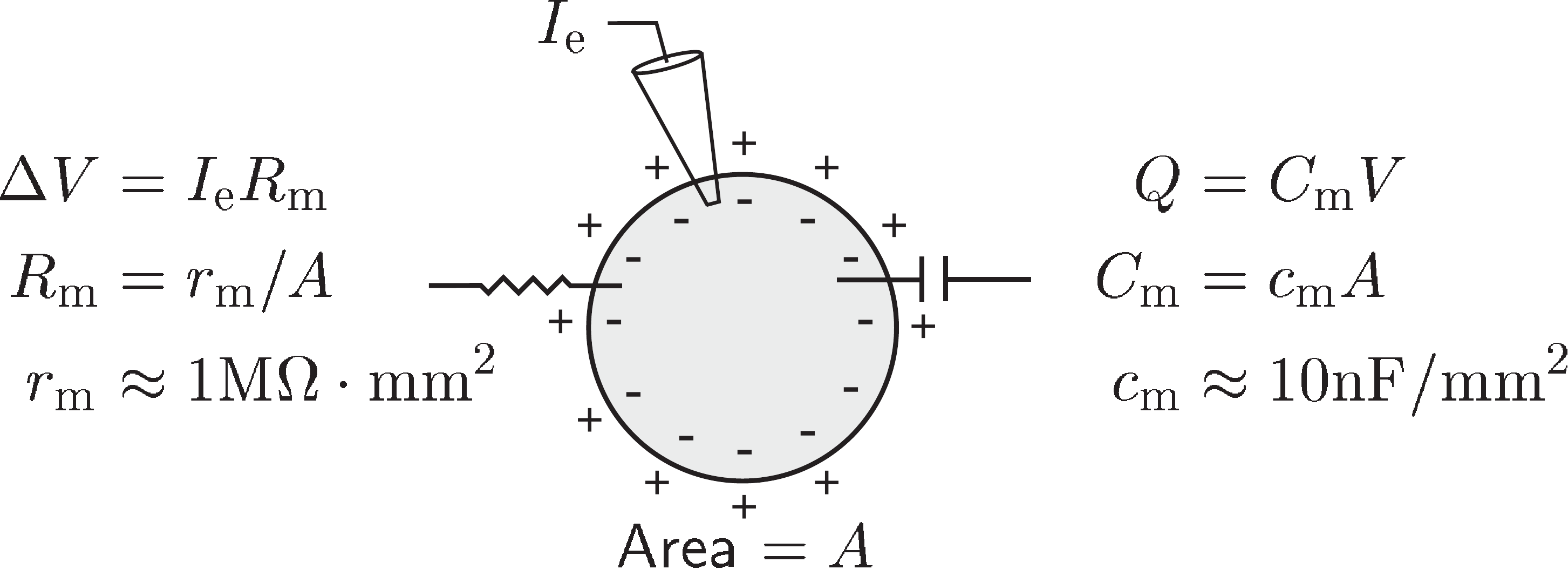}
\par\end{centering}
\centering{}\caption[Single-compartment model]{\textbf{The capacitance and the resistance of a neuron in the single-compartment
model}. Typical values of specific membrane capacitance and resistance,
$c_{m}$ and $r_{m}$, are given. In the equation relating the current
and the potential flowing through the membrane resistance, $R_{m}$,
the restriction to small currents $\text{\textgreek{D}V}$ is needed
to have $R_{m}$ constant over the range \textgreek{D}V, as assumed
by the Ohm's law. (Source: \citealp{Dayan})\label{fig_single_compartment}}
\end{figure}

We have mentioned that there is typically an excess of negative charge
on the inside surface of the cell membrane of a neuron, and a balancing
positive charge on its outside surface (see figure \ref{fig_single_compartment}).
In this arrangement we can model the neuron as a spherical capacitor
(whose potential is kept constant by the ionic pumps) and by introducing
the capacitance, $C_{m}$, we can use the standard equation for a
capacitor to relate the variation of the total charge between the
internal and external surfaces of the membrane, $Q$, to the variation
of the potential: $Q=C_{m}V$ \footnote{For the frequency range that is of interest in physiology studies
(0 to \textasciitilde{}3 kHz, \citet{berens2010local}), the inductive,
magnetic, and propagative effects of the bioelectrical signals in
the extracellular space can be neglected, permitting a quasi-static
description of the electric field for which Ohm\textquoteright s law
applies. \citep{Logothetis03}}. By doing the time derivative of the previous equation we can determine
how much current is required to change the membrane potential at a
given rate:
\begin{equation}
C_{m}\frac{dV}{dt}=\frac{dQ}{dt}.\label{eq_capacitance}
\end{equation}
This is the basic relationship that determine the membrane potential
for a single-compartment model. The specific membrane capacitance
is approximately the same for all neurons: $c_{m}\approx10$nF/mm\textsuperscript{2},
while the surface area, $A$, is usually between 0.01 and 0.1 mm\textsuperscript{2},
so the capacitance, $C_{m}$, is typically in the range 0.1 to 1 nF.

The neuron also shows features that can be modeled by a resistance,
indeed when a current, $I_{e}$, is injected into a neuron through
an electrode, see figure \ref{fig_single_compartment}, the relationship
between the current and the variation of the potential, \textgreek{D}V,
can be modeled by a membrane or input resistance, $R_{m}$. Analogously,
the membrane resistance determines how much current is required to
keep the potential fix to a given value different from its resting
value, while the capacitance of a neuron determines how much current
is required to make the membrane potential change at a given rate
(see equation \ref{eq_capacitance}). The specific membrane resistance
of a neuron at rest is around 1 M\textgreek{W} mm\textsuperscript{2},
but its value is much more variable than the specific membrane capacitance. 

The right-hand side of equation \ref{eq_capacitance} is given by
the total amount of current entering the neuron and it can be split
in two main components: the physiological currents, $i_{m}$, (flowing
through all the membrane and the synapses) and the experimentally
injected currents, $I_{e}$ (if it is present). The former are usually
in units of current per unit area (to facilitate comparisons between
neurons of different sizes), while the latter is the total current
injected through the electrode\footnote{We adopted the usual convention: when a variable is indicated both
with the uppercase and the lowercase letter, the latter case refers
to the measure of the variable related to the surface area. In particular,
$c_{m}$ is the membrane capacitance per unit area {[}nF/mm\textsuperscript{2}{]}
and $r_{m}$is the membrane resistance divided by the surface area
{[}M\textgreek{W} mm\textsuperscript{2}{]}. }, therefore, putting all together we can rewrite the equation \ref{eq_capacitance}
as:
\begin{equation}
C_{m}\frac{dV}{dt}=-i_{m}(t)+\frac{I_{e}(t)}{A},\label{eq_single_compartment}
\end{equation}
where the sign of $i_{m}$ is negative because, by convention, membrane
currents are defined as positive when positive ions flow outward the
neuron (i.e., membrane-hyperpolarizing currents\footnote{A hyperpolarizing current is a current that makes the membrane potential
more negative (that hyperpolarizes the neuron).} are positive) and negative when positive ions flow inward the neuron.
On the other hand, when the current enters through an electrode the
signs of the currents (i.e. electrode-depolarizing currents\footnote{A depolarizing current is a current that makes the membrane potential
less negative (that depolarizes the neuron).} are positive).

\subsection{Nernst equation }

The movement of ions through the channels of the membrane is due to
two distinct effects: electric forces and thermal diffusion. Indeed
the ionic pumps maintain an inhomogeneous concentration of charges
(in particular Cl\textsuperscript{-}, Na\textsuperscript{+} e Ca\textsuperscript{2+}are
more concentrated outside, while K\textsuperscript{+}inside) between
the interior and the exterior of the cells, therefore, both the electric
force (positive ions will be attracted towards negative potentials
and vice versa) and the thermal energy (which tends to homogeneously
diffuse the ions) act on the ions. We can characterize the balance
between these two contributes by means of \textquotedbl{}equilibrium
potentials\textquotedbl{}. The equilibrium potential, $E$, is indeed
defined as the membrane potential at which current flow due to electric
forces cancels the diffusive flow and there is not net movement of
charges through the cell membrane. The equilibrium potential is a
function of the state of the neuron and of to the considered and active
channels. For example, when the neuron is at rest $E$ is around -65
mV and it results by summing contributions from all the active channels
of the neuron.

In the simplest case where the channel $x$ conducts only one type
of ion, $x$, having electric charge $zq$ ($q$ is the charge of
a proton), by using the Boltzmann distribution to evaluate the thermal
energy and by equating the ionic flows due to the thermal and to the
electric contributes, we obtain:
\begin{equation}
E_{x}=\frac{V_{T}}{z}ln\left(\frac{[outside]_{x}}{[inside]_{x}}\right),\label{eq_nernst}
\end{equation}
where $[outside]_{x}$ and $[inside]_{x}$ are the values of the concentration
of the ion $x$ respectively outside and inside the neuron. This is
the Nernst equation, which allows to compute the equilibrium potential
of a ionic channel (that allows only one type of ion to pass through
it) $x$. In the table \ref{table_3.15_bear} are shown some typical
values for the concentration of the four most important ions involved
in the transmission of the neuronal signal.\begin{table}[ht!] 
\centering 
\begin{tabular}{l | p{1.8cm} | p{1.8cm} |l || p{2.5cm}} 
Ion ($x$) & $[outside]_x$ (in mM) & $[inside]_x$ (in mM) & $[outside]:[inside]$ & $E_x$ (at \ang{37}C) [mV]\\ \hline 
K$^+$ & 5 & 100 & 1:20 & -80\\ 
Na$^+$ & 150 & 15 & 10:1 & 62\\ 
Ca$^{2+}$ & 2 & 2$\cdot$10$^{-4}$ & 10$^4$:1 & 123\\ 
Cl$^-$ & 150 & 13 & 11.5:1 & -65
\end{tabular} 
\caption[Concentrations of the principal ions]{\textbf{Approximated concentrations of the principal ions} on both the membrane surfaces. (Adapted from \citealp{Bear2007})} \label{table_3.15_bear} \end{table}

In our modeling framework each channel represents a conductance that
allows the current to flow through the membrane. The direction of
this current depends on the value of the membrane potential with respect
to the equilibrium potential. Indeed a conductance $x$ with an equilibrium
potential $E_{x}$ tends to move the membrane potential of the neuron
toward the value $E_{x}$. When $V>E_{x}$ this means that positive
current will flow outward, and when $V<E_{x}$, positive current will
flow inward. This is the reason why the equilibrium potential is also
called reversal potential and indicated by $V_{x}$. For example,
looking at table \ref{table_3.15_bear} we see that Ca\textsuperscript{2+}
and Na\textsuperscript{+}conductances have positive reversal potentials,
so they tend to depolarize a neuron, while the opposite normally happens
with K\textsuperscript{+} channels.

\subsection{Membrane currents}

In order to model the membrane current flowing through the channel
$x$, $i_{m,x}$, we make a first order approximation obtaining: $i_{m,x}(t)=g_{x}(t)(V(t)-V_{x})$.
Summing over the different types of channels, we obtain the total
membrane current (see equation \ref{eq_single_compartment}):
\begin{equation}
i_{m}(t)=\underset{x}{\sum}g_{x}(t)(V(t)-V_{x}).
\end{equation}
The term $(V(t)-V_{x})$ is called the driving force, because it is
responsible for the intensity and the direction of the net movement
of ions across the channel. In particular, the current flows in the
direction that tends to minimize the driving force. The factor $g_{x}$
is the conductance per unit area due to the channel $x$ and it is
in general a function of the time. Indeed much of the complexity and
richness of neuronal dynamics arises because membrane conductances
change over time (the channels can open and close depending on many
factors, see section \vref{conductance_changes}). Nevertheless some
of the factors which contribute to the total membrane current can
be treated as channels with a relatively constant synaptic conductance
(e.g. the ionic pumps). In the simplest version of the model, they
are grouped together into a single term called leakage current whose
conductance is not a function of time. Therefore the total physiological
current, $i_{m}$, can be split into two contributes: 
\begin{equation}
i_{m}(t)=i_{leak}(t)+\underset{x}{\sum}i_{x}^{active}(t),
\end{equation}
where
\begin{equation}
i_{leak}(t)=g_{leak}(V(t)-V_{leak}).\label{eq_leak_current}
\end{equation}
Since the leak conductance is time-independent, it is also called
passive conductance to distinguish it from the variable conductances,
which are termed active because they interact with the surrounding.
Indeed they can be affected by both the state of the neuron (like
the membrane potential value) and by the environment where the neuron
is placed (like the concentration of a given ion). We can write the
active currents as the product of a maximal conductance, $\bar{g}_{x}$,
times an active probability, $s_{x}$, (that is the probability of
finding the channel $x$ in an open and active state) in the following
way:
\begin{equation}
i_{x}^{active}(t)=\bar{g}_{x}s_{x}(t)(V(t)-V_{x}),\label{eq_active_currents}
\end{equation}
where $V_{x}$ is the reversal potential of the channel and $s_{x}(t)$
a function that describes the channel dynamics.

\subsection{Synaptic currents}

A very important class of active conductances is given by the synapses.
Indeed also the synapses can be modeled as conductances and depending
on the value of their reversal potential they are termed excitatory
or inhibitory. In particular, if a synaptic conductance has a reversal
potential higher than the threshold for action potential generation,
its activation will produce an increase of the membrane potential
(that is a depolarization of the neuron) and the synapse is called
excitatory. On the other hand when the reversal potential is lower
than the threshold, its activation will hyperpolarize the neuron and
the synapse is called inhibitory. This conductances cannot be efficiently
modeled as constants, therefore, we write the synaptic currents as
in equation \ref{eq_active_currents}:
\begin{equation}
i_{syn}(t)=\bar{g}_{syn}s_{syn}(t)(V(t)-V_{syn}),\label{eq_synaptic_current}
\end{equation}
where $V_{syn}$ is the reversal potential of the synapse $syn$ and
$s_{syn}(t)$ is a function that models the synaptic dynamics. In
particular, $s_{syn}(t)$ expresses the probability that the synaptic
channel is open as a consequence of the arrival of an action potential.
In order to write an explicit equation for $s_{syn}(t)$, we firstly
introduce $P_{pre}$, which is the probability that the presynaptic
neuron is activated by the arrival of an action potential at the synaptic
terminal (and the neurotransmitter released in the synaptic cleft).
When an action potential invades the presynaptic terminal, the transmitter
concentration in the synaptic cleft rises extremely rapidly after
vesicle release, remains at a high value for a period of duration
$T$, and then falls rapidly to 0. Therefore, as a simple model of
the presynaptic transmitter release, we assume that $P_{pre}$ is
a square pulse (with pulses located at the spike times) of amplitude
$T$. 

Let's assume now to model the opening an closing of the synaptic channel
as two exponentials and introduce the following coefficients: $\alpha_{syn}$,
which represents the opening rate of the synapse and $\beta_{syn}$,
which is the closing rate of the synapse. In general these two coefficients
are not constant and they can be function, for example, of the neurotransmitter
concentration and of the membrane potential. In particular, since
we are interested in the case where the synaptic channel is open when
an action potential arrives, we take as effective opening rate the
product $\alpha_{syn}P_{pre}$. The probability that a synaptic gate
opens over a short time interval is proportional to the probability
of finding the gate closed, $(1-s_{syn})$, multiplied by the opening
rate $\alpha_{syn}P_{pre}$. Likewise, the probability that a synaptic
gate closes during a short time interval is proportional to the probability
of finding the gate open, $s_{syn}$, times the closing rate $\beta_{syn}$.
Therefore, the equation for the probability that the synaptic channel
is active is:
\begin{equation}
\frac{ds_{syn}}{dt}=\alpha_{syn}P_{pre}(t)(1-s_{syn}(t))-\beta_{syn}s_{syn}(t).\label{eq_ssyn}
\end{equation}
The solution for this equation depends on the spike train impinging
on the neuron (through the presynaptic probability $P_{pre}$). The
contribution to $s_{syn}$ of each post synaptic potential is given
by the difference of two exponentials\label{ssyn}: one models the
opening of the synaptic gates (that is the increase of $s_{syn}$
observed in correspondence of the arrival of an action potential)
with time rise constant $\tau_{r}=1/\alpha_{syn}$, while the other
exponential, which describes the closing of the synaptic gates with
decay time $\tau_{d}=1/\beta_{syn}$, tends to vanish $s_{syn}$.
$\alpha_{syn}$and $\beta_{syn}$ are obtained by fitting experimental
data and typically $\tau_{r}$ is considerably smaller than $\tau_{d}$. 

The synaptic currents can also be written in a simplified way by neglecting
the dependence on the membrane potential $V(t)$:
\begin{equation}
i_{syn}(t)=j_{syn}s_{syn}(t),\label{eq_synaptic_current_CUB}
\end{equation}
where $j_{\ensuremath{syn}}$ is a constant (in units of current per
unit area) which models the synaptic efficacy of the connections.

\subsection{Leaky Integrate-and-Fire model}

The leaky integrate-and-fire (LIF) model \citep{Lapicque} is one
of the simplest single-neuron model that includes the action potential
generation. It is a single-compartment model with some ad hoc assumptions
needed to introduce the action potential in the neuronal dynamic. 

\begin{figure}
\begin{centering}
\includegraphics{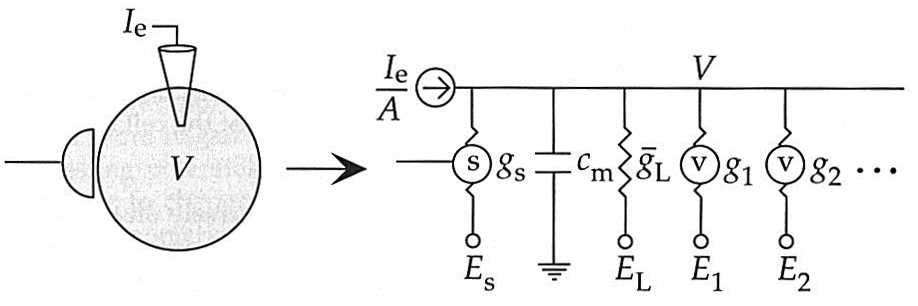}
\par\end{centering}
\centering{}\caption[Equivalent circuit for a single-compartment model]{ \textbf{On the right}, it is shown the \textbf{equivalent circuit
for a single-compartment neuron model}. \textbf{On the left} is represented
the neuron (having surface area $A)$ with a single synapses and a
current-injecting electrode. The equation \ref{eq_single_compartment2}
determines the evolution of the voltage in such a circuit. In particular,
the circled $s$ indicates a synapse conductance (with reversal potential
$E_{s}$), which is a function of the presynaptic neuron activity.
Then we have the capacitance and the leak conductance, which are constant,
and eventually a series of voltage dependent conductances (indicated
by the circled \textit{v}). The dots stand for possible additional
membrane conductances or active currents (such as the spike-rate adaptation
current). All the physiological active currents are included in the
summation over the channels in the equation \ref{eq_single_compartment2}.
(Source: \citealp{Dayan})\label{fig_equivalent_circuit}}
\end{figure}

In particular, by substituting equation \ref{eq_leak_current} into
equation \ref{eq_single_compartment} we obtain a more explicit equation
for the membrane potential of a single-compartment model:
\begin{equation}
c_{m}\frac{dV}{dt}=-g_{leak}(V(t)-V_{leak})-\underset{x}{\sum}g_{x}(t)(V(t)-V_{x})+\frac{I_{e}(t)}{A}.\label{eq_single_compartment2}
\end{equation}
In the simplest case the leak conductance, $g_{leak}$, can be approximated
by the input conductance (that is the inverse of the input resistance,
see figure \ref{fig_single_compartment}): $g_{leak}=1/r_{m}.$ Therefore,
by multiplying both sides by the specific membrane resistance, we
obtain:
\begin{equation}
\tau_{m}\frac{dV}{dt}=-V(t)+V_{leak}-\underset{x}{\sum}\frac{g_{x}(t)}{g_{leak}}(V-V_{x})+R_{m}I_{e},\label{eq_integrate_and_fire}
\end{equation}
where $\tau_{m}=c_{m}r_{m}$ is a constant with units of time. It
is called the membrane time constant and its typical values is between
10 and 100 ms. If there are not input currents (that is $g_{x}(t)=I_{e}=0$),
the membrane potential exponentially decades to the value $V_{leak}$
with time constant $\tau_{m}$. Therefore $V_{leak}$ is the potential
of the cell and it is also called the resting potential of the neuron.
Equation \ref{eq_single_compartment2} is the equation for the potential
in a electrical circuit, called equivalent circuit, consisting of
a capacitor and a set of variable and non variable resistors corresponding
to the different channels of the membrane. Figure \ref{fig_equivalent_circuit}
shows the equivalent circuit for a generic one-compartment model.

A neuron will typically fire an action potential when its action potential
reaches a threshold value of about -55 to -50 mV. The generation and
propagation of an action potential in a neuron are due to a cascade
of events that are very complex and depends on a lot of variables.
In the leaky integrate-and-fire model the description of these biophysical
mechanisms is simply avoided\footnote{Note that the mechanisms by which voltage-dependent conductances produce
action potentials are well understood and they can be modeled quite
accurately, for example, with the well-known Hodgkin-Huxley model.
Nevertheless, in this work, we do not use these biophysically detailed
models, since they require high computational costs.}: the subthreshold dynamics of the membrane potential follow the equation
\ref{eq_integrate_and_fire} and each time the membrane potential
overcomes a fixed threshold, $V_{thresohld}$:
\begin{enumerate}
\item An action potential is instantaneously fired
\item The membrane potential is instantaneously set to the reset value,
$V_{reset}$
\item The firing on another action potential is forbidden for a given absolute
refractory period
\end{enumerate}
Equation \ref{eq_integrate_and_fire} combined with the three rules
just stated define the leaky integrate-and-fire model of a neuron.\label{LIF_model}

If there are not currents injected through an electrode ($I_{e}=0$)
and all the active currents are due to synaptic inputs, equation \ref{eq_integrate_and_fire}
becomes:
\begin{equation}
\tau_{m}\frac{dV}{dt}=-V(t)+V_{leak}-\underset{syn}{\sum}\frac{\bar{g}_{syn}}{g_{leak}}s_{syn}(t)(V(t)-V_{syn}),\label{eq_integrate_and_fire_COB}
\end{equation}
By using the simplified expression of the synaptic current, shown
in equation \ref{eq_synaptic_current_CUB}, the equation for the membrane
potential is:
\begin{equation}
\tau_{m}\frac{dV}{dt}=-V(t)+V_{leak}-\underset{syn}{\sum}\frac{j_{syn}}{g_{leak}}s_{syn}(t).\label{eq_integrate_and_fire_CUB}
\end{equation}

The difference between the single-neuron model described in equation
\ref{eq_integrate_and_fire_COB} and the one in equation \ref{eq_integrate_and_fire_CUB}
relies on the synaptic current. Both of them are LIF neurons, but
in the latter case, the dependence on the membrane potential is neglected
and the synapses are termed ``current-based'', while in the first
case we have ``conductance-based''\footnote{The reason for this terminology will be clear in chapter \ref{chapter_frontiers}}
synapses\label{COB_vs_CUB}. 

\begin{figure}
\begin{centering}
\includegraphics{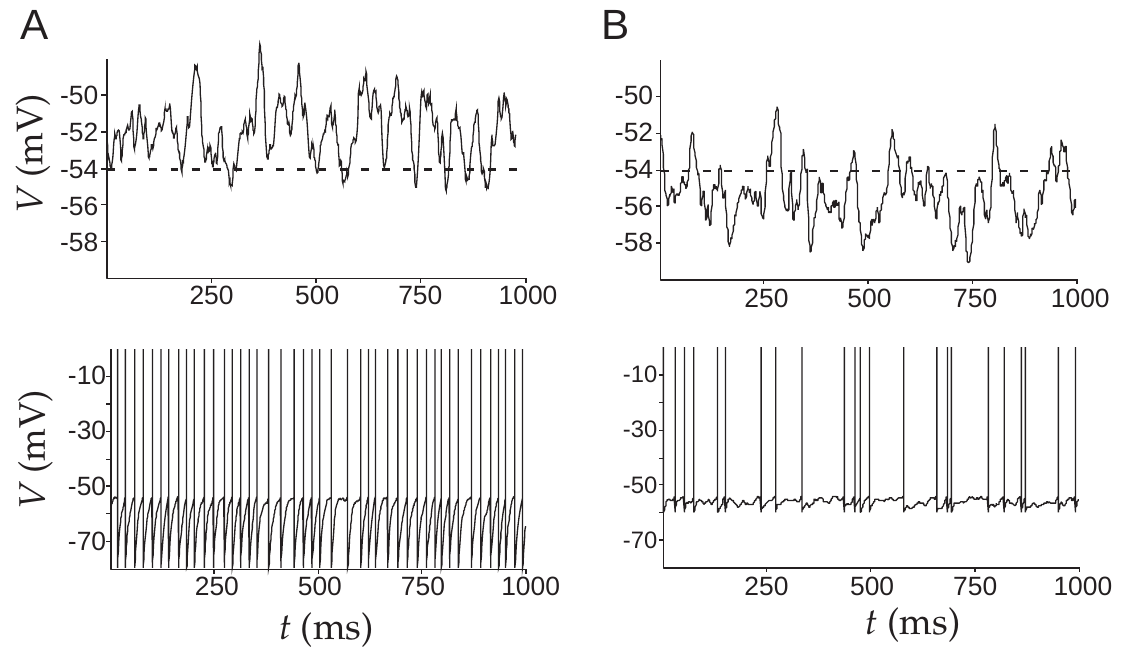}
\par\end{centering}
\centering{}\caption[Regular and irregular firing modes of a LIF neuron]{\textbf{The regular (A) and irregular (B) firing modes} of an integrate-and-fire
model neuron. In the upper panels it is shown the membrane potential
of the neuron when the spike generation mechanism is turned off (the
dashed line is the spike threshold), while in the lower panels the
membrane potential of the same neuron when the spiking mechanism is
active. (Source: \citealp{Dayan})\label{fig_regular_irregular}}
\end{figure}

The leaky integrate-and-fire models is very useful to investigate,
for example, how neurons integrate a high number of synaptic inputs.
A major difference in the way neurons can respond to multiple synaptic
inputs depends on the balance between excitatory and inhibitory inputs.
In figure \ref{fig_regular_irregular}A the excitation is so strong,
with respect to inhibition, to produce an average membrane potential
(when action potential generation blocked) above the spiking threshold
of the model. By turning on the spiking mechanism the neuron fires
in a regular way (that is with a regular pattern of action potentials).
In this case the timing of the action potentials is only weakly related
to the temporal structure of the input currents, since it is mainly
determined by the charging rate of the neuron, which depends on its
membrane time constant. On the other hand, in figure \ref{fig_regular_irregular}B,
the mean membrane potential (in absence of spiking mechanism) is below
the threshold for action potential generation and the resulting spiking
activity is irregular: action potentials are generated only when the
fluctuations in the synaptic input are sufficiently strong to bring
the membrane over the threshold. In this case the degree of variability
of the spiking activity (than can be measured for example with the
coefficient of variation of the interspike interval, CV ISI) is much
higher than in the regular regime and it is more similar to the high
degree of variability observed in the spiking patterns of in vivo
recordings of cortical neurons. Furthermore in the irregular-firing
mode the spiking activity reflect the temporal properties of fluctuation
in the input currents. For these reason the irregular-firing mode
is by far the most investigated and, depending on the context, it
is also termed inhibition-dominated or fluctuation-driven regime.
In particular this is also the regime we will investigate. 

\subsection{Neural networks}

By using different experimental methods has been proved that different
cerebral areas are specialized for single functions \citep{Kandel,Nicholls},
even if no single areas are entirely responsible for a complex mind
faculty. Indeed each area performs only some basic operations. In
particular, all the most complex faculties are due to series and parallel
connections across many different cerebral areas \citep{Nicholls}.
In summary, extensive synaptic connectivity is a hallmark of neural
circuitry. For example, a pyramidal neuron in the mammalian neocortex
receives about 10,000 synaptic inputs where 75\% are excitatory synapses
and 25\% inhibitory (this numbers change across the different structures
of the cortex) \citep{Abeles91,Braitenberg1991}. The merging of a
so high number of synaptic inputs on a single-neuron of the cortex
is indicative of how broad is the integration of the signal that happens
at the single-neuron level and, more in general, of how complex is
the computation underlying each recording site. Network models allow
us to explore the computational potential of such connectivity, using
both analysis and simulation\footnote{In this work we mainly use simulation to investigate network dynamics.}. 

Networks are used to study a broad spectrum of phenomena such as selective
amplification of inputs, short-term memory, gain modulation, input
selection, coding of sensory stimuli and so on. Neocortical circuits
are the focus of our discussion. In the neocortex, neurons lie in
six vertical layers highly coupled within cylindrical columns. Such
columns have been suggested as basic functional units, and stereotypical
patterns of connections (both within a column and between columns)
are repeated across cortex. In particular, we can divide the observed
interconnections within cortex in three main classes \citep{Dayan}:
\begin{itemize}
\item feedforward connections, the input travels in a defined direction
going from a given area (or layer) to another located in a following
stage along the signal pathway 
\item top-down connections, the input travels in a defined direction going
from a given area (or layer) to another located in an earlier stage
along the signal pathway 
\item recurrent connections, the neurons are interconnected within a given
area which is considered to be at the same stage along the processing
pathway
\end{itemize}
There is another major distinction between neural networks: they can
be firing-rate or spiking models. In the former case each neuron-like
unit of the network has output consisting of firing rates rather than
action potential. This simplification is very useful to allow analytical
calculations of same aspects of network dynamics that could not be
treated in the case of spiking neurons. When we are dealing with spiking
neurons networks, it means that the neuron-like unit of the network
implements a model of action potential generation, so the output is
given by the membrane potential and the spike train of each neuron.

The last classification of neural networks models we introduce is
based on the kinds of single neurons that compose the network. In
particular, if all the neurons belong to the same population of excitatory
either inhibitory neurons, then we have a one-population network,
while, when both the populations are present, the network is a two-population
network. Eventually, if all the neuron of a given population have
identical free parameters, the network is homogeneous, while, if the
single-neuron parameters can differ from neuron to neuron (at fixed
population), the network is inhomogeneous. 

In this work we will investigate neural dynamics by means of two-populations
recurrent networks of LIF neurons (that is spiking neurons).

\section{Information Theory}

A major purpose of our investigation of network dynamics by means
of models is to understand the way neuronal networks convey information
about sensory stimuli. Indeed the information calculation allows as
to answer the following important question: ``How much does the neural
response tell us about a stimulus?''; by answering this question
we can also investigate which forms of neural response are optimal
for conveying information about natural stimuli. 

In order to quantify the information transmitted by neurons, we treat
the brain as a communication channel and we assume that the coding
and transmission processes are stochastic and noisy. More precisely,
we compute the Shannon mutual information \citep{Shannon} between
two random variables \citep{Panzeri07,Quiroga,Shannon} to quantify
and analyze the information about the external stimulus obtained from
different neural codes (i.e., different neural responses, as done
in our previous work \citet{Mazzoni2011}) and with different synaptic
current models (see chapter \ref{chapter_frontiers}). 

\subsection{Shannon information and neuroscience\label{section_Shannon_information}}

We introduce now the general concept of mutual information (hereafter
information) of two random variables and we make, for clarity, examples
by referring to our (discretized) case. Each time we run a simulation
of a network model, we are basically computing an output signal as
a function of the (noisy) input we inject to the network during the
time interval $T$. We call that input signal the stimulus, $S$.
In order to compute the information (that is the information of two
random variables, where one is the stimulus $S$ and the other is
the answer $R$\footnote{Capital letters are used to indicate that are random variables.})
we need to define the neural response, $R$, that is the variable
(or the set of variables) we take as output of the model. Note that
this is the most important choice we make when computing information
because it defines the neural code used to convey information, reflecting
our hypotheses about which are the most important aspects of neural
activity. We just mentioned that the response $R$ can be given by
one or more variables; more in general, it can be a scalar quantity
or a vector (``response vector'') and the dimension of the response,
$L$, is the dimension of the code. 

For each presentation of the stimulus $s$ in the time interval $T$,
the response $R$ will assume the value $r$, and the amplitude of
$T$ determines the temporal precision of the code.

A crucial point in this computation relies on the fact that the coding
is a stochastic and noisy process: the value of the response $r$
does not depend on the stimulus $s$ in a deterministic way. Indeed
the response is a stochastic function of the input where the noise
plays a crucial role. This reflects a basic neuronal feature: real
neurons are ``noisy'', that is they can produce different responses
when presenting the same external stimulus. Two recordings (or simulations)
where the stimulus $s$ is the same, which differ only for the stochastic
(noisy) component are called ``trials''\label{noise_role}. By means
of information we want to investigate the relationship between the
stimulus $S$ and the answer $R$ by quantifying which is the average
reduction in the uncertainty of $S$ due to the observation of $R$
(decoding point of view) or, equivalently, which is the average reduction
in the uncertainty of the response $R$ due to the presentation of
the stimulus $S$ (encoding point of view)\footnote{We will see afterward (equation \ref{informazione}) that the two
points of views are equivalent.}.

Let's assume the decoding point of view, and introduce the way to
quantify the average level of uncertainty associated with the stimulus
$S$. We define the probability that stimulus $s$ is presented as:
$P(s)=N_{s}/N_{tot}$, where $N_{s}$ is the number of times the stimulus
$s$ has been presented and $N_{tot}$ the total number of stimuli
presented. We can now introduce the (Shannon's) total stimulus entropy:
\begin{equation}
H(S)=-\underset{s}{\sum}P(s)\log_{2}P(s),\label{eq_shannon_entropy}
\end{equation}
where, by convention, base 2 logarithms are used so that information
can be compared easily with results for binary systems. To indicate
that the base 2 logarithm is being used, information is reported in
units of ``bits''. This quantity is the average uncertainty about
which stimulus $s$ is presented in a time interval $T$. Indeed if
the stimuli $s$ are all equal, $H(S)=0$, while it reaches its maximum
when all the presented stimuli are different and equally likely: $H(S)=-\log_{2}(1/N_{tot})$.

We define similarly the Shannon's total entropy of the stimulus $S$
given the response $R$:\begin{equation}
H(S|R)=-\sum_{r,s}P(r)P(s|r)\log _{2}P(s|r), 
\label{equivocation} 
\end{equation}where $P(r)$ is the probability that response $r$ is observed and
$P(s|r)$ is the conditional (prior) probability that the stimulus
$s$ was presented when the response $r$ is observed. This quantity
represents the average uncertainty about which stimulus $s$ was presented
in a time interval $T$ where the response $r$ is known\footnote{Note that this variability is due to the stochastic nature of the
coding process: if the relationship between stimulus and response
was deterministic, $H(S|R)$ would be 0.}. We can now define the mutual information between the response and
the stimulus as the average reduction in the uncertainty of $S$ due
to the observation of $R$ (in a time interval $T$):\begin{eqnarray} 
I(S;R)&=&H(S)-H(S|R)\nonumber\\ 
&=&\sum_{s,r} P(r)P(s|r)\log _2\frac{P(s|r)}{P(s)}.
\label{informazione_0} 
\end{eqnarray}The total stimulus entropy $H(S)$ represents the maximum information
theoretically available with the given distribution of stimuli (irrespective
of the code chosen). On the other hand, if $S$ and $R$ are independent,
there is no reduction of the stimulus uncertainty due to the knowledge
of the response, $H(S|R)=H(S)$, and the information is 0. The information,
like entropy, is measured in bits; each bit of information corresponds
to an average reduction of the uncertainty about the presented stimulus
of a factor 2 as a consequence of the observation of a response $r$
in the time interval $T$. Note that the information (measured in
bits) is obtained from the observation of the neuronal response over
a time interval $T$, therefore it does depend on this value. In some
cases it is useful to normalize the information by T to obtain units
of bits/sec.

The Bayes theorem relates the prior probability $P(s|r)$ to the current
probability $P(r|s)$ in the following way:
\begin{equation}
P(s|r)=\frac{P(r|s)P(s)}{P(r)}.
\end{equation}
By using the Bayes theorem in the equation \ref{informazione_0},
we obtain:\begin{eqnarray} 
I(S;R)&=&H(S)-H(S|R)\nonumber\\ 
&=&\sum _{s,r}P(s,r)\log _2\frac{P(s,r)}{P(s)P(r)}\nonumber \\ 
&=&\sum_{s,r} P(s)P(r|s)\log _2\frac{P(r|s)}{P(r)}\nonumber \\
&=&H(R)-H(R|S)=I(R;S), 
\label{informazione} 
\end{eqnarray}where $P(s,r)=P(s)P(s|r)=P(r)P(r|s)$ is the joint probability of
stimulus $s$ appearing ans response $r$ being evoked. 

From equation \ref{informazione} we conclude that information is
symmetric with respect to interchange of $S$ and $R$. This is the
reason why the information is mutual: $S$ and $R$ can be inverted
without affecting the information. In the end we demonstrated what
we mentioned above: the average reduction in the uncertainty of $S$
due to the observation of $R$ is equivalent to the average reduction
in the uncertainty of the response $R$ due to the presentation of
the stimulus $S$. 

This last point of view (encoding point of view, corresponding to
the last row of equation \ref{informazione}) represents a different
interpretation of information which is based on the fact that the
more the response is variable, the higher is the theoretical capacity
of a code to convey information. Indeed an higher level of variability
in the response $R$ corresponds to an higher value of the total response
entropy, $H(R)$, which represents the maximum information theoretically
achievable with the given code (distribution of responses) \citep{deRuyter97,Dayan}.
The variability in the response as measured by the total response
entropy includes both the variability due to the presentation of different
stimuli and to the noise (which gives rise to different responses
when the stimulus is fixed). The latter contribution is the entropy
of the response $R$ given the stimulus $S$, $H(R|S)$, that is indeed
called the noise entropy. Therefore $I(S;R)=H(R)-H(R|S)$ is the variability
in the response only due to the presentation of different stimuli.
\begin{figure}
\begin{centering}
\includegraphics[scale=0.185]{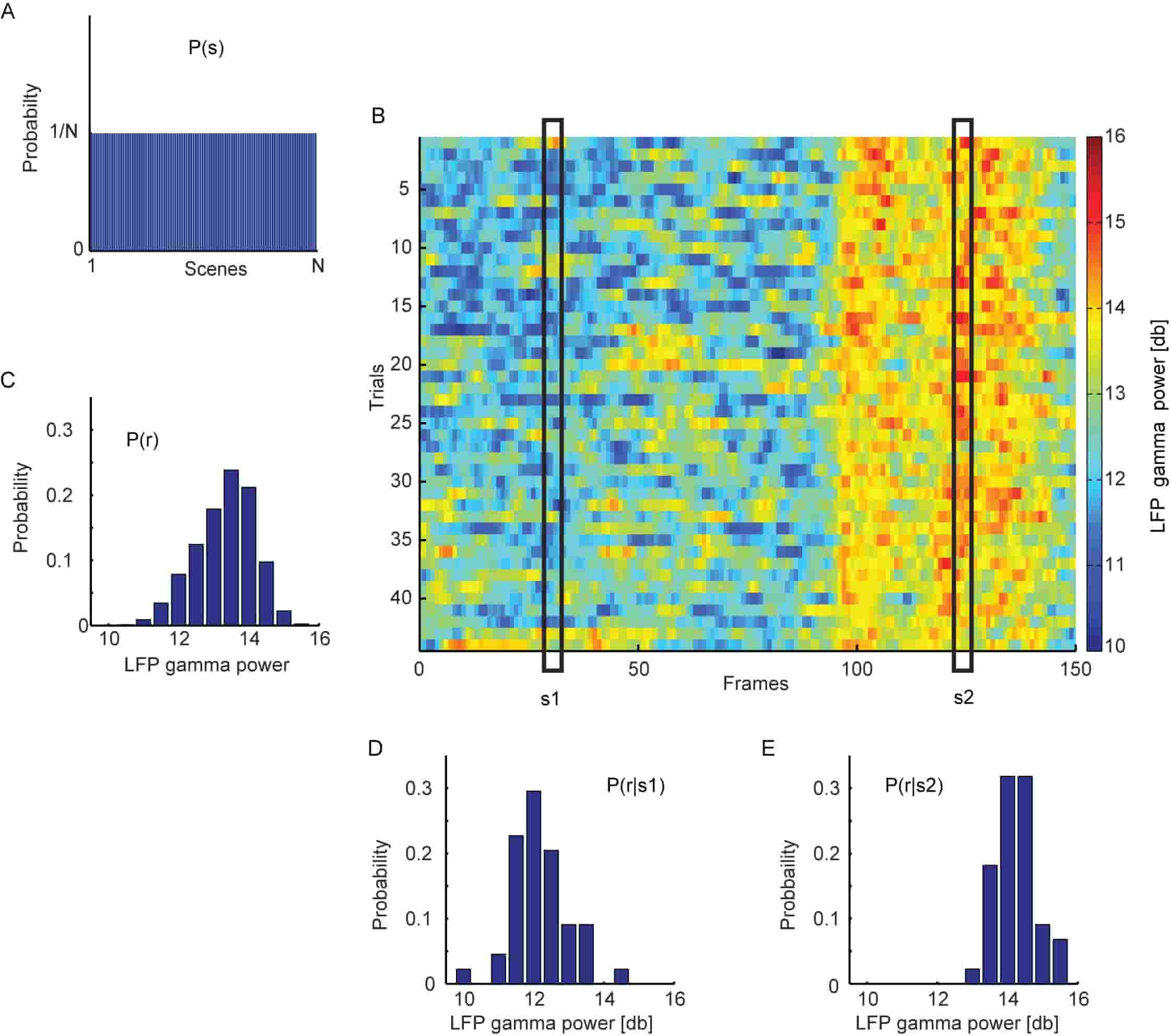}
\par\end{centering}
\centering{}\caption[Computation of information]{\textbf{Schematic representation of the computation of the mutual
information} carried by LFP power about movie scenes of a complex
visual stimulus. The figure illustrates the way to obtain the different
probabilities needed to compute the information $I(S;R)$ (see equation
\label{informazione}) in a specific case where the stimulus, $S$,
is the Hollywood movie presented to a monkey and the response, $R$,
is the LFP power in the gamma band. \textbf{(A)} First the entire
movie presentation time is portioned into non-overlapping window,
each considered a different stimulus $s$ (a ``scene''). The set
of the stimuli is the set of the different scenes, each of which is
presented once every trial, therefore the probability of each scene,
$P(s)$, is the inverse of the number $N$ of the scenes presented
and it is constant. \textbf{(B)} The color plot shows the single-trial
LFP gamma power (in this example, in the {[}72\textendash 76 Hz{]}
frequency range) across all trials and movie scenes. From these data
it is possible to compute the \textbf{(C)} probability distribution
$P(r)$ of the LFP gamma power across all trials and scenes\textbf{
}and\textbf{ (D, E) }the probability distribution $P(r|s)$ of the
LFP gamma power across trials given the presented scenes $s1$ and
$s2$ respectively. The differences between the two distributions
and the distribution $P(r)$ suggest that the LFP gamma power carried
information about which scene is presented. By computing $P(r|s)$
for all scenes and inserting it in equation \label{informazione}
the actual value of the mutual information is obtained. (Adapted from
\citet{Mazzoni2011})\label{fig_info_computation_method}}
\end{figure}
\\
In figure \ref{fig_info_computation_method} is showed a schematic
representation of the computation of the mutual information in an
example case where the stimulus is given by a movie presented to a
monkey and the response is the power of LFP oscillations in a given
frequency band. 

We can extend the information computation to the case where we want
to quantify how much information is conveyed by the simultaneous observation
of two distinct responses $R_{1}$ and $R_{2}$ (for example the power
spectral density of two frequencies of the LFP). In this case the
mutual information is:\begin{equation} 
I(S;R_1,R_2)=\sum_{s,r} P(s)P(r_1, r_2|s)\log _2\frac{P(r_1, r_2|s)}{P(r_1, r_2)} 
\label{informazione_congiunta} 
\end{equation}If the two responses $R_{1}$ and $R_{2}$ were tuned to independent
stimulus features, and they do not share any source of noise, then
we would expect that $I(S;R_{1},R_{2})=I(S;R_{1})+I(S;R_{2})$, which
means that the two responses convey completely independent information
about the same stimulus. Therefore to quantify how independent are
the contributions to information given by the two responses we introduce
the following information redundancy \citep{Gieselmann08,Logothetis02,Logothetis07}:
\begin{equation}
Red(R_{1},R_{2})=I(S;R_{1})+I(S;R_{2})-I(S;R_{1},R_{2}).
\end{equation}
Redundancy is never negative, when it is zero, the two responses convey
completely independent information about the stimulus, otherwise (at
least part of) the information carried by $R_{1}$ and $R_{2}$ is
redundant (is the same).

We conclude this section by pointing out some important features that
underlain information computation when evaluating the relationship
between the stimulus and the neural activity evoked:
\begin{itemize}
\item it is simple and allows a easy comparison between data obtained from
experiments and from models \citep{Mazzoni2011}
\item there are no assumptions about which features of the stimulus shape
the neuronal response and in this way no one is missed \citep{deRuyter97}
\item what matters when computing information is the probability to observe
the answer $r$ when presenting the stimulus $s$, therefore the units
of the answer do not matter. This allows to build codes where the
answer $R$ combines different measurements of the neural activity
(for example the spiking activity and the LFP) observed in time intervals
of amplitude $T$. In the latter case we speak of ``nested'' codes
\citep{Kayser09}, to distinguish them from the case where the response
is given by a single variable (like the firing rate or the spike time).
Furthermore the response $r$ (defined in the time interval $T$)
can include variables measured on different temporal scales, $\varDelta t\leq T$.
For example, it can be represented by the precise timing of individual
spikes on the scale ($\varDelta t$) of milliseconds and by the phase
of the slow oscillations of the concomitant LFP on the scale ($T$)
of hundreds of milliseconds. In these cases we call the code a ``multiplexed''
code \citep{Panzeri2010}.
\end{itemize}

\section{Neural encoding and decoding}

A fundamental issue in neuroscience is the investigation of the link
between stimulus and response. In section \vref{section_Shannon_information},
we saw that by means of the mutual information we can characterize
``how much'' the neural response tells us about the presented stimulus.
An alternative and complementary approach to the same matter focus
on the question: ``What does the response of a neuron tell us about
a stimulus?'' Neural encoding and decoding face precisely this question. 

We already showed that when computing information the stimulus and
the response can be interchanged without affecting the result. Thus,
from a mathematical point of view, there is not an a priori distinction
between the stimulus and the answer... it is just matter of choice.
On the other hand, when you are doing an experiment it is always the
case that it is clear which is the stimulus (if there is) you are
presenting/injecting and which is the response you are recording.
This is the reason why there are the two distinct names: neural encoding
and decoding. Neural encoding refers to the map from stimulus to response,
while neural decoding refers to the reverse map. 

\subsection{Spike trains and firing rates}

\begin{figure}
\begin{centering}
\includegraphics[scale=0.55]{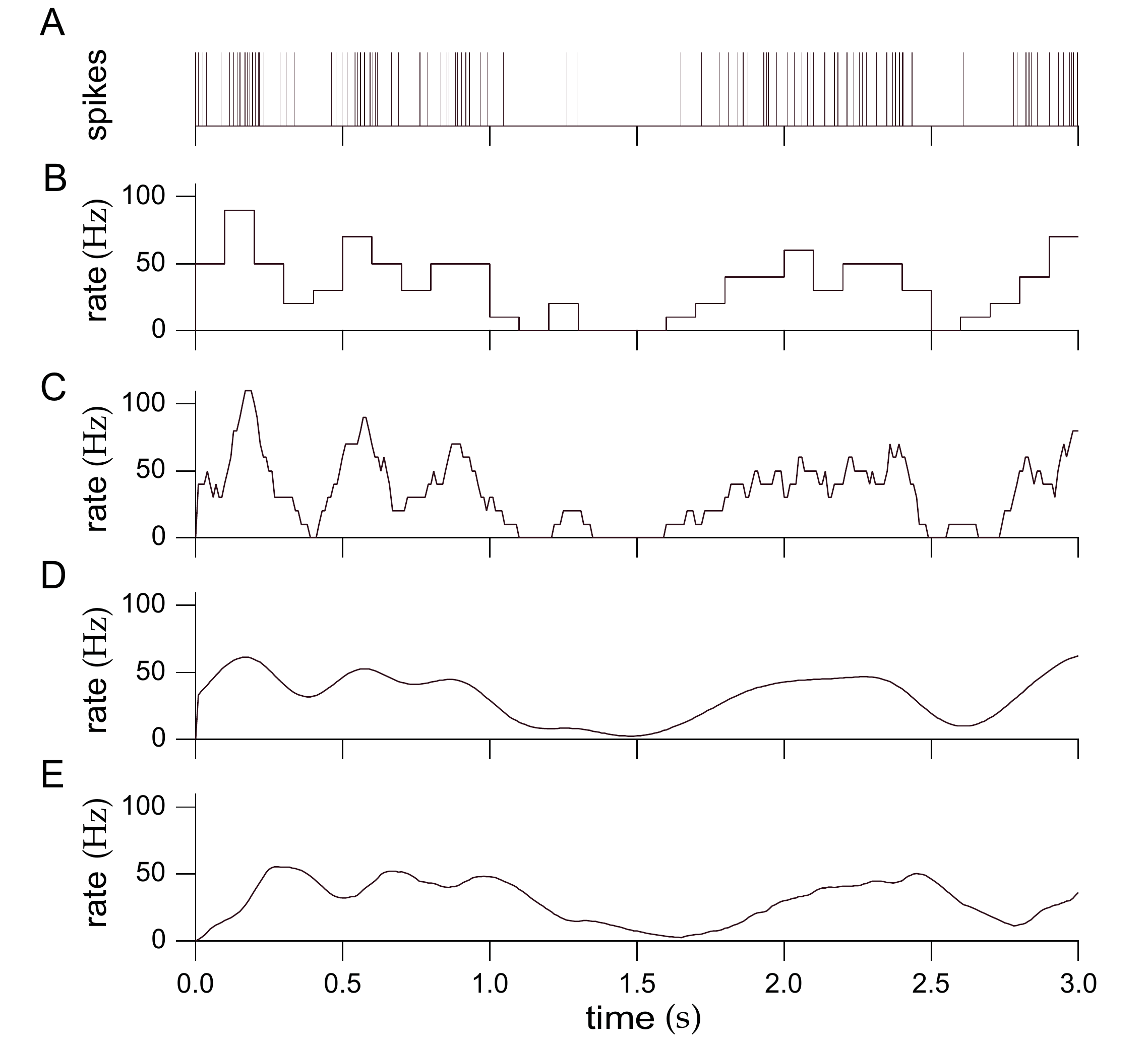}
\par\end{centering}
\centering{}\caption[Different procedures to approximate the firing rate]{\textbf{Different procedures to approximate the firing rate}. \textbf{(A)}
Sampled spike train of a neuron, $\rho(t)$. \textbf{(B)} This is
a discrete time approximation of the firing rate called Post Stimulus
Time Histogram (PSTH) obtained by dividing time into bins of fixed
amplitude (here $\varDelta t=100$ ms) and counting the number of
spikes within each bin. \textbf{(C)} Approximate firing rate obtained
by the discrete version of equation \ref{eq_firing_rate} with $\varDelta t=100$
ms. \textbf{(D)} Approximate firing rate computed using equation \ref{eq_smoothed_FR}
with $w(t)$ is a Gaussian window function with $\sigma_{t}=100$
ms. \textbf{(E)} Approximate firing rate computed using equation \ref{eq_smoothed_FR}
with a causal window function ($\alpha$ function). (Source: \citealp{Dayan})\label{fig_firing_rates_comp}}
\end{figure}

In real neurons, action potentials can vary somewhat in duration,
amplitude, and shape. However, when dealing with neural coding, action
potentials are typically treated as identical stereotyped events and
what matters is only the spike timing. Thus, we ignore the duration
of an action potential (about 1 ms), and characterize the firing activity
of a neuron by means of a list of the times when spikes occurred:
for $n$ spikes, we denoted these times by $t_{i}$ with $i=1,2,...,n$.
From the mathematical point of view, we assume the spike sequence
can be represented as a sum of Dirac $\delta$ functions:
\begin{equation}
\rho(t)=\stackrel[i=1]{n}{\sum}\delta(t-t_{i}).\label{eq_spke_train}
\end{equation}
$\rho(t)$ is the spike train (or neural response function; it represents
the spiking times). Because of the trial-to-trial variability of the
neural response, $\rho(t)$ is typically treated statistically or
probabilistically (see section \vref{section_Shannon_information}).
Thus we use angle brackets, $\langle\rangle$, to denote average over
trials at fixed stimulus and we introduce the trial-averaged spike
train,$\langle\rho(\tau)\rangle.$ Then the ``average firing rate''
over a time window $T$ is given by:
\begin{equation}
\langle r\rangle=\frac{1}{T}\int_{0}^{T}d\tau\langle\rho(\tau)\rangle,
\end{equation}
while the firing (or spiking) rate, $r(t)$, has the following expression:
\begin{equation}
r(t)=\frac{1}{\varDelta t}\int_{t}^{t+\varDelta t}d\tau\langle\rho(\tau)\rangle.\label{eq_firing_rate}
\end{equation}
This is the ``firing rate'' computed on time windows of amplitude
$\varDelta t$. Formally the dependence on $\varDelta t$ can be removed
by taking the limit $\varDelta t\rightarrow0$ on the right hand side
of the equation (that is $r(t)=\langle\rho(t)\rangle$). Actually
the firing rate, $r(t)$, being a probability density, cannot be determined
exactly from the limited amounts of data available from a finite number
of trials. Therefore we need to approximate the true firing rate from
a spike sequence. There are several procedures to do it and some of
them are illustrated in figure \ref{fig_firing_rates_comp}. A very
common way consists in making the convolution of the available spike
train, $\rho(t)$, (or the PSTH, see figure \ref{fig_firing_rates_comp})
with a window function (also called the filter kernel), $w(t)$, in
order to obtain a more smoothed signal (and avoid jagged curve, like
the ones showed in figure \ref{fig_firing_rates_comp}B,C): 
\begin{equation}
r(t)=\int_{-\infty}^{+\infty}d\tau w(\tau)\rho(t-\tau),\label{eq_smoothed_FR}
\end{equation}
where $w(\tau)$ goes to 0 outside a region near $\tau=0$ and has
time integral equal to 1 (in order to not affect units of the firing
rate). The filter kernel specifies how the spike train evaluated at
time $t-\tau$ contributes to the firing rate approximated at time
$t$. Therefore, if we want the approximated firing rate in $t$ depends
only on the spikes occurred before $t$, the window function must
be 0 when its argument is negative. Such a kernel is termed causal. 

\subsection{Spike-triggered average}

A simple and effective way to perform neural encoding (that is to
characterize the average neural response to a given stimulus) is to
count the (trial-averaged) number of action potential fired during
the presentation of different stimuli. By plotting this number as
a function of the parameters $s$ chosen to characterize the stimuli,
we obtain the response tuning curve. The neural response to an external
stimulus is mediated by the interaction between the stimulus and the
sensory surface (e.g. in case of visual stimuli between the presented
image and the retina). The portion of the sensory surface (and, by
extension, of the external stimulus) responsible for the modulation
of the firing activity of a given neuron is called the receptive field
of the neuron.

Response tuning curve characterizes the average neural response to
a given stimulus. The complementary procedure, when performing neural
decoding, consists in computing the average stimulus that elicited
a given response. If the response is the spiking activity, this means
to compute the spike-triggered average (STA). Indeed, the spike-triggered
average is the average value of the stimulus at a time interval $\tau$
from the occurrence of a spike. We describe the stimulus with a parameter,
$s(t)$, that varies over time, and define the STA as:
\begin{equation}
C(\tau)=\frac{1}{\langle n\rangle}\int_{0}^{T}dt\langle\rho(t)\rangle s(t-\tau)=\frac{1}{\langle n\rangle}\int_{0}^{T}dt\:r(t)s(t-\tau),\label{eq_STA}
\end{equation}
where $\langle n\rangle$ is the average number of spikes in each
trial, which is assumed to be constant over trials. Although the range
of $\tau$ values in equation \ref{eq_STA} extends over the entire
trial length time, the response is typically affected only by the
stimulus in a window a few hundred milliseconds wide immediately preceding
and following a spike. To understand the reason of this behavior,
let's introduce the cross-correlation between the firing rate and
the stimulus:
\begin{eqnarray}
Q_{rs}(\tau) & = & \frac{1}{T}\int_{0}^{T}dt\:r^{\ast}(t)s(t+\tau)\label{eq_cross_correlation}\\
 & = & \frac{1}{T}\int_{0}^{T}dt\:r(t)s(t+\tau).\nonumber 
\end{eqnarray}
By substituting equation \ref{eq_cross_correlation} into equation
\ref{eq_STA}, we obtain that 
\begin{equation}
C(\tau)=\frac{T}{\langle n\rangle}Q_{rs}(-\tau)=\frac{Q_{rs}(-\tau)}{\langle r\rangle}.\label{eq_reverse_correlation}
\end{equation}
Now it is clear than the STA will approach to zero for positive $\tau$
values larger than the correlation time between the stimulus and the
response (that is usually in the order of hundred of milliseconds
or smaller). Furthermore the response of a neuron cannot depend on
future stimuli, thus, unless the stimulus has temporal autocorrelation\footnote{If a signal has temporal autocorrelation other than zero on a time
interval $\varDelta t$, it means that the signal in $t\pm t^{\ast}$
($t^{\ensuremath{\ast}}<\varDelta t$) is not independent on the signal
in $t$. }, we expect for $C(\tau)$ to be zero for $\tau<0$.

Because of the minus sign of the argument in the right hand side of
equation \ref{eq_reverse_correlation}, the spike-triggered average
is also called ``reverse correlation function''. 

\subsection{Reverse correlation and Wiener kernels}

When investigating the relationships between network oscillations
such as LFPs and EEGs and single-neuron activity, we cannot use (a
priori) the categories of stimulus and answer. Indeed we do not know
if there are (and in which directions) causal relationships between
the two signals. In this respect, neither we can talk of neural encoding
nor decoding. Our purpose is to estimate the time course of the local
field potentials from the spiking activity of a neuron and vice versa.
We also want to test how robust and general can be this estimation.

If we have a nonlinear system, where the input $x(t)$ and the output
$y(t)$ are functions of the time related by some functional transformation
$y(t)=F[x(t)]$, methods developed by Volterra \citep{volterra2005theory}
and Wiener \citep{wiener1966nonlinear}, provide a power series expansion
of the output function: 
\begin{eqnarray}
y(t) & = & h_{0}+\int_{-\infty}^{+\infty}d\tau\:h_{1}(\tau)x(t-\tau)\label{eq_Volterra_series}\\
 &  & +\int_{-\infty}^{+\infty}d\tau_{1}\int_{-\infty}^{+\infty}d\tau_{2}\:h_{2}(\tau_{1},\tau_{2})x(t-\tau_{1})x(t-\tau_{2})\nonumber \\
 &  & +\int_{-\infty}^{+\infty}d\tau_{1}\int_{-\infty}^{+\infty}d\tau_{2}\int_{-\infty}^{+\infty}d\tau_{3}\:h_{3}(\tau_{1},\tau_{2},\tau_{3})x(t-\tau_{1})x(t-\tau_{2})x(t-\tau_{3})+\ldots\nonumber 
\end{eqnarray}
Under certain conditions, the proper choice of the (Volterra) kernels,
$h_{n}$, will provide a complete description of any transformation
$x(t)\rightarrow y(t)$ \citep{volterra2005theory}. Note that, in
general, this is not a causal reconstruction of the output signal
$y$, indeed the integrals can range over negative values of the time
variable $\tau$, which means that the value of the input $x$ at
time instants later than $t^{\ast}$ can affect the value of $y(t^{\ast})$\footnote{To obtain a causal Volterra's series, the integrals in equation \ref{eq_Volterra_series}
have to range from 0 to $+\infty$. }. The series was rearranged by Wiener to make the terms easier. In
particular, Wiener reformulated Volterra's expansion by making the
successive terms independent, which means that we can compute the
terms individually. In this formulation, the filter kernels are called
Wiener kernels.

Since we are interested in building a linear model of the relationships
between single-neuron activity and network oscillations, let's focus
on the first (i.e., linear) Wiener\footnote{It is called also Wiener-Kolmogorov filter.}
kernel $h_{1}$\label{Wiener_kernel_definition}. To have a clear
intuition of what we are doing, remember that the simplest way to
construct an estimate of a time varying signal $y$ starting from
$x$, $y_{est}$, is to assume that at any given time, $t^{\ast}$,
$y_{est}(t^{\ast})$ can be expressed as a weighted sum of the values
taken by $x$. Let's assume that the weights are constant in time
(i.e., they are not a function of the time instant $t^{\ensuremath{\ast}}$:
the estimation is time invariant), therefore we write the estimated
signal as the convolution between a kernel $h_{x2y}$ and the input
signal (plus a constant $y_{0}$),
\begin{eqnarray}
y_{\mathrm{est}}(t) & = & y_{0}+\int_{0}^{T}d\tau\:h_{x2y}(t-\tau)x(\tau),\label{eq_linear_approximation}\\
 & = & y_{0}+\int_{t-T}^{t}d\tau\:h_{x2y}(\tau)x(t-\tau).
\end{eqnarray}
The Wiener filter $h_{x2y}$\footnote{We use the subscript ``$x2y$'' to specify the direction in which
we are doing the estimation.} gives the weights of the sum (that is the integral on time): it determines
how strongly, and with what sign, the value of the input $x$ in $(t-\tau)$
contribute to the value of the output in $t$. Since we are dealing
with real signals, the integral does not range from minus to plus
infinity (as in equation \ref{eq_Volterra_series}) but it is restricted
over the time interval where the signals are defined (from 0 to $T$,
the length of the trial). Note that in equation \ref{eq_linear_approximation}
(and hereafter) the signal $x(t)$ is defined with its mean value
subtracted out\footnote{This subtraction is needed to simplify the kernel computation, and
does not affect the performance estimation. } (that is $\int_{-\infty}^{+\infty}dt\:x(t)=0$), thus the constant
term $y_{0}$ is the mean value of $y_{est}$\footnote{Indeed, if $\langle x\rangle=0$, the convolution theorem implies
$\langle h_{x2y}\ast x\rangle=0$.} and compensates for the mean subtraction done on $x$ and it also
accounts for any background output activity we could have when $x=0$.

The filter $h_{x2y}$is chosen to minimize the mean (over the duration
of the trial, $T$) squared distance (MSD) between the original signal,
$y$, and the estimated one, $y_{est}$:
\begin{equation}
\mathrm{MSD}(y,y_{\mathrm{est}})=\frac{1}{T}\int_{0}^{T}dt[y(t)-y_{\mathrm{est}}(t)]^{2}.\label{eq_MSD}
\end{equation}
By minimizing this expression it is possible to obtain an explicit
formula for the Fourier transform of the Wiener optimal kernel$:$
\begin{equation}
\tilde{h}_{x2y}(\omega)=\frac{\tilde{Q}_{xy}(\omega)}{\tilde{Q}_{xx}(\omega)}=\frac{\tilde{Q}_{yx}(-\omega)}{\tilde{Q}_{xx}(\omega)}\label{eq_FOURIER_wiener_kernel}
\end{equation}
thus
\begin{equation}
h_{x2y}(t)=\frac{1}{2\pi}\int_{-\infty}^{+\infty}d\omega\:\frac{\tilde{Q}_{xy}(\omega)}{\tilde{Q}_{xx}(\omega)}e^{-i\omega t},\label{eq_wiener_kernel}
\end{equation}

where the $\tilde{f}$ indicates the Fourier transform of $f$. $Q_{xy}(\omega)$
is the cross-correlation between x and y (see equation \ref{eq_cross_correlation})
and $Q_{xx}(\omega)$ is the autocorrelation. The Wiener-Khinchin
theorem assures that if $x$ and $y$ are wide-sense stationary random
processes\footnote{Note that the importance of this theorem relies on the fact that if
a signal is a wide-sense stationary random process its Fourier transform
does not exist.}, $\tilde{Q}_{xy}$ can be computed as the cross power spectral density
of $x$ and $y$, $S_{xy}(\omega)$, and $\tilde{Q}_{xx}$ as the
power spectral density of $x$, $S_{xx}(\omega)$. Since the mean
value of $x$ is zero, the convolution theorem (i.e., $\widetilde{f\ast g}=k\,\tilde{f}\,\tilde{g}$)
implies that the mean value of the kernel, $h_{x2y}$, is zero.\\
Note that the Wiener kernel can be seen also as the transfer function
of the linear time-invariant system made by the input $x$ and the
output $y$. 

To have a more clear idea of what the kernel represents suppose that
the input $x(t)$ is an uncorrelated signal (i.e., its autocorrelation
is a delta function: $Q_{xx}(\tau)=k\delta(\tau)$), as in the case
of white noise, and the output is a firing rate $y=r(t)$. Thus, from
equation \ref{eq_FOURIER_wiener_kernel}, we obtain:
\begin{equation}
h_{x2r}(\tau)=\frac{Q_{rx}(-\tau)}{k}=\frac{\langle r\rangle C(\tau)}{k},
\end{equation}
where $C(\tau)$ is the STA and the last equality follows from equation
\ref{eq_reverse_correlation}. Therefore in case of white-noise input
and spike train output, the first Wiener kernel is proportional to
the spike-triggered average\footnote{More in general, when the input is not a white-noise, it is possible
to demonstrate that the kernel $h_{x2y}$ is proportional to the input
that gives rise to the highest estimated output $y_{est}$.}. On the other hand, if the input is an uncorrelated firing rate,
$r(t)$, (which tends to happen at low rates), $Q_{rr}(\tau)=\langle r\rangle\delta(t)$,
the equation for the Wiener kernel becomes:
\begin{equation}
h_{x2r}(\tau)=\frac{Q_{ry}(\tau)}{\langle r\rangle}=C(-\tau).\label{eq_358_0}
\end{equation}

We conclude this section by noting that comparing equations \ref{eq_358_0}
and \ref{eq_FOURIER_wiener_kernel} we can have a better insight on
the difference between the STA and the Wiener kernel when performing
a decoding task. In particular, the numerator in equation \ref{eq_FOURIER_wiener_kernel}
reproduces the STA in equation \ref{eq_358_0}, thus the role of the
denominator in the expression of the Wiener kernel is to correct for
the autocorrelation in the response spike train. Indeed such autocorrelation
introduce a bias in the decoding, which is removed by using the Wiener
kernel\label{STA_vs_wienerKernel}. Note that, when the input is a
firing rate, the convolution in equation \ref{eq_linear_approximation}
translates into a simple rule: every time a spike appears, we replace
it with the kernel. 

\subsection*{Causality in the estimation\label{section_FILTERcausality}}

The linear estimation performed by using equation \ref{eq_linear_approximation}
is not causal in general, indeed the argument of the kernel can range
over negative values. The simplest way to force the estimation of
$y$ to be causal is to set the kernel equal to 0 for negative values\footnote{Note that, in this case, the restricted kernel is no longer the optimal
kernel.} (i.e., $h_{x2y}(t)=0$ for $t<0$), or, equivalently, to restrict
the interval of integration in equation \ref{eq_linear_approximation}:
\begin{equation}
y_{\mathrm{est}}^{\mathrm{causal}}(t)=y_{0}+\int_{0}^{t}d\tau\:h_{x2y}(t-\tau)x(\tau).\label{eq_linear_approx_causal}
\end{equation}

A complementary procedure useful to implement causality is given by
the introduction of a delay in the filter \citep{Dayan}. In equation
\ref{eq_linear_approximation} we attempt to estimate the signal $y$
in $t$ by using the values of $x$ over the entire trial length,
while in equation \ref{eq_linear_approx_causal} we use only the value
of $x$ prior to the time $t$. We already mentioned that, when we
are dealing with decoding tasks, the signal $y(t)$ we want to estimate
is the stimulus and $x(t)$ is the elicited response (e.g. spike-train
decoding: we attempt to construct an estimation of the stimulus from
the evoked spikes). The stimulus required a finite amount of time
$\tau_{0}$ to affect the response, thus, to make the decoding task
easier we can introduce a prediction delay $\tau_{0}$ and estimate
the stimulus $y$ at time $(t-\tau_{0})$ from the values of the response
$x$ up to time $t$:
\begin{equation}
y_{\mathrm{est}}(t-\tau_{0})=y_{0}+\int_{0}^{t}d\tau\:h_{x2y}(t-\tau)x(\tau).
\end{equation}
The delay $\tau_{0}$ results in the Wiener kernel expression in the
following way:
\begin{equation}
\tilde{h}_{x2y}(\omega)=\frac{\tilde{Q}_{xy}(\omega)}{\tilde{Q}_{xx}(\omega)}e^{i\omega\tau_{0}},
\end{equation}
and equation \ref{eq_358_0} becomes:
\begin{equation}
h_{x2y}(\tau)=\frac{Q_{ry}(\tau-\tau_{0})}{\langle r\rangle}=C(\tau_{0}-\tau).\label{eq_358}
\end{equation}
Note that, if there is not stimulus autocorrelation, $C(\tau)=0$
for $\tau<0$ (i.e., the filter is zero for $\tau>0$). On the other
hand, the causality requires the filter to be zero for $\tau<0$.
Therefore, from equation \ref{eq_358}, it is clear the need for either
stimulus autocorrelation or a nonzero prediction delay $\tau_{0}$
when $x$ is the stimulus and $y$ the response.

\newpage{}

~

\newpage{}

\chapter{How synaptic currents shape network dynamics\label{chapter_frontiers}}

\ohead{\headmark} 
\pagestyle{scrheadings}    

\lettrine[lines=2]{I}{n this} chapter we investigate in a modelling
framework some aspects of the relationship between dynamics at the
single-neuron and at the network level. More precisely, we focus on
how different features of the synaptic input affect network dynamics
as measured by LFPs and average properties across neurons.\\
We already mentioned that models of networks of Leaky Integrate-and-Fire
(LIF) neurons are a widely used tool for theoretical investigations
of brain functions. These models have been used both with current-
and conductance-based synapses (see section \vref{COB_vs_CUB}). However,
the differences in the dynamics expressed by these two approaches
have been so far mainly studied at the single-neuron level. To investigate
how these synaptic models affect network activity, we compared the
single-neuron and neural population dynamics of conductance-based
networks (COBNs) and current-based networks (CUBNs) of LIF neurons.
These networks were endowed with sparse excitatory and inhibitory
recurrent connections, and were tested in conditions including both
low- and high-conductance states. We developed a novel procedure to
obtain comparable networks by properly tuning the synaptic parameters
not shared by the models. The so defined comparable networks displayed
an excellent and robust match of first order statistics (average single-neuron
firing rates and average frequency spectrum of network activity).
However, these comparable networks showed profound differences in
the second order statistics of neural population interactions and
in the modulation of these properties by external inputs. The correlation
between inhibitory and excitatory synaptic currents and the cross-neuron
correlation between synaptic inputs, membrane potentials and spike
trains were stronger and more stimulus-modulated in the COBN. Because
of these properties, the spike train correlation carried more information
about the strength of the input in the COBN, although the firing rates
were equally informative in both network models. Moreover, the network
activity of COBN showed stronger synchronization in the gamma band,
and spectral information about the input higher and spread over a
broader range of frequencies. These results suggest that the second
order statistics of network dynamics depend strongly on the choice
of the synaptic model.

\section{Introduction}

Networks of Leaky Integrate-and-Fire (LIF) neurons are a key tool
for the theoretical investigation of the dynamics of neural circuits.
Models of LIF networks express a wide range of dynamical behaviors
that resemble several of the dynamical states observed in cortical
recordings (see \citet{Brunel2013} for a recent review). An advantage
of LIF networks over network models that summarize neural population
dynamics with only the density of population activity, such as neural
mass models \citep{Deco08}, is that LIF networks include the dynamics
of individual neurons. This allows to investigate at the same time
the single-neuron and the network level, and, for example, LIF networks
can be used to investigate phenomena, such as the relationships among
spikes of different neurons, that are not directly accessible to simplified
mass models of network dynamics. 

A basic choice when designing a LIF network is whether the synaptic
model is voltage-dependent (conductance-based model) or voltage-independent
(current-based model). In the former case the synaptic current depends
on the driving force, while this does not happen in the current-based
model(see section \vref{COB_vs_CUB}). Current-based LIF models are
popular because of their relative simplicity (see e.g. \citet{Brunel2013})
and they have the key advantage of facilitating the derivation of
analytical closed-form solutions. Thus current-based synapses are
convenient for developing mean field models \citep{grabska2014well},
event based models \citep{Touboul11}, or firing rate models \citep{helias2010instantaneous,Ostojic11,Schaffer2013},
as well as in studies examining the stability of neural states \citep{babadi2010intrinsic,mongillo2012bistability}.
Moreover, current-based models are often adopted, because of their
simplicity, to investigate numerically network-scale phenomena \citep{Memmesheimer2010,Renart2012,Gutig2013,Lim2013,zhang2014distribution}.
On the other hand, conductance-based models are also widely used because
they are more biophysically grounded \citep{Kuhn2004,Meffin2004}.
In particular, only conductance-based neurons can reproduce the fact
that when the synaptic input is intense, cortical neurons display
a three- to fivefold decrease in membrane input resistance (thus they
enter a high-conductance state), as observed in intracellular recordings
in vivo \citep{Destexhe2003}. However, an added complication of conductance-based
models is that their differential equations can only be evaluated
numerically or approximated analytically \citep{Rudolph-Lilith2012}
rather than being fully analytically treatable. 

Despite the widespread use of both types of models, the differences
in the network dynamics that they generate has not been yet fully
understood. Previous studies comparing conductance- and current-based
LIF models focused mostly on the individual neuron dynamics \citep{Kuhn2004,Meffin2004,Richardson2004}.
Here we extended these previous works by investigating the network
level consequences of the synaptic model choice. In particular, we
investigated which aspects of network dynamics are independent of
the choice of the specific synaptic model, and which are not. Understanding
this point is crucial for fully evaluating the costs and implications
of adopting a specific synaptic model. 

We compared the dynamics of two sparse recurrent excitatory-inhibitory
LIF networks, a conductance-based network (COBN) with conductance-based
synapses, and a current-based network (CUBN) with current-based synapses.
To properly compare the two networks, we set to equal values all the
common parameters (including the connectivity matrix). Building on
previous works \citep{laCamera2004,Meffin2004}, we devised a novel
algorithm to obtain two comparable networks by properly tuning the
synaptic conductance values of the COBN given the set of values of
synaptic efficacies of the CUBN. Since the differences between the
dynamics of the two synaptic models depend on the fluctuations of
the driving force (i.e., of the membrane potential), they should be
close to zero when the synaptic activity is low. Thus, when decreasing
the background synaptic activity, the Post-Synaptic Currents (PSCs)
of the two models should become more and more similar. Consequently,
our procedure calibrated the conductances so that PSCs became exactly
equal in the limit of zero synaptic input (see section \textcolor{red}{\vref{section_procedure_to_Gsyn}}).
Then we investigated whether this procedure could generate COBNs and
CUBNs with matching average single-neuron stationary firing rates
under a reasonably wide range of parameters and network stimulation
conditions. We then studied how comparable conductance- and current-
based networks differed in more complex characterizations of population
dynamics, such as the cross-neuron correlations of membrane potential
(MP), input current and spike train, as well as the spectrum of network
fluctuations. The latter was investigated not only for total average
firing rates, but also for the simulated Local Field Potential (LFP)
computed from the massed synaptic activity of the networks \citep{mazzoni2008encoding}.
To study the spectrum of network fluctuations it is useful to use
a LFP model (rather than a massed spike rate) mainly because cortical
rhythms are more easily measured in experiments by recording LFPs
rather than the spike rate \citep{Buzsaki2012,Einevoll2013}; therefore
this quantification makes the models more directly comparable to experimental
observations. We then quantified how the external inputs modulate
the firing rate, the LFP spectrum and the spike train correlation
by using information theory \citep{Quiroga,crumiller2011estimating}.
Finally, we discuss the similarities and differences of COBN and CUBN
against recent experimental observations of dynamics of cortical network
correlations \citep{Lampl1999,Kohn2005,deLaRocha2007correlation,Okun2008,Ecker2010,Renart2010}.

\section{Methods}

\subsection{Network structure and external inputs\label{section_network_and_input}}

\begin{figure}
\begin{centering}
\includegraphics[scale=0.6]{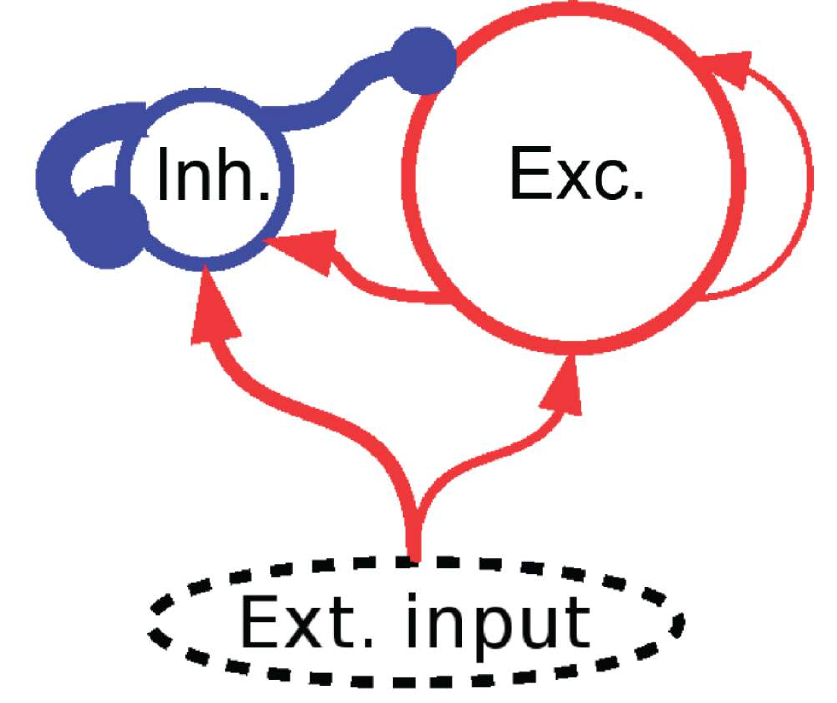}
\par\end{centering}
\begin{centering}
\caption[Network structure]{\textbf{\textcolor{black}{Network structure}}. The network is composed
of 1000 inhibitory (blue) and 4000 excitatory LIF neurons (red). Connectivity
is random, each directed pair of neurons is connected with a probability
of 0.2. The size of the arrows represents schematically the different
synaptic strengths. In addition to recurrent interactions both populations
receive an external excitatory input. Adapted from \citep{mazzoni2008encoding}.\label{fig_network_struct}}
\par\end{centering}
\centering{}
\end{figure}

We considered two networks of LIF neurons with identical architecture
and injected with identical external inputs. The only difference between
the two networks was in the synaptic model: one was composed by neurons
with conductance-based synapses and the other by neurons with current-based
synapses (see section \vref{COB_vs_CUB}). The network structure we
adopt was already used in other works such as \citep{Brunel03,mazzoni2008encoding,Mazzoni2011}.
Each network was composed of 5000 neurons. Eighty percent of the neurons
were excitatory, that is their projections onto other neurons formed
AMPA-like excitatory synapses, while the remaining 20\% were inhibitory,
that is their projections formed (A-type) GABA-like inhibitory synapses.
The 4:1 ratio is compatible with anatomical observations \citep{Braitenberg1991}.
The network had random connectivity with a probability of directed
connection between each pair of neurons of 0.2 \citep{Sjostrom2001,Holmgren2003},
thus any neuron in the network received on average 200 synaptic contacts
from inhibitory neurons and 800 from excitatory neurons \textcolor{black}{(see
figure \ref{fig_network_struct})}. Both populations received a noisy
excitatory external input taken to represent the activity from thalamocortical
afferents, with inhibitory neurons receiving stronger inputs than
excitatory neurons. This simulated external input was implemented
as a series of spike times that activated excitatory synapses with
the same kinetics as recurrent AMPA synapses, but different strengths

The input spike trains activating the model thalamocortical synapses
were generated by a Poisson process, with a time varying rate, $\nu_{ext}(t)$,
identical for all neurons. Note that this implied that the variance
of the inputs across neurons increased with the input rate. $\nu_{ext}(t)$
was given by the positive part of the superposition of a \textquotedblleft signal\textquotedblright ,
$\nu_{signal}(t)$, and a \textquotedblleft noise\textquotedblright{}
component , $n(t)$:
\begin{equation}
\nu_{ext}(t)=[\nu_{signal}(t)+n(t)]_{+}\label{eq_net_input_c3}
\end{equation}
The separation of signal and noise in the input spike rate was to
reproduce the classical experimental design in which a given sensory
stimulus is presented many times, with each presentation (or \textquotedblleft trial\textquotedblright )
eliciting different responses due to variations in intrinsic network
dynamics from presentation to presentation. We achieved this by identifying
the external stimulus with the signal term, $\nu_{signal}(t)$, (which
was thus exactly the same across all trials of the same stimulus)
and by using a noise term, $n(t)$, generated (as explained below)
independently in each trial.

In this study we used three kinds of external signals. \\
For the majority of the simulations we used constant stimuli, $\nu_{signal}(t)=\nu_{0}$,
(with $\nu_{0}$ ranging from 1.5 to 6 spikes/ms). \\
In a second set of simulations we used periodic stimuli made by superimposing
a constant baseline term to a sinusoid: $\nu_{signal}(t)=A\sin(2\pi ft)+\nu_{0}$,
where A = 0.6 spikes/ms; $f$ ranged from 2 to 16 Hz in \textcolor{black}{figure
\ref{fig_12_paperFR}} and from 2 to 150 Hz in \textcolor{black}{figure
\ref{fig_13_paperFR}} and $\nu_{0}$ was set to 1.5 (respectively
5) spikes/ms when studying the low- (respectively high-) conductance
state. \\
We also used a time varying signal, called \textquotedblleft naturalistic\textquotedblright ,
that reproduced the time course of Multi Unit Activity recorded from
the LGN of an anesthetized macaque during binocular presentation of
commercially available color movies\citep{Belitski2008}. More precisely,
the MUA was measured as the absolute value of the high pass filtered
(400\textendash 3000 Hz) extracellular signal recorded from an electrode
placed in the LGN while the monkey was presented binocularly a color
movie (we refer to \citealp{Rasch08} for full details on experimental
methods). The MUA measured in this way is thought to represent a weighted
average of the extracellular spikes of all neurons within a sphere
of <140\textendash 300 mm around the tip of the electrode \citep{Logothetis03},
and thus gives a good idea of the spike rate fluctuations of a patch
of geniculate input to cortex during viewing of natural stimuli. We
took 40 consecutive seconds of LGN MUA recordings during movie presentation,
we divided it into 20 non-overlapping intervals of 2 seconds (ideally
corresponding to different movie scenes) following the procedure used
in \citep{Belitski2008}, and each interval was considered as a different
visual stimulus. For the purposes of the present work, it is mainly
useful to remind that the naturalistic input was a slow signal dominated
by frequencies below 4 Hz. 

The noise component of the stimuli, $n(t)$, was generated by an Ornstein-Uhlenbeck
(OU) process with zero mean:
\begin{equation}
\tau_{n}\frac{dn(t)}{dt}=-n(t)+\sigma_{n}(\sqrt{2\tau_{n}})\eta(t),\label{eq_OU}
\end{equation}
where $\eta(t)$ is a Gaussian white noise. $\sigma_{n}^{2}$=0.16
spikes/ms is the (stationary, that is for $t\rightarrow\infty$) variance
of the noise, while the stationary mean is 0. The time constant $\tau_{n}$
was set to 16 ms to have a cutoff frequency of 10 Hz. The OU process
is a stationary, Gaussian, and Markovian process we chose for the
two following reasons: 
\begin{itemize}
\item the power spectrum, which is flat up to the cutoff frequency and then
it decays as $f^{-2}$. Therefore it does not diverge and it has the
highest power spectral density in the low frequencies, in agreement
with what found in the background activity of the cortex \citep{Mazzoni2011}
\item It is a mean-reverting process. Indeed, the drift term is not constant
(since it depends on the value assumed by the process) and it always
tends to drift the variable towards its long-term mean (0 in our case).
\end{itemize}
Note that the trial-to-trial differences in the stochastic process
generated by equation \ref{eq_OU} were the first and largest source
of trial-to-trial variability in the model (that is the variability
at fixed stimulus, see section \vref{noise_role}), the second and
last being the fact that each neuron received an independent realization
of the Poisson process with rate $\nu_{ext}(t)$.

In a specific set of control stimulations (\textcolor{black}{figure
}\ref{fig_supp4_paperFR}), instead of the OU process described above,
we used a Gaussian white noise with the same variance. Note that,
for low frequencies, the power spectrum of the OU process was higher
than the one of the white noise.

\subsection{Single-neuron models}

Both inhibitory and excitatory neurons were modeled as (LIF) neurons
(see section \vref{LIF_model}). The leak membrane potential, $V_{leak}$,
was set to -70 mV, the spike threshold, $V_{threshold}$, to -52 mV
and the reset potential, $V_{reset}$, to -59 mV. The absolute refractory
period was set to 2 ms for excitatory neurons and to 1 ms for inhibitory
neurons \citep{Brunel03}. Since we had no current injected into the
neuron through an electrode, the equation for the sub-threshold dynamic
of the MP of i-th neuron (see equation \ref{eq_integrate_and_fire})
took the form:
\begin{equation}
\tau_{m}\frac{dV^{i}(t)}{dt}=-V^{i}(t)+V_{leak}-\frac{I_{tot}^{i}(t)}{G_{leak}},\label{eq_3_paperFR}
\end{equation}
where $\tau_{m}$ is the membrane time constant (20 and 10 ms for
excitatory and inhibitory neurons respectively), $G_{leak}$ is the
leak membrane conductance\footnote{Note that here we use the capital letter because this is the absolute
value of the capacitance (not referred to the cell surface).} (25 and 20 nS for excitatory and inhibitory neurons respectively)
\citep{Brunel03} and $I_{tot}^{i}(t)$ is the total synaptic input
current. The latter was given by the sum of all the synaptic inputs
entering the i-th neuron:
\begin{equation}
I_{tot}^{i}(t)=\underset{N_{(i,AMPArec)}}{\sum}I_{AMPArec}^{i}(t)+\underset{N_{(i,GABA)}}{\sum}I_{GABA}^{i}(t)+I_{AMPAext}^{i}(t),\label{eq_4_paperFR}
\end{equation}
the value of $N_{(i,AMPArec)}$ (respectively $N_{(i,GABA)}$) being
the set of excitatory (respectively inhibitory) neurons projecting
into the i-th neuron, and $I_{\ensuremath{AMPArec}}^{i}(t)$, $I_{\ensuremath{GABA}}^{i}(t)$,
$I_{AMPAext}^{i}(t)$ the different synaptic inputs entering the i-th
neuron from: recurrent AMPA, GABA, and external AMPA synapses respectively. 

The difference between current- and conductance-based synapses lied
in the definition of these synaptic input currents and. Current-based
synapses (see equations \ref{eq_synaptic_current_CUB}), $I^{CUBN}$,
were modeled as follows:
\begin{equation}
I_{syn}^{CUBN}(t)=J_{syn}s_{syn}(t),\label{eq_5_paperFR}
\end{equation}
while conductance-based currents (see equation \ref{eq_synaptic_current}),
$I^{COBN}$, as
\begin{equation}
I_{syn}^{COBN}(t)=G_{syn}s_{syn}(t)(V(t)-V_{syn}).\label{eq_6_paperFR}
\end{equation}
Both models had the same synaptic kinetics, that is the same functions
$\varDelta s_{syn}(t)$ described the time course of the synaptic
currents: every time a presynaptic spike occurred at time $t^{*}$,
$s_{syn}(t)$ of the postsynaptic neuron was incremented by an amount
described by a delayed\footnote{The delay models the fact that the transmission of the action potential
from the pre- to the post-synaptic neuron requires a finite time.} difference of exponentials (see section \vref{ssyn}) \citep{Brunel03}:
\begin{equation}
\varDelta s_{syn}(t)=\frac{\tau_{m}}{\tau_{d}-\tau_{r}}\left[\exp\left(-\frac{t-\tau_{l}-t^{*}}{\tau_{d}}\right)-\exp\left(-\frac{t-\tau_{l}-t^{*}}{\tau_{r}}\right)\right],\label{eq_7_paperFR}
\end{equation}
where the latency $\tau_{l}$, the rise time $\tau_{r}$ and the decay
time $\tau_{d}$ are shown in table \vref{tab_tau_syn}.
\begin{table}[h]
\begin{centering}
\begin{tabular}{|c|c|c|c|}
\hline 
\textbf{Synaptic time constants (ms)} & \textbf{$\tau_{l}$} & \textbf{$\tau_{r}$} & \textbf{$\tau_{d}$}\tabularnewline
\hline 
\hline 
GABA & 1 & 0.25 & 5\tabularnewline
\hline 
AMPA on inhibitory & 1 & 0.2 & 1\tabularnewline
\hline 
AMPA on excitatory & 1 & 0.4 & 2\tabularnewline
\hline 
\end{tabular}
\par\end{centering}
\medskip{}

\centering{}\caption[Synaptic time constants of network models]{\textbf{Synaptic time constants of both models.\label{tab_tau_syn}}}
\end{table}

\begin{table}[h]
\begin{centering}
\begin{tabular}{|c|c|}
\hline 
\multicolumn{2}{|c|}{\textbf{\textit{CURRENT-BASED NETWORK}}}\tabularnewline
\hline 
\hline 
\multicolumn{2}{|c|}{\textbf{Synaptic efficacies, $J_{syn}$ (pA)}}\tabularnewline
\hline 
GABA on inhibitory & 54\tabularnewline
\hline 
GABA on excitatory & 42.5\tabularnewline
\hline 
AMPA\textsubscript{recurrent} on inhibitory & -14\tabularnewline
\hline 
AMPA\textsubscript{recurrent} on excitatory & -10.5\tabularnewline
\hline 
AMPA\textsubscript{external} on inhibitory & -19\tabularnewline
\hline 
AMPA\textsubscript{external} on excitatory & -13.75\tabularnewline
\hline 
\end{tabular}
\par\end{centering}
\medskip{}

\centering{}\caption[Synaptic efficacies of the CUBN]{\textbf{Synaptic efficacies of the current-based network}.\label{tab_Jsyn}}
\end{table}
The current-based synapses (see equation \ref{eq_synaptic_current_CUB})
were characterized by the synaptic efficacies $J_{syn}$ whose value
are reported in table \ref{tab_Jsyn}.On the other hand, the parameters
shaping the conductance-based synapses (see equation \ref{eq_synaptic_current})
were the conductances, $G_{syn}$, and the reversal potential of the
synapses, $V_{\ensuremath{syn}}$ (see table \ref{tab_gsyn}).

\begin{table}[h]
\begin{centering}
\begin{tabular}{|c|c|}
\hline 
\multicolumn{2}{|c|}{\textbf{\textit{CONDUCTANCE-BASED NETWORK}}}\tabularnewline
\hline 
\hline 
\multicolumn{2}{|c|}{\textbf{Synaptic conductances (nS)}}\tabularnewline
\hline 
GABA on inhibitory & 2.70\tabularnewline
\hline 
GABA on excitatory & 2.01\tabularnewline
\hline 
AMPArecurrent on inhibitory & 0.233\tabularnewline
\hline 
AMPArecurrent on excitatory & 0.178\tabularnewline
\hline 
AMPAexternal on inhibitory & 0.317\tabularnewline
\hline 
AMPAexternal on excitatory & 0.234\tabularnewline
\hline 
\multicolumn{2}{|c|}{\textbf{Synaptic reversal potential (mV)}}\tabularnewline
\hline 
VGABA & -80\tabularnewline
\hline 
VAMPA & 0\tabularnewline
\hline 
\end{tabular}
\par\end{centering}
\medskip{}

\centering{}\caption[Synaptic parameters of the COBN]{\textbf{Reference values of the synaptic parameters in the conductance-based
model.\label{tab_gsyn}}}
\end{table}
A useful parameter for conductance-based neuron analysis is the effective
membrane time constant, $\tau_{eff}$ . Following a standard procedure,
we computed the total effective membrane conductance for the i-th
neuron as:
\begin{alignat}{1}
G_{tot}^{i}(t) & =G_{leak}+\underset{N_{(i,AMPArec)}}{\sum}G_{AMPArec}s_{AMPArec}^{i}(t)+\underset{N_{(i,GABA)}}{\sum}G_{GABA}s_{GABA}^{i}(t)+\nonumber \\
 & +G_{AMPAext}s_{AMPAext}^{i}(t)\label{eq_8_paperFR}
\end{alignat}
and we rewrote equation \ref{eq_3_paperFR} as follows:
\begin{equation}
\tau_{eff}^{i}(t)\frac{dV^{i}(t)}{dt}=-V^{i}(t)+\frac{G_{leak}V_{leak}+\underset{N(i,syn)}{\sum}G_{syn}s_{syn}^{i}(t)V_{syn}}{G_{tot}^{i}(t)},\label{eq_9_paperFR}
\end{equation}
where \textquotedblleft syn\textquotedblright{} indicates: recurrent
AMPA; GABA; external AMPA synapses and
\begin{equation}
\tau_{eff}^{i}(t)=\frac{\tau_{m}G_{leak}}{G_{tot}^{i}(t)}\label{eq_10_paperFR}
\end{equation}
is the effective membrane time constant. In particular, for the i-th
neuron, the effective AMPA conductance is defined as $\sum_{N(i,AMPArec)}G_{AMPArec}s_{AMPArec}^{i}(t)+G_{AMPAext}s_{AMPAext}^{i}(t)$
and the effective GABA conductance as $\sum_{N(i,GABA)}G_{GABA}s_{GABA}^{i}(t)$
\textcolor{black}{(see figure \ref{fig_3_paperFR}).}

By looking at equation \ref{eq_9_paperFR} we can have a new insight
about the differences between the two synaptic models. Indeed, by
comparing equation \ref{eq_9_paperFR} (for the conductance-based
neurons) with equation \ref{eq_integrate_and_fire_CUB} (for the current-based
model), we see that the former case differs from the latter essentially
because of:
\begin{itemize}
\item the leak conductance $G_{\ensuremath{leak}}$ has been replaced by
the total conductance $G_{tot}(t)\geq G_{leak}$, which is a function
of the synaptic input to the neuron
\item the membrane time constant $\tau_{m}$ has been replaced by the effective
membrane time constant $\tau_{eff}(t)\leq\tau_{m}$, which is a function
of the synaptic input to the neuron
\end{itemize}
As a consequence of the variability in the total conductance, $G_{tot}(t)$,
the conductance-based neurons can switch from low- to high-conductance
states \citep{Destexhe2003} and vice versa. In particular, when the
network activity increases, the neurons tend to move towards high-conductance
states (see equation \ref{eq_8_paperFR} and \textcolor{black}{figure
\ref{fig_3_paperFR}}) and vice versa.

\subsection{Numerical methods\label{section_numerical_methods}}

Network simulations were done using a finite difference integration
scheme based on the second-order Runge Kutta algorithm \citep{Press1992},
also known as the midpoint method, with time step $\varDelta t=0.05$
ms. 

The noise, $n(t)$, was obtained from equation \ref{eq_OU} by implementing
an exact numerical simulation of the Ornstein-Uhlenbeck process\citep{Gillespie1996}.
The temporal durations of the simulations varied from 4.5 s to 100.5
s, and they are specified in the figure captions. The regimes we investigated
displayed average firing rates relatively low (0.4\textendash 13 Hz),
thus, when computing the Inter-Spike Interval (ISI) and the pairwise
spike train correlation, we used the longest simulation times (25.5
and 100.5 s) to obtain larger spike datasets. Since we studied stationary
responses, the first 500 ms of the simulations were never included
in any analysis. Analysis and simulations (the latter implemented
using MEX file) were performed in Matlab. Both COBN and CUBN model
source codes are available on the ModelDB sharing repository (\url{http://senselab.med.yale.edu/ModelDB/ShowModel.asp?model=152539})
with accession number 152539. 

\subsection{Spectral analysis\label{section:Spectral-analysis}}

To compute the power spectrum we used the Fast Fourier Transform with
the Welch method (pwelch function in Matlab), dividing the time window
under investigation into eight subwindows with 50\% overlap. 

For the entrainment analysis showed in\textcolor{red}{{} }\textcolor{black}{figure
\ref{fig_13_paperFR}} in case of periodic inputs with frequency $f$,
we bandpassed the LFP at the correspondent frequency f with a Kaiser
filter with zero phase lag and 2 Hz bandwidth, very small passband
ripple (0.05 dB) and high stopband attenuation (60 dB). We extracted
then the instantaneous phase by means of the Hilbert transform of
the signal. To quantify entrainment, we computed the phase coherence
between the phase of the input signal and of the LFP at the corresponding
frequency \citep{mormann2000mean}. Phase coherence, which we computed
using the CircStat toolbox \citep{Berens2009}, ranges from zero (no
relationships between phases) to 1 (perfect phase locking between
the two signals).

\subsection{LFP as a measure of network-level dynamics\label{section_LFP}}

A very common way to track network dynamics, that are dynamics due
to the overall integrative processes of networks of neurons, is by
means of a signal called local field potential. LFPs are obtained
by low pass filtering an extracellularly recorded signal (called extracellular
field potentials), which represents the electric activity resulting
from the neuronal processes of cells close to the recording site \citep{Belitski2008}.
More precisely, from a theoretical point of view, the neurons are
placed in an extracellular medium, which acts as a conductor, with
a specific impedance\footnote{The impedance of the cerebral cortex is isotropic and independent
on the signal frequency \citep{Ranck63,Logothetis07}, therefore it
should not affect the oscillations and the power spectral density
of the recorded signal. } ($\sim200-400$ $\Omega/$cm depending on the neuronal site, \citet{Ranck63,Ranck66,Mitzdorf85,Nicholson75})
higher than the one of a saline solution ($\sim65$ $\Omega/$cm).
This high impedance reflects the fact that ion moves around cells
in a very limited space. In the extracellular recording, the inflow
of positive ions (mainly Na\textsuperscript{+}) inside a neuron,
through an active regions of the membrane, corresponds to an inward
current and the active region to a current sink. When the current
flows inside the neurons, for the continuity equation of electric
charge, the inactive regions of the membrane act as sources of charge
for the active regions (that is, through the inactive regions outward
currents will flow). The superposition of currents from all sinks
and sources, due to the impedance of the extracellular medium, elicits
an electric field called extracellular field potential (EFP). To obtain
EFPs actually useful to investigate network dynamics (i.e., dynamics
due to a population of neurons), the measurement is done with an electrode
(or pipette) with a sufficiently low impedance and whose tip is not
too close to the spike generation site of a single neuron (in order
to to avoid that the action potentials from the single neuron prevail
on the overall neuronal signal). The EFPs recorded in this way collect
both integrative processes due to the dendritic/synaptic activity
of neurons and spikes fired from a groups of neurons in the proximity
of the recording site. These two different contributions can be reliably
segregated by frequency band separation. In particular, with a low
pass filter (with a cutoff around 200 Hz) we obtain LFP, while, a
high pass filter cutoff of $\sim500$ Hz is used in most recordings
to obtain the multi-unit activity. From MUA we can then extract the
spiking activity of small neural populations in a sphere of 100\textendash 300
$\mu$m radius, and, by performing a spike sorting, even of single
(or few) neuron (i.e., the single-unit activity) \citep{Logothetis08}.

The LFP reflects the perisynaptic activity of a neural population,
which, to have an idea can be within 0.5-3 mm from the electrode tip.
The size of the neural population is debated and depends on both the
method used to measure it and on the kind of electrode used. In general,
the slow oscillations seem to be correlated on higher distances than
the fast oscillations, thus depending on neurons located in a larger
area. LFP is thought to be given by a weighted sum of all the potential
changes close to the electrode, which depends on the current flows
in the extracellular space. The latter, in turn, are related to all
the integrative subthreshold processes. These processes are not only
due to synaptic activity (i.e., synaptic potentials), but also to
other types of slow oscillations, like voltage-dependent membrane
oscillations\footnote{They are variations of the membrane potential due to the opening/closing
of membrane channel, which, in turn, is regulated by the membrane
potential value.} and spike afterpotentials\footnote{More precisely, the soma-dendritic spike afterpotential indicates
a brief depolarization, followed by a longer lasting hyperpolarization.
It generally happen after a soma-dendritic spike in the neurons of
central nervous system and have a duration on the orders of 10s of
milliseconds.}, in areas such as the dendritic trees, not accessible by the spiking
activity of few neurons. In conclusion, the LFP does not reflect the
output of a cortical area, but rather the synaptic and dendritic processes
and the local processing of the signal in the cortex \citep{Logothetis08}.

\subsubsection{Computation of simulated LFP\label{section_Computation-of-LFP}}

\begin{figure}
\begin{centering}
\includegraphics[scale=0.3]{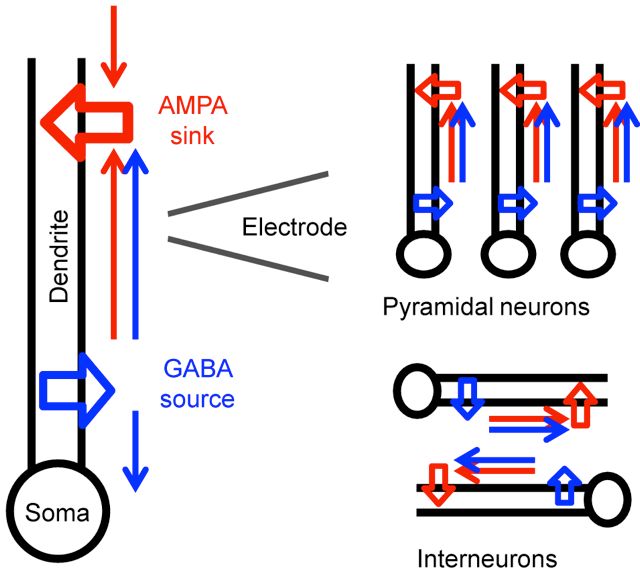}
\par\end{centering}
\centering{}\caption[Simulated LFP computation]{\textbf{Schematic of the computation of the simulated LFP.} The arrows
indicate the direction of the flow of positive charges (i.e., cations)
in the extracellular medium due to GABA (blue) and AMPA synaptic currents
(red arrows). \textbf{Left side}: representation of a pyramidal neuron
in an open field configuration with excitatory synapses (AMPA) on
apical dendrites and inhibitory synapses (GABA) close to the soma.
We computed the simulated LFP as the GABA currents minus the AMPA
currents because the pyramidal neurons are usually in an open field
configuration thus the dipoles generated by excitatory and inhibitory
currents sum with the same sign along the dendrite (remember that,
by convention, inhibitory currents are positive and excitatory currents
negative, see equation \ref{eq_single_compartment}). \textbf{Right
side}: we summed only currents from synapses of pyramidal neurons
because, due to their approximate open field arrangement, they contribute
to LFP more than interneurons, which instead have a much less regular
dendritic spatial organization. Therefore, the contribution from different
interneurons tend to cancel out each other. (Source: \citep{Mazzoni2011})
\label{fig_campo_aperto}}
\end{figure}
We computed from network activity the LFP by using a procedure that
has been proposed in previous works \citep{mazzoni2008encoding,Mazzoni10}.
More precisely, we computed the simulated LFP as the difference between
the sum of the GABA currents and the sum of the AMPA currents (both
external and recurrent) that enter all excitatory neurons. This quantity
was then divided by the leak membrane conductance to obtain units
of mV:
\begin{equation}
LFP=\frac{1}{G_{leak}}\left(\underset{i_{exc}}{\sum}I_{GABA}^{i}-\underset{i_{exc}}{\sum}I_{AMPAtot}^{i}\right).\label{eq_LFP}
\end{equation}
As explained above in detail, LFPs are experimentally obtained by
low-pass filtering the extracellularly recorded neural signal, and
are thought to reflect to a first approximation the current flow due
to synaptic activity around the tip of the recording electrode \citep{Buzsaki2012}.
The simple recipe in equation \ref{eq_LFP}, we used to model that
current flow, was motivated by two well-known geometrical properties
of cortical circuits (see figure \ref{fig_campo_aperto}). First,
AMPA synapses tend to be apical, i.e., they contact the dendrites
away from the soma, while GABA synapses tend to be peri-somatic, i.e.,
they contact the soma or the dendrites close to the soma. Because
of this spatial arrangement, the sink and sources of the flow of cations
resulting from the activation of both AMPA and GABA synapses will
tend to produce in the extracellular field a dipole oriented from
apical dendrites toward soma; hence we computed the LFP by subtracting\footnote{Remember that, by convention, inhibitory currents are positive and
excitatory currents negative, see equation \ref{eq_single_compartment}.} the AMPA currents from the GABA currents (divided by the leak membrane
conductance). Second, pyramidal neurons contribute more than interneurons
to generation of LFPs in cortex because (i) they are bigger than interneurons
eliciting stronger action potentials, furthermore (ii) their apical
dendrites are organized in an approximate open field configuration
\citep{Johnston95,Logothetis03}, thus the contribution of each pyramidal
neuron sum up to each other (see figure \ref{fig_campo_aperto}).
On the other hand, in the interneurons, due to their star-shaped dendrites
and their geometrical disorder, contributions from each cell are smaller
and tend to cancel out each other \citep{LorentedeNO1947,murakami2006contributions,Linden2011}.
Therefore, we computed LFPs by considering only input currents to
excitatory neurons (taken here to correspond to cortical pyramidal
neurons). \\
Note that this model neglects all the contribution to the LFP not
due to synaptic potentials and does not assume any dependencies of
the contributions from different neurons on the topology of the network
(indeed there are not weights in the summation in equation \ref{eq_LFP}).
Nevertheless, though simple, it proved to be an effective way to generate
a realistic LFP signal that match many characteristics of LFPs in
sensory cortex \citep{Mazzoni10,Mazzoni2011,mazzoni2008encoding}.

\subsection[Procedure to determine comparable CUBNs and COBNs]{Procedure to determine comparable current- and conductance-based
models\label{section_procedure_to_Gsyn}}

As mentioned above all the parameters that were directly shared between
the two models were set equal; also the connectivity matrix was the
same in the CUBN and in the COBN. The starting point of our comparison
was to completely define the CUBN, by specifying the synaptic efficacies,
$J_{syn}$ (reported in table \eqref{tab_Jsyn}), as well as the values
of the common set of parameters. Then, we computed the synaptic parameters
of the COBN that made it comparable to the given CUBN. To simplify
the problem, we first set the reversal potentials of the COBN to biophysically
plausible values: $V_{AMPA}$ = 0 mV and $V_{GABA}$ = \textminus 80
mV (as reference values, but we also tested other values, see\textcolor{black}{{}
figure \ref{fig_6_paperFR}C,D, \ref{fig_7_paperFR}D}). The \textquotedblleft free\textquotedblright{}
parameters now left to set were only the COBN conductances ($G_{syn}$
in equation \ref{eq_6_paperFR}). 
\begin{figure}
\begin{centering}
\includegraphics[scale=0.85]{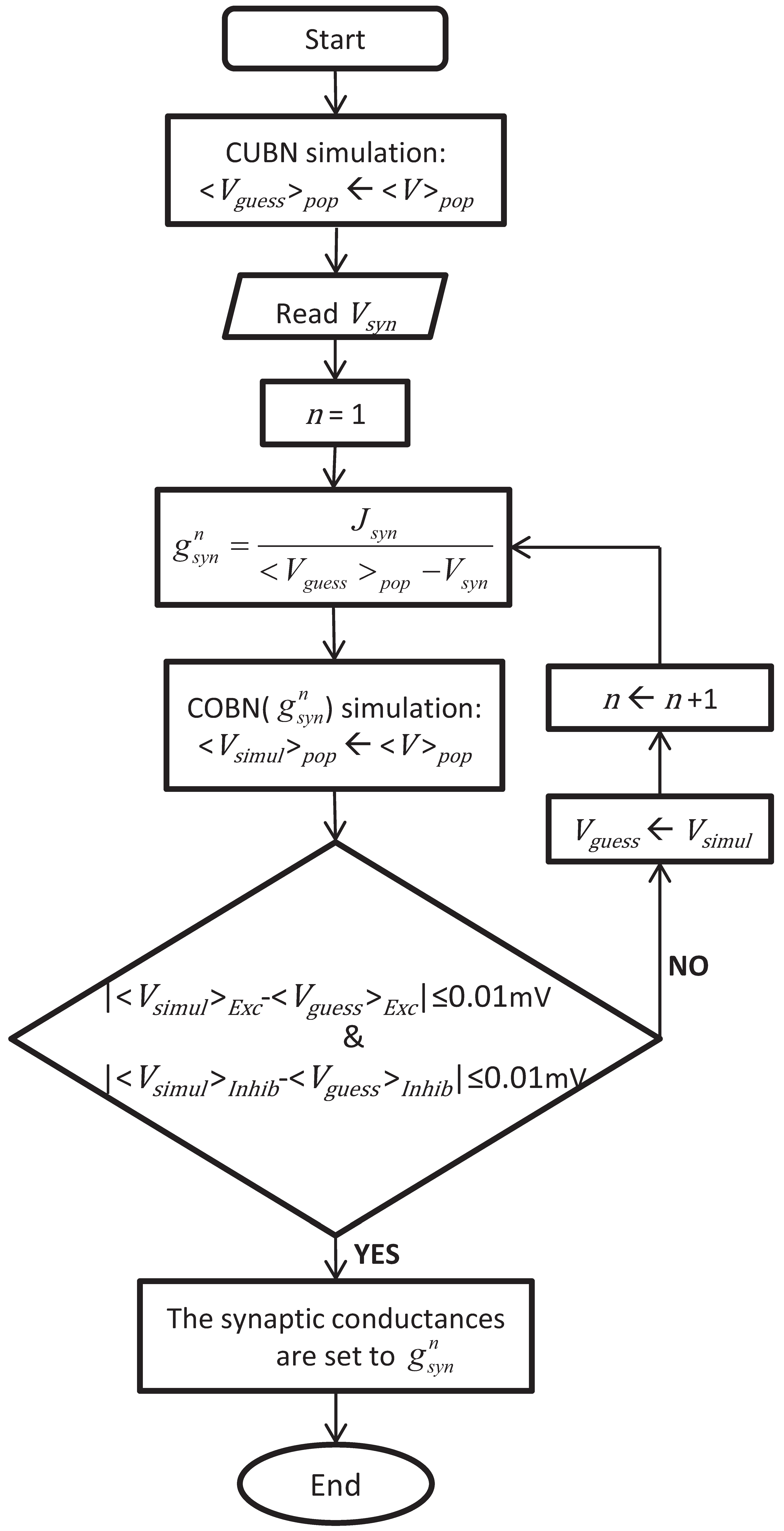}
\par\end{centering}
\raggedright{}\caption[Procedure to set the synaptic conductances of the COBN]{\textbf{Procedure to set the synaptic conductances of the COBN}.
The flowchart illustrates the iterative algorithm we used to set the
synaptic conductances, $G_{syn}$, such in a way to obtain a COBN
comparable with the given CUBN. The two networks shared all the common
parameters, so, once the CUBN was given, the synaptic conductances
depended only on the synaptic reversal potentials of the COBN, $V_{syn}$.\label{fig_procedure_to_Gsyn}}
\end{figure}

The procedure used to obtain the conductance values leading to comparable
COBN and CUBN is illustrated in \textcolor{black}{figure}\textcolor{red}{{}
\ref{fig_procedure_to_Gsyn}} and described in the following. Consistent
with the fact that the effective membrane time constant of the COBN
is equal to the membrane time constant of the CUBN only in absence
of synaptic input (see equation \ref{eq_10_paperFR}), we set the
conductances of each synapse type to obtain the same PSCs as in the
corresponding current-based synapse in the limit of no synaptic activity.
Explicitly, for each synapse type:
\begin{equation}
G_{syn}=\frac{J_{syn}}{(\langle V\rangle_{pop}-V_{syn})},\label{eq_11_paperFR}
\end{equation}
where $\langle V\rangle_{pop}$ was the average (over time and neurons)
MP of excitatory and inhibitory populations obtained from network
simulation of 4.5 s with a constant external input of 1.5 (spikes/ms)/cell.
This last value was chosen because it was the lowest stimulus used
throughout the paper, i.e., the one that induced the lowest synaptic
activity. Since $\langle V\rangle_{pop}$ depended on $G_{syn}$,
we determined both values numerically and recursively. We used as
first guess the average MP obtained with the CUBN, we computed the
associated conductances with equation \ref{eq_11_paperFR}, we ran
a COBN simulation with those conductances and then we used the resulting
$\langle V\rangle_{pop}$ to compute the updated conductances, until
$\langle V\rangle_{pop}$ (and consequently the conductances) reached
a stable value (see \textcolor{black}{figure}\textcolor{red}{{} \ref{fig_procedure_to_Gsyn}}).
Note that convergence was very fast: stability within a tolerance
on average MPs of 0.01 mV was achieved usually in less than 10 steps.
By using equation \ref{eq_11_paperFR}, we rewrote the equation \ref{eq_6_paperFR}
as follows:
\begin{equation}
I_{syn}^{COBN}(t)=J_{syn}s_{syn}(t)\left[1+\frac{V(t)-\langle V\rangle_{pop}}{\langle V\rangle_{pop}-V_{syn}}\right].\label{eq_12_paperFR}
\end{equation}
Comparing equation \ref{eq_12_paperFR} with equation \ref{eq_5_paperFR}
it is clear that the synaptic currents of the two networks are the
same only when $V(t)=\langle V\rangle_{pop}$, that is in the limit
of no synaptic input.

Conductance-based neurons can undergo transitions from low- to high-conductance
states \citep{destexhe2001fluctuating} and the simulations performed
in this work included both states. However, current-based neurons
cannot undergo such transitions and their membrane time constant is
close to the effective membrane time constant of conductance-based
neurons in a low-conductance state (see\textcolor{black}{{} figure \ref{fig_3_paperFR}A}).
Therefore, the correspondence between the two models that we defined
is consistent with the physiologically-meaningful requirement that
the differences between the two synaptic models decrease with synaptic
activity \citep{Destexhe2003}.

\newpage{}

\subsection[Computation of the average PSPs in the COBN]{Computation of the average post-synaptic potentials in the conductance-based
network\label{section_averagePSP}}

Modeling the synaptic input as conductance transients produces an
activity-dependent increase of membrane conductance (that is a reduction
of effective membrane time constant, see equation \ref{eq_10_paperFR})
which attenuates and shortens the Post-Synaptic Potentials (PSPs)
\citep{Destexhe1999}. In order to extract the average (activity-dependent)
PSPs of the COBN we used a procedure similar to the one used in \citep{Kumar2008a}:
for each synapse type (see table \ref{tab_gsyn}) we randomly selected
300 neurons from the network and we made a copy of them. These \textquotedblleft cloned\textquotedblright{}
neurons received the synaptic input of the original ones and had exactly
the same spiking activity. The only difference with respect to the
original is that the cloned neurons received an extra spike, from
the synapse under investigation, each 100 ms (except for the first
500 ms), for a total of 100 PSPs for each cloned neuron (i.e., simulations
lasted 10.5 s). We subtracted then the MP of the original neurons
from the one of the cloned neurons and, by doing a spike triggered
average over time and selected neurons, we obtained the average effective
PSP.

\subsection{Computation of correlations among signals in the networks\label{section_computation_correl}}

We quantified the effects of the choice of the synaptic model on the
cross-neuron correlation in time. We computed the cross-neuron pairwise
Pearson\textquoteright s correlation coefficient of the time course
of AMPA currents and of GABA currents entering the neurons, MPs and
spike trains. The spike trains were binned in non-overlapping time
windows of 5 ms and their correlation coefficients were averaged over
all neuron pairs of the network (\textcolor{black}{figure \ref{fig_10_paperFR}A}-C).
Time courses of the other variables were expressed with the original
time steps of 0.05 ms and the correlation was estimated averaging
the correlation coefficients over all neurons\textquoteright{} pairs
obtained from two randomly selected subpopulations of 200 excitatory
and 200 inhibitory neurons (\textcolor{black}{figure }\ref{fig_9_paperFR}). 

We measured also the average correlation between the time course of
AMPA and GABA currents entering each single-neuron. In particular,
we computed the normalized cross-correlation between AMPA and GABA
currents entering each neuron belonging to the two subpopulations
of 200 neurons above mentioned. Then we averaged (over the neurons)
the peak value and the peak position, i.e., the time lag for which
the correlation was strongest (\textcolor{black}{figure }\ref{fig_8_paperFR}).

\subsection{Computation of information about the external inputs\label{section_Computation-of-information}}

We introduced the notion of mutual information in section \vref{section_Shannon_information}.
Here we only specify some details of the information computation performed
in this context. As explained above, we used three kinds of external
input signals: constant input (figures \ref{fig_2_paperFR}-\ref{fig_11_paperFR}),
periodic input (figures \ref{fig_12_paperFR}, \ref{fig_13_paperFR})
and a naturalistic input (figure \ref{fig_14_paperFR}). In the constant
input case, each input rate, $\nu_{0}$, was considered a different
stimulus (with simulations lasting 25.5 s), while, for the periodic
stimuli, each stimulus corresponds to a frequency f (with simulations
lasting 10.5 s). In the naturalistic case, the stimulus presentation
time (80 s) was divided into 2 s long non-overlapping windows and
each window was considered as a different \textquotedblleft stimulus\textquotedblright{}
for the information calculation, following the procedure described
in \citep{Belitski2008}. We discarded an interval at the beginning
of the simulations (500 ms both for constant and periodic case and
2 s for the naturalistic case) to avoid artifacts due to initial conditions.
When computing information we considered three different response
sets $R$: the average network firing rate, the average cross-neuron
spike train correlation, and the LFP power of each single frequency
\citep{Belitski2008} in the (1\textendash 150) Hz range. To facilitate
the sampling of response probabilities, the whole range of response
values was divided into six consecutive intervals. Each of these intervals
contained the same number of responses (i.e., they were equi-populated).
All the responses belonging to a given interval assumed then the same
interval-specific discrete value. In summary, we discretized the responses
into six equi-populated bins. Then conditional probabilities $P(r|s)$
were evaluated empirically by using the results from 50 trials per
each stimulus $s$. We corrected information estimations for the limited
sampling bias \citep{Panzeri07} by using the \textquotedblleft quadratic
extrapolation procedure\textquotedblright{} described in \citep{strong1998application}
implemented in the Information Breakdown Toolbox \citep{magri2009toolbox}.

\section{Results}

We investigated the differences in the dynamics of neural populations
between conductance-based LIF networks (COBNs) and current-based LIF
networks (CUBNs), with particular emphasis in understanding how the
neural population activity of these two types of network is modulated
by external inputs. We first introduced an iterative procedure to
determine synaptic parameter values so that the CUBN and the COBN
were placed on a fair common ground, and could therefore be legitimately
compared. We then analyzed similarities and differences of single-neuron
dynamics and of interactions among neurons in the two networks as
a function of strength and nature of the external stimuli. 

\subsection[Determining synaptic parameter values to build comparable CUBNs and
COBNs]{Determining synaptic parameter values to build comparable current-
and conductance-based networks}

\begin{figure}
\begin{centering}
\includegraphics[scale=0.8]{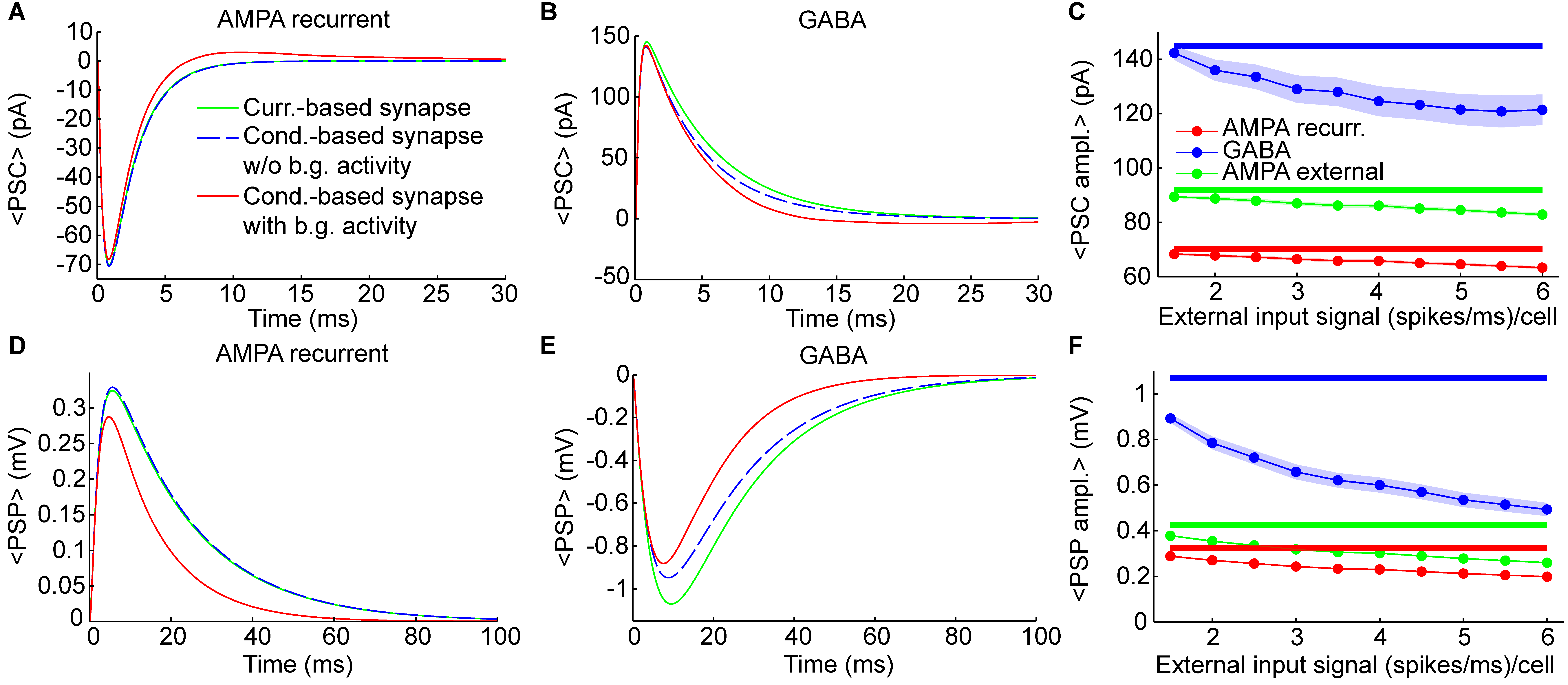}
\par\end{centering}
\centering{}\caption[Individual synaptic events in network models]{\textbf{\textcolor{black}{Individual synaptic events in both models}}.
\textcolor{black}{\small{}Dynamics of single synaptic events on excitatory
neurons (see section \vref{section_averagePSP}). Results were qualitatively
very similar when considering synaptic inputs impinging on inhibitory
neurons (see \textquotedblleft PSP peak amplitude\textquotedblright{}}\textbf{\textcolor{red}{\small{}
}}\textcolor{black}{\small{}in table }\textbf{\textcolor{red}{\small{}\ref{table_suppl_paperFR}}}\textcolor{black}{\small{}).
}\textbf{\textcolor{black}{\small{}(A,B)}}\textcolor{black}{\small{}
Shape of Post-synaptic Currents (PSCs, top) for individual synaptic
events in case of recurrent AMPA (A) and GABA (B) connection (thalamic
AMPA case is not shown because it is qualitatively very similar to
the recurrent AMPA case). The origin of the time axis corresponds
to the arriving time of the spike. Green lines represent the kinetics
in current-based neurons, which is independent from background synaptic
activity. Dashed blue lines indicate the kinetics of an isolated conductance-based
neuron (thus without background activity), having starting membrane
potential equal to $\langle V\rangle_{exc}$ = \textminus 58.8 mV
, that is the average potential of the excitatory neurons of the network
when the external input signal is 1.5 (spikes/ms)/cell. Red lines
indicate the average PSCs in conductance-based neurons embedded in
the network (thus with background activity) when the external input
signal is 1.5 (spikes/ms)/cell (see Methods for details). Blue and
green lines are superimposed in (A). }\textbf{\textcolor{black}{\small{}(C)}}\textcolor{black}{\small{}
Absolute average values of the PSC peaks as a function of the external
input rate for neurons embedded in the network. Results are relative
to recurrent AMPA (red) external AMPA (green), and GABA (blue) synapses
for current- (thick lines) and conductance-based (thin lines with
markers) neurons. Shaded areas for the conductance-based neurons correspond
to the standard deviation across neurons (for AMPA connections the
shaded areas are not visible because they are too small). }\textbf{\textcolor{black}{\small{}(D\textendash F)}}\textcolor{black}{\small{}
Same as (A\textendash C) for Post-Synaptic Potentials (PSPs). PSPs
are more relatively affected by the choice of the synaptic model with
respect to the PSCs, because, in the COBN, the PSCs depend on the
driving force, while the PSPs both on the driving force and on the
effective membrane time constant.}\label{fig_2_paperFR}}
\end{figure}
A necessary requirement to compare the activity of two different network
models is to define a meaningful and sound correspondence between
them. Our first step was thus to define a procedure to achieve comparable
networks. In brief, we set all the common parameters to exactly equal
(and biologically plausible) values in both models. In this way the
two models differed only because of the different synaptic model adopted:
voltage-independent for CUBN (see equation \ref{eq_5_paperFR}) and
voltage-dependent for COBN (see equation \ref{eq_6_paperFR}). In
particular, the expression of the Post-Synaptic Currents (PSCs) in
the COBN depended on conductances $G_{syn}$ and on reversal potentials
($V_{AMPA}$ and $V_{GABA}$), while in the CUBN the PSCs depended
only on synaptic efficacies $J_{syn}$. We set $V_{AMPA}$ and $V_{GABA}$
at 0 and \textminus 80 mV respectively (but importantly our results
were robust to changes in these parameters, see fi\textcolor{black}{gures
\ref{fig_6_paperFR}C,D, \ref{fig_7_paperFR}D)}. We then used an
iterative algorithm (detailed in section \vref{section_procedure_to_Gsyn}
and illustrated in figure \ref{fig_procedure_to_Gsyn}) to set the
values of the conductances $G_{syn}$ of the COBN in such a way to
obtain a COBN comparable to the CUBN with the given synaptic efficacies
$J_{syn}$.

The PSCs and the Post-Synaptic Potentials (PSPs) of recurrent AMPA
and GABA synapses in the comparable networks are shown in figures
\ref{fig_2_paperFR}A,B,D,E for three different cases: current-based
synapse, conductance-based synapse of a single neuron without background
synaptic activity and conductance-based synapse of neurons embedded
in the COBN network (that thus received background synaptic activity).
The post-synaptic kinetics of conductance-based neurons is activity
dependent. The terms that mediate this dependency are: the driving
force (see equation \ref{eq_6_paperFR}) and the increase of the total
effective membrane conductance (see equation \ref{eq_8_paperFR}).
Both these terms tend to reduce the post-synaptic stimulus, but the
PSCs are affected only by the driving force, while the PSPs by both
the driving force and the effective membrane conductance. To understand
how these two terms shape the post-synaptic stimulus, it is important
to compare post-synaptic responses of conductance-based neurons, with
and without background activity. Firstly, we compared PSCs and PSPs
of the current-based synapse with those of the conductance-based synapse
in the absence of background activity. In this condition the shape
of excitatory PSCs and PSPs was almost identical for the two models
when considering AMPA synapses (figures \ref{fig_2_paperFR}A,D),
while, for GABA synapses, differences between the two models were
visible (figures \ref{fig_2_paperFR}A,D). This asymmetry was due
to the fact that the value of the average MP (see figure caption)
was much closer to the reversal potential of GABA synapses than to
the one of AMPA synapses (see equation \ref{eq_12_paperFR}). Consequently
the relative reduction of driving force during the post-synaptic event
was higher for GABA synapses, provoking a stronger reduction of both
PSCs and PSPs, with respect to the AMPA synapses (figures \ref{fig_2_paperFR}B,E).
Moreover, the PSPs of fast synapses (that is synapses with short $\tau_{d}$)
are less affected by synaptic bombardment \citep{Koch1999,Kuhn2004},
so, being the AMPA $\tau_{d}$ shorter than the GABA ones (see table
\ref{tab_tau_syn}), the asymmetry was even stronger when looking
at the PSPs (figures \ref{fig_2_paperFR}D,E). Secondly, we considered
the conductance-based neurons embedded in the COBN and we found that
in this case both AMPA and GABA synapses displayed a reduction in
the amplitude and in the timescale, because the background network
activity affected the time course of the MP (thus of the driving force)
and increased the total effective membrane conductance. 

As stated above, differences between the two synaptic models were
expected to increase with input strength because the background synaptic
activity increases. We measured this effect by injecting in the network
constant inputs ranging from 1.5 to 6 (spikes/ms)/cell. Figures \ref{fig_2_paperFR}C,F
show the amplitude of the different PSCs and PSPs as a function of
the external input rate. Note that the PSCs (figure \ref{fig_2_paperFR}C)
and PSPs (figure \ref{fig_2_paperFR}F) in the CUBN were activity-independent
by construction, while, in the COBN, both PSCs and PSPs decreased
substantially when input rate was increased; furthermore the relative
reduction was the strongest for the slowest PSPs of GABA synapses
(as stated above). Table \ref{table_suppl_paperFR} reports average
PSP amplitude values on both inhibitory and excitatory neurons.
\begin{figure}
\begin{centering}
\includegraphics[scale=0.8]{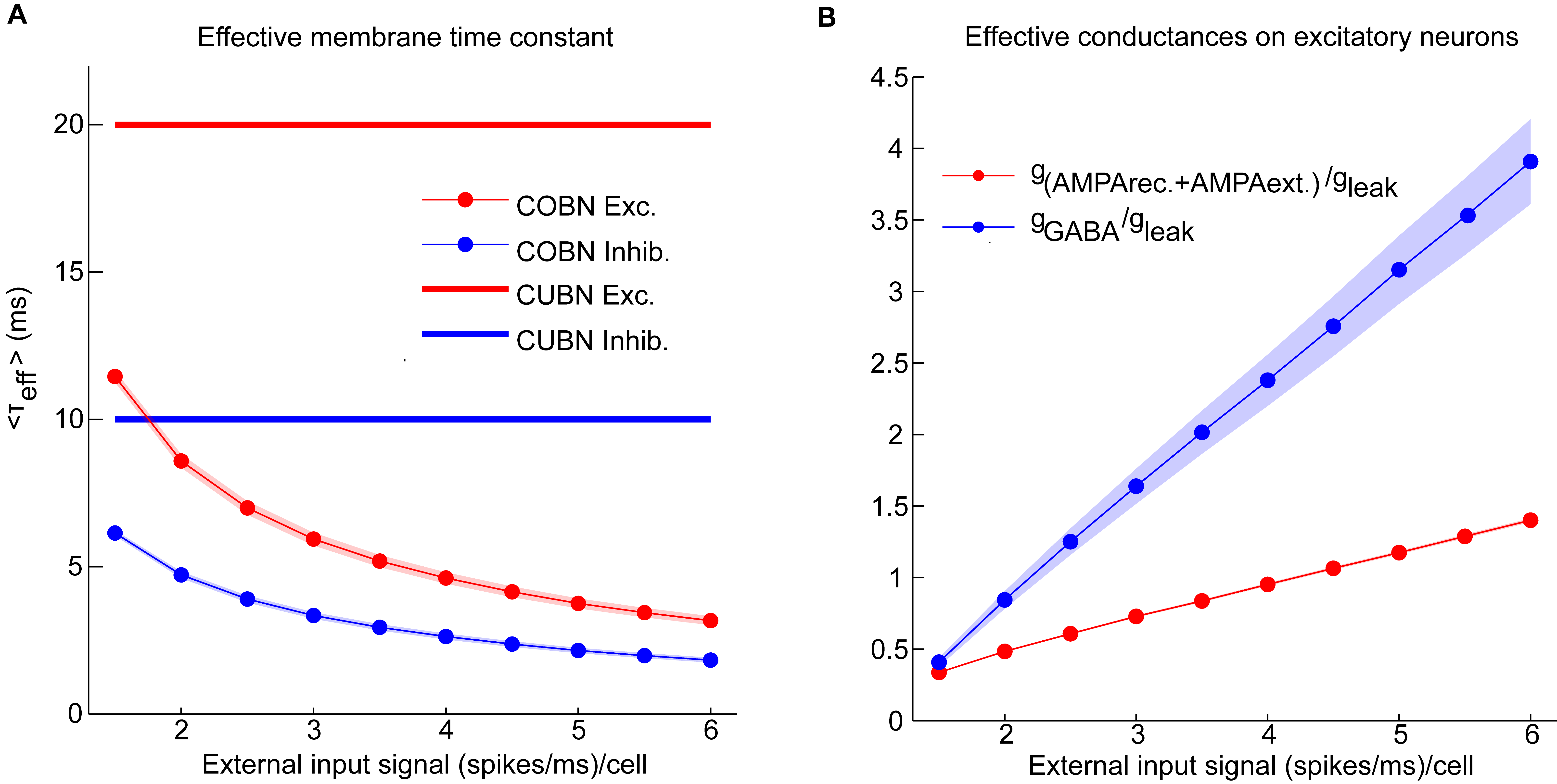}
\par\end{centering}
\begin{centering}
\caption[Effective parameters in conductance-based networks]{\textbf{\textcolor{black}{\small{}}}\textbf{\textcolor{black}{Effective
parameters in conductance-based networks}}\textcolor{black}{. Input
rate modulations of COBN-specific parameters. }\textbf{\textcolor{black}{(A)}}\textcolor{black}{{}
Average effective membrane time constant for conductance-based excitatory
neurons (red markers) and inhibitory neurons (blue markers) as a function
of the external input rate. Membrane time constants of the current-based
neurons are shown for reference as thick lines. Results show that
conductance-based membrane timescale is much faster than current-based
one and that it decreases with input strength. }\textbf{\textcolor{black}{(B)}}\textcolor{black}{{}
Average effective AMPA (red) and GABA (blue) conductances on excitatory
neurons as a function of the external input rate. Results show that
the COBN goes from low- to high-conductance states in the range of
external stimuli considered. Same color code as (A). Shaded areas
represent standard deviation across neurons {[}in (A) for inhibitory
time constant and in (B) for AMPA conductances they are not visible
because too small{]}. Values are computed from a simulation of 10.5
s per stimulus and are averaged over time and neurons.}\textcolor{black}{\small{}\label{fig_3_paperFR}}}
\par\end{centering}
\centering{}
\end{figure}

Figure \ref{fig_2_paperFR} shows that, in the COBN, PSPs were not
only smaller but also faster than in the CUBN, consistently with previous
results \citep{Kuhn2004,Meffin2004}. This reflected the decrease
of the effective membrane time constant, $\tau_{eff}$, of the COBN,
whose average value is shown in figure \ref{fig_3_paperFR}A as a
function of the input rate. When injecting stimuli with high input
rates, we found that for both neuron populations the effective time
constant, $\tau_{eff}$, was in the 1\textendash 5 ms range, matching
experimental observations relative to the high-conductance states
\citep{Destexhe2003}. 

We then asked how the effective conductances associated with the AMPA
and GABA currents varied in the COBN as a function of the input rate.
We found (figure \ref{fig_3_paperFR}B) that the average conductances
grew linearly with input rate, as observed in single-neuron case\citep{Kuhn2004}.
Crucially, for high input rates, the relative conductances $G_{AMPA}/G_{leak}$
and $G_{GABA}/G_{leak}$ displayed values respectively close to 1
and 3.5, in the range of those found experimentally in high-conductance
states \citep{Destexhe2003}. This suggested that our input range
was suited to investigate the whole continuum going from low- to high-conductance
states.

\subsection{Average single-neuron properties}

\begin{figure}
\begin{centering}
\includegraphics[scale=0.8]{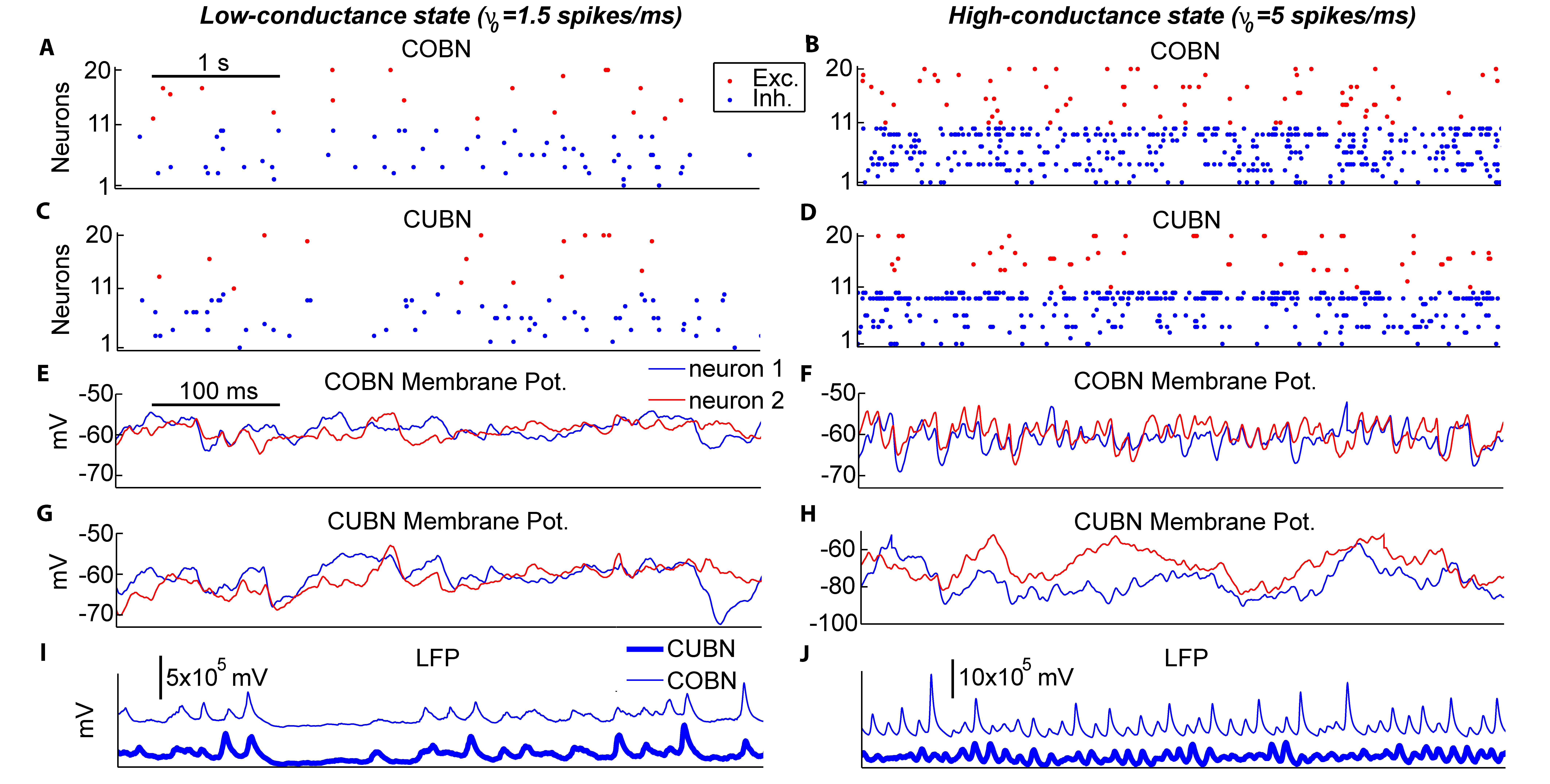}
\par\end{centering}
\centering{}\caption[Examples traces]{\textbf{\textcolor{black}{Example traces}}. Examples of 5 s (A\textendash D)
and 500 ms (E\textendash J) of data traces generated by the two networks
when using constant stimuli. The left column shows the activity in
response to an input rate $\nu_{0}$ set to 1.5 spikes/ms generating
a low-conductance state. The right column shows the activity in response
to an input rate $\nu_{0}$ set to 5 spikes/ms generating a high-conductance
state. \textbf{\textcolor{black}{(A\textendash D)}} Raster plot of
10 excitatory and 10 inhibitory neurons taken from the COBN (A,B)
and from the CUBN (C,D). The selected neurons and the color code are
the same across panels (A\textendash D). \textbf{\textcolor{black}{(E\textendash H)}}
Membrane potential of two neurons taken from the COBN (E,F) and from
the CUBN (G,H). The neurons displayed and the color code are the same
across the panels (E\textendash H). \textbf{\textcolor{black}{(I,J)}}
Simulated LFP obtained from the COBN (thin line) and from the CUBN
(thick line).\label{fig_4_paperFR}}
\end{figure}
After having examined the properties of PSPs and conductances in the
two comparable networks, we began investigating how these properties
affect the dynamics of neural activity in the networks. To gain some
visual intuition about this, we plotted (figure \ref{fig_4_paperFR})
example traces of how variables reflecting single-neuron and network
activity evolve over time for the two types of network both in the
low- and high-conductance state. The overall spike rate of individual
neurons was similar for the two networks in both low- and high-conductance
state (compare panels \ref{fig_4_paperFR}A with \ref{fig_4_paperFR}C
and panels \ref{fig_4_paperFR}B with \ref{fig_4_paperFR}D) suggesting
that the level of network firing was only mildly dependent on the
synaptic model adopted. On the other hand, single-neuron MP traces
were similar in the two networks in the low-conductance regime (compare
panels \ref{fig_4_paperFR}E with \ref{fig_4_paperFR}G), but different
in many aspects in the high-conductance regime (compare panels \ref{fig_4_paperFR}F
with \ref{fig_4_paperFR}H). In particular, in the high-conductance
state, the COBN MPs had rapid gamma-range variations which were correlated
across neurons and whose amplitude was more prominent than that of
the gamma oscillations in the CUBN MPs, suggesting that the oscillation
regime in the high-conductance state was tighter in the COBN than
in the CUBN. Finally, we considered the traces of the LFP (which can
potentially capture both supra- and sub- threshold massed neural dynamics).
LFP traces were relatively similar across networks in the low-conductance
state (figure \ref{fig_4_paperFR}I). However, there was an interesting
qualitative difference in the LFP traces in the high-conductance state:
the COBN LFP had transient peaks of very high amplitude, which were
not observed in the CUBN. At fixed level of overall firing rate, the
amplitude of the LFP is modulated by the relative timing of the synaptic
events contributing to it. Therefore this observation suggests that
the COBN may undergo larger fluctuations in synchronization than the
CUBN. The visual inspection of example traces suggests that, while
some network properties such as overall firing rate are consistently
close in the two networks, other more subtle aspects of network dynamics
(such as the ability of the network to transiently synchronize its
activity) may not be entirely equivalent in the two networks, especially
in the high-conductance state. In the following we will systematically
quantify this intuition. 
\begin{figure}
\begin{centering}
\includegraphics[scale=0.75]{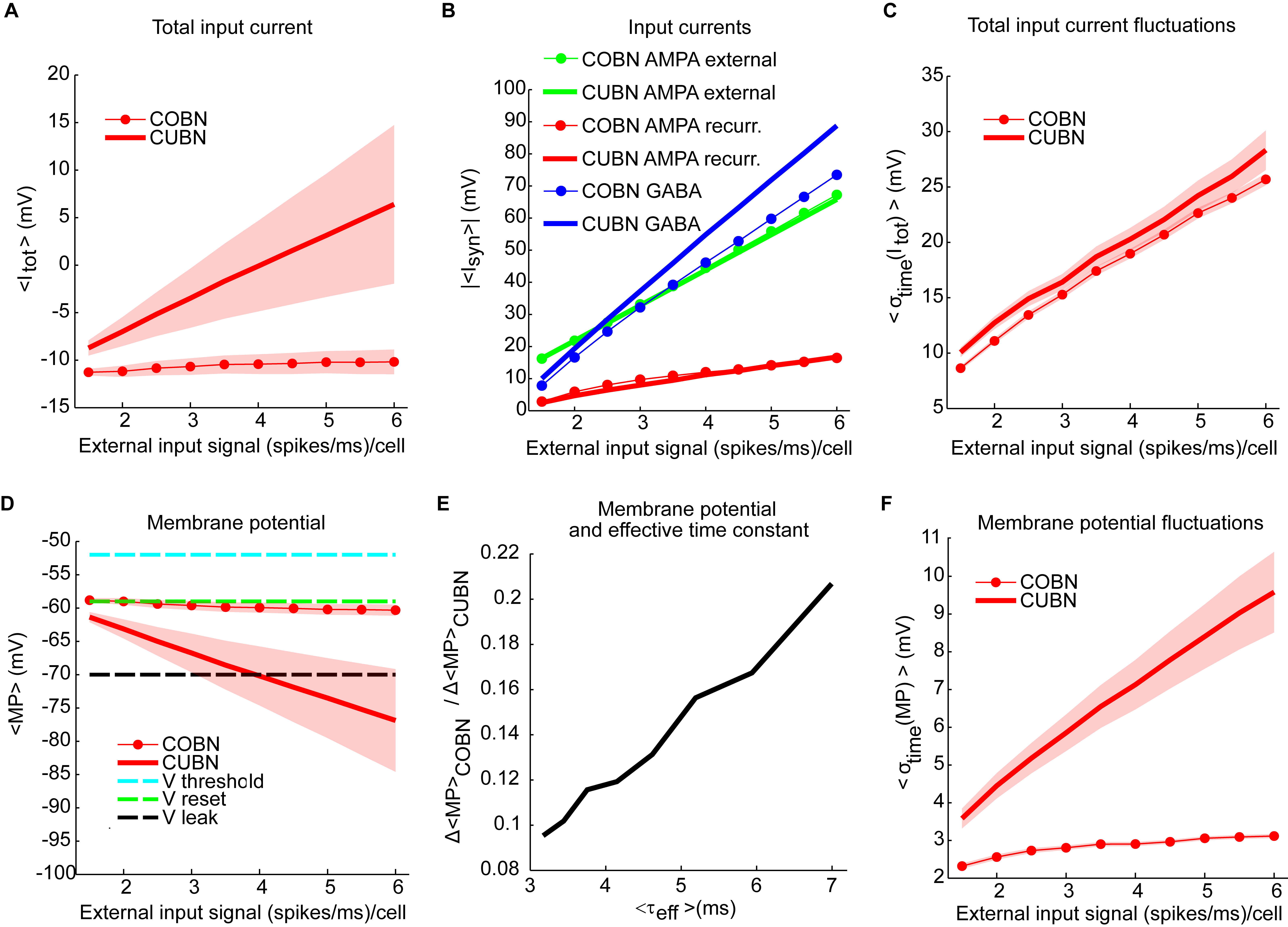}
\par\end{centering}
\centering{}\caption[MP and synaptic currents as a function of the external input]{\textbf{\textcolor{black}{}}\textbf{\textcolor{black}{\scriptsize{}MP
and synaptic currents as a function of the external input rate}}\textcolor{black}{\scriptsize{}.
Effects of external input rate modulation on the net synaptic input
currents and the membrane potential of excitatory neurons. The synaptic
currents in panels (A\textendash C) are divided by the leak membrane
conductance to obtain units of mV . Results are qualitatively very
similar when considering inhibitory neurons {[}see \textquotedblleft MP\textquotedblright{}
and \textquotedblleft $\sigma_{time}$(MP)\textquotedblright{} in
table }\textbf{\textcolor{red}{\scriptsize{}\ref{table_suppl_paperFR}}}\textcolor{black}{\scriptsize{}{]}.
We studied separately the average over time and the standard deviation
over time of the variables by using a simulation of 10.5 s per stimulus.
Shaded areas correspond to standard deviation across neurons. }\textbf{\textcolor{black}{\scriptsize{}(A)}}\textcolor{black}{\scriptsize{}
Average total synaptic input current in CUBN (thick line) and COBN
(thin line with markers) as a function of the external input rate.
}\textbf{\textcolor{black}{\scriptsize{}(B)}}\textcolor{black}{\scriptsize{}
Different input currents in the two networks. Blue/red/green lines
represent respectively the average GABA/recurrent AMPA/external AMPA
currents in CUBN (thick lines) and in COBN (thin lines with markers).
}\textbf{\textcolor{black}{\scriptsize{}(C)}}\textcolor{black}{\scriptsize{}
Average (over neurons) standard deviation in time of the total input
current in the two networks as a function of the input rate. }\textbf{\textcolor{black}{\scriptsize{}(D)}}\textcolor{black}{\scriptsize{}
Average membrane potential in the two networks as a function of the
external input rate. For reference, the panel shows also threshold
potential (cyan), reset potential (green) and leak membrane potential
(black). }\textbf{\textcolor{black}{\scriptsize{}(E)}}\textcolor{black}{\scriptsize{}
Ratio of the decrease of the average MP observed in the two networks
when increasing the external inputs as a function of the effective
membrane time constant (see figure \ref{fig_3_paperFR}A). The decrease
in MP is computed for external inputs greater than 2 (spikes/ms)/cell
with respect to the average MP obtained with an external input of
2 (spikes/ms)/cell. }\textbf{\textcolor{black}{\scriptsize{}(F)}}\textcolor{black}{\scriptsize{}
Average (across neurons) standard deviation over time of the membrane
potential in the two networks as a function of the input rate. Shaded
area for COBN is not visible because it is too small. Results show
that for the COBN both average total input current and membrane potential
are almost constant across stimuli, while in the CUBN both quantities
change dramatically for different input strengths. Cross-neuron variability
of both variables is much higher in the CUBN. In both networks net
input current fluctuations become larger when input rate is increased.
This is reflected in larger fluctuations in the membrane potential
in the CUBN, but not in the COBN. In panels (A,B,D,E) the average
values of MP and input currents are computed over time and neurons.}{\footnotesize{}\label{fig_5_paperFR}}}
\end{figure}

An important feature of the models is the dynamics of the average
(over time and neurons) of the total synaptic input current $I_{tot}$
(equation \ref{eq_4_paperFR}). We observed in both networks (figure
\ref{fig_5_paperFR}A) an increase of $I_{tot}$ with the input rate
(Pearson correlation test, $p<10^{-5}$). However, $I_{tot}$ was
significantly higher for the CUBN over all inspected inputs (t-test
$p\ll10^{-10}$). The net input current $I_{tot}$ was also less modulated
by the input rate in the COBN: the difference between the current
(divided by the leak membrane conductance) at maximum and minimum
input was 1 mV for COBN and 15 mV for CUBN. Even if the firing rate
was very similar in the two networks (see figure \ref{fig_6_paperFR}A),
average GABA currents were weaker in COBN, while average AMPA currents
were very similar (see figure \ref{fig_5_paperFR}B). This discrepancy
in the dynamics of the net input current was due to the fact that
individual PSCs of GABA currents were more affected (i.e., reduced)
by the change from CUBN to COBN with respect to the AMPA PSCs, as
pointed out in figure \ref{fig_2_paperFR}. Note also that in the
case of external AMPA current, the spike trains that activated the
synapses (more precisely the function $s(t)$ in equations \ref{eq_5_paperFR}
and \ref{eq_6_paperFR}) are exactly the same in the two models, while
they were different for the other currents. 

Consistent with the sample traces shown in figures \ref{fig_4_paperFR}G,H,
the average MP of the CUBN decreased steeply when we increased the
input (\textminus 15 mV between maximum and minimum input, figure
\ref{fig_5_paperFR}D). This is due to the fact that, in the CUBN,
the net input current strongly increased when increasing the external
inputs (figure \ref{fig_5_paperFR}A). Conversely, and consistently
with the sample traces in figures \ref{fig_4_paperFR}E,F, the decrease
in COBN MP was smaller (\textminus 2 mV between maximum and minimum
input, figure \ref{fig_5_paperFR}D), consistent with previous results
\citep{Meffin2004}. It is important to note that an increase of the
input current led to an increase the voltage fluctuations in both
models. However in the COBN, it caused also a concomitant increase
of the membrane conductance, which in turn decreased the membrane
voltage fluctuations. The dynamics of MP in COBN thus resulted from
the competition between these two effects, which overall produced
a suppression of both fluctuations and mean of the MP \citep{Meffin2004,Kuhn2004,Richardson2004}.
We found that, for external inputs higher than 2 (spikes/ms)/cell,
there was a linear relation ($R^{2}$ = 0.98, $p\ll10^{-10}$) between
the ratio of the average MP changes induced by the external inputs
in the two networks and the effective membrane time constant of the
COBN (see figure \ref{fig_5_paperFR}E). This result confirmed and
extended what found for a single-neuron model in a high-conductance
state in \citep{Richardson2004}. Shaded areas in figures \ref{fig_5_paperFR}A,D
indicate standard deviation across neurons, and show that the cross-neuron
variability in both net input currents and MP was much larger in the
CUBN than in the COBN, suggesting a more coherent activity for the
latter (see section \textcolor{red}{\vref{section_Cross-neuron-correlations}}). 

When we looked at the variability over time of the input currents,
we found that it grew almost linearly and with very similar values
for both COBN and CUBN (figure \ref{fig_5_paperFR}C), while the increase
of the variability over time of the MP was much more pronounced in
the CUBN than in the COBN (figure \ref{fig_5_paperFR}F). This result
is still consistent with the suppression of voltage fluctuations typical
of conductance-based model with respect to the current-based one. 

In sum, our findings so far confirmed that dynamics previously observed
in simpler conditions were valid also over a more extended range of
conditions, proved that the range of input rates considered encompassed
both low- and high-conductance regimes, and highlighted some of the
differences between the dynamics of COBNs and CUBNs.

\subsection{Firing rate modulations}

Having established a procedure that computes comparable CUBN and COBN
parameters, and having investigated the synaptic responses in these
comparable networks, we next compared the average firing rates of
single neurons in the two networks, and studied how they are modulated
by the strength of the input to the networks. 
\begin{figure}
\begin{centering}
\includegraphics[scale=0.8]{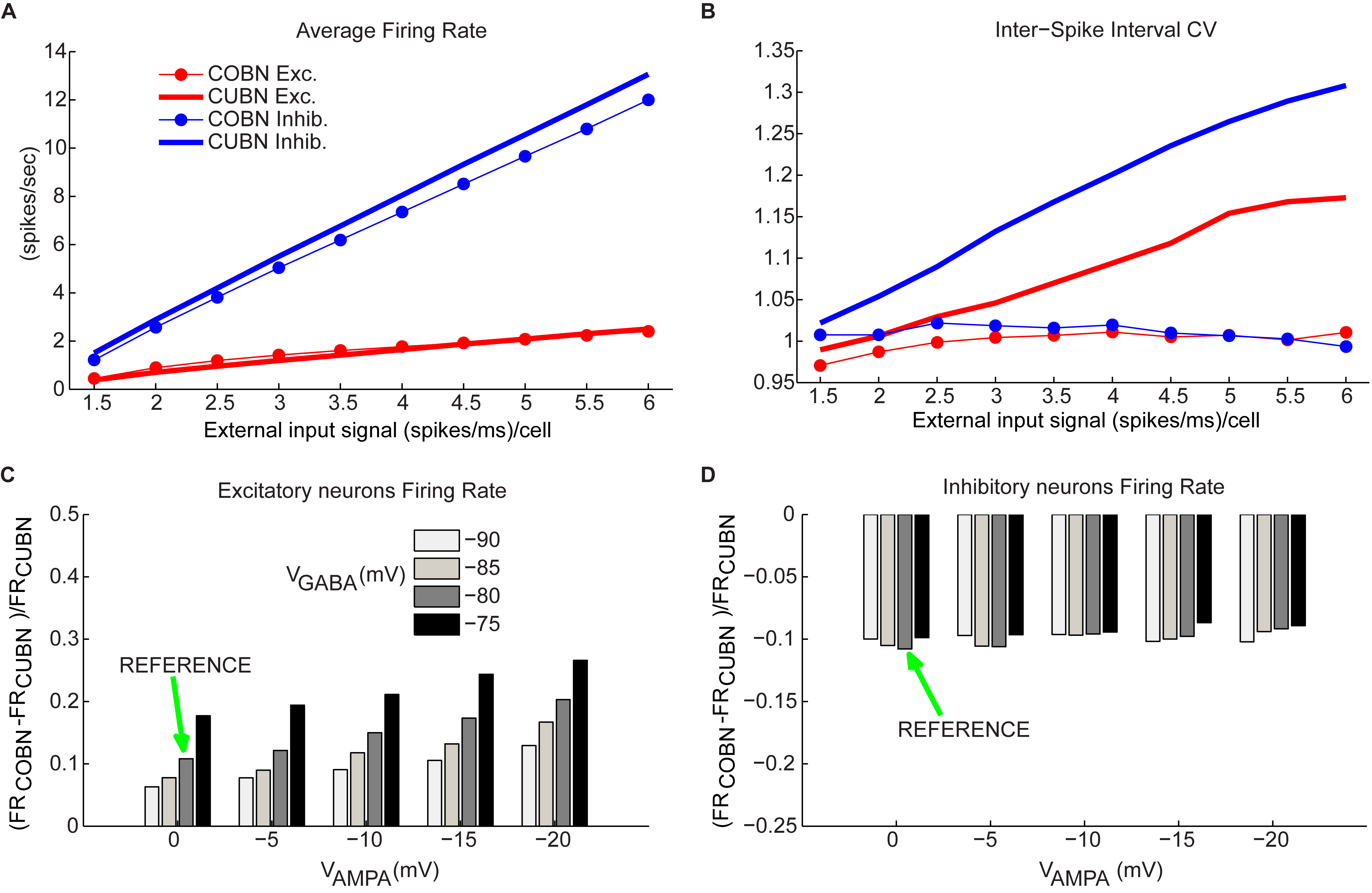}
\par\end{centering}
\centering{}\caption[Firing rates in comparable COBNs and CUBNs]{\textbf{\textcolor{black}{Firing rates comparison}}.\textbf{\textcolor{black}{{}
(A) }}Comparison between average firing rate (FR) of inhibitory (blue)
and excitatory neurons (red) for COBN (thin lines with markers) and
CUBN (thick lines) as a function of the external input rate.\textbf{\textcolor{black}{{}
(B)}} Average Coefficient of Variation of the Inter-Spike Interval
in the two networks. Same color code as (A). \textbf{\textcolor{black}{(C)}}
Relative difference between the average FR of excitatory neurons in
COBN and CUBN computed for different AMPA and GABA reversal potentials.
The relative difference is averaged over the whole stimuli set ranging
from 1.5 to 6 (spikes/ms)/cell. Green arrow indicates reference value
of reversal potentials that were used in all the analysis (see table
\ref{tab_gsyn}). \textbf{\textcolor{black}{(D)}} Same as (C) for
inhibitory neurons. In (A,C,D) the results are obtained from 50 trials
of 4.5 s per stimulus, while for the panel (B) we used a single trial
of 100.5 s per stimulus (see section \vref{section_numerical_methods}).
Results show that the two models have similar firing rates over the
whole input range. This agreement is stable over a wide range of network
parameters. On the other hand, the CV of the ISI increases with the
input rate in the CUBN, while it does not in the COBN.\label{fig_6_paperFR}}
\end{figure}

We considered individually the excitatory and inhibitory neural populations
since they fired at very different rates \citep{Brunel03}. Figure
\ref{fig_6_paperFR}A shows the way inhibitory and excitatory firing
rates increase with the input rate in the two networks. Consistently
with the qualitatively intuition gained form the visual inspection
of the raster plots in figure \ref{fig_4_paperFR}A\textendash D,
we found that the discrepancies between COBN and CUBN firing rates
were extremely small (average difference over external inputs of 10\%),
though significant (t-test $p<0.05$ except for excitatory neurons
with external input rates greater than 4 spikes/ms). This shows that
the algorithm used to set comparable networks produces networks whose
neurons have similar average firing rates with a similar dependence
on the input strength, both in low- and high-conductance states.

To verify if the agreement of the firing rate in the two comparable
networks was robustly achieved over a wide range of parameters, we
computed the COBN synaptic conductances for a set of 20 different
COBN networks (obtained by using the setting procedure illustrated
in figure \ref{fig_procedure_to_Gsyn} with 20 different combinations
of the synaptic reversal potentials, $V_{AMPA}$, ranging from 0 to
\textminus 20 mV , and $V_{GABA}$, ranging from \textminus 75 to
\textminus 90 mV). We then computed the average firing rates for each
resulting network. We found that even when $V_{AMPA}$ was \textminus 20
mV and $V_{GABA}$ \textminus 75 mV , and hence the discrepancies
between the two models were stronger, the excitatory neurons firing
rate differed between COBN and CUBN at most by 25\%, but usually the
difference was much smaller, on the order of 10\% (figure \ref{fig_6_paperFR}C).
Note that, given the very low firing rate of excitatory neurons, the
relative difference corresponded always to small values of absolute
difference ($<0.4$ spikes/ms). The difference in the firing rate
of the inhibitory neurons between COBN and CUBN were of the order
of 10\% for all reversal potentials combinations inspected (figure
\ref{fig_6_paperFR}D). 

These results show that our procedure determines COBNs with firing
rates similar to the compared CUBN for a wide range of parameters.
In current-based neurons the firing rate is modulated only by the
increase in the MP fluctuations (figure \ref{fig_5_paperFR}F), while
in conductance-based neurons, the firing rate activity is the result
of two different competing effects: the shortening of the timescales
(figure \ref{fig_3_paperFR}A) and the increase of the membrane fluctuations
(figure \ref{fig_5_paperFR}F), that tend to facilitate the firing
activity, and the increase of the effective membrane conductance,
that acts in the opposite direction (figure \ref{fig_3_paperFR}B)
\citep{Kuhn2004,Meffin2004,Richardson2004}. It is therefore quite
interesting that these underlying different dynamics compensate to
produce, in the two corresponding network models, very similar firing
rates over a wide range of inputs and parameters.

We then considered how the coefficient of variation (CV) of the inter-spike
interval (ISI) changed with the strength of the input rate. We found
(figure \ref{fig_6_paperFR}B) that the two networks showed a very
different dependence of CV on input rates. The ISI CV of neurons of
the COBN was close to one for all considered input rates (indicating
near-Poisson firing statistics). In contrast, in CUBN, the ISI CV
was higher than 1 (i.e., the firing was more variable than that of
a Poisson process) and increased with the input rate, reaching values
up to 1.33 and 1.16 for inhibitory neurons and excitatory neurons
respectively, confirming results of \citep{Meffin2004}. The difference
between the CVs of neurons in COBN and CUBN was highly significant
(t-test, $p<10^{-7}$) for all input rates above 1.5 spikes/ms. The
larger ISI CV of neurons in COBN was consistent with our finding of
larger MP fluctuations in time in the COBN (figure \ref{fig_5_paperFR}F).
ISI CV values were within the experimentally observed range 0.5\textendash 1.5
for both networks, but only the COBN reproduced the experimental result
that the ISI CV of cortical neurons is not affected by the firing
rate \citep{Maimon2009}. 

The discrepancy between the similarity of the firing rates and the
dissimilarity of the ISI CVs suggests that the first order statistics
of the two networks were close to match, but the second order statistics
differed significantly.

\subsection{Spectral modulations in simulated LFPs}

We investigated then the differences in the spectral modulations of
network activity, as measured by the simulated LFP and by the total
excitatory and inhibitory firing rate generated by the two networks.
LFP models can offer interesting insights into the dynamics of cortical
networks \citep{Einevoll2013} because they offer an insight in both
supra- and sub-threshold dynamics that can be compared with experimental
recordings; however the differences in LFPs computed from networks
with either current- or conductance-based synapses have not been investigated
yet. We expected significant differences to arise because, as detailed
above, the sub-threshold dynamics of COBNs and CUBNs were quite different.
\begin{figure}
\begin{centering}
\includegraphics[scale=0.8]{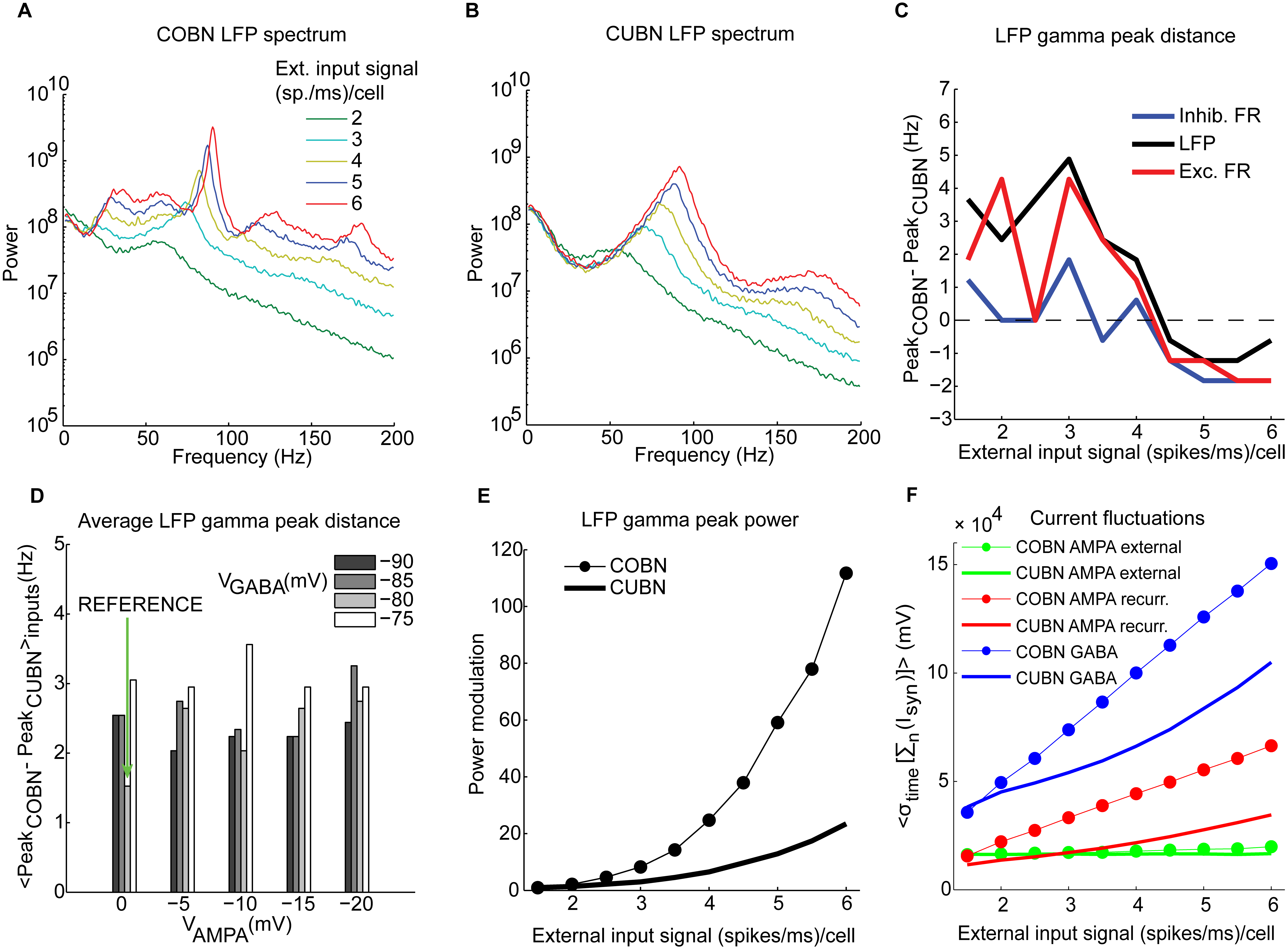}
\par\end{centering}
\centering{}\caption[Spectral dynamics of LFP and firing rate]{\textbf{\textcolor{black}{}}\textbf{\textcolor{black}{\footnotesize{}Spectral
dynamics of LFP and firing rate}}{\footnotesize{}. Input rate-dependent
modulations of the LFP , studied focusing on position and amplitude
of the gamma frequency peak. }\textbf{\textcolor{black}{\footnotesize{}(A)}}{\footnotesize{}
LFP power spectra in COBN as a function of the external input rate.
Data are averaged over trials. }\textbf{\textcolor{black}{\footnotesize{}(B)}}{\footnotesize{}
Same as (A) for CUBN.}\textbf{\textcolor{black}{\footnotesize{} (C)}}{\footnotesize{}
Difference in the position of the gamma band {[}(30\textendash 100
Hz){]} peak of the power between the two networks. The analysis was
performed for the LFP signal (black), and for the total firing rate
of excitatory (red) and inhibitory neurons (blue). }\textbf{\textcolor{black}{\footnotesize{}(D)}}{\footnotesize{}
Difference in the position of the LFP gamma peak averaged over the
constant external inputs used (ranging from 1.5 to 6 (spikes/ms)/cell
with steps of 0.5 (spikes/ms)/cell) as a function of AMPA and GABA
reversal potentials. Green arrow indicates reference values (see table
\ref{tab_gsyn}). }\textbf{\textcolor{black}{\footnotesize{}(E)}}{\footnotesize{}
Modulation of the LFP gamma peak power for the two networks. Power
modulation is defined as the difference of the power of a frequency
at a given input signal and its power at the input signal of 1.5 (spikes/ms)/cell,
normalized to the latter power. }\textbf{\textcolor{black}{\footnotesize{}(F)}}{\footnotesize{}
Average (over trials) amplitude of the fluctuations of the sum of
the currents entering the excitatory neurons for the two networks
as a function of the input rate. The currents are divided by the leak
membrane conductance to obtain units of mV. Blue, red, and green lines
represent GABA, recurrent AMPA and external AMPA respectively. These
are the currents we used to compute LFP. Note that the external AMPA
currents ($I_{AMPAext}$ in equation \ref{eq_4_paperFR}) are almost
identical between the two networks because their synapses are activated
by the same spike trains in COBN and CUBN. Results are computed by
using 50 trials of 4.5 s per stimulus and show that (i) the gamma
peak position across stimuli is similar for the two networks and this
agreement is robust to change in the network parameters, (ii) the
amplitude of the peak power is more modulated in the COBN because
of the stronger fluctuations of the synaptic currents at the network
level.}\label{fig_7_paperFR}}
\end{figure}

The dependence of LFP spectrum on the input rate (figures \ref{fig_7_paperFR}A,B)
shows that, consistent with previous results \citep{Brunel03,mazzoni2008encoding,Mazzoni2011},
both networks develops gamma range (30\textendash 100 Hz) oscillations
that become stronger and faster as the input is increased. Figures
\ref{fig_4_paperFR}I,J illustrate this effect in the time domain.
Figures \ref{fig_7_paperFR}A,B show the LFP input rate-driven modulation
in COBN and CUBN. The dependence of response to variations in input
rate in the two networks was qualitatively similar. There was no modulation
for frequencies below 5 Hz (Pearson correlation test, $p>0.1$); there
was strong modulation in the gamma band and above (Pearson correlation
test, $p<0.01$). The difference between the position of the COBN
and CUBN gamma peak was always below 5 Hz (figure \ref{fig_7_paperFR}C).
For comparison, we also computed the power spectrum of the total firing
rate of excitatory or inhibitory neurons (figure \ref{fig_7_paperFR}C).
The spectral peaks of COBN and CUBN were very close also in this case. 

We tested the robustness of the agreement between spectral peaks of
CUBNs and COBNs by measuring the average (over stimuli) gamma-peak
distance between the two networks for different AMPA and GABA reversal
potentials (similarly to what was done in the analysis represented
in figures \ref{fig_6_paperFR}C,D), and we found that the two networks
always displayed almost identical positions of the gamma frequency
peaks (figure \ref{fig_7_paperFR}D). 

Note that we did not build the comparable networks to obtain robustly
similar firing rates and similar dominant frequencies in the gamma
band, as we used other constraints to select comparable parameters.
The equivalence and robustness of rates and gamma peaks arose from
network dynamics, and, in particular, the robustness corroborates
the notion that our procedure indeed produces a meaningful comparison.
We also tested other kinds of procedures to set the COBN synaptic
conductances, $G_{syn}$, given the CUBN synaptic efficacies, $J_{syn}$.
In particular we define $G_{syn}$ such in a way to maximize the similarity
of PSCs (in one case) or PSPs (in another case) between the two networks
at the single-neuron level, to compensate for the post-synaptic stimulus
reduction that is peculiar of the COBN with respect to the CUBN (figure
\ref{fig_2_paperFR}). When using these procedures the results were
both less robust to change in the synaptic reversal potentials and
less similar between CUBN and COBN (data not shown). 

On the other hand, differences between the LFP spectra of the two
networks are also apparent in figures \ref{fig_7_paperFR}A,B. First,
the COBN gamma peak was larger and was modulated by the input rate
in a much stronger way than the CUBN gamma peak (figure \ref{fig_7_paperFR}E).
Given the fact that the net input current in the COBN was smaller
(figure \ref{fig_5_paperFR}A) and also fluctuated slightly less than
in CUBN (figure \ref{fig_5_paperFR}C), at first we found this result
surprising. However, the phenomenon can be understood after measuring
the AMPA and GABA fluctuations. As reported in figure \ref{fig_7_paperFR}F,
the size of recurrent AMPA and GABA current fluctuations was larger
in COBN than in CUBN, and the difference increased with the input
rate. Indeed, while the simultaneous increases of AMPA and GABA fluctuations
compensated each other in the COBN net input current (figures \ref{fig_5_paperFR}A,B),
the contributions of these two currents to the computed LFP have the
same sign (see equation \ref{eq_LFP}), and this led to a stronger
spectral peak in the COBN. Second, the CUBN displayed a broad LFP
spectral peak in the high gamma region ($>60$ Hz), and small fluctuations
in the low gamma region ($<60$ Hz), while, in the COBN, for inputs
greater than 3 (spikes/ms)/cell there was a sharp peak in the high
gamma band and also a pronounced plateau in the low gamma. Third,
since the power associated with this plateau was modulated by the
input rate, for the COBN all frequencies above 20 Hz were significantly
modulated, while in the CUBN significant modulation occurred only
for frequencies above 60 Hz. As we will see in the next section, the
narrower gamma peak indicates a stronger synchronization in the COBN
than in the CUBN, while the stronger modulation in the gamma power
makes the amount of information conveyed by the COBN larger than in
the CUBN (see section \vref{section_Information-about-external}).

For both networks, the spectra of the total firing rate were qualitatively
very similar to the spectra of the LFP for all input rates considered
(data not shown). Therefore all the aforementioned differences were
present also when comparing the COBN and CUBN total firing rate power
spectra.

\subsection{Correlation between AMPA and GABA currents}

\begin{figure}
\begin{centering}
\includegraphics[scale=0.8]{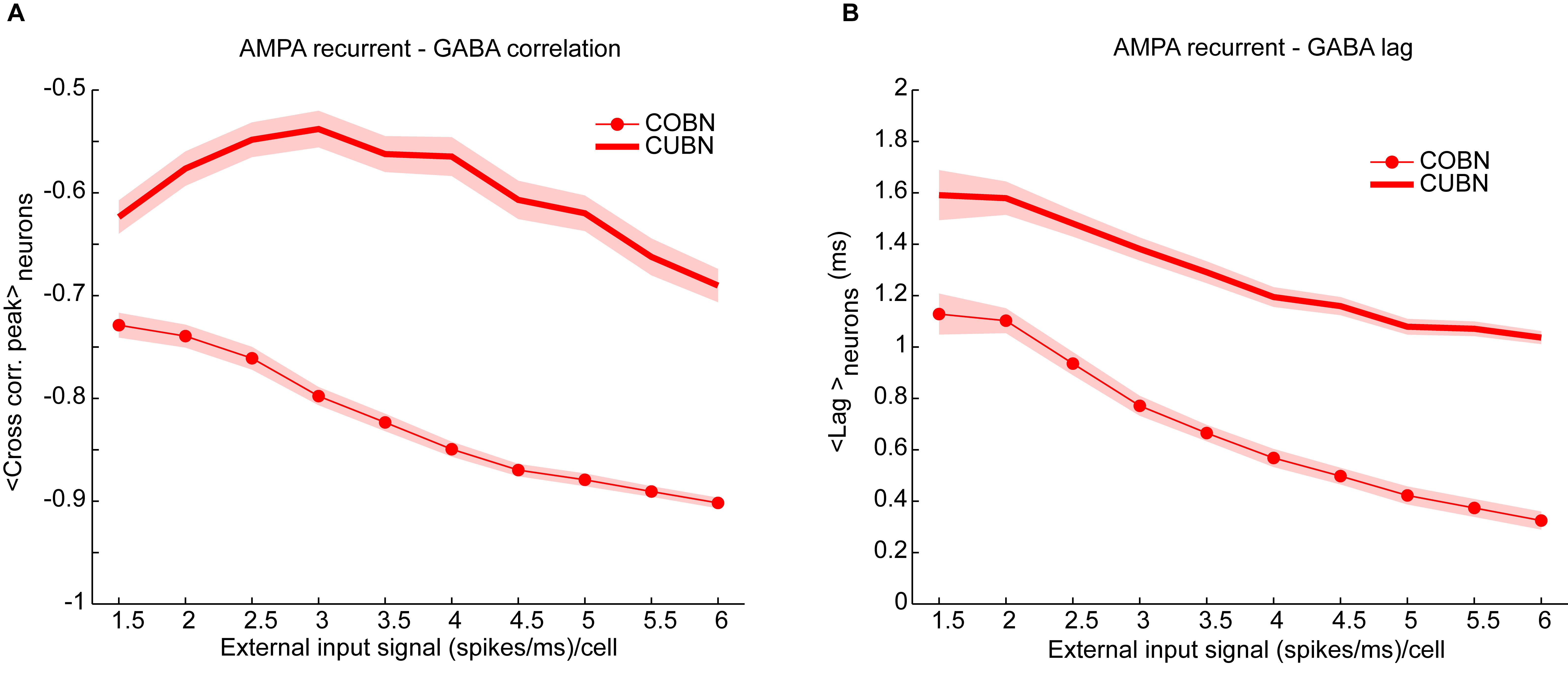}
\par\end{centering}
\centering{}\caption[Cross-correlation between AMPA and GABA inputs to excitatory neurons]{\textbf{\textcolor{black}{}}\textbf{\textcolor{black}{\small{}Cross-correlation
between AMPA and GABA synaptic currents}}{\small{}. Cross-correlation
between the time course of recurrent AMPA and GABA currents entering
excitatory neurons. }\textbf{\textcolor{black}{\small{}(A)}}{\small{}
Average peak value of cross-correlation between AMPA and GABA input
currents into excitatory neurons (see section \vref{section_computation_correl}
for details) for CUBN (thick line) and COBN (thin line with markers).
Note that, AMPA and GABA currents having opposite sign, the correlation
is negative. Shaded areas correspond to standard deviation across
neurons. }\textbf{\textcolor{black}{\small{}(B)}}{\small{} Cross correlation
average peak position. This measure quantify how much AMPA inputs
lags behind GABA ones. Same color code as (A). Results are computed
by using a simulation of 10.5 s per stimulus and show that (i) correlation
between recurrent AMPA and GABA input currents is stronger in the
COBN than in the CUBN, (ii) input correlation decreases monotonously
with input rate in COBN, while it does not in CUBN, (iii) GABA inputs
lags behind AMPA inputs by few milliseconds in both networks.}\label{fig_8_paperFR}}
\end{figure}
\begin{figure}
\begin{centering}
\includegraphics[scale=0.8]{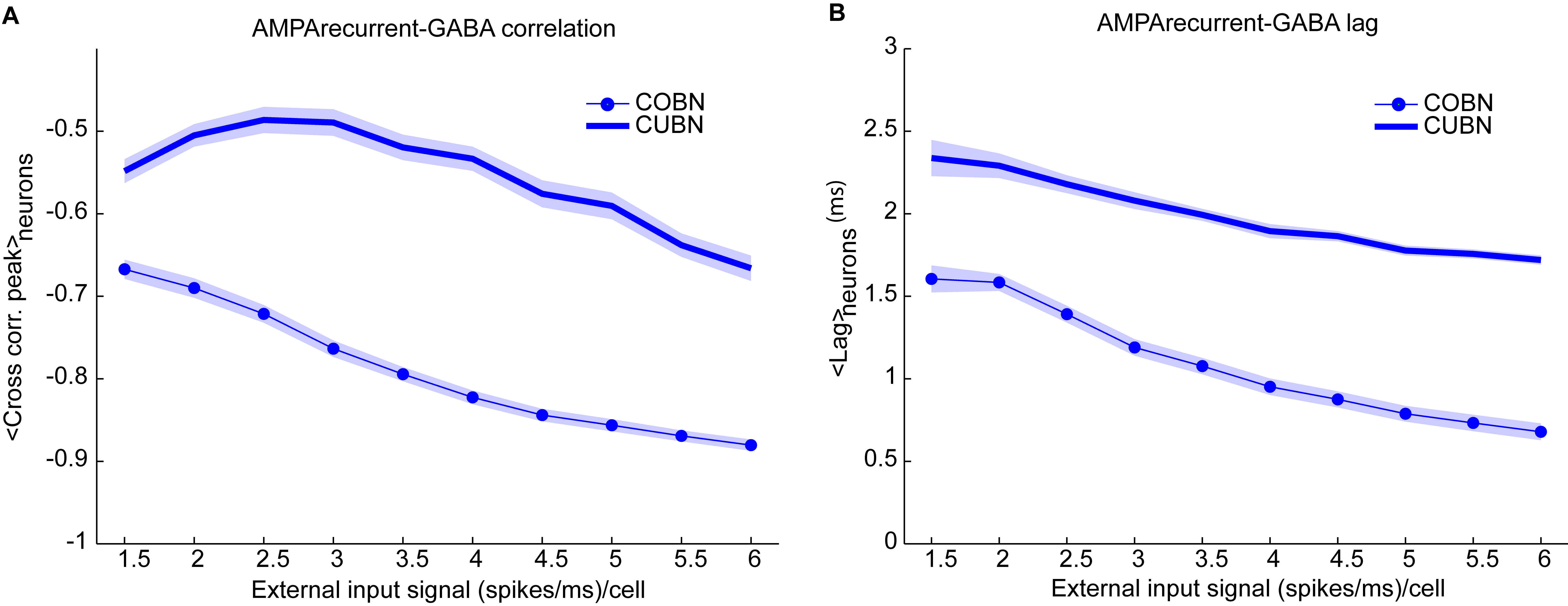}
\par\end{centering}
\centering{}\caption[Cross-correlation between AMPA and GABA inputs to inhibitory neurons]{\textbf{\textcolor{black}{}}\textbf{\textcolor{black}{\small{}Correlation
between AMPA and GABA inputs to inhibitory neurons}}{\small{}. }\textbf{\textcolor{black}{\small{}(A)}}{\small{}
Same as figure \ref{fig_8_paperFR}(A) for inhibitory neurons. }\textbf{\textcolor{black}{\small{}(B)}}{\small{}
Same as figure \ref{fig_8_paperFR}(B) for inhibitory neurons.}\label{fig_supp2_paperFR}}
\end{figure}

The correlation between AMPA and GABA synaptic currents is known to
play a very important role in determining the network dynamics and
in particular the spike train variability \citep{Isaacson2011}. A
negative correlation of AMPA and GABA input currents leads to sparse
and uncorrelated firing events, while positive values lead to strong
bursty synchronized events \citep{Renart2010}. We thus compared the
cross correlation between recurrent AMPA and GABA currents impinging
on single neurons in COBN and CUBN. We found that the correlation
between GABA and AMPA inputs was stronger (i.e., more negative) in
the COBN for all external input rates (figure \ref{fig_8_paperFR}A).
Moreover, in both networks, AMPA currents led GABA currents with lags
shorter than 5 ms, of the order of those observed in \citep{Okun2008}.
However, for all external input rates, AMPA-GABA lags were smaller
in the COBN (figure \ref{fig_8_paperFR}B). Although figure \ref{fig_8_paperFR}
shows results only for excitatory neurons, similar results held for
inhibitory neurons (figure \ref{fig_supp2_paperFR}). Finally, these
results held also when using as external noise a white noise process
instead of an Ornstein-Uhlenbeck process (figure \ref{fig_supp4_paperFR}C).

\subsection{Cross-neuron correlations\label{section_Cross-neuron-correlations}}

The fact that the cross-neuron variability in average current inputs
and MPs was much smaller (figures \ref{fig_5_paperFR}A,D) and high
gamma frequency peaks were narrower in the COBN (figures \ref{fig_7_paperFR}A,B)
suggested that the activity was more coherent in the COBN than in
the CUBN. This view was further corroborated by the finding that the
sum of the recurrent currents was larger in the COBN (figure \ref{fig_7_paperFR}F)
and suggested that, in this network, input currents may be more correlated
across different neurons. 
\begin{figure}
\begin{centering}
\includegraphics[scale=0.8]{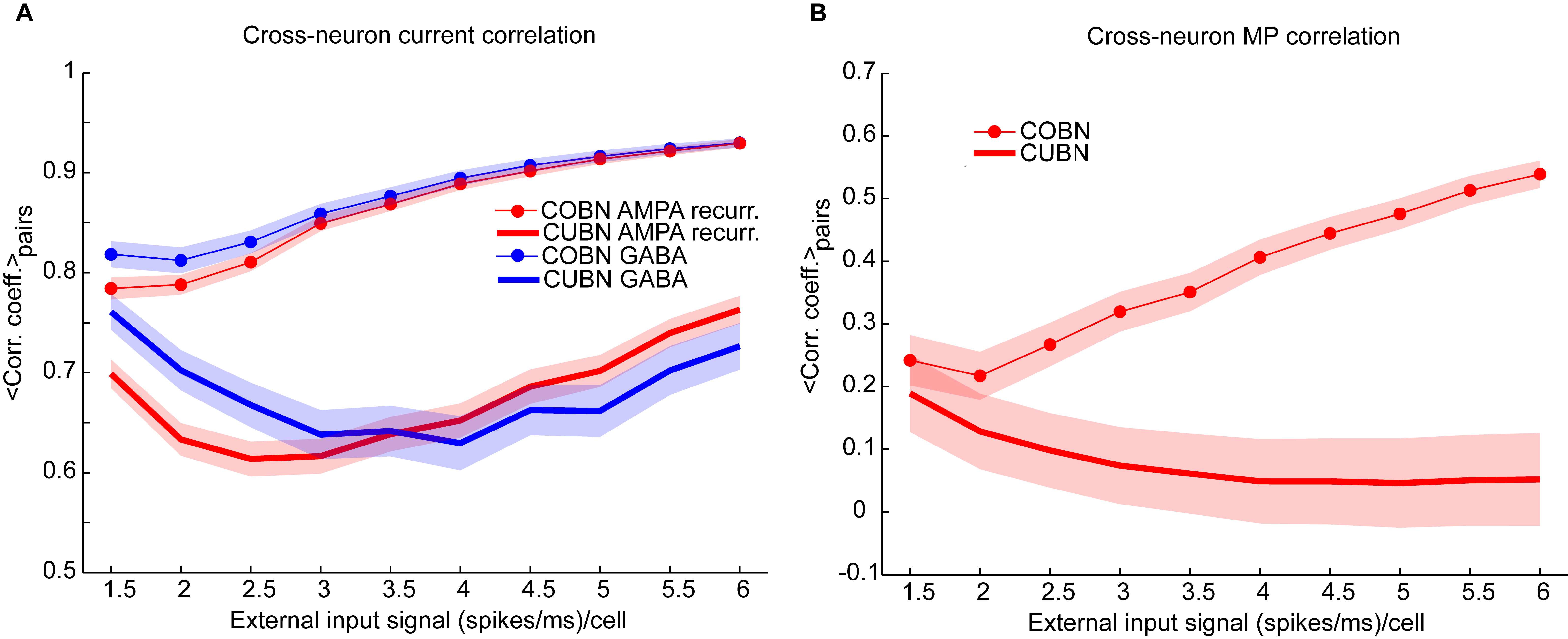}
\par\end{centering}
\begin{centering}
\caption[Cross-neuron correlation properties]{\textbf{\textcolor{black}{}}\textbf{\textcolor{black}{\small{}Correlation
of the synaptic inputs and of membrane potentials across neurons}}{\small{}.
}\textbf{\textcolor{black}{\small{}(A)}}{\small{} Average cross-neuron
correlation coefficient between the time course of recurrent AMPA
currents (red lines) and GABA currents (blue lines) on excitatory
neurons, for CUBN (thick lines) and COBN (thin line with markers),
as a function of the external input rate. Similar results hold for
inhibitory neurons (see \textquotedblleft Rec. AMPA-Rec. AMPA \textquotedblright{}
and \textquotedblleft GABA-GABA \textquotedblright{} }\textcolor{black}{\small{}in
table }\textbf{\textcolor{red}{\small{}\ref{table_suppl_paperFR}}}{\small{}).
}\textbf{\textcolor{black}{\small{}(B)}}{\small{} Average correlation
coefficient between the membrane potential (MP) time courses of pairs
of excitatory neurons as a function of the external input rate. While
in the COBN the MP correlation increases with input rate, the opposite
occurs in the CUBN. Shaded areas correspond to standard deviation
across neuron pairs. Results are computed by using a simulation of
10.5 s per stimulus and show that in COBN the cross-neuron correlations
between membrane potentials and between input currents are stronger
than in CUBN}.\label{fig_9_paperFR}}
\par\end{centering}
\centering{}
\end{figure}

We verified this hypothesis by measuring the average Pearson correlation
coefficient between the time evolution of the recurrent AMPA and of
the GABA input currents over neuron pairs (see section \vref{section_computation_correl}),
Figure \ref{fig_9_paperFR}A shows that for both AMPA and GABA currents
the average cross-neuron correlation coefficient was indeed significantly
stronger (t-test, $p\ll10^{-10}$) in the COBN for all external input
rates. Figure \ref{fig_9_paperFR}A shows also that, in the COBN,
the cross-neuron correlation grew with the external input rate for
both currents (Pearson correlation test, $p<10^{-5}$). In the CUBN
the AMPA currents were linearly correlated to the input rate (Pearson
correlation test, $p<0.05$), while GABA currents varied with the
input rate in a non-monotonic way. However, if we used white noise,
instead of the Ornstein-Uhlenbeck noise (see section \vref{section_network_and_input}),
the cross-neuron current correlation was again higher in the COBN
(t-test, $p\ll10^{-10}$), but grew monotonously with the input rate
for both networks (Pearson correlation test, $p<10^{-5}$), as shown
in figure \ref{fig_supp4_paperFR}A. The increase in the difference
between the cross-neuron current correlation in COBN and CUBN with
the input rate (figure \ref{fig_9_paperFR}A) led to the increase
of the difference in AMPA and GABA total fluctuations in the two networks,
shown in figure \ref{fig_7_paperFR}F. To fully appreciate the key
role played by correlations note that, if the correlations were similar
in COBN and CUBN, fluctuations would be expected to be larger in CUBN
since the firing rate was similar for the two networks (figure \ref{fig_6_paperFR}A)
and the single PSC amplitude was larger for the CUBN (figure \ref{fig_2_paperFR}).
\begin{figure}
\includegraphics[scale=0.8]{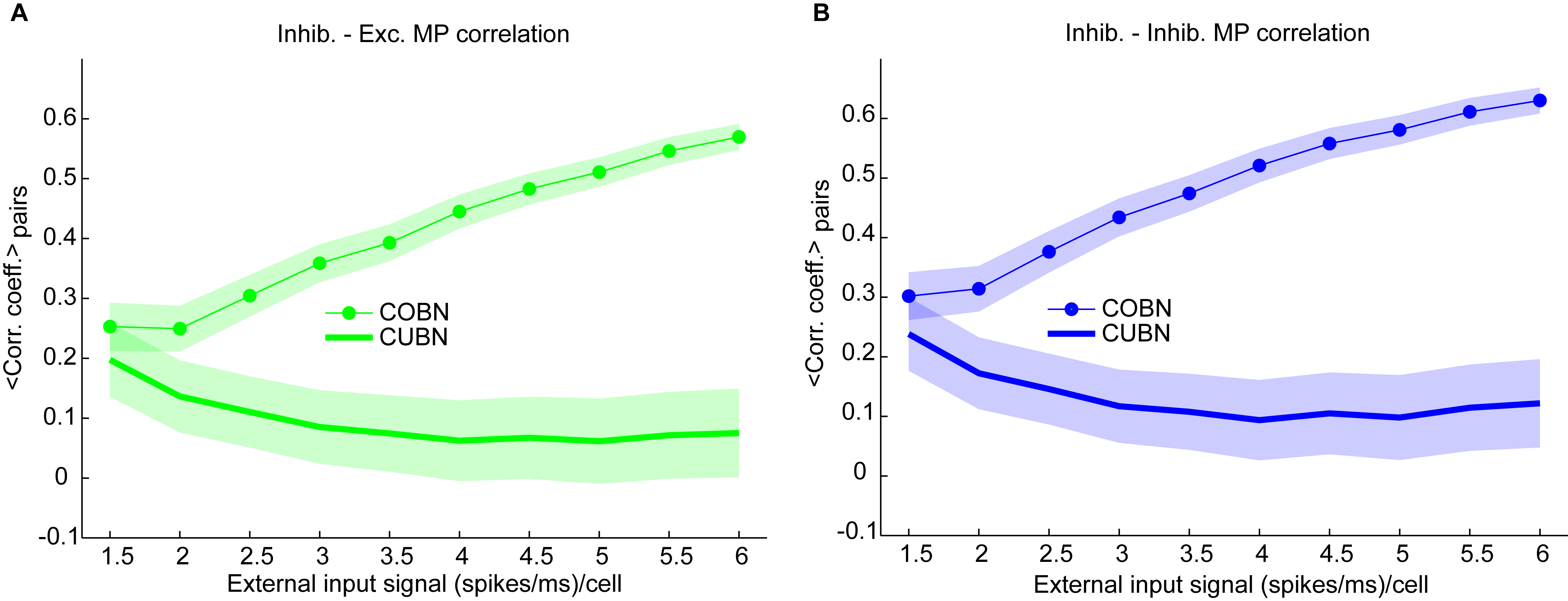}

\caption[MP correlation across neurons]{\textbf{\textcolor{black}{Membrane potential correlation across neurons.
(A)}} Same as figure \ref{fig_9_paperFR}(B) for pairs composed by
an inhibitory and an excitatory neuron. \textbf{\textcolor{black}{(B)}}
Same as figure \ref{fig_9_paperFR}(B) for pairs composed by two inhibitory
neurons.\label{fig_supp3_paperFR}}
\end{figure}
 Cross-neuron correlation of the input currents should be reflected
in cross-neuron MP correlation. The previously shown sample traces
of the MP of neuron pairs (figures \ref{fig_4_paperFR}E,H) suggested
that the correlation was indeed similar for COBN and CUBN in the low-conductance
state, but much stronger for the COBN in the high-conductance state.
We thus analyzed the average correlation of the MP time courses of
pairs of excitatory neurons (figure \ref{fig_9_paperFR}B). Over the
whole external input range considered, MP correlation in the COBN
was significantly stronger than in the CUBN (t-test, $p\ll10^{-10}$).
Cross-neuron MP correlation in the COBN increased with external input
rate (Pearson correlation test, $p<10^{-8}$), while it was only mildly
affected in the CUBN (Pearson correlation test, $p<0.02$). These
results held for all considered neuron pairs (figure \ref{fig_supp3_paperFR})
and also when considering white noise, instead of Ornstein-Uhlenbeck
noise (figure \ref{fig_supp4_paperFR}B). 
\begin{figure}
\begin{centering}
\includegraphics[scale=0.8]{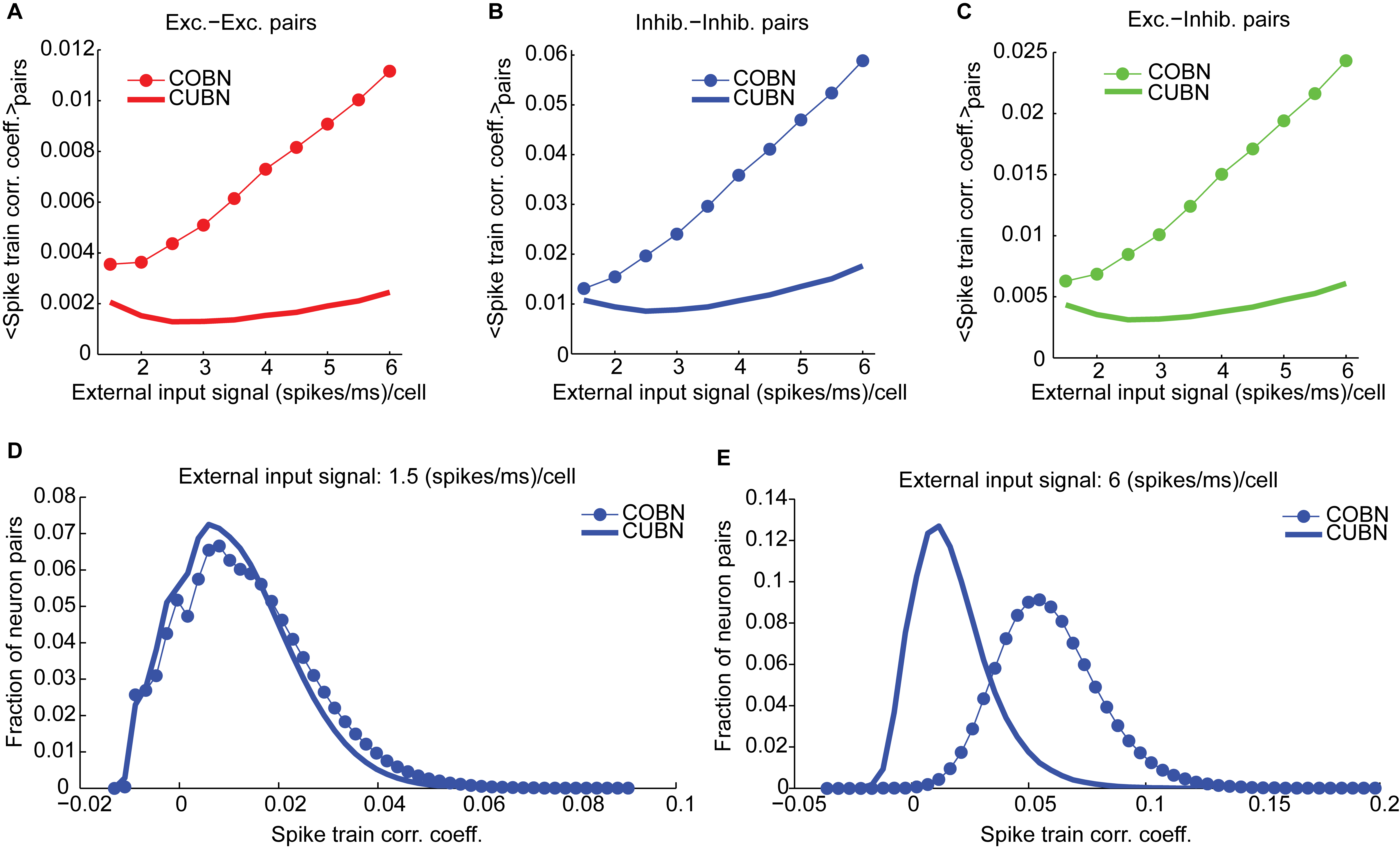}
\par\end{centering}
\begin{centering}
\caption[Spike train correlation]{\textbf{\textcolor{black}{}}\textbf{\textcolor{black}{\small{}Spike
train correlation}}{\small{}. Spike train pairwise coefficient of
correlation between neurons belonging to the same (A,B) or to different
(C) populations. }\textbf{\textcolor{black}{\small{}(A)}}{\small{}
Average spike train correlation between pairs of excitatory neurons
as a function of the external input rate for CUBN (thick line) and
COBN (thin line with markers). }\textbf{\textcolor{black}{\small{}(B)}}{\small{}
Same as (A) for correlation between pairs of inhibitory neurons. }\textbf{\textcolor{black}{\small{}(C)}}{\small{}
Same as (A) for correlations between pairs composed by an inhibitory
and an excitatory neuron. }\textbf{\textcolor{black}{\small{}(D)}}{\small{}
Distribution of the correlation coefficient across inhibitory neurons
pairs for an input of 1.5 (spikes/ms)/cell for the two networks. }\textbf{\textcolor{black}{\small{}(E)}}{\small{}
Same as (D) for an input of 6 (spikes/ms)/cell. Note that panels (A\textendash C)
do not have error bars for clarity, but the range of correlation values
is similar to the one displayed in panels (D,E). Results are computed
by using a simulation of 100.5 s per stimulus and show that firing
rate correlation is very low for both networks, and it increases with
input rate in the COBN, but not in the CUBN.}\label{fig_10_paperFR}}
\par\end{centering}
\centering{}
\end{figure}

We finally computed the cross-neuron spike train correlation. We expected
it to be related to the MP correlation displayed in figure \ref{fig_9_paperFR}B,
even if, since both networks were in a fluctuation-driven state, the
spike train correlation should be close to zero \citep{Brunel03,Renart2010}.
We found indeed a very low average spike train correlation (figures
\ref{fig_10_paperFR}A\textendash C) such that, for low input rates,
a significant fraction of pairs displayed negative correlation (figure
\ref{fig_10_paperFR}D).
\begin{figure}
\begin{centering}
\includegraphics[scale=0.8]{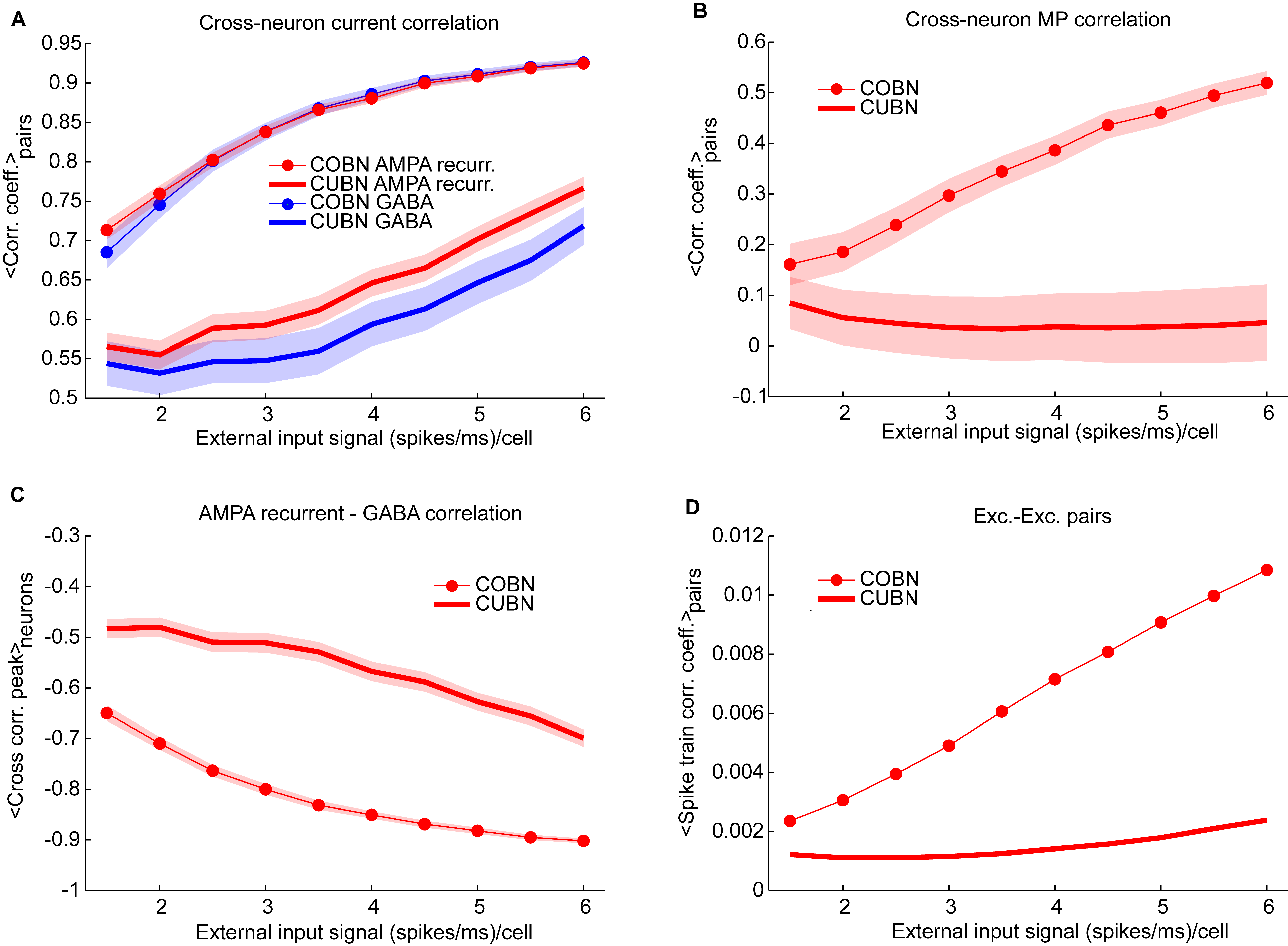}
\par\end{centering}
\centering{}\caption[Correlations in presence of white noise]{\textbf{\textcolor{black}{}}\textbf{\textcolor{black}{\small{}Correlations
in presence of white noise.}}{\small{} Same correlation analysis already
performed. The difference lies in the fact that here we model the
external input noise, n(t), as a Gaussian white noise instead of as
an Ornstein-Uhlenbeck process (see section \vref{section_network_and_input}).
The white noise has the same variance of the OU process used in the
main text. }\textbf{\textcolor{black}{\small{}(A)}}{\small{} Same
as figure \ref{fig_9_paperFR}A. }\textbf{\textcolor{black}{\small{}(B)
}}{\small{}Same as figure \ref{fig_9_paperFR}(B). }\textbf{\textcolor{black}{\small{}(C)}}{\small{}
Same as figure \ref{fig_8_paperFR}(A). }\textbf{\textcolor{black}{\small{}(D)}}{\small{}
Same as figure \ref{fig_10_paperFR}(A).}\label{fig_supp4_paperFR}}
\end{figure}
 However, in the CUBN, the spike train correlation was weaker and
less sensitive to input rate changes than in the COBN (see figures
\ref{fig_10_paperFR}A\textendash C and compare figures \ref{fig_10_paperFR}D,E).
These results did not change if we injected white noise, instead of
Ornstein-Uhlenbeck noise, in the network (figure \ref{fig_supp4_paperFR}D). 

\newpage{}

\begin{table}[H]
\includegraphics[scale=0.82]{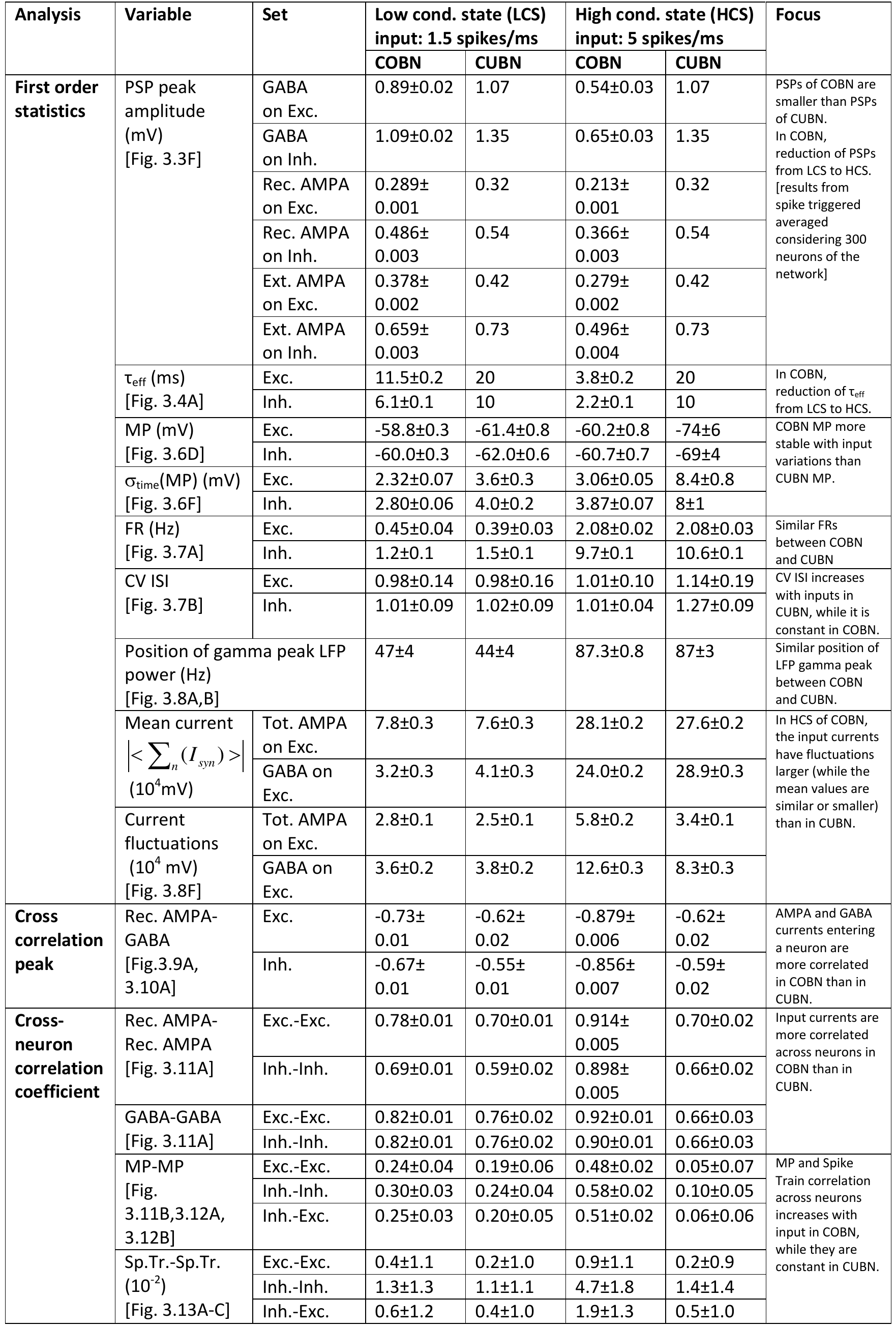}

\caption[Summary of differences between two comparable COBN and CUBN]{\textbf{\textcolor{black}{}}\textbf{\textcolor{black}{\small{}Summary
of differences between two comparable COBN and CUBN models}}\textcolor{black}{\small{}.\label{table_suppl_paperFR}}}

\end{table}

\newpage{}

\fbox{\begin{minipage}[t]{1\columnwidth - 2\fboxsep - 2\fboxrule}%
\textbf{\textcolor{black}{\footnotesize{}Table \ref{table_suppl_paperFR}.
Summary of differences between two comparable COBN and CUBN models}}\textcolor{black}{\footnotesize{}.
This table summarizes our main findings by comparing the values of
different features in the COBN used as reference (see table \ref{tab_gsyn})
and in the CUBN, when using two constant stimuli: $\nu_{0}$=1.5 and
5 (spikes/ms)/cell. These inputs cause respectively a low-conductance
state (LCS) and a high-conductance state (HCS). Values are reported
as mean \textpm{} standard deviations. }\textbf{\textcolor{black}{\footnotesize{}PSP
peak amplitudes}}\textcolor{black}{\footnotesize{} of the COBN are
computed by using a spike triggered averaged over 300 neurons from
the network in a simulation of 10.5 s (see section \ref{section_averagePSP}).
The effective membrane time constant of the COBN, $\mathbf{\boldsymbol{\tau}_{eff}}$,
the membrane potential, }\textbf{\textcolor{black}{\footnotesize{}MP}}\textcolor{black}{\footnotesize{},
the fluctuations on time of the membrane potential, $\boldsymbol{\sigma_{time}}$}\textbf{\textcolor{black}{\footnotesize{}(MP)}}\textcolor{black}{\footnotesize{}
and the coefficient of variation of the ISI, }\textbf{\textcolor{black}{\footnotesize{}CV
ISI}}\textcolor{black}{\footnotesize{}, are computed for each neuron
and then averaged across neurons by using data from a single trial
(of 10.5 s for $\tau_{eff}$, MP and $\sigma_{time}$(MP) and of 100.5
s for CV ISI); the standard deviations are computed across neurons.
The average firing rate, }\textbf{\textcolor{black}{\footnotesize{}FR}}\textcolor{black}{\footnotesize{},
the }\textbf{\textcolor{black}{\footnotesize{}position of the gamma
peak of the LFP power}}\textcolor{black}{\footnotesize{} spectrum,
the }\textbf{\textcolor{black}{\footnotesize{}current mean}}\textcolor{black}{\footnotesize{}
and the }\textbf{\textcolor{black}{\footnotesize{}current fluctuations}}\textcolor{black}{\footnotesize{}
are computed for each trial (of 4.5 s) considering the activity of
all the (excitatory or inhibitory) neurons of the network and then
are averaged over 50 trials (the standard deviations are computed
thus across trials). The current mean and the current fluctuations
refer to the sum of the (AMPA or GABA) currents entering all the excitatory
neurons, as indicated by the summation over neurons, $\sum_{n}$,
which are exactly the variables used to simulate the LFP. The sum
of external AMPA (Ext. AMPA) and recurrent AMPA (Rec. AMPA) is stated
as Tot. AMPA. Correlations are computed by using a single trial of
10.5 s. In particular, the }\textbf{\textcolor{black}{\footnotesize{}cross
correlation peak}}\textcolor{black}{\footnotesize{} is averaged over
the neurons obtained from two randomly selected subpopulations of
200 excitatory and inhibitory neurons (see section \vref{section_computation_correl}),
while the }\textbf{\textcolor{black}{\footnotesize{}cross-neuron correlation
coefficient}}\textcolor{black}{\footnotesize{} is averaged over all
the couples of neurons obtained from the same subpopulations (see
section \vref{section_computation_correl}).}%
\end{minipage}}

\subsection{Information about external inputs\label{section_Information-about-external}}

In the previous subsections we investigated how the average level
of spike rate, LFP and spike train correlation depends on the external
input to the network, finding a more pronounced stimulus modulation
of LFP gamma power and of cross-neural correlation in COBN. To quantify
these stimulus modulations of network activity, we computed the mutual
information between the stimuli to the network and various aspects
of network activity (see for section \vref{section_Computation-of-information}
details).

\begin{figure}
\begin{centering}
\includegraphics[scale=0.8]{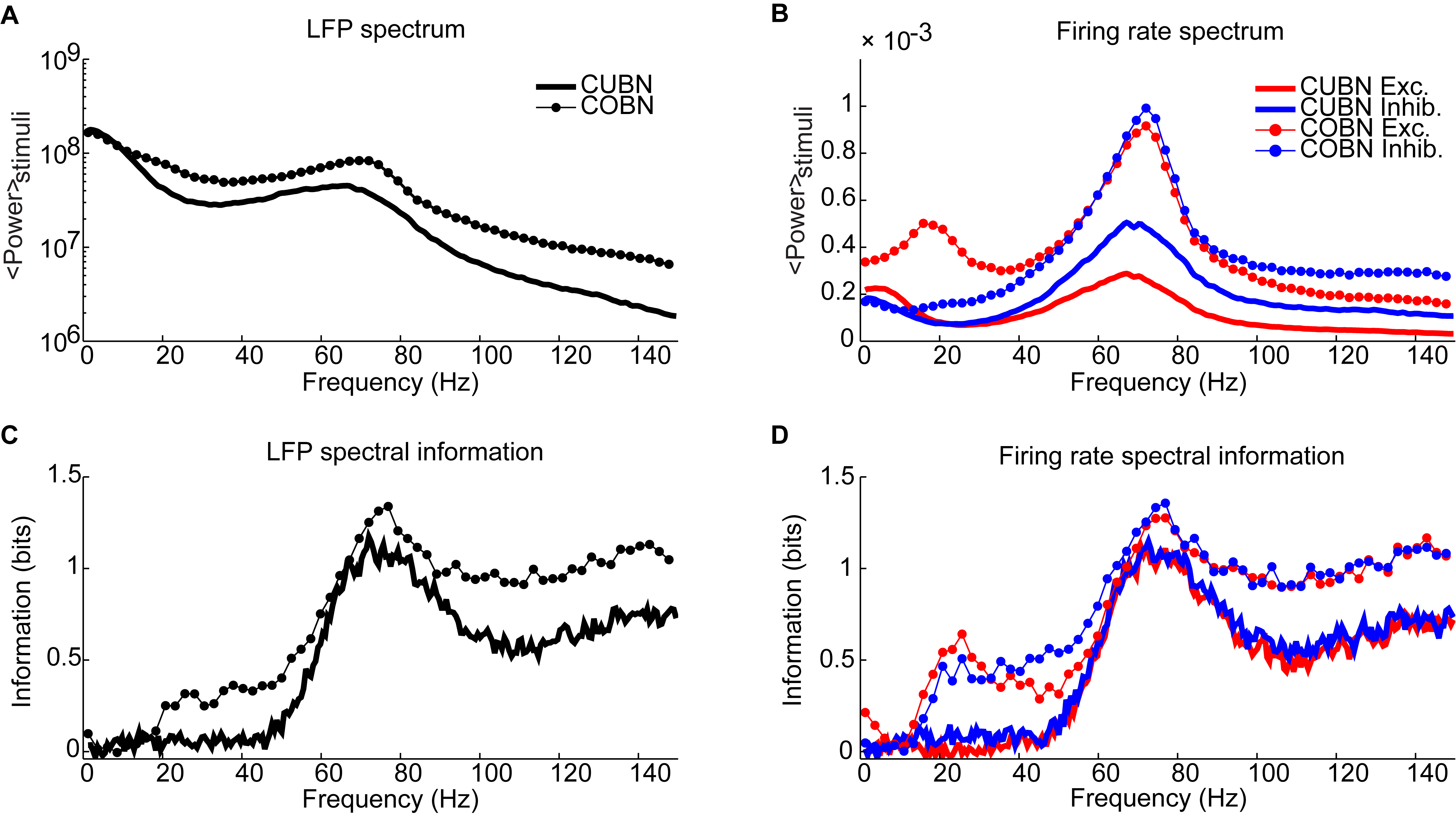}
\par\end{centering}
\centering{}\caption[Spectral information relative to the input rate]{\textbf{\textcolor{black}{Spectral information relative to the input
rate}}. Information carried by LFP power spectrum (left column) and
population firing rates power spectra (right column) about constant
inputs ranging from 1.5 to 3 (spikes/ms)/cell with steps of 0.1 (spikes/ms)/cell.
Data are obtained by using 50 trials of 4.5 s per stimulus. \textbf{\textcolor{black}{(A)}}
Average power spectrum of LFP over the entire stimulus range for the
COBN and the CUBN (thin line with markers and tick line respectively).\textbf{\textcolor{black}{{}
(B)}} Average power spectrum of the total firing rate of excitatory
and inhibitory neurons (red and blue respectively) for the two networks
{[}same line code as (A){]}. \textbf{\textcolor{black}{(C)}} Spectral
information carried by LFP about the input rate (see section \vref{section_Computation-of-information}
for details). Same color code as (A). \textbf{\textcolor{black}{(D)}}
Spectral information carried by total excitatory and inhibitory firing
rate about the input rate. Same color code as (B). Results show that
the COBN carries more information about constant stimuli for all considered
frequencies, both in LFP and in firing rates.\label{fig_11_paperFR}}
\end{figure}
We first measured the information carried by the average firing rate,
both of excitatory and inhibitory neurons, in the two networks by
using constant stimuli in the range 1.5\textendash 3 (spikes/ms)/cell
with steps of 0.1 (spikes/ms)/cell. We found that, consistently with
the results shown in figure \ref{fig_6_paperFR}A, the information
carried by the average firing rate had the same value of 2.3 bits
for both neural populations in both network models. Given that the
modulation of spike train correlation with external input is greater
in the COBN than in the CUBN, we expected that also the mutual information
between the spike train correlation and the input rate was greater
in the COBN than in the CUBN. Indeed this was the case: information
in spike train correlation was much larger in the COBN (1.6 and 2.0
bits for excitatory and inhibitory neurons respectively) than in the
CUBN (1.4 and 0.9 bits for excitatory and inhibitory neurons respectively).

We measured then the information content of the LFP power spectrum.
The LFP power spectrum averaged over all the presented constant stimuli
was higher for the COBN than for the CUBN for all frequencies above
15 Hz (figure \ref{fig_11_paperFR}A). We found that, at all frequencies
above 20 Hz, the COBN LFP spectrum carried more information about
input rate than the CUBN LFP spectrum (figure \ref{fig_11_paperFR}C).
Most notably, the peak information increased by about 20\%, and the
(20\textendash 45) Hz frequency range was informative in the COBN,
but not in the CUBN. We repeated the analysis considering the power
spectra of the total inhibitory and excitatory firing rate in the
two networks. Excitatory neurons in the COBN had stronger power than
excitatory neurons in the CUBN for all frequencies (figure \ref{fig_11_paperFR}B,
note that here the y-scale is linear, while in figure \ref{fig_11_paperFR}A
is logarithmic) and showed a secondary peak at about 20 Hz. For inhibitory
neurons, instead, the COBN power spectrum was higher only for frequencies
above 15 Hz, as in the LFP . 

So far we have investigated only the information carried about the
strength of a time-independent input to the network. In a previous
work on CUBN \citep{mazzoni2008encoding} it has been shown that when
the input to the CUBN is dominated by low frequency fluctuation, the
network oscillations (captured by both LFP and massed firing rate
measures) form two largely independent frequency information channels.
A gamma-range information channel is generated by recurrent interactions
of inhibitory and excitatory neurons and conveys information about
the mean input rate. A low-frequency information channel is generated
by entrainment of the low frequency network activity to the slow fluctuations
of the input stimulus and carries information about the stimulus time
course on such slow time scales. We wanted to test how these two information
channels, developed when presenting the network with time-varying
stimuli, depended on the choice of the synaptic model.

\begin{figure}
\begin{centering}
\includegraphics[scale=0.8]{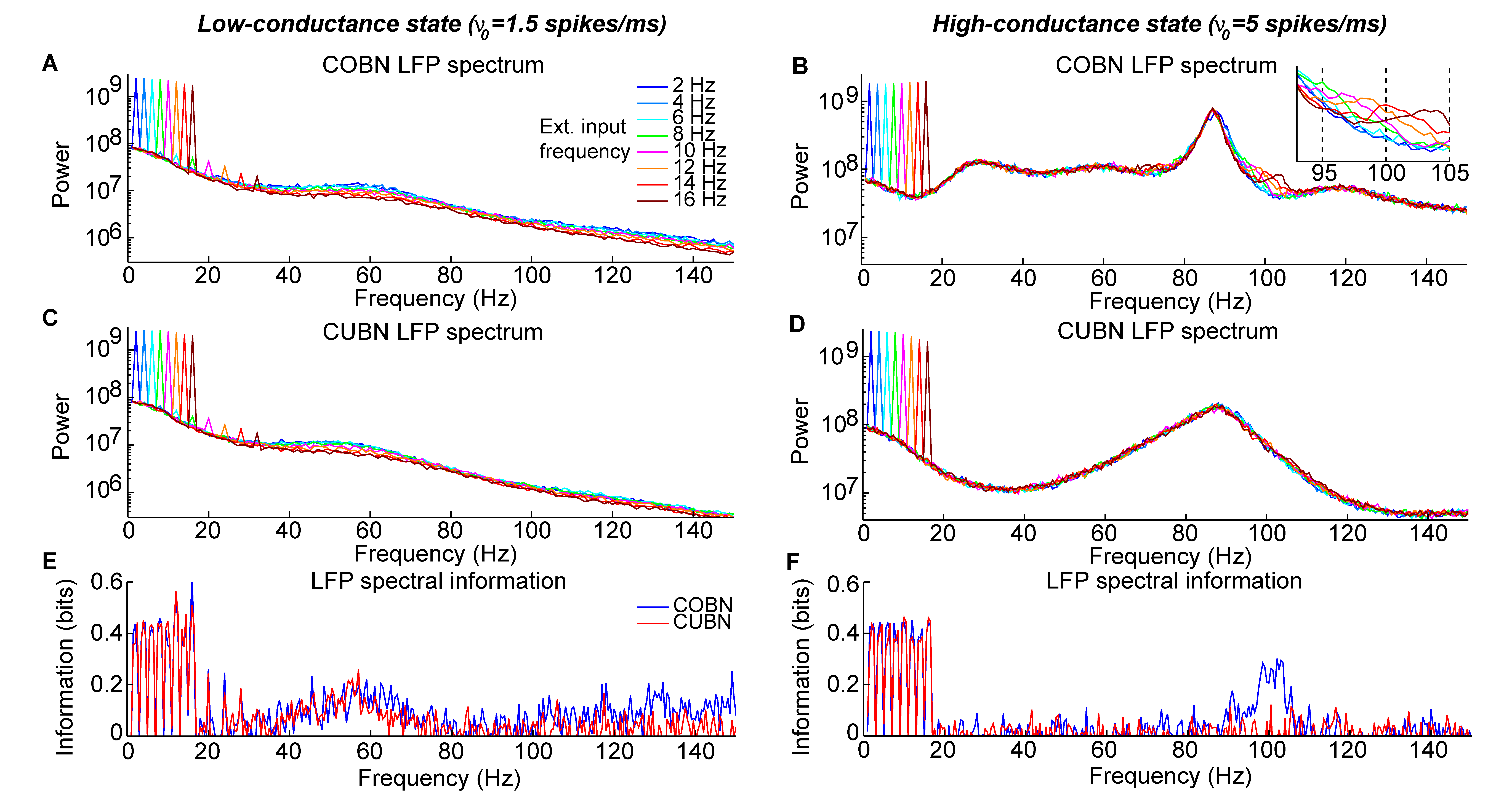}
\par\end{centering}
\begin{centering}
\caption[Spectral information relative to periodic input oscillations]{\textbf{\textcolor{black}{Spectral information relative to periodic
low frequency inputs}}. Dynamics of the COBN and CUBN when injected
with slowly oscillating inputs. The input signals are sine curves
with amplitude A = 0.6 spikes/ms and frequency $f$, from 2 to 16
Hz, superimposed to a baseline of $\nu_{0}$ = 1.5 spikes/ms in the
left column and $\nu_{0}$ = 5 spikes/ms in the right column. The
first baseline value produces a low-conductance state, while the second
originates a high-conductance state. Data are obtained from 50 trials
of 10.5 s per stimulus. \textbf{\textcolor{black}{(A,B)}} LFP power
spectrum in the COBN as a function of the external signal frequency.
The power spectrum is averaged over trials. (B) Same color code as
in (A). \textbf{\textcolor{black}{(C,D)}} Same as (A,B) for the CUBN.
The inset in (B) shows a detail of the panel in the frequency range
where beats are displayed. \textbf{\textcolor{black}{(E,F)}} Spectral
information carried by the LFP about the frequency of the stimulus
presented (see section \vref{section_Computation-of-information}
for details) for COBN (blue line) and CUBN (red line). Results show
that the information due to the entrainment of the LFP to the slow
input oscillations is almost the same in COBN and CUBN. The only difference
is due to the beats that appear in the high-conductance state of the
COBN {[}inset in (B){]}, which result in a peak of information around
100 Hz (F).\label{fig_12_paperFR}}
\par\end{centering}
\centering{}
\end{figure}
To investigate this point, we injected into the two networks periodic
stimuli with fixed amplitude and frequency varying between 2 and 16
Hz. These input frequencies below 16 Hz were taken to represent the
slow naturalistic fluctuations present in natural input signals \citep{Luo07,Chandrasekaran10,Gross2013}.
Since we wanted to investigate potential differences between models
separately in low- and high- conductance states, we generated two
kinds of input signals: a low-input regime (corresponding to a low-conductance
state) and a high-input regime (corresponding to a high-conductance
state). Thus the periodic input was made of a sinusoidal signal at
a given frequency superimposed to a constant baseline that was set
to a low value ($\nu_{0}$ = 1.5 spikes/ms) to induce a low-conductance
state and to a high value ($\nu_{0}$ = 5 spikes/ms) to induce a high-conductance
state. The amplitude , $A$, of the sinusoidal component of the input
was 0.6 spikes/ms across all simulations. Results are reported in
figure \ref{fig_12_paperFR}. 

We examined first the low-conductance state (left column of figure
\ref{fig_12_paperFR}). We considered the LFP power spectra of the
two networks in response to periodic stimuli of different frequencies
(figures \ref{fig_12_paperFR}A,C). With respect to the previously
examined constant input case (figures \ref{fig_7_paperFR}A,B), the
LFP power spectrum of both networks had an additional high narrow
peak exactly at the same frequency of the periodic input. This peak
signaled the entrainment of the network to the periodic input \citep{mazzoni2008encoding}.
The ability of the two networks to entrain their dynamics to the low-frequency
stimuli suggested that the power of the LFP at such low frequencies
could discriminate which of these periodic inputs was being presented.
We tested this suggestion quantitatively by using mutual information,
and we found that the slow LFP frequencies conveyed indeed information
about the stimuli, approximately in the same amount in both networks
(figure \ref{fig_12_paperFR}E). Note that, in the low-conductance
state, there was also a slight modulation with the input frequencies
of the power in the gamma band (40\textendash 70) Hz, with slightly
lower gamma power for stimuli of faster frequency (figures \ref{fig_12_paperFR}A,C).
These modulations of gamma-range power resulted in moderate amounts
of stimulus information in the same range, (40\textendash 70) Hz,
(figure \ref{fig_12_paperFR}E), and were likely due to the time taken
by the networks to develop gamma oscillations following the very low
input values occurring at the trough of the sinusoidal input. 

\begin{figure}
\begin{centering}
\includegraphics[scale=0.8]{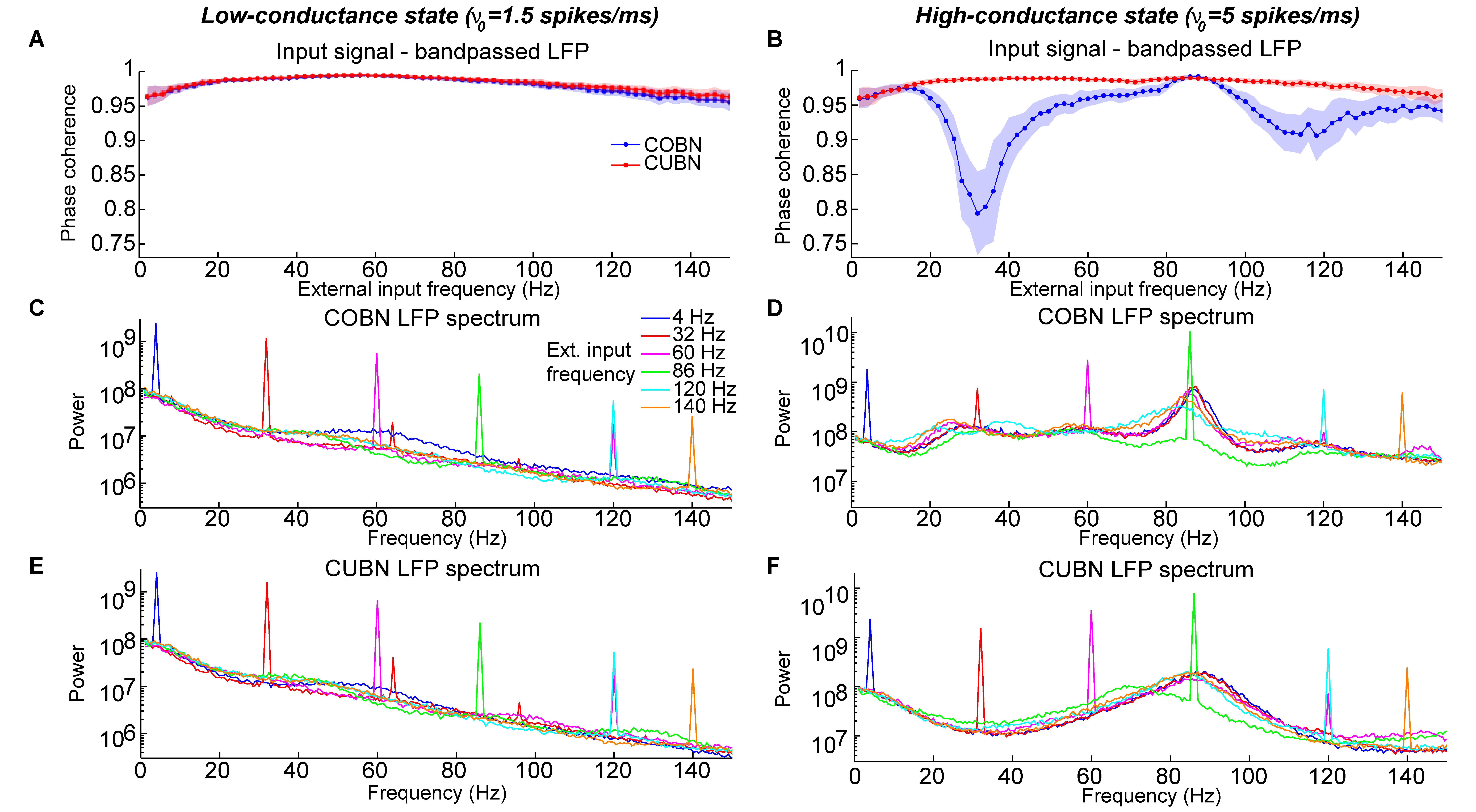}
\par\end{centering}
\centering{}\caption[Entrainment of LFP to input oscillations]{\textbf{\textcolor{black}{Entrainment of LFP to input oscillations}}.
Entrainment of the network oscillations to the frequencies of the
periodic input in COBN and CUBN. The input signals are periodic curves
as in figure \ref{fig_12_paperFR}, but with frequency $f$ from 2
to 150 Hz. \textbf{\textcolor{black}{(A,B)}} Average (over trials)
coherence between the phase of the input signal, with frequency $f$,
and the phase of the LFP bandpassed in the corresponding frequency
range $(f-1,f+1)$ Hz (see section \vref{section:Spectral-analysis}
for details). Note that the phase coherence lies in the interval (0,
1). Data are obtained from 50 trials of 10.5 s per stimulus; shaded
areas represent standard deviations across trials. Blue lines display
results from COBN and red lines from CUBN. \textbf{\textcolor{black}{(C,D)}}
LFP power spectrum in the COBN as a function of some selected external
signal frequencies. The power spectrum is averaged over 50 trials.
(D) Same color code as in (C). \textbf{\textcolor{black}{(E,F)}} Same
as (C,D) for the CUBN. In the low-conductance state both networks
entrain very well to the external stimulus, whereas in the high-conductance
regime the COBN entrains less well than the CUBN in the middle and
in the highest frequency regimes.\label{fig_13_paperFR}}
\end{figure}
We then investigated the high-conductance state (right column of figure
\ref{fig_12_paperFR}). Figures \ref{fig_12_paperFR}B,D shows that
entrainment of both networks to low frequencies (signaled by the high
narrow peak of LFP spectrum at the same frequency as the input) occurred
strongly in the high-conductance state. The information about which
of these periodic inputs was being presented, carried by the low frequency
LFP power, was still identical in the two networks (figure \ref{fig_12_paperFR}F).
Moreover, and consistently with the above results obtained with constant
inputs (figures \ref{fig_7_paperFR}A,B), the gamma peak in the high-conductance
states was much stronger and narrower in the COBN than in the CUBN.
Probably because of this, the COBN (but not the CUBN) developed beats
of the low-frequency peaks into the frequency range around 100 Hz
(inset figure \ref{fig_12_paperFR}B). Since the low-frequency peak
varied with the input, these beats led to an amount of information
in the COBN LFP power around 100 Hz. The moderate gamma-range information
peak, observed in the (40\textendash 70) Hz range for the low-conductance
state (figure \ref{fig_12_paperFR}E), was absent in both networks
for the high-conductance regime (figure \ref{fig_12_paperFR}F), because
the input rate was always high at any time point. Thus gamma oscillations
in the range (80\textendash 94) Hz were always strong, with relatively
small fluctuations over time, leading to not discernible modulation
across the set of input frequencies considered (figure \ref{fig_12_paperFR}B,D). 

We then investigated the ability of the network to entrain to a wider
range of input frequencies, in particular including frequencies as
fast as or faster than the gamma oscillations intrinsically generated
by the network. We did so by testing the network with periodic stimuli
over the 2\textendash 150 Hz range of input frequencies (figure \ref{fig_13_paperFR}).
Again, to investigate differences between models separately in low-
and high-conductance regimes, we generated two kinds of input signals
that only differed for the value of the baseline, as described above.
We quantified entrainment by computing the coherence between the phase
of the input signal and the phase of the LFP bandpassed in a narrow
band (with 2 Hz bandwidth) centered at the frequency of the periodic
input. In the low-conductance state both networks were strongly entrained
to the input over the whole range of frequencies examined, as indicated
by the high phase coherence (figure \ref{fig_13_paperFR}A). However,
when injecting the same input frequencies with the highest baseline
(i.e., making the network operate in a high-conductance state), the
behavior of the two networks was very different. The CUBN could still
entrain extremely well over the entire input frequency range tested.
The COBN entrained extremely well to inputs in the (80\textendash 94)
Hz input frequency range, but less well to inputs with frequency between
16 Hz and 80 Hz, and above 94 Hz. The reason for the presence in the
COBN of frequency regions with lower phase coherence (and thus less
accurate entrainment to the periodic input) may be because, in the
high-conductance state, the COBN had stronger internally generated
recurrent oscillations (of higher power than the CUBN, see figures
\ref{fig_13_paperFR}D,F) whose dynamics likely did not interfere
constructively with the dynamics of the entrainment to the input.
This resulted in peaks of less high amplitude in the COBN LFP spectrum
at the exact frequency of the periodic input (figures \ref{fig_13_paperFR}D,F).
It is interesting to note that the COBN still entrained very well
in the (80\textendash 94) Hz input frequency range (figure \ref{fig_13_paperFR}B),
despite this was also the frequency range exhibiting the strongest
recurrent oscillations. Indeed, this range coincided with the peak
amplitude of the internally generated gamma oscillations (figure \ref{fig_12_paperFR}B).
The ability of the network to entrain well in this gamma range can
be understood by observing that this was also the range more strongly
modulated by the input rate (figure \ref{fig_7_paperFR}A). Thus,
due to their particularly strong responsiveness to the input, external
and internal oscillation in this range could interfere constructively,
resulting in large peaks of the network LFP at the input frequency
(figure \ref{fig_13_paperFR}D). 

\begin{figure}
\begin{centering}
\includegraphics[scale=0.8]{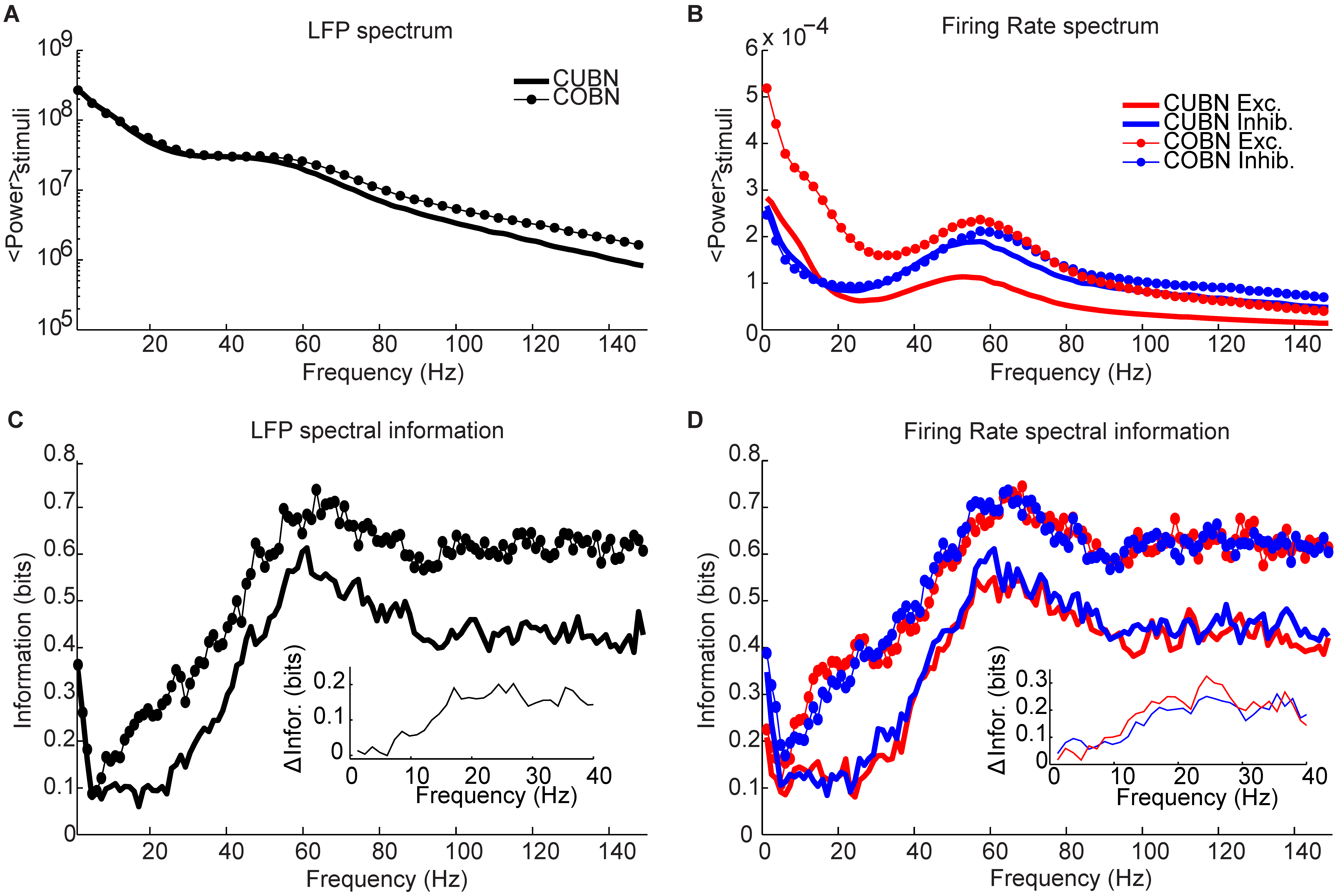}
\par\end{centering}
\begin{centering}
\caption[Spectral information relative to naturalistic stimuli]{\textbf{\textcolor{black}{Spectral information relative to naturalistic
stimuli}}. Information carried by LFP power spectrum (left column)
and population firing rates power spectra (right column) about intervals
of naturalistic stimulation based on LGN recordings in monkeys watching
a movie. Recording time (80 s) is divided into 40 intervals, considered
as different stimuli and the information is computed over 50 trials
(see section \vref{section_Computation-of-information} for details).
\textbf{\textcolor{black}{(A)}} Average power spectrum of LFP over
the entire naturalistic input for COBN and CUBN (thin line with markers
and thick line respectively). \textbf{\textcolor{black}{(B)}} Average
power spectrum for the total firing rate of excitatory and inhibitory
neurons (red and blue respectively) for the two networks. Same line
code as in (A). \textbf{\textcolor{black}{(C)}} Spectral information
carried by LFP (see Methods for details). Same color code as in (A).
In the inset, it is shown the difference between COBN and CUBN information
in the low frequency band. \textbf{\textcolor{black}{(D)}} Spectral
information carried by total excitatory and inhibitory firing rates.
Same color code as (B). In the inset, it is shown the difference between
COBN and CUBN information in the low frequency band. Results show
that, also considering complex stimuli, the information relative to
the mean value of the input {[}that here is the information carried
by the frequencies above the delta band, (1\textendash 4) Hz{]} is
higher and carried on a broader range of frequencies in the COBN,
both in LFP and in firing rates. The information conveyed by delta
band frequencies is instead almost identical in the two networks.\label{fig_14_paperFR}}
\par\end{centering}
\centering{}
\end{figure}
To study the differences in the responses of the two networks to stimuli
more complex and more biologically relevant than periodic functions,
we finally compared the information carried by the LFP and firing
rate spectra in COBN and CUBN when using the naturalistic time-varying
inputs. We injected then in the networks naturalistic stimuli based
on MUA recordings from the LGN of an anesthetized macaque presented
with a commercial 80 s color movie clip. The average LFP and total
firing rate power spectra for both networks with this set of stimuli
are displayed respectively in Figures 14A and B. All these spectra
had higher power at low frequencies (as the input signal had), and
the gamma peaks were low because the average stimulus rates were in
the range 1.2\textendash 2 spikes/ms. We computed information about
which part of the time-varying naturalistic signal was being presented
(see \vref{section_Computation-of-information} for details). We found
that both LFP and firing rates spectra carried more information in
the COBN than in the CUBN, for all frequencies (figures \ref{fig_14_paperFR}C,D).
The difference in spectral information between COBN and CUBN for frequencies
below 5 Hz was almost zero for the LFP and very low for the firing
rates (see insets of figures \ref{fig_14_paperFR}C,D). 

Our findings therefore confirm that the two independent information
channels (one in the low frequencies due to the entrainment to the
input, and one in the gamma band due to internally generated oscillations),
which were previously reported for the CUBN \citep{mazzoni2008encoding},
also exist in the COBN. Moreover, our results show that the information
about the input conveyed by low frequencies, both in low- and high-conductance
states, does not depend on the details of the synaptic model adopted,
while the information encoded in the gamma range is larger in the
COBN than in the CUBN.

\newpage{}

~

\newpage{}

\chapter[Relationship between EEGs/LFPs and single-neuron activity]{Relationship between EEGs/LFPs and cell-specific single-neuron firing
during slow wave oscillations\label{chapter_Fellin} }

\ohead{\headmark} 
\pagestyle{scrheadings}    

\lettrine[lines=2]{I}{n this} chapter we investigate how to characterize
empirically the relationship between mesoscopic or macroscopic network
dynamics and the firing activity of identified single neurons (i.e.,
neurons belonging to specific classes such as the classes of pyramidal
neurons or of interneurons of different types). We address the issue
by developing mathematical methods to estimate the linear component
of the relationship between firing activity and mass signals and by
applying them to concurrent recordings of single-unit firing and of
mass circuit activity (LFPs, EEGs) in the neocortex of anesthetized
mice in a regime of slow wave oscillations.

\section{Introduction\label{PFsec:Introduction}}

EEGs and LFPs are measures of mass neural dynamics that are easier
and more stable to perform than measures of single-neuron spiking
activity \citep{hall2014real}. In particular, LFPs are invasive measures
that capture mostly postsynaptic potentials (for a full description
see \vref{section_LFP}) typically collecting the activity of populations
of neurons located a few hundred micrometers from the recording site
\citep{Einevoll2013}. EEG is the extracranial counterpart of the
LFP and (like the LFP) captures the mass postsynaptic potentials of
large populations of neurons (that are less localized than in the
LFP) and it can be measured non-invasively; therefore EEG can be used
to monitor neural activity with high temporal precision in healthy
humans during cognitive tasks \citep{da2013eeg}. Practical advantages
of EEGs and LFPs over recordings of spiking activity are that (i)
LFPs can be recorded more stably for longer periods (ii) their recording
requires less power consumption. This is due to the fact the highest
power spectral density values of mass signals are found for the lowest
frequencies, thus the sampling rates are not required to be high,
while spiking activity always needs an high sampling frequency \citep{hall2014real}.
However, a difficulty in interpreting these mass (i.e., circuit-level)
signals in terms of neural computation is their intrinsic ambiguity:
in absence of specific information of how mass signals arise from
the individual components that contribute to their generation, it
is unclear how they relate to the time course of the underlying spiking
activity of neurons in the proximity of the electrode \citep{Einevoll2013}. 

Being able to develop simple, yet reasonably accurate, mathematical
expressions that relate EEGs or LFPs to spiking activity of single
identified cell would be important for several reasons. The estimation
of the dynamics of firing of specific classes of neurons from EEGs/LFPs
would allow us to correctly infer the underlying neural computations
from mass recordings, a feat which is not possible with current computational
technologies. With this expression, for example, we would be able
to tell, from non-invasive electrical recordings only, that some specific
neural classes of interneurons increase or decrease their firing when
a subject is performing a certain task, thereby giving quantitative
information that we can put into models\footnote{This could be very useful, for example, in brain-machine interfaces
applications, such as spike-based neuroprostheses \citep{hall2014real}.}. The ability of estimating the dynamics of EEGs/LFPs from the spiking
activity of individual cell types (the inverse problem of the one
just described above) would be valuable to understand how the firing
of single cells relates to the circuit ``context'' which led the
neuron to fire, therefore giving precious insights about the way circuits
shape single-neuron activity.

Before discussing in more details these topics, it is useful to point
out that two different variables can be used to define spiking activity:
\begin{itemize}
\item the spike (or firing) rate, FR, (i.e., $r(t)$, see equation \ref{eq_smoothed_FR}),
which is the average number of spikes in windows of a given amplitude
\item the spike times or spike train (i.e., $\rho(t)$, see equation \ref{eq_spke_train}),
which is the position in time of each spike with a given sampling
frequency
\end{itemize}
Furthermore, as stated above, it also useful to consider the relationship
between single-neuron and network-level activities in two opposite
directions:
\begin{itemize}
\item the direction that goes from single-neuron activity towards mass circuit
activity (that we will denote in the following as ``spk2EEG/LFP'')
\item the direction that goes from mass circuit activity towards single-neuron
activity. The latter, in turn, can be measured by the firing rates
(``EEG/LFP2FR'') or by the spike trains (``EEG/LFP2spk'')
\end{itemize}
A series of recent studies has investigated the relationship between
mass signals, measured with LFPs/EEGs, and firing activity in different
cortical areas and during both stimulation and absence of stimulus
\citep{hall2014real,musall2014effects,ng2013eeg,zanos2012relationships,bansal2011relationships,okun2010subthreshold,nauhaus2009stimulus,whittingstall2009frequency,rasch2009neurons,Rasch08,mukovski2007detection}.
In the following, we briefly review the progress made by these studies,
and we highlight the questions that these studies left open and that
we have tried to address in this thesis. 

Whittingstall and Logothetis \citep{whittingstall2009frequency} recorded
simultaneously multi-unit activity and EEGs/LFPs from primary visual
cortex during visual stimulation. Spectral analysis of EEGs/LFPs recordings
reveals that cortical activity presents a reach spectrum of oscillatory
activity spread over a wide range of frequencies. Thus they performed
a frequency decomposition and focused on the relationship between
the network oscillations obtained from LFPs/EEGs and the multi-unit
activity in the direction from EEG/LFP to firing rate (i.e., EEG/LFP2FR).
In both cases (EEG and LFP), they found that the time course of MUA
on a scale of 10-50 ms related statistically both to the delta-band
(2-4 Hz) phase and to the gamma-band (30-100 Hz) amplitude of mass
signals, with a linear combination of gamma power and delta phase
affording more predictability of the multi-unit spike rates than either
signal alone. In a second study \citep{musall2014effects}, they extended
this study by investigating how synchrony between different multi-unit
sites relates to EEG amplitude. 

Panzeri and colleagues \citep{Mazzoni10} developed a network model
that provides a detailed mechanistic explanation (both during visual
stimulation and during spontaneous activity) of the relationships
between activity of excitatory neurons and the phase and amplitude
of EEG/LFP rhythms at different frequencies. This modeling accounted
well many of the aspects of EEG/spikes relationships observed experimentally
by Whittingstall and Logothetis. 

In alternative to studying relationships between EEG/LFP networks
rhythms and spike times in the frequency domain, Kreiman and colleagues
\citep{rasch2009neurons} investigated the relationship between multi-unit
spikes and LFPs in the time domain during both visual stimulation
and absence of stimulus. Their approach was to approximate the LFP
time series as a linear convolution of the spike train time series
with a temporal \textquotedblleft kernel\textquotedblright{} (i.e.,
spk2LFP investigation) that describes the spike-field relationship
(this \textquotedblleft kernel\textquotedblright{} is similar, though,
not identical, to the spike triggered LFP average). This time domain
approach is of interest for two reasons. First, if the \textquotedblleft kernels\textquotedblright{}
are found to be relatively constant across cells, this approach could
be in principle used to estimate the spike rate from the LFP by \textquotedblleft deconvolving\textquotedblright{}
the LFP time series with the inverted kernels. Second, these kernels
are also useful to describe empirically how LFPs/EEGs are generated
within the cortical circuit. Indeed, they are a measure of the spike-LFP/EEG
relationship more accurate with respect to the spike-triggered average
of the local filed potential, since they discount the confounding
factors (in spike-LFP/EEG relationship) that can be given by spatiotemporal
correlation of spikes (see section \vref{STA_vs_wienerKernel}). Thus
they can be used to estimate, for example, whether the LFP or EEG
generation depends on dynamical network parameters such as the cortical
excitability \citep{Einevoll2013}. \\
In another study \citep{Rasch08} Logothetis and colleagues estimated
the spike trains of MUA (with a sampling frequency of 200Hz) from
the LFP (i.e., LFP2spk direction) in monkey primary visual cortex
during both visual stimulation and absence of stimulus. They used
a support vector machine (SVM) to perform binary classification. The
learning algorithm selected a certain number of features (up to 116)
of the LFP in order to maximize the estimation performances. The LFP
features were obtained both from time and frequency domain and the
preferred one is the amplitude of the LFP power fluctuations in the
high gamma band (40-90Hz).

Finally, another study \citep{ng2013eeg} estimated the similarity
between the stimulus selectivity of the firing of auditory cortical
neurons in monkeys and the stimulus selectivity that is obtained by
EEGs recorded in humans using the same auditory stimuli in both species.
The results of this investigation showed that the delta phase is the
parameter of auditory cortex EEG that gives decoding and stimulus
selectivity closer to the firing of cortical neurons, suggesting that
EEG delta phase may be a good proxy for inferring how neurons encode
information. 

All the above mentioned studies investigate the relationship between
LFPs/EEGs and spiking activity of small neuronal populations (or by
means of single-unit activity obtained by applying a spike sorting
algorithm to MUA signal). A very recent paper \citep{hall2014real}
introduced an important novelty: they examined the effect of using
multiple\footnote{Obtained from multichannel (up to 20) recordings.}
(instead of individual) LFPs to estimate individual firing rate activity
and vice versa. However all these works evaluated the firing activity
starting from the multi-unit activity. We note that MUA (see section
\vref{section_LFP}) depends on the (extracellular) action potentials
of a small group of neurons and does not permit any discrimination
of the neural cell types contributing to it, although presumably such
recordings capture mostly spikes from pyramidal neurons, because of
their larger size and higher number \citep{Logothetis03}. Thus:
\begin{itemize}
\item previous investigations of the relationship between EEGs/LFPs and
firing activity lack a characterization of the cell type from which
the firing activity is recorded, and thus cannot reveal the relationship
between network-level activity and the firing of specific cell types.
\item given that the previous investigations used MUA recordings of spiking
activity, and given that MUA has a bias toward measuring pyramidal
neurons, these previous investigations cannot tell anything about
the relationship between the EEGs/LFPs and the activity of neurons
that are not pyramidal cells
\end{itemize}
As a consequence\textcolor{black}{, previous work failed to show light
on how the activity of specific classes of interneurons relate to
the EEGs/LFPs. W}hether or not such relationships can be detected
is made difficult by the fact that inhibitory interneurons, due to
their approximately symmetric shape (i.e., star-shaped dendrites),
generate a dipole for each spike or synaptic event that is approximately
10 times smaller than that of pyramidal neurons\citep{murakami2006contributions}. 

In our work we focus precisely on the relationship between the firing
activity of individual genetically-identified interneurons and EEGs/LFPs. 

\section{Materials and methods}

Note that we performed the same analysis for LFP and EEG signal. In
the materials and methods section (text and equations), we reported
the LFP case, but it is understood that each time you can replace
LFP with EEG.

\subsection{In vivo LFP, EEG and two-photon guided juxtasomal recordings\label{PF_section_exp_setup}}

All experiments were performed in mice (25-30 days old) under urethane
anaesthesia by Stefano Zucca, Tommaso Fellin and other colleagues
in the laboratory of Tommaso Fellin at IIT. These data were kindly
provided to me for the present analysis. Details of the experiments
are concisely reported below.

PV-Cre (B6;129P2- Pvalbtm1(cre)Arbr/J, Jackson Laboratory, Bar Harbor,
USA) and SST-Cre (Ssttm2.1(cre)Zjh/J , Jackson Laboratory, Bar Harbor,
USA) transgenic mice were crossed with the TdTomato (B6;129S6-Gt(ROSA)26Sortm14(CAG-tdTomato)Hze/J)
reporter line and used for simultaneous recordings of the electroencephalogram
(EEG), local field potential (LFP) and single-cell spiking activity.
EEG recordings were obtained by placing two epidural stainless steel
wires unilaterally at about 3.5 mm distance (in the rostro-caudal
direction) from one another. The EEG signal was amplified using an
AM-amplifier (AM-system, Carlsborg, WA), sampled at 10 kHz and stored
with PatchMaster software. For juxtasomal recordings, patch pipettes
(resistance: 4 \textendash{} 9 M\textgreek{W}), filled with artificial
cerebrospinal fluid solution mixed with Alexa Fluor 488 (20 \textgreek{m}M),
were lowered through a small craniotomy placed ipsilaterally and in
between the two EEG recording sites. Parvalbumin-positive (PV-pos)
and Somatostain-positive (SOM-pos) interneurons were identified based
on their TdTomato fluorescence in double transgenic mice under the
two-photon microscope ($\lambda_{exc}$ = 720 nm). Single-cell spiking
activity was recorded with an ELC-01X amplifier (NPI electronic instruments),
the signal was sampled at 10 kHz and stored in the computer via PatchMaster
software. Simultaneous LFP recording was performed by placing a low
resistance glass pipette (0.8 \textendash{} 1 M\textgreek{W}) at a
distance < 500 \textgreek{m}m from the recorded cel\textcolor{black}{l.
The LFP signal was amplified, sampled and stored in the same way as
the EEG signal (datasets: 4 mice, 21 cells for PV-pos and 8 mice,
18 cells for SOM-pos cells). }
\begin{figure}
\begin{centering}
\includegraphics{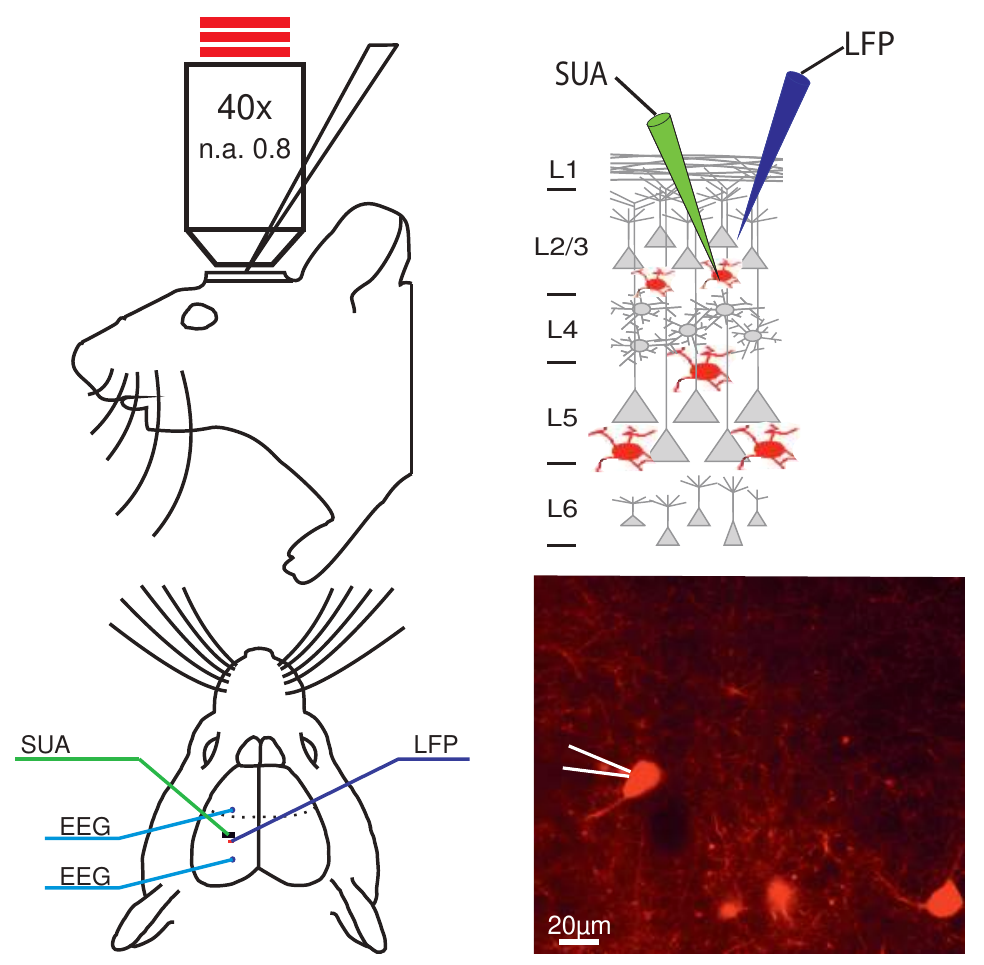}
\par\end{centering}
\centering{}\caption[Experimental setup]{\textbf{In vivo two-photon interneuron identification. }\textbf{\textcolor{black}{On
the left}}\textcolor{black}{: schematic representation of the experimental
setup. Headfixed anesthetized mice were placed under the two-photon
microscope and cells were visualized looking at their fluorescence
(\textgreek{l}\protect\textsubscript{\textcolor{black}{exc}} = 720
nm) through a 40x water-immersion objective (upper part). LFP and
SUA recordings were performed by lowering glass pipettes in two small
craniotomies closed to each other and indicated by respectively violet
and green lines. Cyan lines show the two recording sites for EEG.
}\textbf{\textcolor{black}{Right Upper}}\textcolor{black}{: representation
of cortical architecture and experimental configuration: red cells
indicate Parvalbumin-positive interneurons; green and violet pipettes
display respectively SUA and LFP recording sites. }\textbf{\textcolor{black}{Right
Lower}}\textcolor{black}{: fluorescence imaging showing in-vivo Parvalbumin-positive
interneurons identified under the two-photon microscope and a glass
pipette (white) placed in close contact for juxtasomal recordings.}\textbf{
\label{PFfig_exp_setup}}}
\end{figure}

\subsection{In vivo LFP and Patch-Clamp recordings\label{PF_section_PYR5_LFP}}

Experiments were performed in PV-Cre mice (25-30 days old) under urethane
anaesthesia. For single cell whole cell recordings a patch pipette
filled with an intracellular solution (composition in mM: K-gluconate
140, MgCl2 1, NaCl 8, Na2ATP 2, NaGTP 0.5, HEPES 10, phosphocreatine
10 to pH 7.2 with KOH) was lowered through the tissue (depth: from
450 \textmu m to 700 \textmu m) while applying a small positive pressure.
Once the cell was reached, a negative pressure was imposed to achieve
a stable seal (G\textgreek{W}) and the membrane was then carefully
broken. Pyramidal neurons were identified in current clamp configuration
by looking at their spiking activity during depolarizing steps of
current injections \citep{contreras2004electrophysiological,cauli2000classification}.
Single cell membrane potential values were amplified, sampled and
stored as the extracellular juxtasomal signals. By positioning a second
glass pipette at the same depth of the recorded cell, the LFP was
acquired in same conditions as reported above (dat\textcolor{black}{aset:
3 mice, 7 cells).}

\subsection{Data preprocessing}

The recordings are composed by temporal sessions of variable length
($\leq750$ sec long) which are preceded and followed by a break in
the data acquisition. Each cell can have multiple temporal sessions
(from 1 to 4) that we divided in segments 240 sec long we call ``trials''
after discarded an initial transient (from 5 to 20 sec) to avoid any
potential border effect. For the PV-pos dataset, this results in a
total number of 120 trials. Then we divided each session in segments
240 sec long we call \textquotedblleft trials\textquotedblright .
Eventually we split each trial into two segments of equal length and
in the whole analysis the first 120 sec are used as training set (to
compute the Wiener filters and the coefficients of the GLM), while
the second half of each trial belonged to the test set and is used
to perform the estimation.

All the analysis has been performed using MATLAB (MathWorks). In order
to facilitate the computation, LFP signals were decimated to 500 Hz.\footnote{The decimation (``decimate'' function in Matlab) performed an intrinsic
low-pass filter with cutoff frequency at 200 Hz. Since we are interested
in the LFP spectrum below 100 Hz, the decimation does not affect the
results.} To detect spike times, firstly we applied a high-pass filter to the
mean-subtracted 10 kHz juxtasomal signal (Kaiser filter with zero
phase lag and 0.5 Hz bandwidth, very small passband ripple (0.05 dB)
and high stopband attenuation (60 dB), cutoff frequency of 100 Hz)
and then we applied a detection threshold. Depending on the noise
level, the thresholds could vary across temporal series and the median
value was 9 SDs of the filtered juxtasomal signal (min=4.5, max=12).
Eventually the spike times are downsampled to 500 Hz (like the LFP
signal). Note that in this way we obtain the spikes emitted only by
the two-photon targeted cell, without need of applying any spike sorting
algorithm (because the recordings are juxtasomal).

\begin{figure}
\begin{centering}
\includegraphics[scale=0.5]{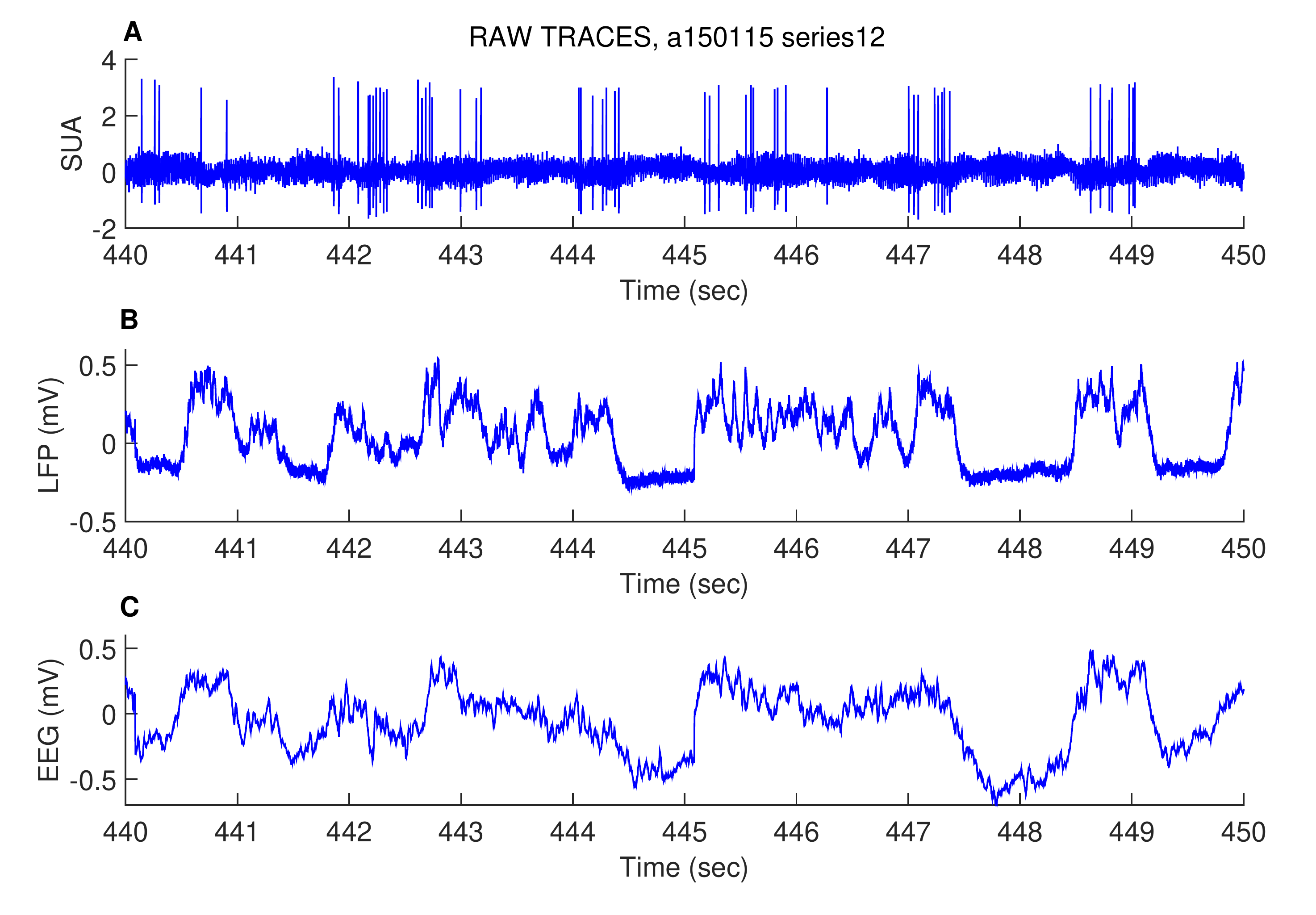}
\par\end{centering}
\centering{}\caption[Raw example traces]{ 10 seconds \textbf{raw traces of simultaneous recordings} obtained
from three different recording sites. \textbf{(A)} Juxtasomal recording
from a PV-pos interneuron. \textbf{(B)} LFP recorded from a glass
pipette (at a distance < 500 \textgreek{m}m from the recorded cell)
and decimated to 500Hz. \textbf{(C)} EEG decimated to 500Hz. \label{PFfig_raw_traces} }
\end{figure}

\begin{figure}
\begin{centering}
\includegraphics[scale=0.5]{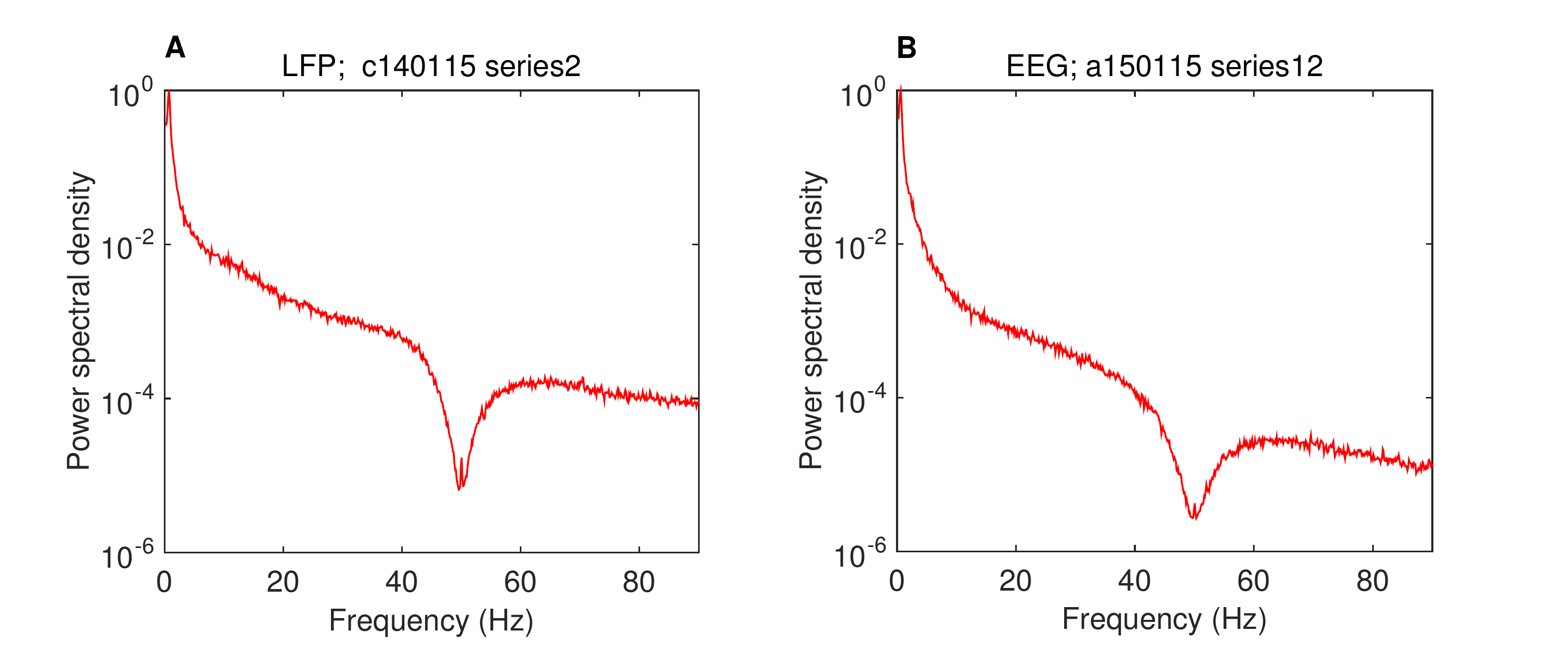}
\par\end{centering}
\centering{}\caption[LFP/EEG power spectral density]{\textbf{LFP and EEG power spectral densit}y. \textbf{(A)} Typical
LFP power spectral density obtained from a temporal series of 750sec.
The power density reduction observed around 50 Hz is due to a band-pass
filter performed by the amplifier in order to remove artefacts due
to electrical equipments.\textbf{ (B)} Same as (A) for EEG signal.
\label{PFfig_example_power_LFP_EEG}}
\end{figure}

Traces of the LFP (decimated to 500 Hz) recorded concurrently with
the juxtasomal signal is displayed in figure \ref{PFfig_raw_traces},
while a typical example of the LFP power spectral density is shown
in figure \ref{PFfig_example_power_LFP_EEG}.

\begin{figure}
\begin{centering}
\includegraphics[scale=0.5]{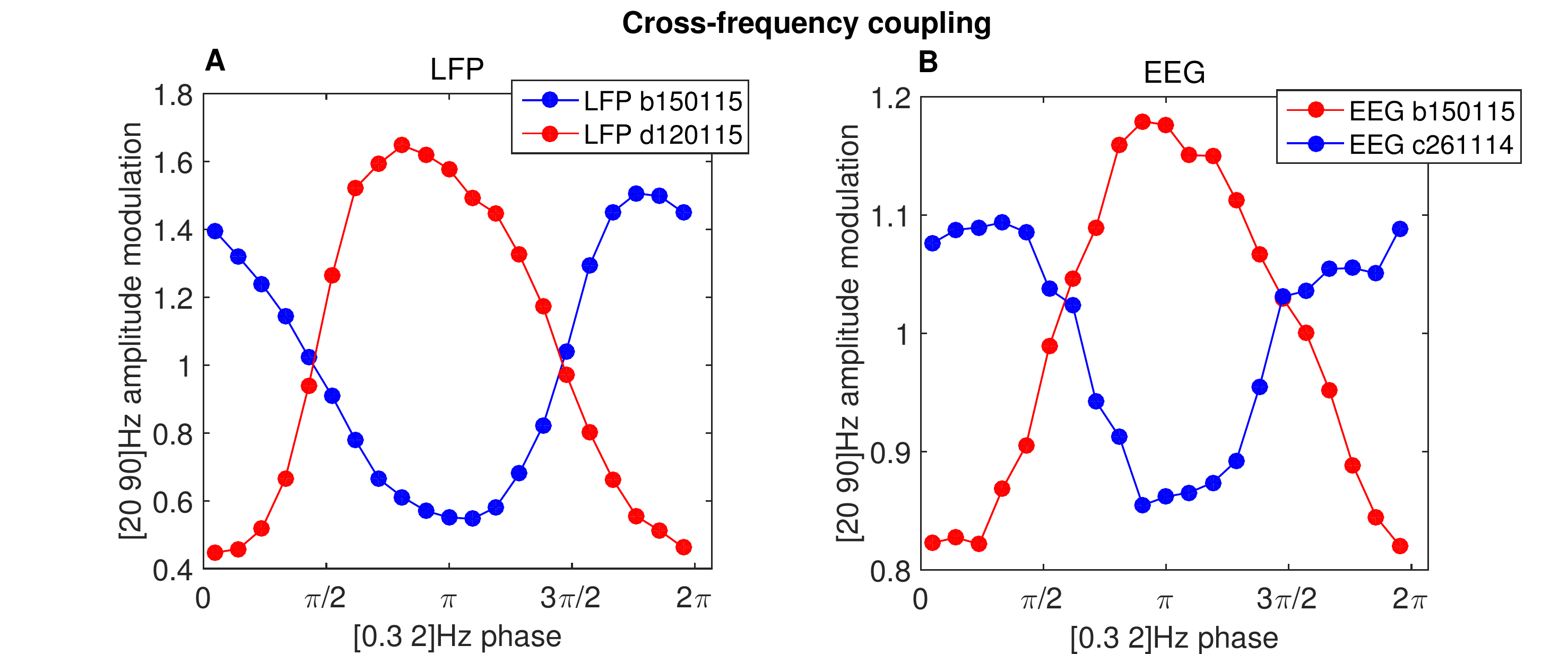}
\par\end{centering}
\centering{}\caption[LFP/EEG cross-frequency coupling]{\textbf{LFP and EEG cross-frequency coupling.} Gamma, {[}20 90{]}Hz,
amplitude modulation of the LFP \textbf{(A)} and EEG \textbf{(B)}
signals plotted as a function of their own low delta, {[}0.3 2{]}Hz,
phases. In particular, cross-frequency coupling is computed as follows.
First, we binned into 21 equispaced intervals the range of low delta
phase angles from 0 to 2$\pi$, and then in each phase interval we
computed the mean gamma amplitude over all data points whose phase
belonged to that phase interval: this value is the mean gamma amplitude
associated with the given phase interval. Finally, in order to obtain
the amplitude modulation, we divided this value by the global mean
gamma amplitude (over the whole temporal session, usually 750sec long).
When the cross-frequency coupling resulted in a downward valley (blue
lines), we inverted the sign of the signal.\label{PFfig_CFC}}
\end{figure}

Depending on the channel takes as reference, both the LFP\footnote{Note that the LFP is recorded always from layer 2 (see figure \ref{PFfig_exp_setup}),
thus its polarity does not depend on the depth of the recording.} and the EEG could have inverted signs. In order to detect the sign,
we computed the dependence of the high frequency power (20 90)Hz on
the low frequency (0.3 2)Hz phase by means of the cross-frequency
coupling of the signal (see figure \ref{PFfig_CFC}). We then aligned
all signals in the same way by reversing the sign of those signals
whose cross-frequency coupling had a downward valley at phase value
$\pi$ (blue lines in figure \ref{PFfig_CFC}).

Finally we low-pass filtered the LFP (Kaiser filter with zero phase
lag and 0.5 Hz bandwidth, very small passband ripple, 0.05 dB, and
high stopband attenuation, 60 dB) with a cutoff frequency of 90 Hz.

\subsection{Linear estimation of the time-varying signals }

We investigated whether we could linearly estimate the LFP and the
EEG from the spiking activity of a single genetically-identified neuron
(a PV-pos neuron, with the exception of figure \ref{PFfig_spk2LFP_SOM_PYR5_datasets}),
and whether we could estimate the firing activity of a neuron from
the mass signals (that it, the ``inverse'' estimation, see figure
\ref{PFfig_schematic_analysis.}). We also asked if the estimation
is general and robust across cells and animals. 

\subsubsection{Estimation of LFP and EEG from single-unit spike train}

The linear estimation was performed by using the Wiener kernel (see
section \vref{Wiener_kernel_definition}). In particular, when we
estimated mass signals from spiking activity (i.e., ``spk2LFP''),
we defined the spike train (see equation \ref{eq_spke_train}) with
its mean value, $\rho_{0}$ , subtracted out:
\begin{equation}
\rho(t)=\underset{i}{\sum}\delta(t-t_{i})-\rho_{0}.
\end{equation}
According to equation \ref{eq_linear_approximation}, the linear estimation
of the (zero-mean) LFP\footnote{Remember that we performed the linear estimation of LFP and EEG by
using the same method. Therefore, if estimating the EEG signal, you
simply need to substitute ``EEG'' to ``LFP'' in the equations
from \ref{PFeq_LFPest} to \ref{PFeq_spk2LFP_mean_filter}. } signal had the following expression:
\begin{equation}
\mathrm{LFP_{est}}(t)=\int_{0}^{T}d\tau\:h_{spk2LFP}(t-\tau)\rho(\tau),\label{PFeq_LFPest}
\end{equation}
where the (trial-specific) Wiener filter see equation \ref{eq_wiener_kernel})
was computed as follows:
\begin{equation}
h_{spk2LFP}(t)=\frac{1}{2\pi}\int_{-\infty}^{+\infty}d\omega\:\frac{\tilde{Q}_{spkLFP}(\omega)}{\tilde{Q}_{spkspk}(\omega)}e^{-i\omega t}=\frac{1}{2\pi}\int_{-f_{cut}}^{+f_{cut}}d\omega\:\frac{\tilde{Q}_{spkLFP}(\omega)}{\tilde{Q}_{spkspk}(\omega)}e^{-i\omega t},\label{PFeq_spk2LFP_filter}
\end{equation}
where the $\tilde{f}$ indicates the Fourier transform of $f$ and
$f_{cut}$ is the cutoff frequency of the LFP. This filter is called
``trial-specific'' because the cross-power spectral density between
the spike train and the LFP, $\tilde{Q}_{spkLFP}(\omega)$, and the
power spectral density of the spike train, $\tilde{Q}_{spkspk}(\omega)$,
are relative to (the training set of) a single trial. In this way
each trial has its own filter. We also considered two other types
of Wiener filter with increasing generality: the ``cell-specific''
filters and the ``general'' filter. The first filter was computed
from and applied to all the trials belonging to a given cell\footnote{Note that each cell can have a different number of trials, resulting
in a median of 6 trials per cell with values from 1 to 9.}, while the general filter was obtained from and allied to the entire
dataset. These filters were computed using the same mathematical procedure
used to obtain the trial-specific filter, that is the minimization
of the sum of the mean squared errors in the estimation over all the
considered trials, as done in \citep{rasch2009neurons}. Thus, when
computing a mean filter, $h^{mean}$, over $N$ trials, we minimized
the following quantity: 
\begin{equation}
\mathrm{MSD}(\mathrm{LFP,LF}\mathrm{P}_{\mathrm{est}})=\sum_{j=1}^{N}\int_{0}^{T}dt\left[\mathrm{LF}\mathrm{P}_{j}(t)-\int_{0}^{T}d\tau\:h_{spk2LFP}^{mean}(t-\tau)\rho_{j}(\tau)\right],\label{PFeq_spk2LFP_MSD}
\end{equation}
where $LFP_{j}$ and $\rho_{j}$ are respectively the LFP and the
spike train (with the mean subtracted out) of the j-th trial. The
explicit expression of the Fourier transform of the mean Wiener kernel
takes the following form (see \citet{Rasch08} for a derivation):
\begin{equation}
\tilde{h}_{spk2LFP}^{mean}(\omega)=\frac{\sum_{j=1}^{N}\tilde{Q}_{spkLFP}^{j}(\omega)}{\sum_{j=1}^{N}\tilde{Q}_{spkspk}^{j}(\omega)}=\frac{\sum_{j=1}^{N}\tilde{Q}_{LFPspk}^{j}(-\omega)}{\sum_{j=1}^{N}\tilde{Q}_{spkspk}^{j}(\omega)}.\label{PFeq_spk2LFP_mean_filter}
\end{equation}

\subsubsection{Estimation of single-unit firing rate from LFP and EEG }

\subsubsection*{Firing rate computation\label{PF_section_FRcomputation}}

We performed a linear estimation also in the opposite direction, that
is we estimated single-unit firing rate from the mass signal (i.e.,
``LFP2FR''). A signal linearly estimated from a continuous (i.e.,
analogue) signal (such as LFP and EEG) will be in turn continuous.
Thus, in order to increase the estimation efficiency, we smoothed
the original sequence of spikes in order to obtain a continuous signal
we called ``firing rate'', FR. This smoothing procedure was performed
as follows: first, we averaged the spike train over a rectangular
sliding window (i.e., a spike smoothing window, SSW), then we convolved
the obtained signal with an Hann window 26 ms wide\footnote{In order not to affect the FR units, the Hann window had unitary integral.}
\citep{theunissen2000spectral}. Unless otherwise stated, the SSW
amplitude was 10 ms\footnote{Note that we varied this parameter in the range from 6 to 50 ms and
we chose 10 ms because it maximized the spike train estimation performances.}.

\subsubsection*{Wiener Filter}

Once we obtained the original firing rate (on the training set), we
estimated the FR (on the test set) from the LFP (and in the same way
also from the EEG) by using the equation we previously adopted to
estimate the LFP from the spiking activity, that is:
\begin{equation}
FR_{\mathrm{est}}(t)=\int_{0}^{T}d\tau\:h_{LFP2spk}(t-\tau)\mathrm{LFP}(\tau),\label{PFeq_FRest}
\end{equation}
where
\begin{equation}
h_{LFP2spk}(t)=\frac{1}{2\pi}\int_{-f_{cut}}^{+f_{cut}}d\omega\:\frac{\tilde{Q}_{LFPspk}(\omega)}{\tilde{Q}_{LFPLFP}(\omega)}e^{-i\omega t}.
\end{equation}

Analogously to the mass signals estimation, the above equation represents
the trial-specific filter. Also in this case we considered the cell-specific
and the general filter, which were defined as in equation \ref{PFeq_spk2LFP_mean_filter}
by substituting ``spk'' with ``LFP'' and ``LFP'' with ``FR''.
\\
Note that in order to evaluate how effective was the filter in performing
the estimation, we computed also the estimation performances obtained
by assuming simply $FR_{est}=LFP$ (without performing any filtering
procedure)

\subsubsection*{General Linear Model (GLM)\label{PF_section_GLM}}

\begin{figure}
\begin{centering}
\includegraphics[scale=0.5]{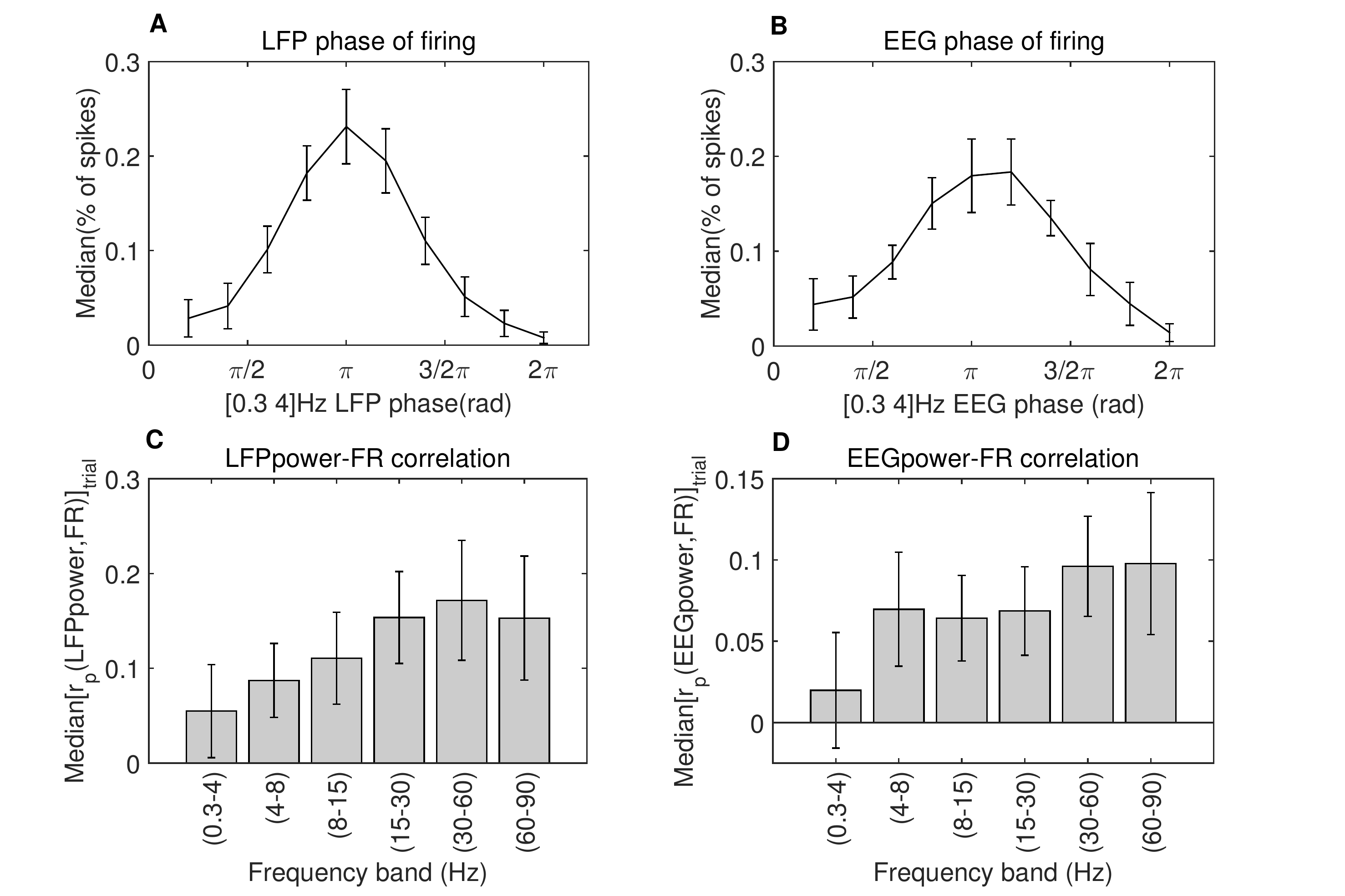}
\par\end{centering}
\centering{}\caption[Setting GLM parameters]{\textbf{Setting general linear model parameters}. We compared the
FR estimation obtained by means of a Wiener kernel with the estimation
obtained by using a general linear model, as done in \citep{whittingstall2009frequency}.
This GLM consisted of three regressors: the phase of the slow network
oscillations, the power of the gamma network oscillations and a constant
term. \textbf{(A)} LFP delta phase of firing. The LFP is bandpassed
in the delta band {[}0.3 4{]}Hz and the instantaneous phase of the
bandpassed signal is computed by means of an Hilbert transform. Then
the phase values are binned into 10 equispaced intervals and each
spike is assigned to the bin corresponding to the LFP phase assumed
when it was fired. \textbf{(B)} Same as (A) for EEG phase of firing.
\textbf{(C)} Median Pearson's correlation between the instantaneous
power (extracted by the Hilbert transform) of the bandpassed LFP and
the FR. The band used to build the power regressor (i.e., the one
that gives the highest performance, data not shown) is the low gamma
band, {[}30 60{]}Hz. \textbf{(D)} Same as (C) for EEG power; the band
used to build the power regressor is the high gamma band, {[}60 90{]}Hz.
In all the panels median values are computed over the training set
and the error bars display the interquartile ranges.\label{PFfig_LFP/EEG2spk_Setting-GLM-parameters.}}
\end{figure}

In order to evaluate better the results of the FR estimation, we compared
the estimation performances obtained by using the kernel with the
ones obtained by adopting a general linear model based on frequency
decomposition. This model had been used to estimate the MUA firing
rate from LFP in a previous work \citep{whittingstall2009frequency},
to which we refer for a full description. Briefly, the GLM performs
a linear estimation of the true FR by using three regressors: the
time resolved power of a given frequency band of the LFP, $Pow_{\mathrm{band1}}$,
the instantaneous phase of a given frequency band of the LFP, $Ph_{\mathrm{band2}}$,
and a constant term, $k$:
\begin{equation}
FR_{\mathrm{est}}(t)=\beta_{1}Pow_{\mathrm{band1}}(t)+\beta_{2}Ph_{\mathrm{band2}}(t)+k.\label{PFeq_GLM}
\end{equation}
The coefficients $\beta_{1}$, $\beta_{2}$ and the constant term,
$k$, were computed by minimizing the mean squared difference between
the true and estimated FRs (as in the Wiener filter case). The frequency
bands used to compute the power and the phase of the oscillations
were chosen to maximize the estimation performance (data not shown)
among the six traditional EEG bands: delta (<4 Hz), theta (4\textendash 8
Hz), alpha (8\textendash 15 Hz), beta (15\textendash 30 Hz), low (30\textendash 60
Hz), and high (60\textendash 100 Hz) gamma. In particular, band1 and
band2 correspond respectively to the band that had the highest correlation
between the oscillatory power and the FR (i.e. {[}30 60{]}Hz for LFP
and {[}60 90{]}Hz for EEG, see panels C,D figure \ref{PFfig_LFP/EEG2spk_Setting-GLM-parameters.})
and the highest phase of firing (<4Hz, see panels A,B figure \ref{PFfig_LFP/EEG2spk_Setting-GLM-parameters.},
where only the band with the highest phase of firing is displayed).
The band-passing procedure was performed by using a Kaiser filter
with zero phase lag and 0.1 Hz bandwidth, very small passband ripple
(0.05 dB) and high stopband attenuation (60 dB). Then respectively
the oscillatory power and phase were obtained as the magnitude and
the phase of the Hilbert transform of the band passed signal. Eventually
both phase and power were normalized resulting in values between 0
and 1. In particular, the oscillatory power was normalized to its
peak value in each single trial. The phase regressor was created by
normalizing the instantaneous phase to its peak phase of firing probability
in each single trial.

The computation of the GLM was carried out in each single trial, resulting
in a ``trial-specific'' GLM. Analogously to what has been done when
performing the estimation by means of the Wiener filter, we also computed
the ``cell-specific'' and ``general'' GLM by averaging the weights
$\beta_{1}$, $\beta_{2}$ and the constant term $k$ across the trials
belonging to a given cell and all the trials, respectively.

\subsubsection{From firing rate to spike times\label{PF_section_thresholds}}

From the estimated FR, we extracted the spike times of the neuron
by using a non-linear threshold. The FR represents a probability of
firing, thus the simplest way to extract spike times relied in detecting
a spike each time the FR\textsubscript{est} had a local maximum that
overcame a given threshold. As a consequence, the value of the threshold
determined the number of spikes in the estimated spike train, $\langle FR\rangle_{\mathrm{est}}$.
As reference value we used thresholds set to get a number of estimated
spikes equal to the number of true spikes; we identified the spike
train estimation performed by using this threshold by saying that
the $\langle FR\rangle_{\mathrm{est}}$ was ``exact'' (see for example
figure \ref{PFfig_LFP/EEG2spk_Example_ottimo}). Therefore, in this
case, we need to know the true average firing rate in each trial,
$\langle FR\rangle_{\mathrm{trial}}$, to obtain the estimated spike
train.\\
In order to quantify if and to which extent the estimation depends
on the exact knowledge of $\langle FR\rangle_{\mathrm{trial}}$, we
used also a less specific (\textquotedblleft general\textquotedblright )
threshold (data not shown). In particular, when using the general
threshold, the estimated average firing rate takes (no longer the
same value as the recorded firing rate but) one out of three possible
values which in turn depends on the recorded FR,$\langle FR\rangle_{\mathrm{trial}}$.
More precisely these values corresponded to the 17th, 50th and 83rd
percentile of the average firing rates distribution (see figure \ref{PFfig_avFR})
and represented respectively a low, medium and high firing activity
for the considered dataset.. The neurons whose$\langle FR\rangle_{\mathrm{trial}}$
fell in the percentile interval (0-33.3) were assigned to the low
FR class (with $\langle FR\rangle_{\mathrm{est}}=2.7$ Hz, corresponding
to the 17th percentile), the neurons with average firing rate in the
percentile interval (33.3-66.6) was assigned to the medium firing
rate class ($\langle FR\rangle_{\mathrm{est}}=4.5$ Hz, that was the
median) and the other to the high class ($\langle FR\rangle_{\mathrm{est}}=9.4$
Hz, that was the 83rd percentile). In this case, we labeled the spike
train estimation by saying that $\langle FR\rangle_{\mathrm{est}}$
was ``similar'' (to the original one; see figures \ref{PFfig_LFP/EEG2spk_rec_spec_perf_VS_FRclass}
and \ref{PFfig_LFP/EEG2spk_general_perf_VS_FRclass}).\\
Eventually, in order to investigate more in general the dynamics of
the estimation as a function of the threshold, we also analyzed three
distinct cases where we adopted a threshold to get respectively the
low, medium and high firing activity defined above for all the neurons,
irrespective of the original firing rate. We identified these three
cases by saying that $\langle FR\rangle_{\mathrm{est}}$ was respectively
``low'', ``medium'' and ``high'' (see figures \ref{PFfig_LFP/EEG2spk_rec_spec_perf_VS_FRclass},
\ref{PFfig_LFP/EEG2spk_general_perf_VS_FRclass} and \ref{PFfig_LFP/EEG2spk_perf_VS_dtACC}).

\subsection{Analysis of cortical datasets}

\begin{figure}
\begin{centering}
\includegraphics[scale=0.6]{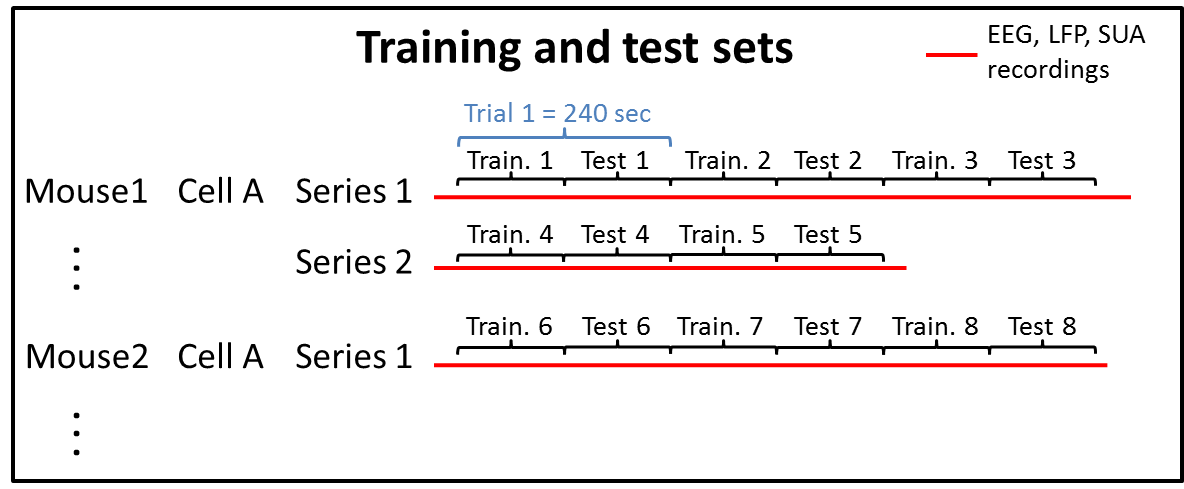}
\par\end{centering}
\centering{}\caption[Training and test sets]{\textbf{Training and test set division}. The original temporal series
(red lines) are divided in trials of 240 sec, and then each trial
is divided in two equal segments: the first 120 sec are used as training
set, while the second half belongs to the test set.\label{PFfig_training_test_set}}
\end{figure}

\begin{figure}
\begin{centering}
\includegraphics[scale=0.5]{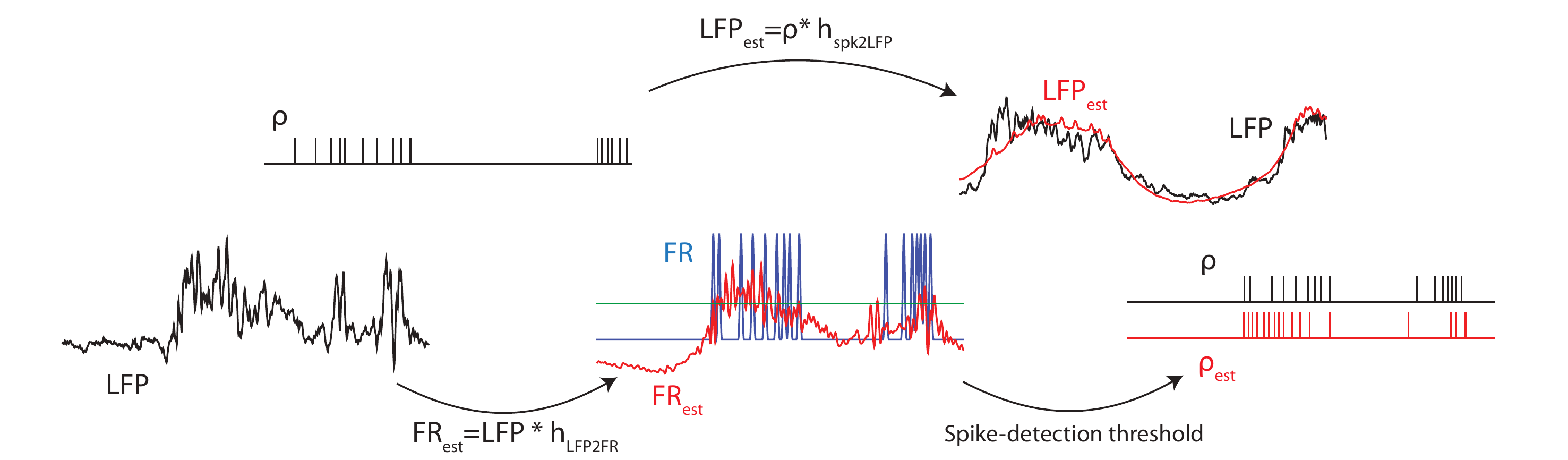}
\par\end{centering}
\centering{}\caption[Schematic signals' estimations]{\textbf{Schematic of the analysis performed. }We linearly estimated
the signals by means of Wiener kernels and we quantified how robust
the estimation is by comparing the performances obtained by trial-specific,
cell-specific filters and also when using a unique filter for all
the mice. \textbf{Top}: the estimated LFP/EEG is obtained by convolving
the relative Wiener filter, $h_{spk2LFP/EEG}$ with the spiking activity,
$\rho$, of single (both excitatory and inhibitory) neurons. \textbf{Bottom}:
firstly, the FR is estimated by convolving the LFP (or the EEG) with
the Wiener kernel, $h_{LFP/EEG2FR}$, then a spike-detection threshold
is applied to detect spike times. We compared this estimation with
the ones performed when (i) avoiding the convolution with the filter
(i.e., by applying the threshold directly to the LFP/EEG) and (ii)
estimating the FR by means of a general linear model based on frequency
decomposition of the mass signals used in \citep{whittingstall2009frequency}.\label{PFfig_schematic_analysis.}}
\end{figure}

Note that we analyzed all the recordings available without performing
any selecting based on the power of slow rhythms.\\
We divided each trial (240 sec long) into two segments of equal length:
the first 120 sec were used as training set to compute the filter,
while the second half of each trial belonged to the test set and was
used to evaluate the estimation performances (see figure \ref{PFfig_training_test_set}).

To compute the Wiener filter we estimated respectively the cross-power
spectral density and the power spectral density (see equation \ref{eq_FOURIER_wiener_kernel})
by using the ``cpsd'' and the ``pwelch'' Matlab functions, which
performed a Welch's averaged periodogram method with Bartlett windowing
and using nfft = 4096\footnote{With a sampling frequency of 500 Hz (as in our case), nfft=4096 results
in a filter 8192 ms wide.} for both the estimation directions. We also performed the analysis
using nfft = 2048 and the results were almost the same (data not shown),
since the filters are significant in an interval about 3000 ms wide
(see figure \ref{PFfig_spk2LFP/EEG-general-filters}). Eventually
we estimated the inverse Fourier transform needed to obtain the time
resolved filter (see equation \ref{eq_wiener_kernel}) with the fast
Fourier transform algorithm implemented in the ``ifft'' Matlab function.\\
We performed the convolution needed to obtain the estimated signal
(see equations \ref{PFeq_LFPest} and \ref{PFeq_FRest}) in the frequency
domain by using the fast Fourier transform implemented in the ``fftfilt''
Matlab function.

\subsection{Quantification of estimation performance\label{PF_section_Performance-measures}}

To quantify the estimation performance of the LFP, EEG and FR estimation
we computed the rank Spearman's correlation\footnote{Very similar results were obtained when considering the Pearson's
correlation (data not shown).} between the estimated, $y_{\mathrm{est}}$, and the true signals,
$y$, and the normalized mean squared distance, defined as follows:
\begin{equation}
\mathrm{NMSD}(y,y_{\mathrm{est}})=\frac{\mathrm{MSD}(y,y_{\mathrm{est}})}{\sigma_{y}^{2}},\label{eq_NMSD}
\end{equation}
where $\sigma_{y}^{2}$ is the variance of the true signal and the
mean squared distance, MSD, is defined in equation \ref{eq_MSD}.
NMSD takes values between 0 and 1 with 0 corresponding to perfect
estimation and 1 estimation not better than chance level \citep{gabbiani1998principles}.
This is true if $y_{est}$ is the optimal linear estimator of $y$
in the mean square sense. Thus, before computing NMSD, we multiplied
$y_{est}$ for the coefficient $k=\sum_{i}[y(t_{i})\cdot y_{\mathrm{est}}(t_{i})]$
/$\sum_{i}y_{\mathrm{est}}^{2}(t_{i})$, which minimizes the MSD between
$y$ and $y_{\mathrm{est}}$. In particular, when evaluating the FR
estimation performances, we rectified the estimated FR before computing
Spearman's correlation and NMSD.\\
To assess the statistical significance of the LFP\footnote{Remember that the same procedure was adopted also in case of EEG estimation.}
estimation, we compared the performance distribution with the one
obtained under the null hypothesis where the Wiener kernel, $h_{spk2LFP}^{\mathrm{rand}}$,
conveyed no information about the relationships between the temporal
structure of spike trains and LFP time courses. More specifically,
for each training trial, we generated a Poisson spike train with the
same average firing rate as the true spike train. We then computed
the Wiener filter $h_{spk2LFP}^{\mathrm{rand}}$ (that could be trial-specific,
cell-specific or general depending on the original filter we were
testing) as in equation \ref{PFeq_spk2LFP_filter} (or \ref{PFeq_spk2LFP_mean_filter})
by minimizing the MSD between the true LFP and the one estimated from
the Poisson spike train. Eventually we used the random filter to estimate
the LFP: 
\begin{equation}
\mathrm{LFP}_{est}^{\mathrm{rand}}(t)=\int_{0}^{T}d\tau\:h_{spk2LFP}^{\mathrm{rand}}(t-\tau)\rho(\tau).
\end{equation}
We repeated this procedure performing 50 different realizations of
the Poisson spike train for each trial and then we averaged the estimation
performance over the 50 realizations to obtain the average random
performance distribution used as null hypothesis. If the filter conveys
no information about the relationship between the temporal structure
of spike trains and LFP time courses, we would expect the estimation
performance of the estimated LFP to be close to the ones obtained
from the random filter.

To quantify the performances when estimating the spike times we need
to introduce a parameter, dt\textsubscript{accuracy}, that we use
to resample the spike trains. Note that we count the number of spikes
in each time window dt\textsubscript{accuracy}, thus the value of
the resampled spike trains in each dt\textsubscript{accuracy} can
be higher than one. We then compared the number of spikes in each
dt\textsubscript{accuracy} window in the true and estimated spike
trains by measuring the sensitivity and the precision of the estimation.
In particular, sensitivity is defined as follows 
\[
\mathrm{Sensitivity}=\frac{TP}{TP+FN},
\]
while precision is 
\begin{equation}
\mathrm{Precision}=\frac{TP}{TP+FP}.
\end{equation}
TP (true positive) is the number of estimated spikes that fitted true
spikes, FN (false negative) is the number of true spikes that do not
have corresponding estimated spikes and finally FP (false positive)
is the number of estimated spikes that do not have corresponding true
spikes. Note that this computation/comparison is performed step by
step in each dt\textsubscript{accuracy}. Thus the sensitivity measures
the percentage of the true spikes that are correctly estimated within
time windows dt\textsubscript{accuracy}wide, while the precision
is the percentage of the estimated spikes that correspond to true
spikes. Note that if the true and the estimated spike trains have
the same number of spikes (as in the case where $\langle FR\rangle_{est}$
is exact, see section \vref{PF_section_thresholds}), $FN=FP$ therefore
$\mathrm{Sensitivity}=\mathrm{Precision}$.\\
To assess the statistical significance of the spike train estimation,
we considered two different null hypotheses. The first is given by
the estimation performances obtained from a Poisson spike train having
the same average FR of the estimated spike train. (The Poisson process
is a point-like process that generates spike times with a given probability
at any given time, independently of spikes emitted at earlier or successive
times). For each trial we performed 50 different realizations of the
Poisson spike train and we took the average performance. This null
hypothesis is labeled as ``Poisson'' in figures \ref{PFfig_LFP/EEG2spk_rec_spec_perf_VS_FRclass}
and \ref{PFfig_LFP/EEG2spk_general_perf_VS_FRclass} and it quantifies
the performance only due to the knowledge of the average firing rate
(independently on the relationship between firing activity and LFP).
The second null hypothesis is obtained by the estimation performances
obtained randomly placing the estimated spikes in the intervals where
the estimated FR is above the spike-detection threshold (see section
\vref{PF_section_thresholds}). For each trial we repeated this procedure
50 times and we took the average performance distribution. This null
hypothesis is labeled as ``Shuffled'' in figures \ref{PFfig_LFP/EEG2spk_rec_spec_perf_VS_FRclass}
and \ref{PFfig_LFP/EEG2spk_general_perf_VS_FRclass} and it analyzes
if the estimated spikes are placed at random level where the FR\textsubscript{est}
is above the spike threshold.

All the performance measurements we adopted do not depend on the magnitude
of the signals. Nevertheless, when computing mean Wiener filters (see
equation \ref{PFeq_spk2LFP_mean_filter}) and mean GLM, the involved
signals had to have the same units across trials. In particular, the
firing activity was in units of (spikes/dt\textsubscript{sampling},
where dt\textsubscript{sampling}= 2 ms) and the LFP were in standard
deviations units (s.d.u.).

\newpage{}

\section{Results}

\begin{figure}
\begin{centering}
\includegraphics[scale=0.5]{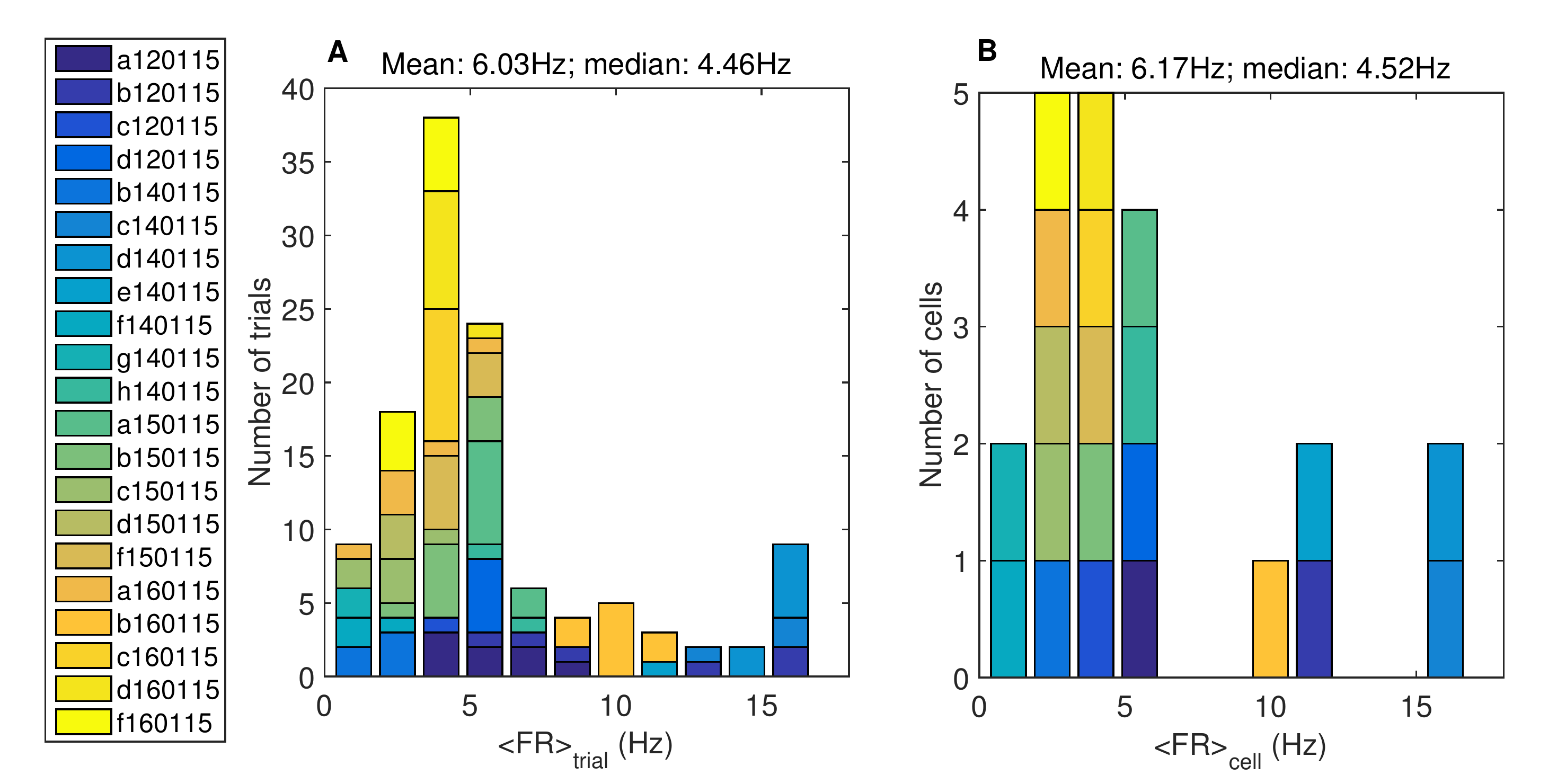}
\par\end{centering}
\centering{}\caption[Distribution of average FRs across trials and cells]{\textbf{Distribution of the average firing rates} across trials (240
sec)\textbf{ (A)} and cells \textbf{(B)}. The legend shows the cell
the data belong to. Mean and median values of the distributions are
displayed in the panel's titles.\label{PFfig_avFR} }
\end{figure}

While mass measures of circuit activity (as EEGs and LFPs) are analogue
signals, spike can be considered as point-like processes. Linear estimation
methods of analogue signals from point-like processes rely on some
kind of filtering operation on the sequence of times when the point-like
is present. Thus, the number of available spikes is crucial when performing
such a kind of estimation of mass signals. In particular, in our datasets
the spiking activity comes from single genetically-identified (i.e.,
PV-pos) interneurons (see section \vref{PF_section_exp_setup}), thus
the firing activity of the identified population of neurons is very
important. In figure \ref{PFfig_avFR}, we showed the distribution
of the average firing rates of the single PV-pos interneurons. These
neurons are fast-spiking neurons and this is the reason why we chose
them, with respect to the SOM-pos interneurons, which had an extremely
low firing activity (see figure \ref{PFfig_spk2LFP_SOM_PYR5_datasets}).

\subsection{Estimating LFP and EEG from SUA }

We performed a linear estimation of both the EEG and the LFP recorded
in layer 2 of neocortex of mice under anesthesia from the concurrently
recorded spiking activity of a single PV-pos interneuron (in layer
2) placed at a distance <500 \textgreek{m}m from the LFP pipette.
We then extended this analysis (see figure \ref{PFfig_spk2LFP_SOM_PYR5_datasets})
by also including spiking activity recorded from either a SOM-pos
interneuron in layer 2 and a pyramidal neuron in a deep layer (i.e.,
5 or 6). The estimation was done by means of the (first order) Wiener
kernel, as described in equation \ref{PFeq_LFPest}. In order to investigate
if and to which extent the estimation algorithm could be generalized,
we considered three kinds of filters, with increasing generality:
trial-specific, cell-specific and general filters. 

\begin{figure}
\begin{centering}
\includegraphics[scale=0.5]{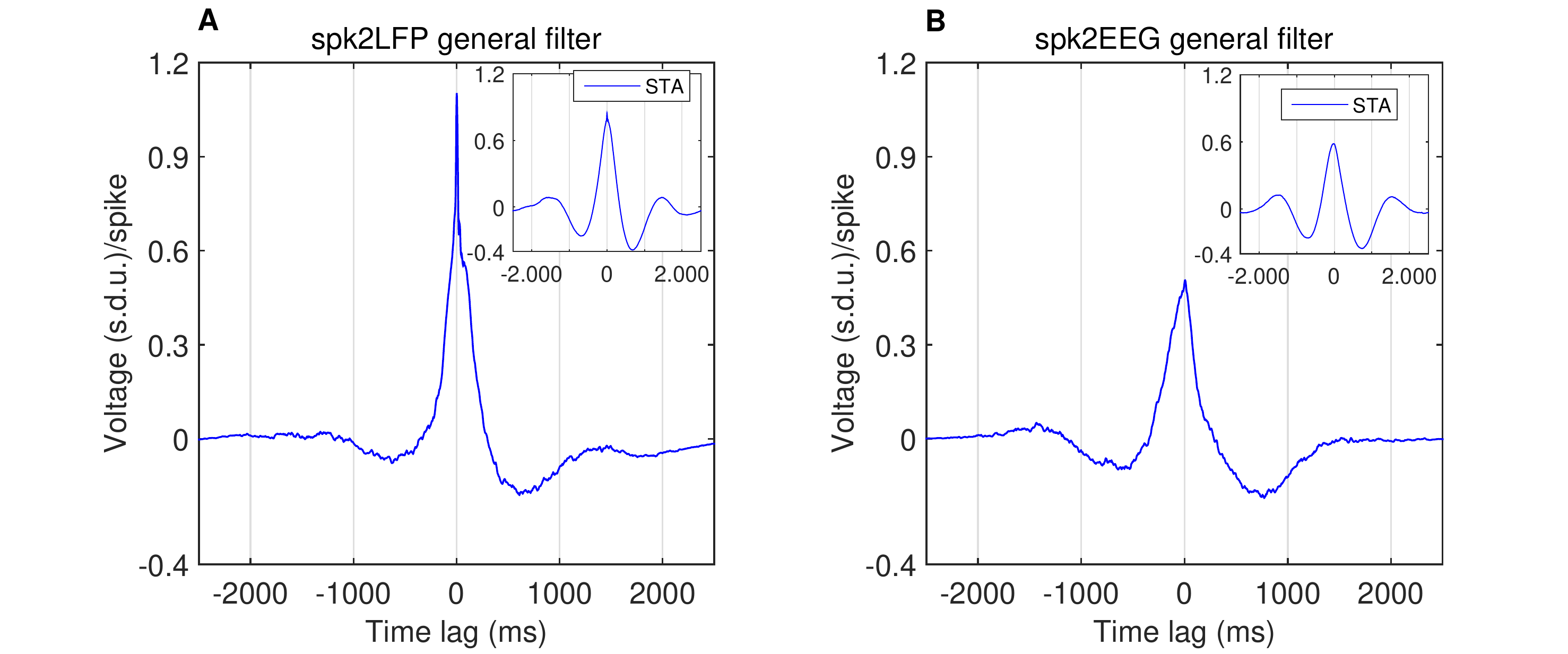}
\par\end{centering}
\begin{centering}
\caption[Spk2LFP/EEG general filter]{\textbf{General Wiener kernels for LFP and EEG estimation}. \textbf{(A)}
Mean filters (over all the trials) used to estimate the LFP from the
spiking activity of a PV-pos interneuron. The peak is at 2 ms lag.
The inset, which has the same axes than the main panel, shows the
LFP spike-triggered average as term of comparison (see section \vref{STA_vs_wienerKernel}).
\textbf{(B)} Same as (A) for EEG estimation. The peak is at 6 ms lag.\label{PFfig_spk2LFP/EEG-general-filters}}
\par\end{centering}
\end{figure}

\begin{figure}
\begin{centering}
\includegraphics[scale=0.5]{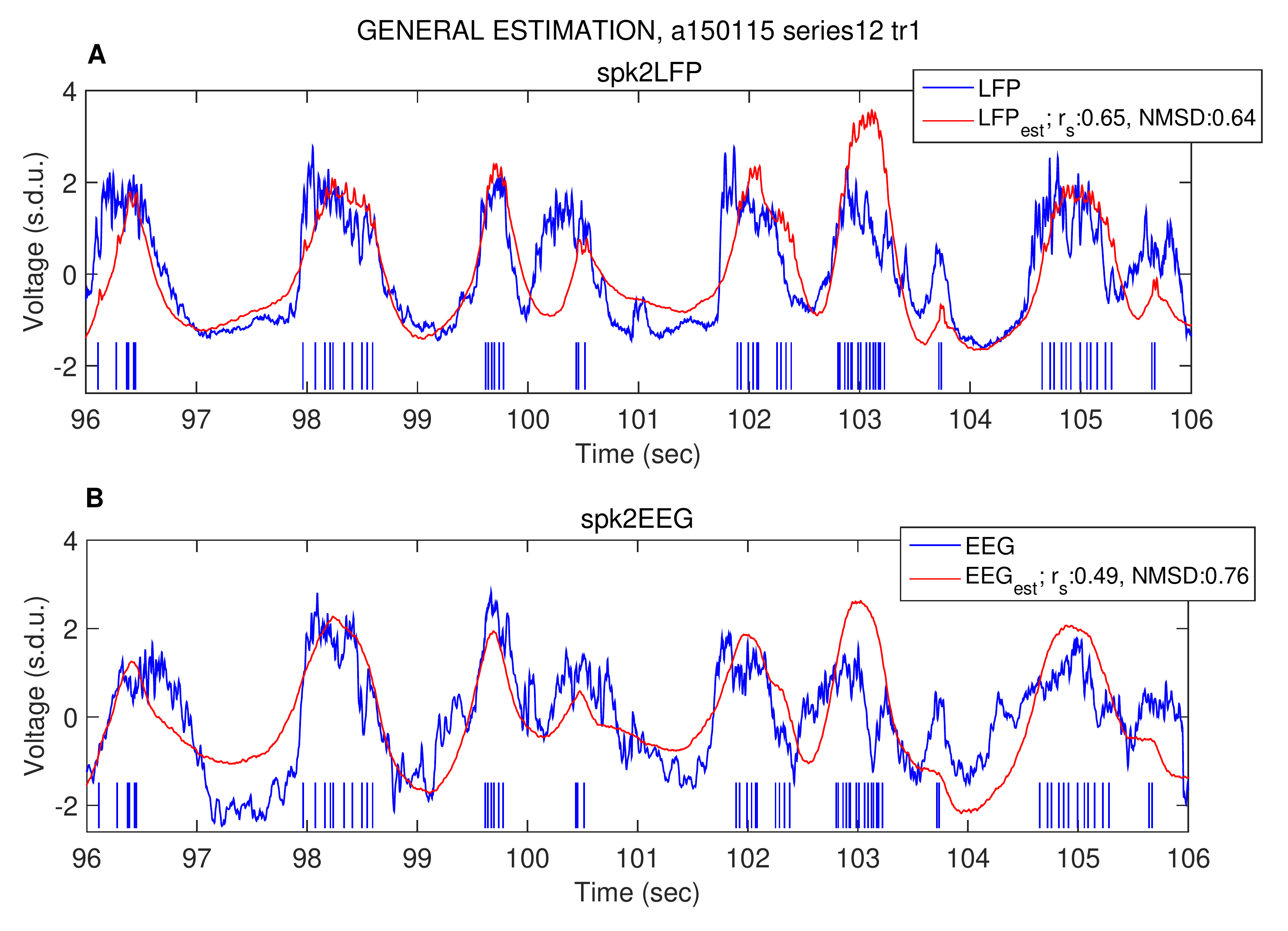}
\par\end{centering}
\centering{}\caption[Spk2LFP/EEG estimation example]{\textbf{LFP/EEG estimation example} when using the general filters
(showed in figure \ref{PFfig_spk2LFP/EEG-general-filters}). \textbf{(A)}
10 seconds trace of the recorded LFP (blue) compared with the estimated
one (red). The latter is obtained by convolving the spike train (blue
vertical lines) with the (general) Wiener kernel. The estimation performances
(i.e., Spearman's correlation and normalized mean squared distance
between LFP and LFP\protect\textsubscript{est}) on the whole test
set are displayed in the legend. \textbf{(B)} Same as (A) for EEG
estimation. Note that the examples in panels (A) and (B) are taken
from the same trial and their estimation performances (and $\langle FR\rangle=4.9$Hz)
are close to the median performances over the entire dataset.\label{PFfig_spk2LFP/EEG_Example}}
\end{figure}

The general filter obtained for respectively the LFP and EEG estimation
is showed in figure \ref{PFfig_spk2LFP/EEG-general-filters} panels
(A) and (B), where also the spike-triggered averages are displayed.
 The LFP (and EEG) estimation is obtained by placing a series of filters
centered on the spike times and then summing them up. Therefore, the
time lag associated with the filter peak (2 ms for LFP and 6 ms for
EEG) indicates the delay where in average there is the strongest relationship
between mass signal fluctuations and firing activity. Thus, for example,
the average strongest effect of a spike will be observed on the LFP
2 ms after the spike emission. Furthermore, the higher and tighter
value of the filter peak when estimating the LFP is an indication
of the fact that the relationship between firing activity and LFP
is closer and more stable than in the EEG case. Note that these filters
are acausal\footnote{To be causal they should be equal to 0 for negative time lags (see
section \vref{section_FILTERcausality}).} (indeed we do not know whether spikes cause directly the mass signal,
for example because the spikes generate a dipole directly captured
by the mass signal, or whether the network oscillations cause the
spike times, for example because the network oscillations suppress
or enhance the likelihood of an individual cell firing at given phase
of network oscillation \citealt{Einevoll2013}). This means that each
spike affects the estimated LFP in time steps both preceding and following
the spike emission. In particular, the values of the filter for respectively
positive (negative) time lags shows the contributions to the LFP\textsubscript{est}
of each spike after (prior to) its emission. In figure \ref{PFfig_spk2LFP/EEG_Example}
we show a representative example of both the LFP and EEG estimation
obtained from the same spiking activity by using the filters displayed
in figure \ref{PFfig_spk2LFP/EEG-general-filters}. 

\begin{figure}
\begin{centering}
\includegraphics[scale=0.5]{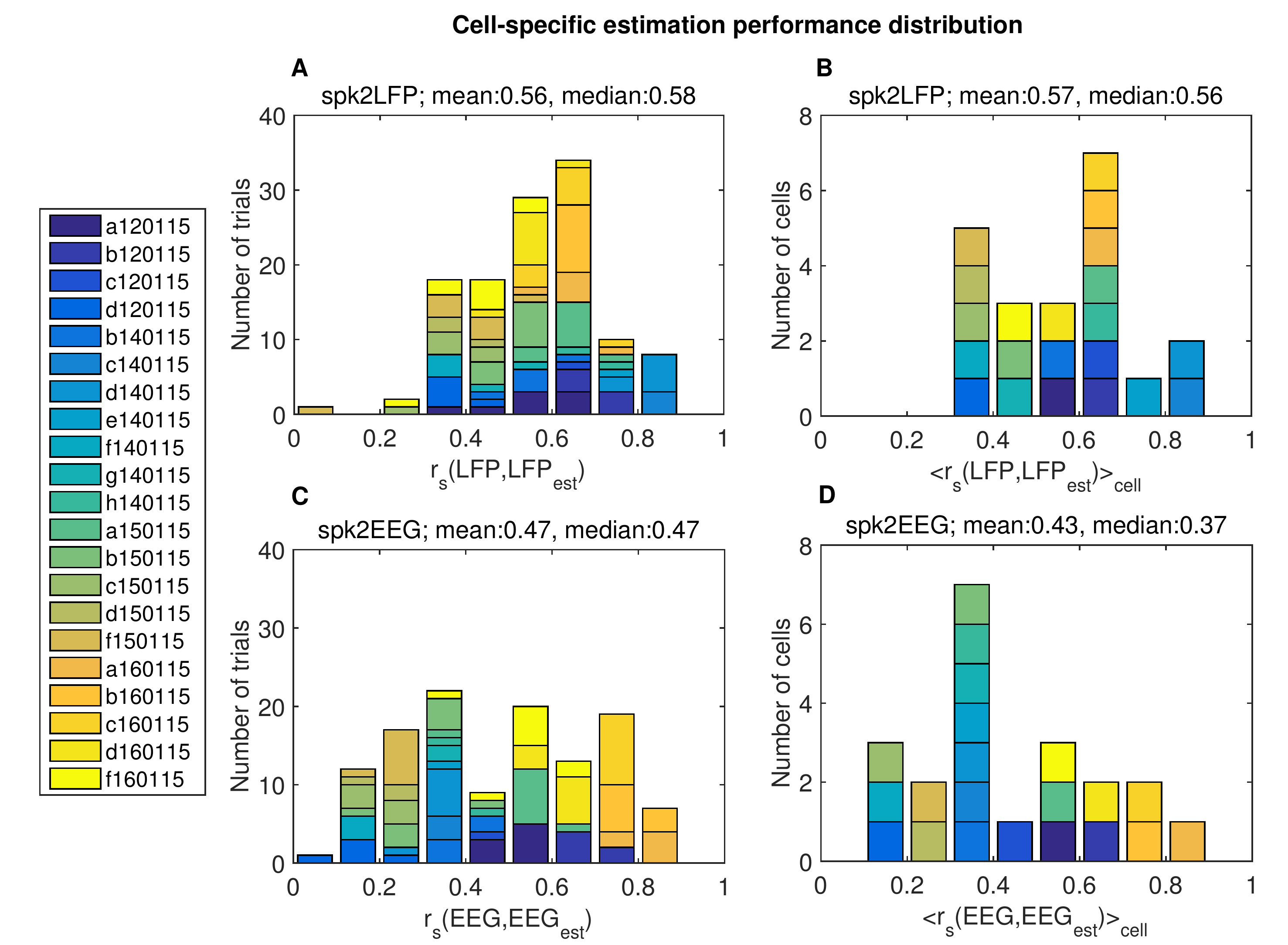}
\par\end{centering}
\centering{}\caption[Spk2LFP/EEG performance distribution]{\textbf{Distribution across trials and cells of LFP and EEG estimation
performances}. \textbf{(A)} Distribution across trials of the LFP
estimation performances (as measured by Spearman's correlation), when
using the cell-specific filter; mean and median values are displayed
in the panel's title. \textbf{(B)} Same as (A) for the distribution
of the average values across cells. \textbf{(C,D)} Same as respectively
(A,B) in case of EEG estimation. The legend indicates the cells the
data belong to.\label{PFfig_spk2LFP/EEG_perf_distribution}}
\end{figure}

We measured the estimation performance by means of the Spearman's
correlation, $r_{s}$, and of the NMSD between the original and estimated
signals on the test set (see section \vref{PF_section_Performance-measures}).
In figure \ref{PFfig_spk2LFP/EEG_perf_distribution} we show the distribution
of the Spearman's correlations across trials and cells both for LFP,
panels (A,B), and EEG estimation, panels (C,D). We found that the
performances vary on a broad range being relatively similar for trials
belonging to the same cell (both for LFP and EEG\footnote{More specifically, in figure \ref{PFfig_spk2LFP/EEG_perf_distribution}
the average (over cells) amplitude of the interval of $r_{s}$ values
found for each cell (with more than one trial) is 0.14 for LFP and
0.13 for EEG estimation.}, compare panels (A,C) with (B,D) in figure \ref{PFfig_spk2LFP/EEG_perf_distribution});
nevertheless, the ranked performances differed when comparing LFP
and EEG estimation (data not shown), suggesting that the relationship
between FR and mass signals does not only depend on the average firing
rate of the neuron, but it also depends on the nature and signal to
noise ratio of the mass signal.
 When using cell-specific filters, the median values of $r_{s}$ for
LFP\textsubscript{est} across all the trials is 0.58$\pm$0.12 (median$\pm$interquartile
range/2, n=120 trials) and NMSD=0.70$\pm0.12$, while, for EEG\textsubscript{est},
$r_{s}$=0.47$\pm$0.18 and NMSD=0.78$\pm$0.17. 
\begin{figure}
\begin{centering}
\includegraphics[scale=0.5]{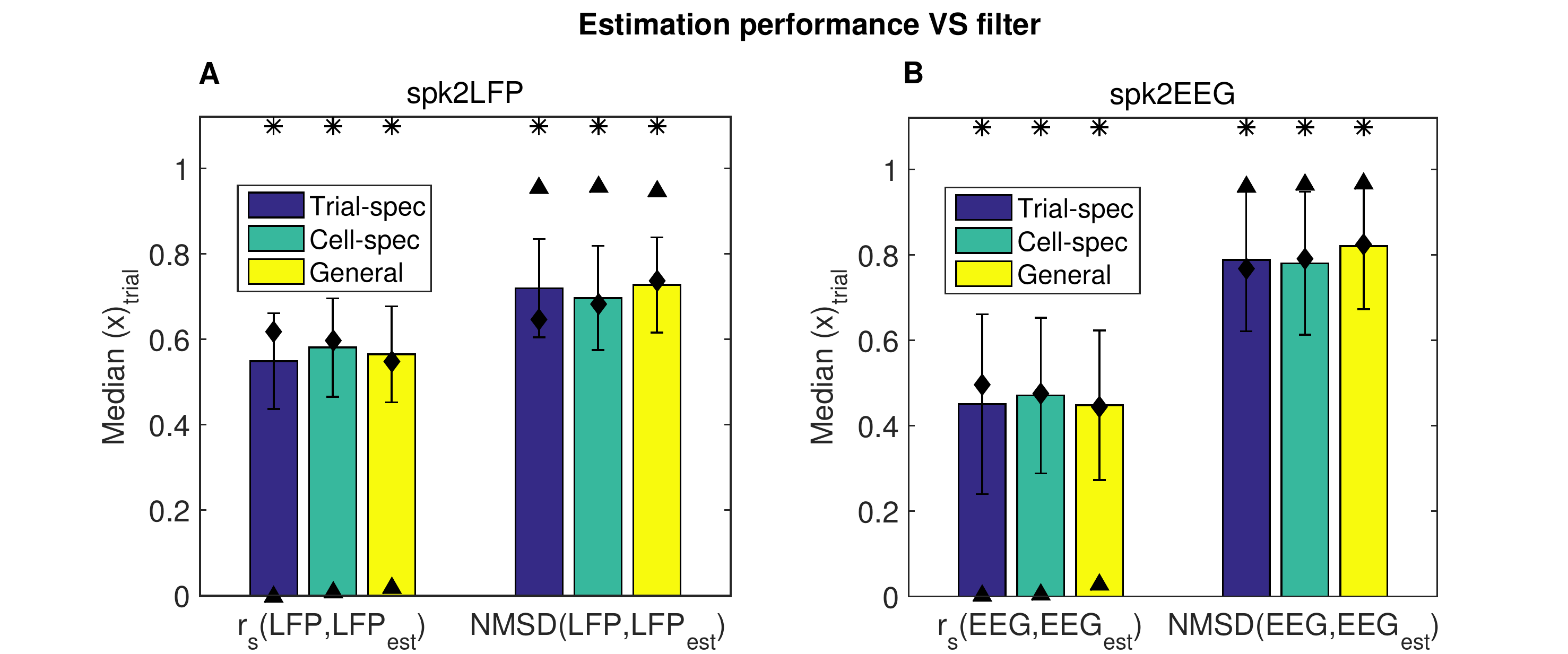}
\par\end{centering}
\centering{}\caption[Spk2LFP/EEG performances VS filters]{\textbf{LFP and EEG estimation performance VS filter specificity.
(A)} LFP estimation performances as a function of the kind of kernels
used; the colored bars indicate the median values (and the error bars
the interquartile ranges) over the test set. The diamonds are the
median values over the training set; the triangles represent the median
estimation performances (on the test set) under the null hypothesis,
that is when using the random filter, $h^{\mathrm{rand}}$ (see section
\vref{PF_section_Performance-measures}). {*} $p\ll10^{-10}$ based
on a one-tailed Kolmogorov-Smirnov test comparing the estimation performances
against the null hypothesis performances. \textbf{(B)} Same as (A)
for EEG estimation. \label{PFfig_spk2LFP/EEG_perf_VS_filter}}
\end{figure}
 Note that the performance distributions obtained from both the trial-specific
and general filters are never significantly different from the cell-specific
one, (according to two-tailed Kolmogorov-Smirnov tests where respectively
p>0.37 (0.29) when comparing Spearman's correlation for LFP (EEG)
estimation), as summarized in figure \ref{PFfig_spk2LFP/EEG_perf_VS_filter}.
Therefore the estimation is robust with respect to the kernel's generality,
suggesting that the relationships underlying the estimation reflect
general and robust network phenomena under the experimental conditions
considered and that the algorithm does not express any kind of overfitting
of trial-specific features.
\begin{figure}
\begin{centering}
\includegraphics[scale=0.5]{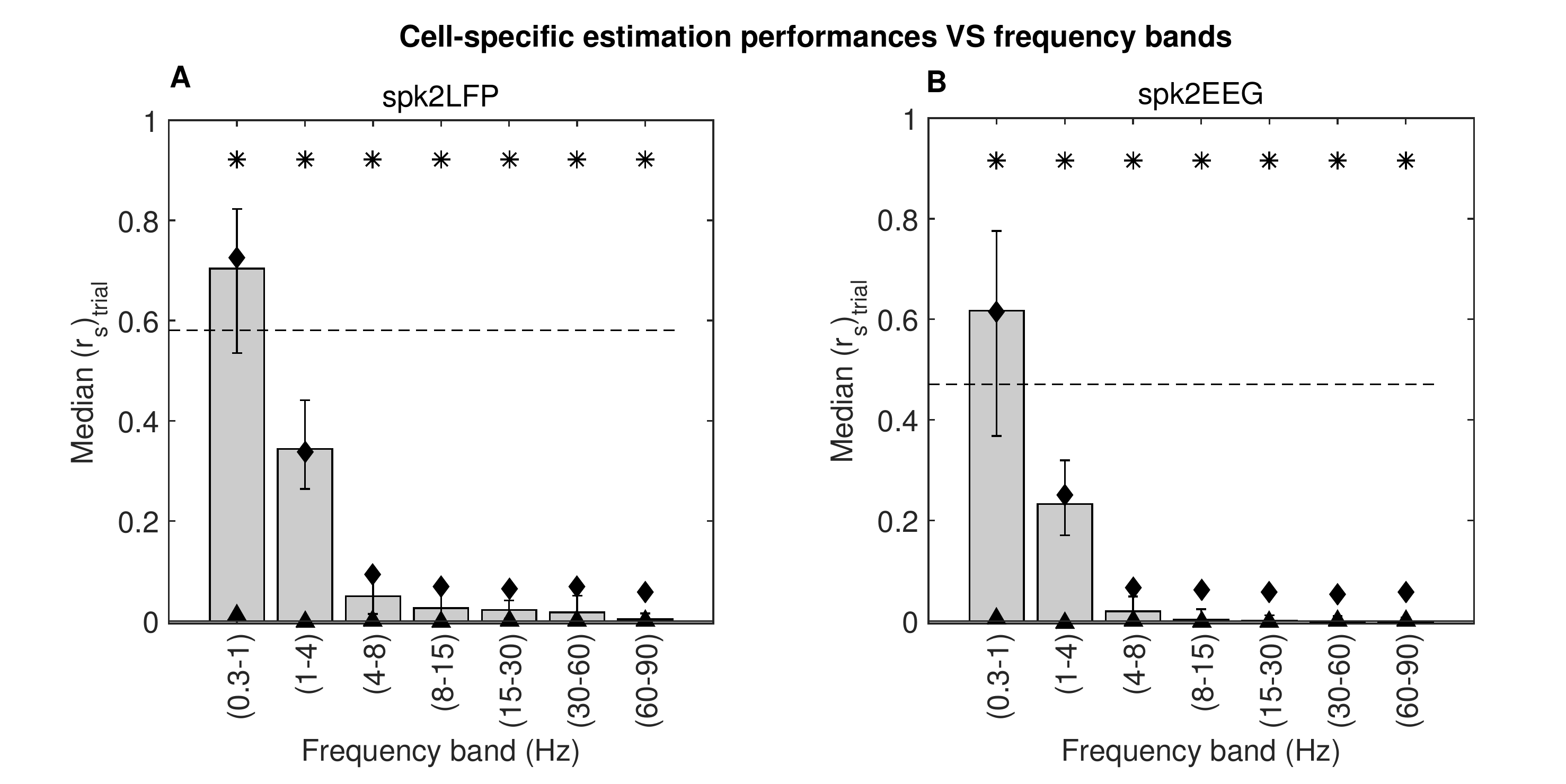}
\par\end{centering}
\centering{}\caption[Spk2LFP/EEG performances VS frequencies]{\textbf{LFP and EEG estimation performance VS frequency bands. (A)}
LFP estimation performance as a function of the frequency bands; First,
both the LFP and the LFP\protect\textsubscript{est} have been filtered
in the specified frequency band, then the Spearman's correlation between
the two filtered signals has been computed. The gray bars indicate
the median values over the test set, while the error bars represent
the interquartile ranges. The diamonds are the median values over
the training set, while the triangles represent the estimation performances
(on the test trial) under the null hypothesis (as in figure \eqref{PFfig_spk2LFP/EEG_perf_VS_filter}).
The horizontal dashed line indicates the median value for the unfiltered
LFP. {*} $p\ll10^{-10}$ based on one-tailed Kolmogorov-Smirnov test
comparing the estimation performances against the null hypothesis
performances. \textbf{(B)} Same as (A) for EEG estimation; {*} $p\ll10^{-6}$.
The showed performances are obtained by using a cell-specific kernel;
similar results are obtained when using both trial-specific and general
kernels (data not shown).\label{PFfig_spk2LFP/EEG_perf_vs_FREQ}}
\end{figure}
\begin{figure}
\begin{centering}
\includegraphics[scale=0.45]{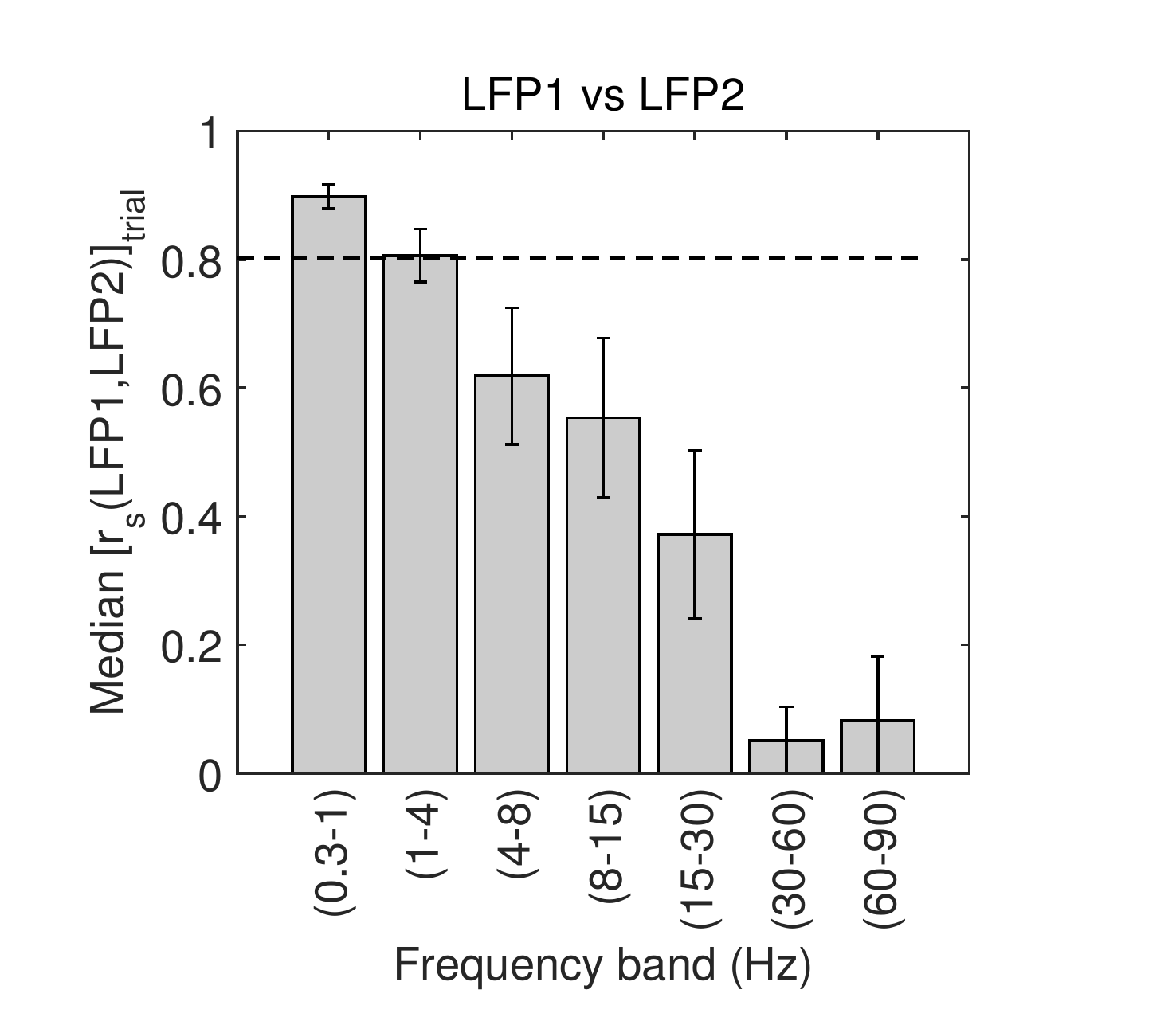}
\par\end{centering}
\centering{}\caption[Spatial synchrony of LFPs]{{\small{}}\textbf{\small{}Comparing two LFP traces recorded at the
SUA-LFP pipette distance}{\small{}. The same analysis done in figure
\ref{PFfig_spk2LFP/EEG_perf_vs_FREQ} is applied here to evaluate
the frequency-dependent similarity between two LFP traces recorded
at a distance < 500 \textgreek{m}m. Note that this is the same distance
present between the pipettes used to record the LFP and the SUA analyzed
in all the other figures. The horizontal dashed line indicates the
value for the unfiltered LFP (that is 0.80$\pm$0.05 (median$\pm$interquartile
range/2), while for the NMSD is 0.36$\pm$0.05, datum not shown).
Each trial lasts 100 sec (3 mice, 31 trials).\label{PFfig_Comparing-two-LFPs}}}
\end{figure}

To understand which components of network oscillations are better
captured by our model, we analyzed how the performances are distributed
across the frequency spectrum of LFP\textsubscript{est} and EEG\textsubscript{est}.
We found \textcolor{black}{(figure \ref{PFfig_spk2LFP/EEG_perf_vs_FREQ})}
that our linear estimation reconstructed better the low frequency
components of the signals, \textcolor{black}{even if the performances
remains significantly higher than random level for the whole spectrum.
This is intuitively expected for two reasons. The first is that the
estimation (especially a linear one) will tend to reconstruct better
the oscillations with largest amplitudes, which are in the lowest
frequency band (see figure \ref{PFfig_example_power_LFP_EEG}). The
second reason is due to the experimental setup. Indeed, the pipette
used to record the spiking activity is placed at a given distance
(<500 }\textgreek{m}m\textcolor{black}{) from the pipette used to
record LFP (and also from the wires used to record EEG), thus we use
the SUA recorded at a given place to reconstruct a LFP recorded some
hundreds micrometers away. As a consequence, the LFP performance estimation
depended also on the spatial synchrony of LFP oscillations, which
is higher for lower frequencies, as shown in figure \ref{PFfig_Comparing-two-LFPs},
where we compared two LFP traces recorded at the distance usually
present between LFP and SUA pipettes}. 

\begin{figure}
\begin{centering}
\includegraphics[scale=0.5]{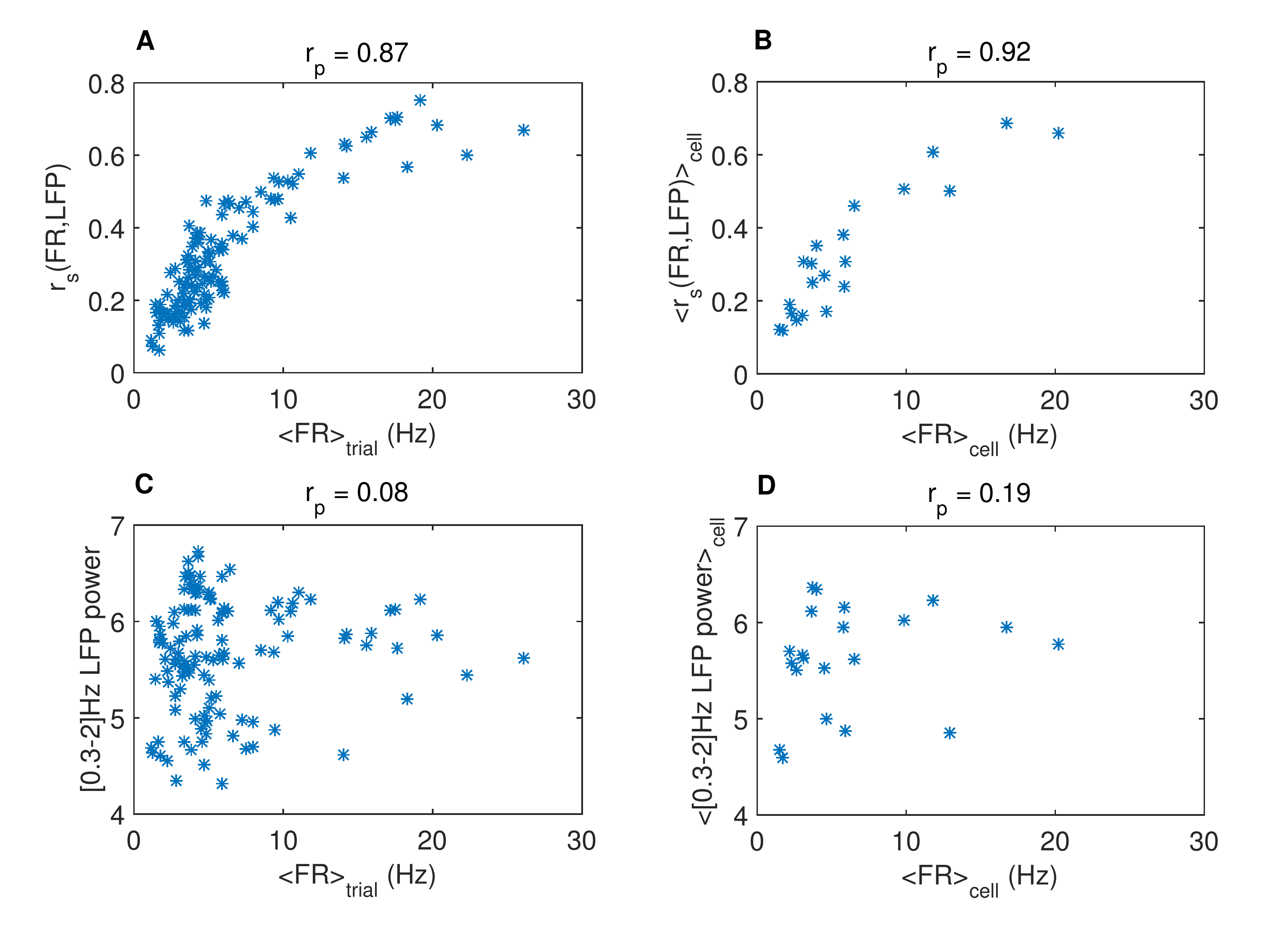}
\par\end{centering}
\centering{}\caption[FR-LFP correlations]{\textbf{Relationships between firing rate} \textbf{and LFP} in the
test set. \textbf{(A)} Spearman's correlation between the LFP and
the concomitant firing rate, $r_{s}(\mbox{FR,LFP})$, as a function
of the average firing rate. The firing rate is computed by using a
spike smoothing window of 50ms (see section \vref{PF_section_FRcomputation}).
The median value of $r_{s}(\mbox{FR,LFP})$ over all the trials is
0.28. \textbf{(B)} Same as panel (A) for the average values in each
cell. \textbf{(C)} Scatter plot between the average FR and the power
spectrum of the low LFP delta band {[}0.3 2{]}Hz; each point represents
the values in a trial. \textbf{(D)} Same as panel (C) when each point
represents the average values over the trials of a cell. The Pearson's
correlations between the plotted variables are displayed in the panels'
titles. \label{PFfig_avFR_LFP_dispersion}}
\end{figure}

\begin{figure}
\begin{centering}
\includegraphics[scale=0.5]{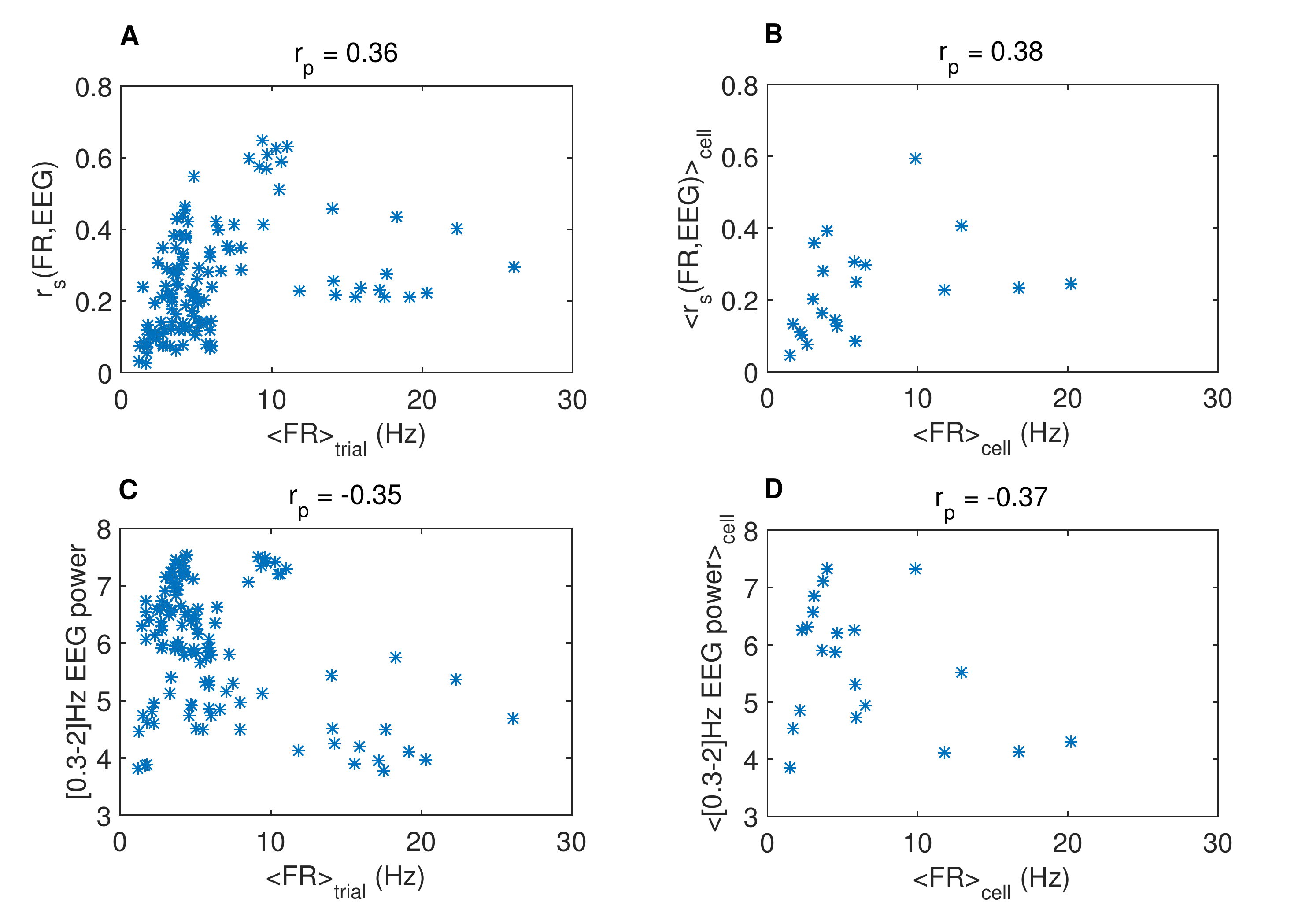}
\par\end{centering}
\centering{}\caption[FR-EEG correlations]{\textbf{Relationships between firing rate} \textbf{and EEG} in the
test set. Same analysis performed in figure \ref{PFfig_avFR_LFP_dispersion}.\textbf{
(A)} Spearman's correlation between the EEG and the concomitant firing
rate, $r_{s}(\mbox{FR,EEG})$, as a function of the average firing
rate. The median value of $r_{s}(\mbox{FR,EEG})$ over all the trials
is 0.22. \textbf{(B)} Same as panel (A) for the average values in
each cell. \textbf{(C)} Scatter plot between the average FR and the
power spectrum of the low EEG delta band. \textbf{(D)} Same as panel
(C) when each point represents the average values over the trials
of a cell.\label{PFfig_avFR_EEG_dispersion}}
\end{figure}

\begin{figure}
\begin{centering}
\includegraphics[scale=0.75]{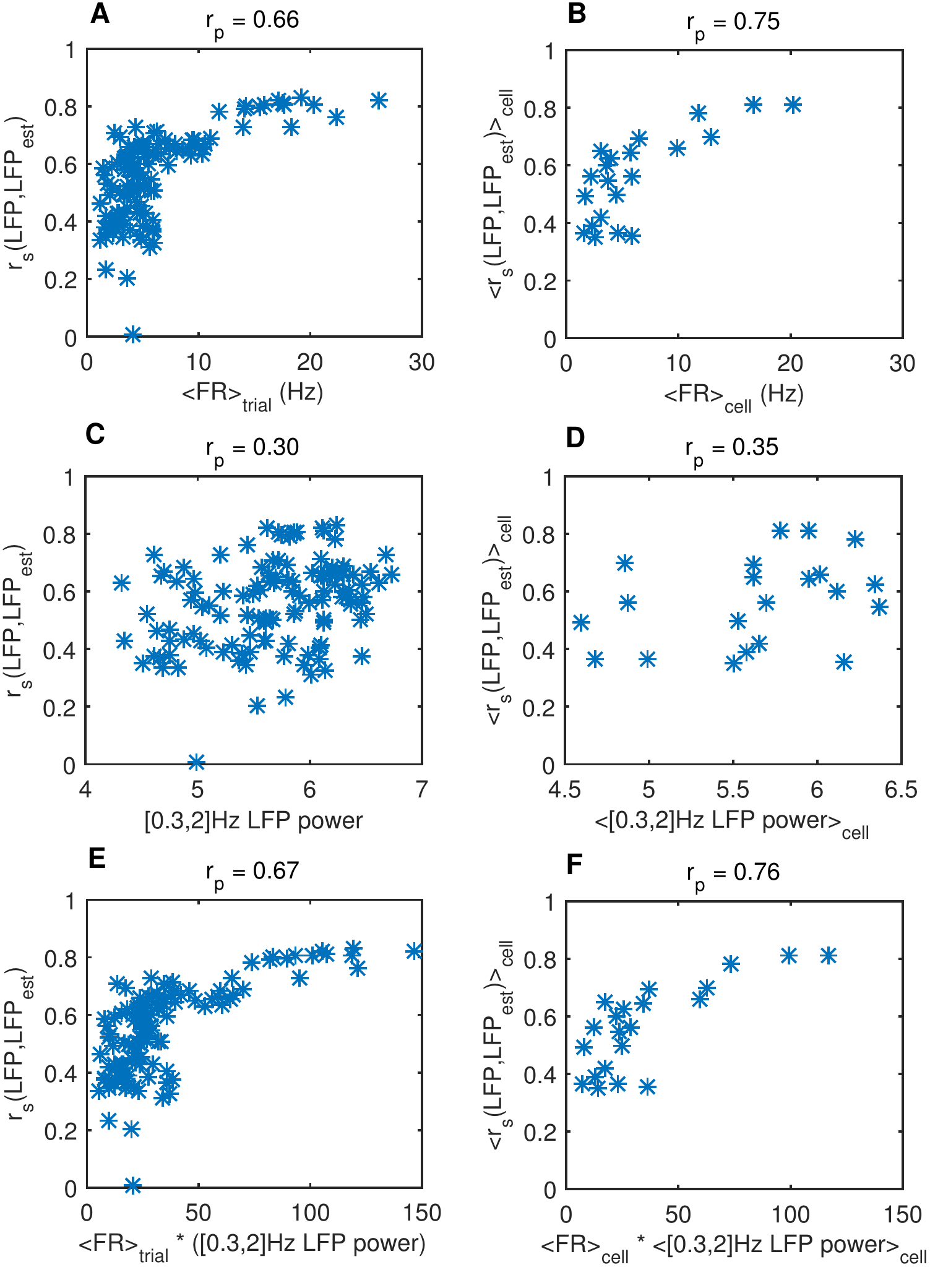}
\par\end{centering}
\begin{centering}
\caption[Spk2LFP performance scatter plots]{\textbf{LFP estimation performance scatter plots.} \textbf{(A)} LFP
estimation performance in each trial (as measured by Spearman's correlation)
as a function of the average FR. \textbf{(C)} LFP estimation performance
of each trial as a function of the true LFP power spectrum in the
low delta band, {[}0.3 2{]}Hz. \textbf{(E)} LFP estimation performance
of each trial as a function of the product between the true low LFP
power spectrum and the average FR. \textbf{(B,D,F)} Same as respectively
(A,C,E) when each variable is averaged over the trials belonging to
a cell. The Pearson's correlations between the plotted variables are
displayed in the panel's titles. The showed performances are obtained
by using a cell-specific kernel and similar results are obtained when
using both trial-specific and general kernels (data not shown). \label{PFfig_spk2LFP_perf_dispersion}}
\par\end{centering}
\centering{}
\end{figure}

\begin{figure}
\begin{centering}
\includegraphics[scale=0.75]{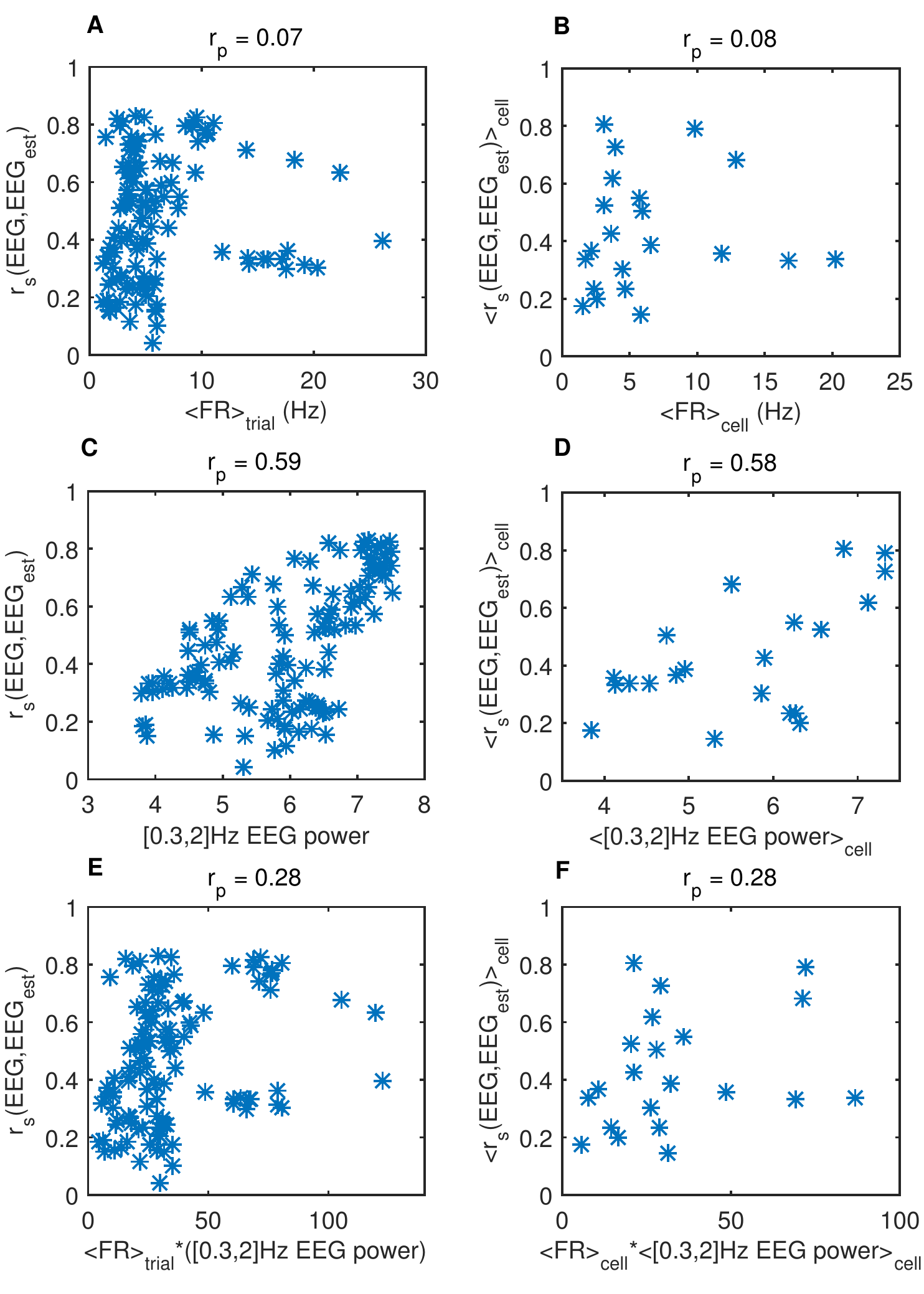}
\par\end{centering}
\centering{}\caption[Spk2EEG performance scatter plots]{\textbf{EEG estimation performance scatter plots. }Same analysis
performed in figure \ref{PFfig_spk2LFP_perf_dispersion}.\textbf{
(A)} EEG estimation performance of each trial as a function of the
average FR. \textbf{(C)} EEG estimation performance of each trial
as a function of the true EEG power spectrum in the low delta band,
{[}0.3 2{]}Hz. \textbf{(E)} EEG estimation performance of each trial
as a function of the product between the true low EEG power spectrum
and the average FR. \textbf{(B,D,F)} Same as respectively (A,C,E)
when each variable is averaged over the trials belonging to a cell.
The Pearson's correlations between the plotted variables are displayed
in the panel's titles. The showed performances are obtained by using
a cell-specific kernel and similar results are obtained when using
both trial-specific and general kernels (data not shown).\label{PFfig_spk2EEG_perf_dispersion}}
\end{figure}

\begin{figure}
\begin{centering}
\includegraphics[scale=0.5]{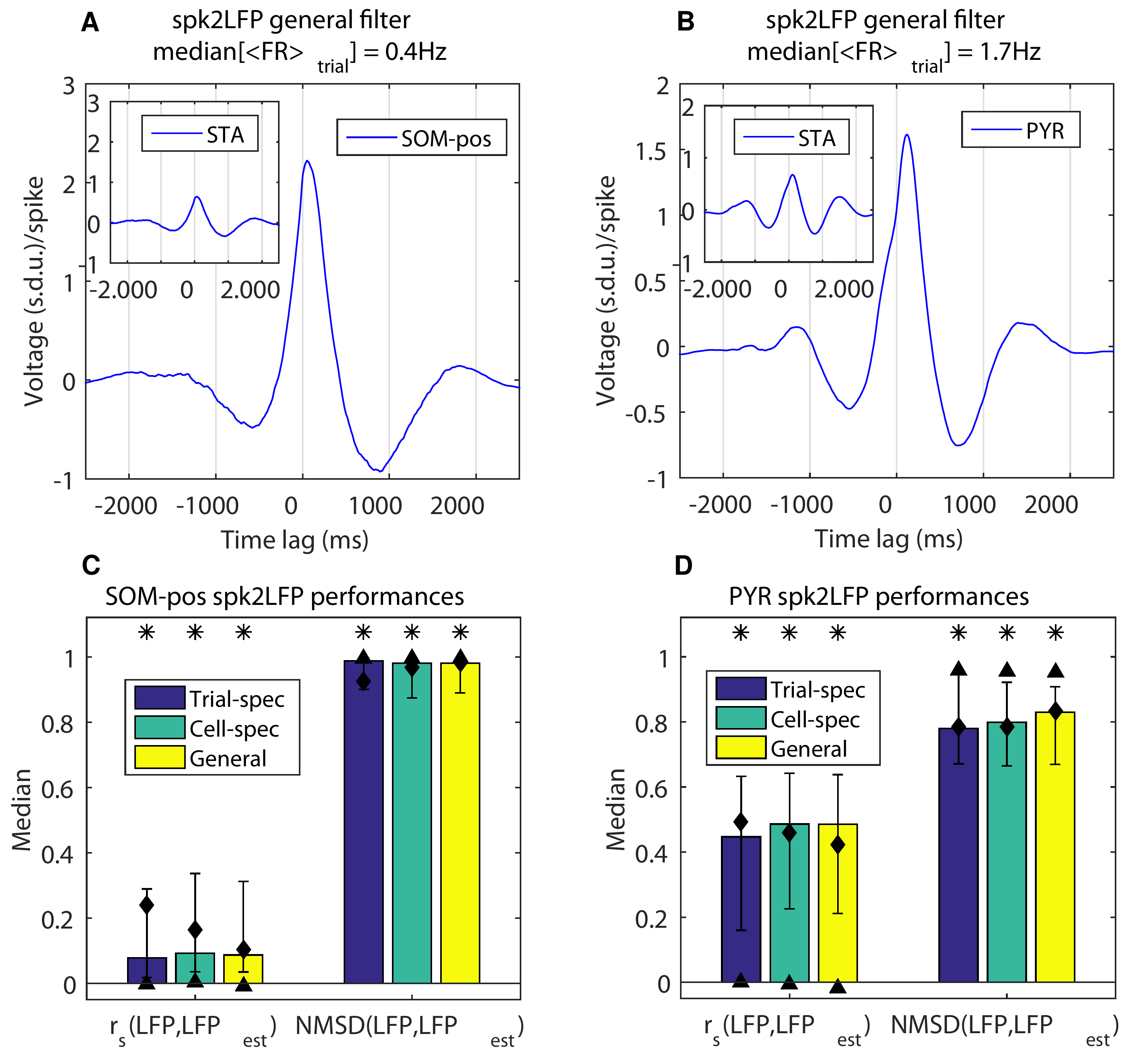}
\par\end{centering}
\begin{centering}
\caption[LFP estimation from SOM-pos and PYR neuron activity]{\textbf{LFP estimation from the firing activity of SOM-pos interneurons
and pyramidal neurons}. Results from the analysis of other two datasets
are shown. \textbf{(A)} General Wiener filter used to estimate the
LFP from the spiking activity of an individual SOM-pos interneuron
(8 mice, 18 cells and 99 trials). The peak is at 52 ms time lag. As
in figure \eqref{PFfig_spk2LFP/EEG-general-filters}, the inset displays
the LFP STA. Note that the median firing activity of this kind of
neurons is very low: median average FR equal to 0.4 Hz (see figure
\eqref{PFfig_avFR} for comparison with the PV-pos activity).\textbf{
(B)} Same as (A) when the firing activity comes from a single pyramidal
neuron of deep layers (3 mice, 7 cells and 23 trials). Filter peak
located at 118 ms time lag. \textbf{(C,D)} As done in figure \eqref{PFfig_spk2LFP/EEG_perf_VS_filter},
we show the performances and their significance against the null hypothesis
as a function of the filter used when estimating the LFP from the
firing activity of a SOM-pos (C; {*} $p<0.004$) and of a pyramidal
neuron (D; {*} $p<10^{-5}$). \label{PFfig_spk2LFP_SOM_PYR5_datasets}}
\par\end{centering}
\centering{}
\end{figure}

To gain a deeper insight about the parameters shaping the relationships
between single-neuron spiking activity and mass signals, we investigated
how the correlations between original firing rates and network oscillations
(figures \ref{PFfig_avFR_LFP_dispersion} and \ref{PFfig_avFR_EEG_dispersion})
affect the estimation performance dynamics. \\
To assess how the FR of single cells is synchronized with the mass
signal, we computed the Spearman\textquoteright s correlation between
FR and mass signal time courses. We found that the median correlation
between the FR and the LFP (0.28) is higher than the one between FR
and EEG (0.22); this is not surprising since the EEG is a signal integrated
over a broader area than the LFP. Nevertheless, what is more important
is that, in the LFP case, there is an high positive Pearson\textquoteright s
correlation between the average firing rate and the synchronization
between FR and LFP ($r_{p}=0.87$), which is strongly attenuated in
the EEG case ($r_{p}=0.36$, compare panels (A,B) in figures \ref{PFfig_avFR_LFP_dispersion}
and \ref{PFfig_avFR_EEG_dispersion}). This means that, in the LFP
case, the higher the average firing rate, the higher is the synchronization
of firing activity with the mass signal, whereas, in the EEG case,
this is not as clear. This is the crucial point to understand the
differences in the way the estimation performances are shaped in the
LFP and EEG case throughout this analysis and, in particular, it is
the reason why we observed a strong correlation between the average
firing rate and the LFP estimation performances (see panels (A,B)
in figure \ref{PFfig_spk2LFP_perf_dispersion}), which is absent in
the EEG estimation (see panels (A,B) in figure \ref{PFfig_spk2EEG_perf_dispersion}).
\\
We also investigated if the average firing rate is related to the
power of slow frequencies ({[}0.3 2{]}Hz) of the LFP and EEG. We found
weak correlations between these two variables both for LFP and EEG
(see panels (C,D) in figures \ref{PFfig_avFR_LFP_dispersion} and
\ref{PFfig_avFR_EEG_dispersion}): the average firing rate is only
weakly dependent on the power of the slow network oscillations. However,
note that our recordings are performed in a regime of slow wave oscillations,
where the slowest frequencies are always the largest in amplitude
(see figure \ref{PFfig_example_power_LFP_EEG}) and their power variation
across trials is smaller than the variation observed in the average
firing rates\footnote{More precisely, while the largest average FR is 26 times the smallest,
the largest respectively LFP (EEG) low power is 1.6 (2.0) times the
smallest, see figures \eqref{PFfig_avFR_LFP_dispersion} and \eqref{PFfig_avFR_EEG_dispersion}.}. 

As a consequence of the fact that the oscillations estimated better
are the slowest ones (see figure \ref{PFfig_spk2LFP/EEG_perf_vs_FREQ}),
we could expect to find a positive correlation between the estimation
performance and the power of the lowest frequencies. On the other
hand, another crucial variable is the number of spikes available to
reconstruct the signal, that is the average firing rate. Therefore,
we will focus on the contributions of slow network oscillations and
average firing rates in shaping the estimation performances\footnote{In extended analysis we also investigated the contributions of the
coefficient of variation of the inter-spike interval (CV ISI) and
of the index of synchronization of network oscillations \citep{cheng2009burst},
which is measured as the ratio between the power of the low {[}0.3
2{]}Hz and of the high {[}30 60{]}Hz frequencies. We found that the
correlations of respectively CV ISI and index of synchronization with
the estimation performances were always lower than the ones obtained
with the average FR and the power of the low frequencies of mass signals
(data not shown), thus we decided to focus on the results relative
to these two latter variables.}. Interestingly, when looking at the scatter plots of the performances
with respect to those two variables, we found that the highest correlation
is with the average firing rate for LFP estimation ($r_{p}=0.66$,
see figure \ref{PFfig_spk2LFP_perf_dispersion}) and with the low
frequency power for EEG estimation ($r_{p}=0.59$, see figure \ref{PFfig_spk2EEG_perf_dispersion}).
This asymmetry is due to the differences in the relationships between
FR and LFP with respect to FR and EEG pointed out in figures \ref{PFfig_avFR_LFP_dispersion}
and \ref{PFfig_avFR_EEG_dispersion}. Indeed, the estimation is enhanced
(i) by high average firing rates, but, importantly, only if the firing
activity is synchronized with the mass signals and (ii) by large amplitudes
of slow oscillations. For what concerns the contributions of the average
firing rate, the correlation between average firing rate and (mass
signal-FR) synchronization is stronger for LFP ($r_{p}=0.87$) than
for EEG ($r_{p}=0.36$). Thus, when a neuron has an high firing activity,
this activity is strongly synchronized with LFP, but much less with
EEG. In addition, the average FR tend to decrease when increasing
the EEG low frequency power ($r_{p}=-0.35$), therefore, in case of
EEG estimation, the performances are almost independent on the average
firing rate (see panels (A,B) figure \ref{PFfig_spk2EEG_perf_dispersion}).
As a consequence, the EEG estimation performances are mainly determined
by the amplitude of slow oscillations (see panels (C,D) figure \ref{PFfig_spk2EEG_perf_dispersion}).
On the other hand, in our regime (i.e., slow wave oscillation), we
have always large amplitude values of the slowest network oscillations
thus (as stated above) their power is relatively stable across trials
while the variation observed in the average firing rates is wider.
The larger variability of the average FRs, combined with the fact
that LFPs are strongly synchronized with high FR activities (as stated
above), results in an higher correlation between performances and
average FRs with respect to the one between performances and LFP low
power (compare panels (A,B) with (C,D) in figure \ref{PFfig_spk2LFP_perf_dispersion}).
\\
In panels (E,F) of figures \ref{PFfig_spk2LFP_perf_dispersion} and
\ref{PFfig_spk2EEG_perf_dispersion}, we show the correlation between
the performances and the average FR multiplied by the low frequency
power. In the EEG case, the resulting correlation is smaller than
the correlation between the performances and the variables taken individually,
because FRs and mass signals are not synchronized and thus an high
firing rate does not improve the performances. On the other hand,
in the LFP case, the correlation in panels (E,F) is (slightly) larger
than in the other panels confirming that, when FRs and mass signals
are synchronized, both slow network oscillations and FR enhance estimation.

\label{PF_section_spk2LFP_avFR_VS_lowFreq}In summary, we found that
the performance of LFP and EEG estimation depends mainly on two distinct
features: (i) the amplitude of low frequencies in the mass signal
and (ii) the number of spikes available to reconstruct the signals
(i.e., the average firing rate of the cell). Both of them, when increasing,
tend to facilitate the estimation. Since the recordings are performed
during slow wave oscillations, we remain always in a regime where
the slowest frequencies are largest and their power is relatively
stable across trials\footnote{Even if mass signals can be more or less synchronized, depending on
the level of the anesthesia.}. On the other hand, the average firing rate can span a broad spectrum
of values, thus its fluctuations are wider than the variations of
LFP and EEG low power. As a result, when the firing activity is synchronized
with the mass signals, as happens with the LFP, the average FR will
prevail in shaping the estimation performances, otherwise, the level
of low frequency oscillations will mainly determine the performances
(EEG case). 

We conclude this analysis by summarizing some preliminary results
of the application of our analysis to other two datasets where the
mass signal is measured only as LFPs and the SUAs come respectively
from Somatostatin-positive interneurons in layer 2 and excitatory
pyramidal neurons from a deep layer (i.e., 5 or 6) in mouse neocortex.
In the first case, the firing activity is extremely low (median{[}$\langle FR\rangle${]}=0.4
Hz, indeed this interneurons are no longer fast-spiking), and this
leads to a fall in the estimation performances (panel (C) in figure
\ref{PFfig_spk2LFP_SOM_PYR5_datasets}). For the excitatory neurons,
even if the average firing rates decrease considerably with respect
to PV-pos (median{[}$\langle FR\rangle${]} from 4.5 Hz to 1.7 Hz),
the performances remain quite high (panel (D) figure \ref{PFfig_spk2LFP_SOM_PYR5_datasets})
suggesting the existence of a strong lock between pyramidal firing
activity and mass signal. Finally, when looking at the correlations
of the estimation performances with average firing rates and powers
of the slow LFP oscillations, we found results very similar to the
ones showed in figure \ref{PFfig_spk2LFP_perf_dispersion} for the
PV-pos interneurons (data not shown). %

\subsection{Estimating SUA from LFP or EEG}

In this section we reverse the direction of the estimation, as we
attempt to estimate SUA from from mass signals. This analysis is important
to understand how we can infer the changes in firing rates of specific
cell types in cases (such as those in human cognitive experiments
with EEGs) when it is only possible record mass signals. We performed
this estimation in two steps: first, the linear estimation of the
FR, through convolution with a Wiener kernel, second, a non-linear
threshold to detect the estimated spikes. We evaluated how the performances
depend on the specificity of the filter (as done for the estimation
in the opposite direction), on the spike-detection threshold and we
also compared the performances obtained by using the Wiener kernel
with the ones obtained by considering other two models of the estimated
FR.

\subsubsection*{Spike train estimation}

\begin{figure}
\begin{centering}
\includegraphics[scale=0.55]{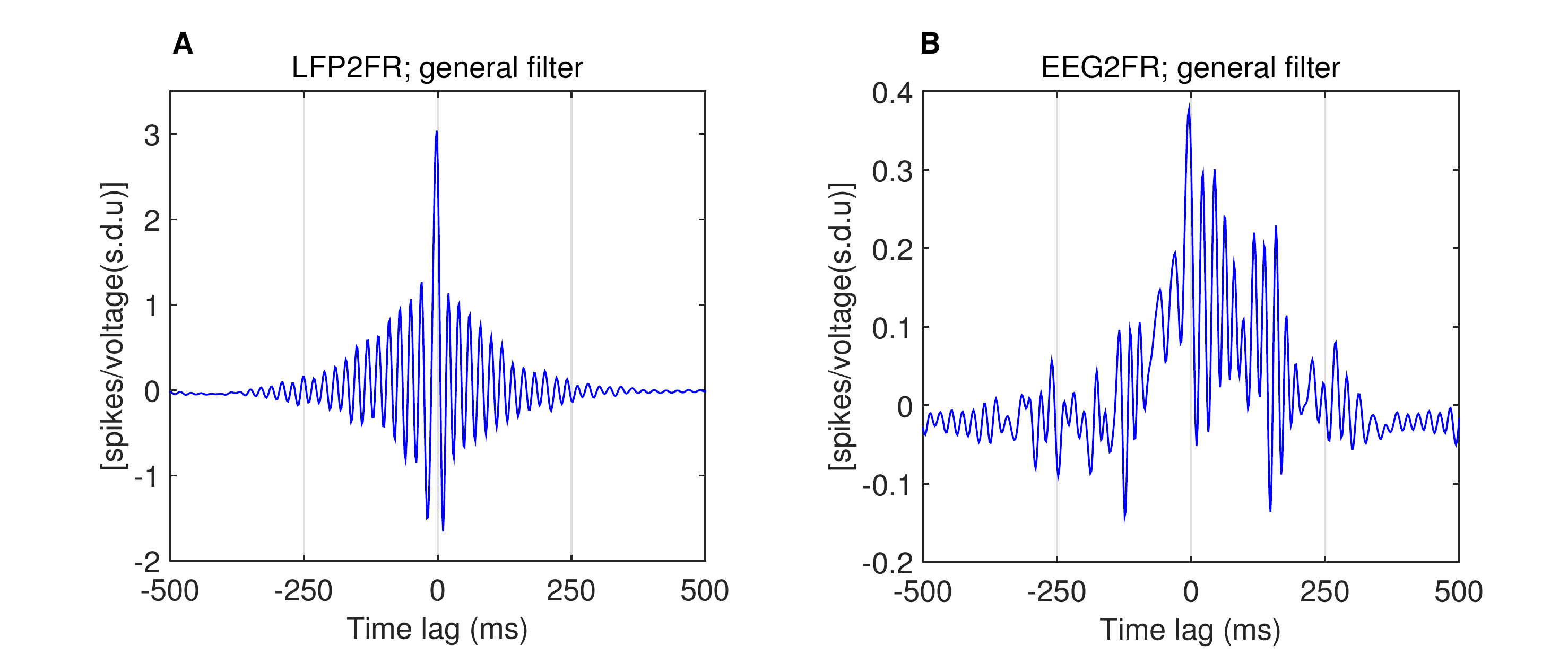}
\par\end{centering}
\centering{}\caption[LFP/EEG2FR general filter]{\textbf{General Wiener kernels for FR estimation from LFP and EEG}.
Mean filters (over all the trials) used to estimate the FR (spike
smoothing window of 10 ms) of a PV-pos interneuron starting from the
LFP \textbf{(A)} and from the EEG \textbf{(B) }signals. In panel (A)
the filter peak is at -2 ms and in panel (B) at -4 ms.\label{PFfig_LFP/EEG2FR_general-filters}}
\end{figure}
\begin{figure}
\begin{centering}
\includegraphics[scale=0.5]{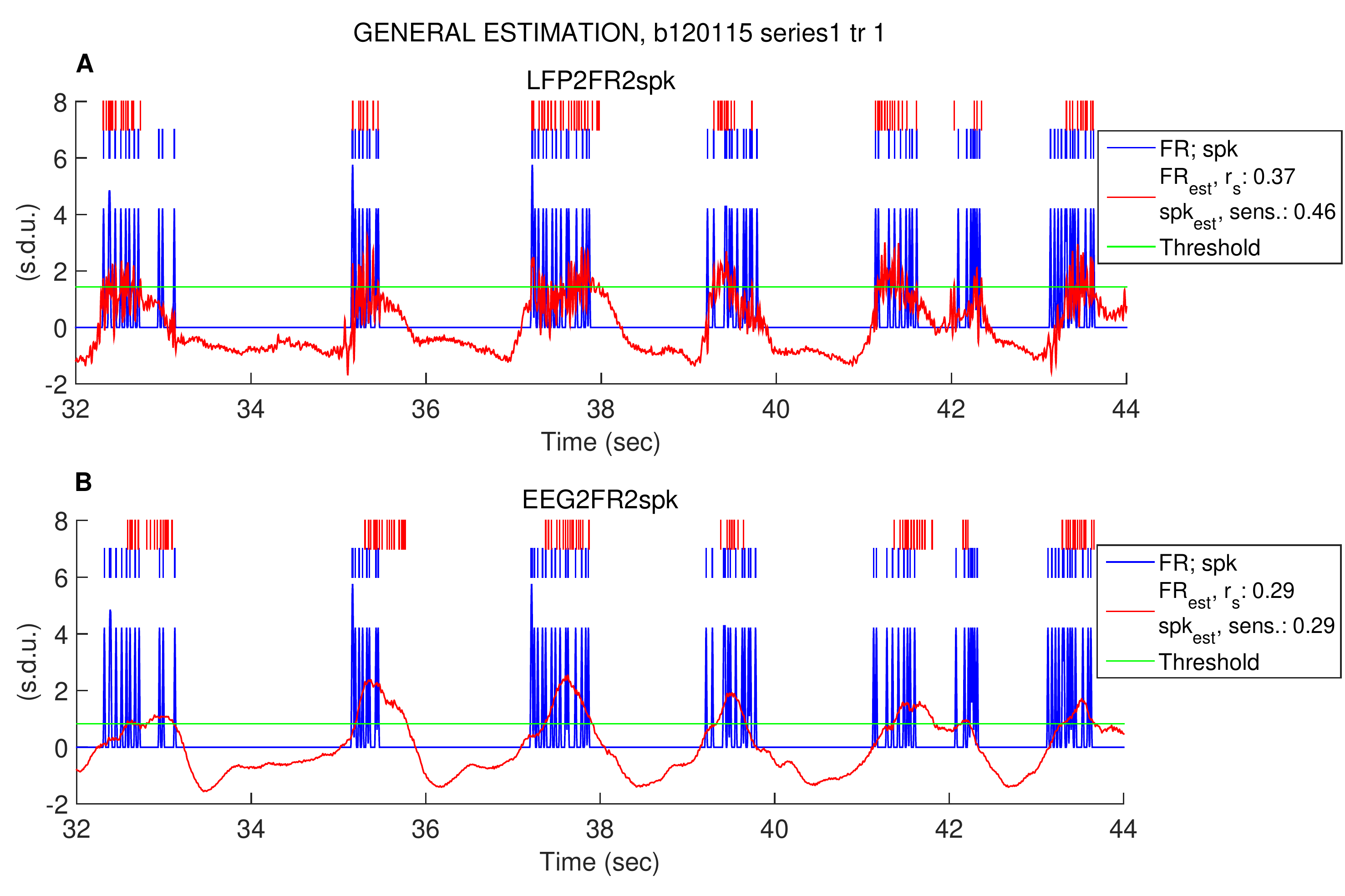}
\par\end{centering}
\centering{}\caption[LFP/EEG2spk estimation example]{\textbf{Example of spike train estimation from LFP and EEG} when
using the general filters (showed in figure \ref{PFfig_LFP/EEG2FR_general-filters}).
\textbf{(A)} We firstly compute the estimated FR, FR\protect\textsubscript{est},
by convolving the recorded LFP with the (general) filter, then we
estimate the spike train, spk\protect\textsubscript{est}, by detecting
a spike each time the FR\protect\textsubscript{est} has a local maximum
that overcomes a given threshold. The threshold (green line) is set
such in a way to obtain the same number of spikes in spk\protect\textsubscript{est}
as in the true spike train (note that we used also other thresholds
throughout the work). 12 seconds trace of the original FR (blue) compared
with the estimated one (red) are shown; in the upper part of the panel
is displayed the original spike train (blue vertical lines) and the
estimated one (red vertical lines). In the legend are displayed the
estimation performances (i.e., Spearman's correlation for the FR\protect\textsubscript{est}
and sensitivity for the spike train estimation) of the trial from
which the traces are taken. \textbf{(B)} Same as (A) for spike train
estimation from EEG signal. The examples in panels (A) and (B) are
taken from the same trials and their estimation performances (and
$\langle FR\rangle=5.9$Hz) are close to the median performances over
the entire dataset.\label{PFfig_LFP/EEG2spk_Example_ottimo}}
\end{figure}

The mean Wiener kernel over all the dataset is shown in figure \ref{PFfig_LFP/EEG2FR_general-filters}
for both LFP and EEG cases. Each point of the filter represents the
weight given to the LFP (EEG) signal in $(t^{*}+\mathrm{time\:lag)}$
when estimating the FR in $t^{*}$. The filters have oscillations
narrower with respect to the filters used to estimate mass signals
(figure \ref{PFfig_spk2LFP/EEG-general-filters}) and this reflects
the fact that the FR itself displayed very narrow oscillations (which
identify the spikes position, see figure \ref{PFfig_LFP/EEG2spk_Example_ottimo}).
Note that the EEG2FR filter is much smaller than the LFP2FR one; this
indicates a weaker synchronization between firing activity and the
EEG oscillations. As expected, the time lags associated with the filter
peaks have inverted sign with respect to the case where the estimations
were performed in the opposite direction. 
 A representative example of the method used to perform the firing
activity estimation is showed in figure \ref{PFfig_LFP/EEG2spk_Example_ottimo},
where the estimated FR, FR\textsubscript{est}, the spike-detection
threshold and the estimated spike train, spk\textsubscript{est},
are displayed.%

\begin{figure}
\begin{centering}
\includegraphics[scale=0.5]{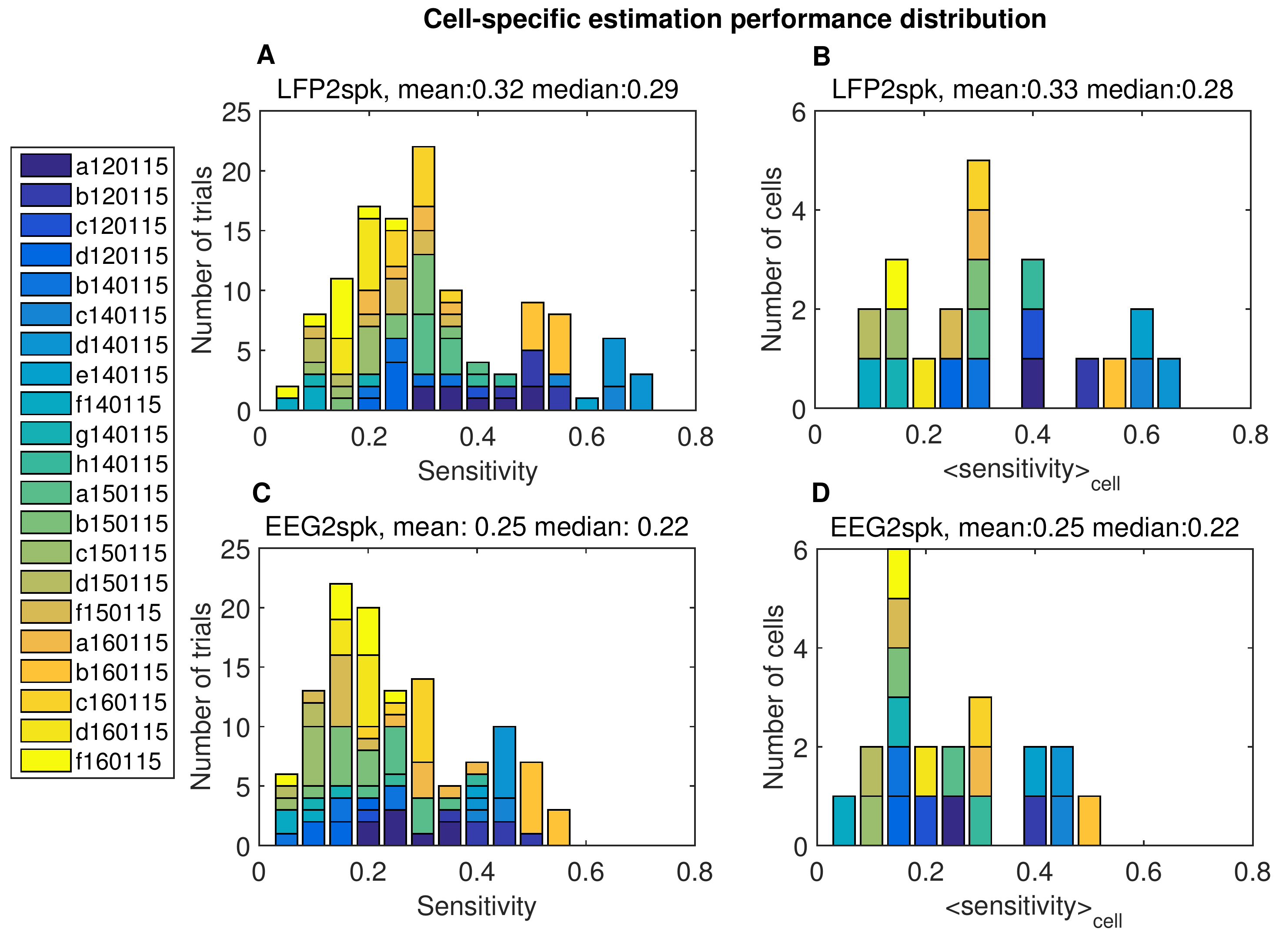}
\par\end{centering}
\centering{}\caption[LFP/EEG2spk performance distribution across trials and cells]{\textbf{Distribution across trials and cells of spike train estimation
performances} when using a spike-detection threshold to obtain the
same number of spikes in spk\protect\textsubscript{est} as in the
true spike train (in this case sensitivity and precision of spk\protect\textsubscript{est}
are equal). \textbf{(A)} Distribution of the sensitivities in the
spike train estimation over the trials. \textbf{(B)} Same as (A) when
the distribution is across the average sensitivities per cell. \textbf{(C,D)}
Same as respectively (A,B) in case of spike estimation from EEG. The
legend specifies the cells the data belong to. Mean and median values
of the distributions are displayed above each panel. The estimation
has been performed by using cell-specific filters and similar results
are obtained when using trial-specific and general filters (data not
shown).\label{PFfig_LFP/EEG2spk_perf_distribution}}
\end{figure}
In figure \ref{PFfig_LFP/EEG2spk_perf_distribution} we show the distribution
across trials of the spike train estimation performances performed
with cell-specific filters. Analogously to what reported in figure
\ref{PFfig_spk2LFP/EEG_perf_distribution}, we found that the performances
vary broadly from cell to cell but are relatively stable for any given
cell\footnote{More specifically, in figure \ref{PFfig_LFP/EEG2spk_perf_distribution}
the average (over cells) amplitude of the interval of sensitivity
values found for each cell (with more than one trial) is 0.11 for
LFP and 0.09 for EEG estimation.}. The median values of the sensitivity of the number of estimated
spikes within time windows of 26 ms is 0.29$\pm$0.11 (median$\pm$interquartile
range/2) for estimation from LFP and 0.22$\pm$0.09 from EEG when
using cell-specific filters. Very similar results were obtained also
with the trial-specific and general filters. In fact the sensitivity
distribution across filters never differed when comparing sensitivity
for estimation from LFP and EEG (two-tailed Kolmogorov-Smirnov tests,
p>0.45 for LFP and p > 0.21 for EEG). 

\begin{figure}
\begin{centering}
\includegraphics[scale=0.5]{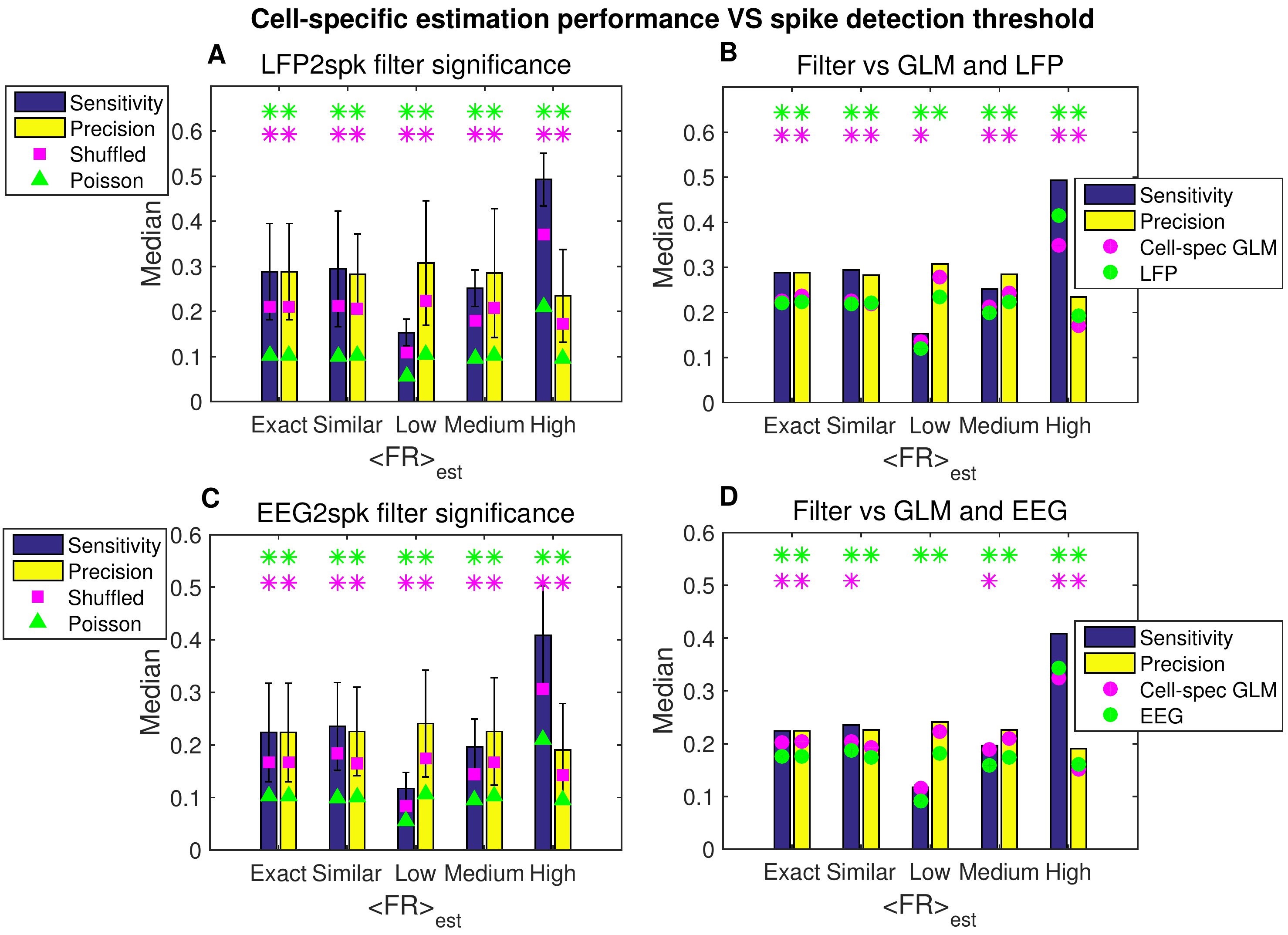}
\par\end{centering}
\centering{}\caption[LFP/EEG2spk cell-spec performance summary]{{\footnotesize{}}\textbf{\footnotesize{}Significance of the spike
train cell-specific estimation performance VS spike-detection threshold.
}{\footnotesize{}We analyze the significance of the Wiener-based estimation
and we also compared the performances with the ones obtained by using
other two different models that are (i) directly the LFP/EEG and (ii)
the GLM to estimate the FR.}\textbf{\footnotesize{} (A)}{\footnotesize{}
Spike train estimation performances and their significance as a function
of the threshold used to detect the estimated spikes (see section}\textcolor{red}{\footnotesize{}\vref{PF_section_thresholds}}{\footnotesize{}).
To evaluate the significance of the spike estimation, we consider
two different null hypotheses (see section \ref{PF_section_Performance-measures}):
(i) the triangles represent the median estimation performances obtained
by taking as spk\protect\textsubscript{est} a Poisson spike train
with the same average FR; (ii) the squares indicate the median estimation
performances obtained by randomly placing the estimated spikes in
the intervals where the estimated FR was above the spike-detection
threshold. The colored bars indicate the median values over the trials,
while the error bars represent the interquartile ranges. {*}$p<10^{-4}$
based on a one-tailed Kolmogorov-Smirnov test comparing the estimation
performances against the null hypothesis performances. }\textbf{\footnotesize{}(B)}{\footnotesize{}
Comparison with the performances obtained by using other two FR estimation
methods as a function of the threshold used to detect the estimated
spikes. In particular, we compare the performances given by the FR\protect\textsubscript{est}
computed as the convolution between the LFP and the (cell-specific)
Wiener filter (colored bars) with the ones obtained from taking directly
FR\protect\textsubscript{est}=LFP (green circles) and by approximating
the FR\protect\textsubscript{est} through a (cell-specific) GLM \citep{whittingstall2009frequency}.
{*}$p<0.03$ based on a one-tailed Kolmogorov-Smirnov test comparing
the estimation performances against the null hypothesis performances
represented by the LFP (green asterisks) and by the GLM (magenta asterisks).
}\textbf{\footnotesize{}(C,D)}{\footnotesize{} Same as respectively
(A,B) when the spike estimation is done from the EEG signal; {*}$p<0.002$
in (C) and {*}$p<0.03$ in (D). Very similar results are obtained
when using trial-specific filters (and trial-specific GLM, data not
shown). \label{PFfig_LFP/EEG2spk_rec_spec_perf_VS_FRclass}}}
\end{figure}
\begin{figure}
\begin{centering}
\includegraphics[scale=0.5]{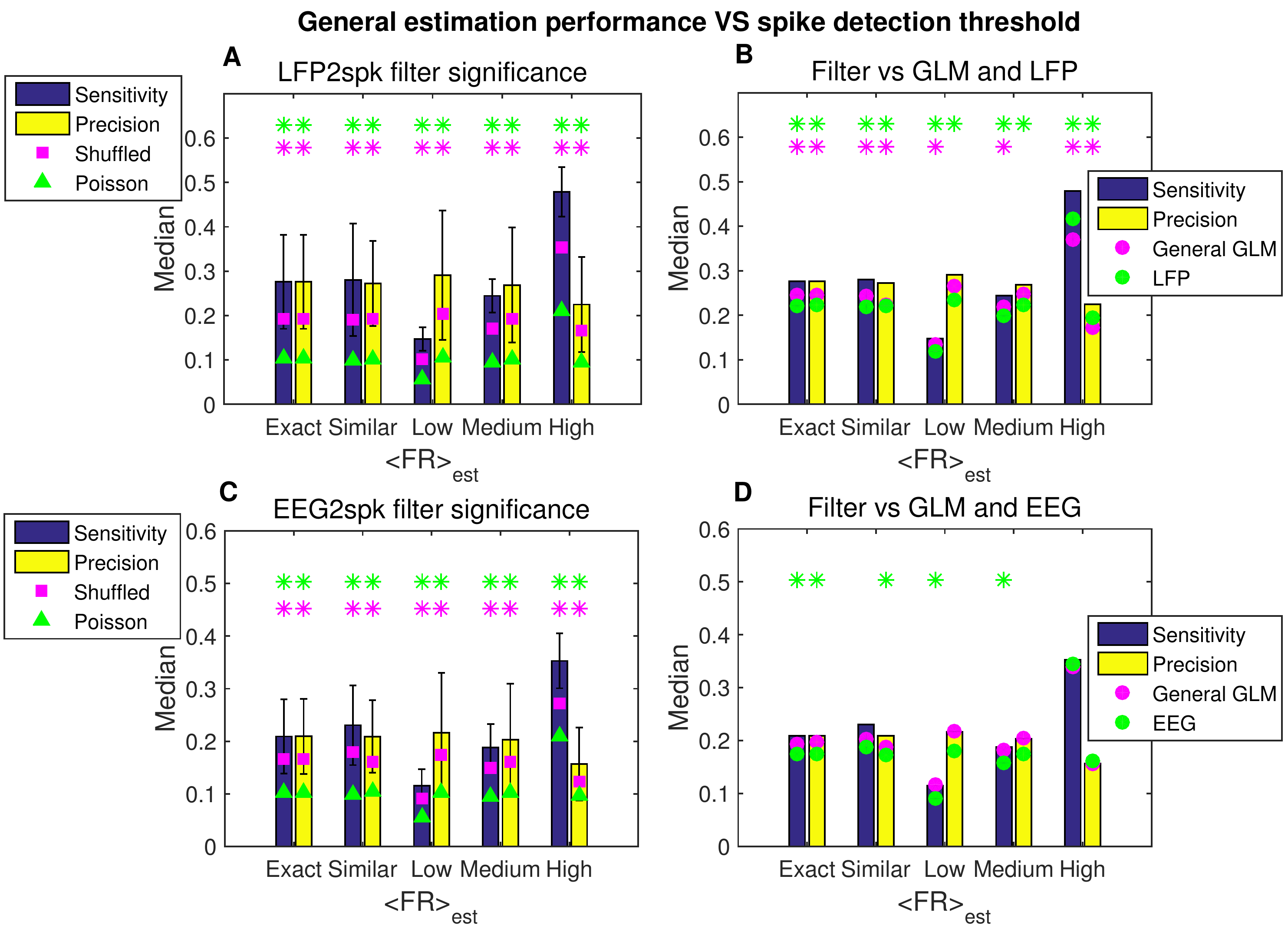}
\par\end{centering}
\centering{}\caption[LFP/EEG2spk general performance summary]{\textbf{Significance of the spike train general estimation performance
VS spike-detection threshold. }Same as figure \ref{PFfig_LFP/EEG2spk_rec_spec_perf_VS_FRclass}
for a general Wiener filter (and a general GLM). \textbf{(A)}{*}$p<10^{-4}$.\textbf{
(B)} {*}$p<0.05$. \textbf{(C)} {*}$p<0.03$. \textbf{(D)} {*}$p<0.05$.\label{PFfig_LFP/EEG2spk_general_perf_VS_FRclass}}
\end{figure}
\begin{figure}
\begin{centering}
\includegraphics[scale=0.75]{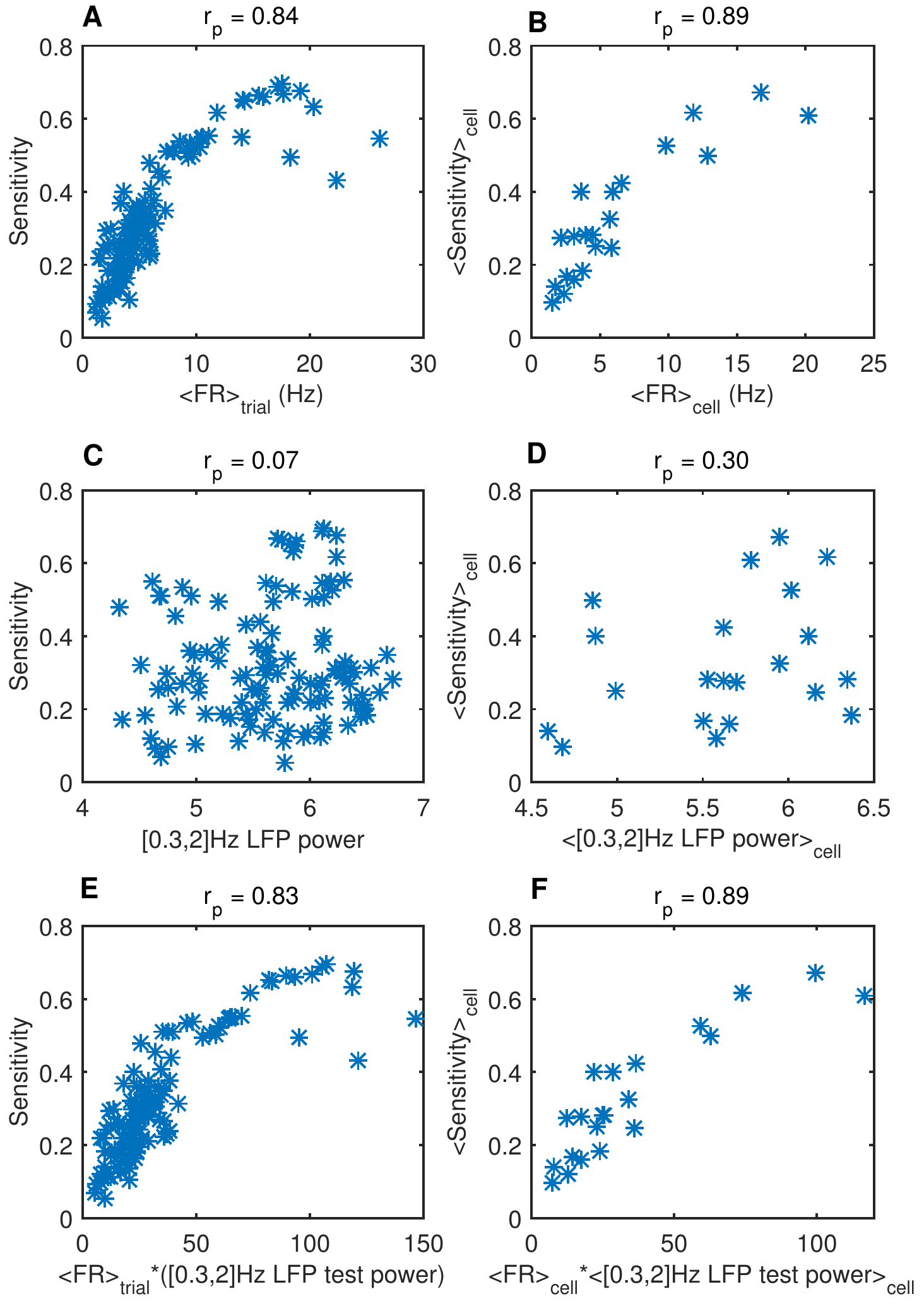}
\par\end{centering}
\centering{}\caption[LFP2spk performance scatter plots]{{\small{}}\textbf{\small{}Scatter plots of the performances of the
spike trains estimated from LFPs. }{\small{}Spike train estimation
performed by using a spike-detection threshold to obtain the same
$\langle FR\rangle_{est}$ as in the true spike train (i.e., ``exact''
case).}\textbf{\small{} (A)}{\small{} Sensitivity of the estimated
spike train of each trial as a function of the average FR. }\textbf{\small{}(C)}{\small{}
Sensitivity of the estimated spike train of each trial as a function
of the LFP power spectrum in the low delta band, {[}0.3 2{]}Hz. }\textbf{\small{}(E)}{\small{}
Sensitivity of the estimated spike train of each (test) trial as a
function of the product between the low LFP power spectrum and the
average FR. }\textbf{\small{}(B,D,F)}{\small{} Same as respectively
(A,C,E) when the data represents the average values over all the trials
belonging to a given cell. The values of the Pearson's correlation
between the two plotted variables are reported in the titles of each
panel. The showed performances are obtained by using a cell-specific
kernel, but similar results are obtained when using both trial-specific
and general kernels (data not shown).\label{PFfig_LFP2spk_perf_dispersion}}}
\end{figure}
\begin{figure}
\begin{centering}
\includegraphics[scale=0.75]{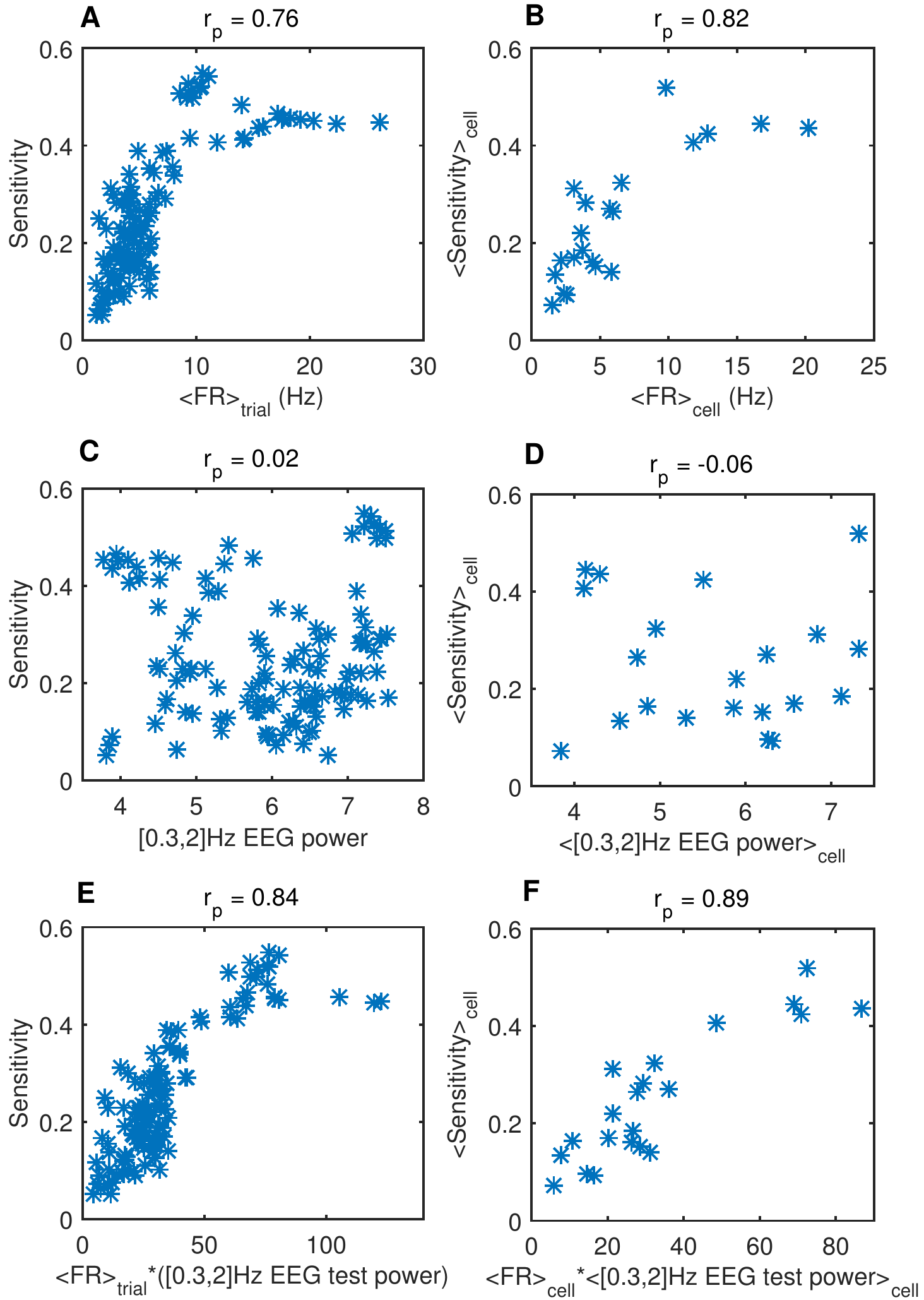}
\par\end{centering}
\centering{}\caption[EEG2spk performance scatter plots]{\textbf{Scatter plots of the performances of the spike trains estimated
from EEGs. }Same analysis as in figure \ref{PFfig_LFP2spk_perf_dispersion}
when estimating the spike trains from EEG signals.\label{PFfig_EEG2spk_perf_dispersion}}
\end{figure}

In order to evaluate the goodness of the performances obtained, we
performed both statistical significance tests and comparisons with
the performances obtained by using other methods. We found that the
spike train estimation performed with the Wiener kernel is always
significant and not only with respect to the chance level (for details
see panels (A,C) in figures \ref{PFfig_LFP/EEG2spk_rec_spec_perf_VS_FRclass}
and \ref{PFfig_LFP/EEG2spk_general_perf_VS_FRclass}). Furthermore,
when evaluating the performances obtained by taking directly the mass
signals as the estimated FR, we found that the Wiener kernel actually
produces an enhancement in the spike train estimation performances
(p<0.05, according to one-tailed Kolmogorov-Smirnov tests, see panels
(B,D) in figures \ref{PFfig_LFP/EEG2spk_rec_spec_perf_VS_FRclass}
and \ref{PFfig_LFP/EEG2spk_general_perf_VS_FRclass}). Therefore,
we can conclude that the filtering procedure is effective also after
the application of the non-linear threshold to detect spikes. \\
Finally, we compared the results of our method to those obtained with
a general linear model (see section \vref{PF_section_GLM}) constructed
on frequency decomposition of network oscillations\footnote{Note that the Wiener filter is built instead considering the whole
spectrum of the network oscillations. } which was used in \citep{whittingstall2009frequency} to estimate
the firing rate of MUA from both EEG and LFP signals. We estimated
the FR by using the GLM and then we applied the spike-detection threshold
to estimated spike times (as done in the previous cases). Note that
we set the GLM parameters in order to maximize its performances (see
figure \ref{PFfig_LFP/EEG2spk_Setting-GLM-parameters.}) and we still
found that, in all the cases, the performances of the filter were
significantly higher (p<0.05, according to one-tailed Kolmogorov-Smirnov
tests, see panels (B,D) in figure \ref{PFfig_LFP/EEG2spk_rec_spec_perf_VS_FRclass}
and panel (B) in figure \ref{PFfig_LFP/EEG2spk_general_perf_VS_FRclass}),
excepting when estimating spikes from the EEG signal using a general
filter (where the performances were not statistically different, see
panel (D) in figure \ref{PFfig_LFP/EEG2spk_general_perf_VS_FRclass}).\\
In figures \ref{PFfig_LFP/EEG2spk_rec_spec_perf_VS_FRclass} and \ref{PFfig_LFP/EEG2spk_general_perf_VS_FRclass}
we report the performances obtained when using cell-specific and general
estimation methods, respectively. Importantly, we found that the results
were stable across the three kinds of filters we considered. In more
detail, we found that the distributions of the performance values
associated with the different filters could never\footnote{We performed the statistical test by considering the ``exact'' and
``similar'' classes of estimation.} be statistically distinguished (respectively p>0.36 (0.21) for estimation
from LFP (EEG), according to two-tailed Kolmogorov-Smirnov tests).
\\
We also looked at the spike train estimation performances as a function
of the cutoff frequency (in a range between 10 and 90 Hz) of the LFP
and of the EEG and we found that the performances monotonically decreased
when decreasing the cutoff frequency in a similar way for all estimation
considered above (data not shown).\\
We then investigated how the estimation depends on the spike-detection
threshold. We found that, when the estimated FR is exactly equal to
the original one, the performances are never distinguishable from
those obtained when $\langle FR\rangle_{est}$ was similar to the
original one (for details see section \vref{PF_section_thresholds}),
for both LFP and EEG cases (p>0.55 according to two-tailed Kolmogorov-Smirnov
tests). This means that the performances are robust to a jitter in
the spike-detection threshold and that, in particular, we can estimate
the spiking activity from mass signals in a blind way, where the only
free parameter is given by the range (low, medium or high) of the
firing activity. Interestingly, we observed that by augmenting the
number of estimated spikes (that is going form the ``low'' to the
``high'' $\langle FR\rangle_{est}$), the sensitivity increase is
greater than the concurrent precision decrease\footnote{This dynamics is always observed and, in particular, when estimating
spikes from EEG with trial-specific (data not shown) or cell-specific
filters (see panel (C) in figure \ref{PFfig_LFP/EEG2spk_rec_spec_perf_VS_FRclass}),
the precision sensitivity in the ``exact'' and ``high'' are not
distinguishable (p>0.16, according to two-tailed Kolmogorov-Smirnov
tests), while the sensitivity increase is clearly significant (p$\ll10^{-10}$).}. This means that, when increasing the number of estimated spikes,
the majority of the added spikes will correctly predict true spikes.
This fact, together with the already observed positive correlation
between average FRs and the synchronization of FR and mass signals
(panels (A,B) in figures \ref{PFfig_avFR_LFP_dispersion} and \ref{PFfig_avFR_EEG_dispersion}),
suggests that there should be a positive correlation between the spike
estimation performances and the average FR. This is indeed what we
found, for both LFP and EEG signals (see panels (A,B) in figures \ref{PFfig_LFP2spk_perf_dispersion}
and \ref{PFfig_EEG2spk_perf_dispersion}). 


By comparing the scatter plots of the performances of spike train
estimation (figures \ref{PFfig_LFP2spk_perf_dispersion} and \ref{PFfig_EEG2spk_perf_dispersion})
with the same plots when estimating mass signals (figures \ref{PFfig_spk2LFP_perf_dispersion}
and \ref{PFfig_spk2EEG_perf_dispersion}), we note that the Pearson's
correlations with the average firing rate increase, while the correlations
with the power of the low frequencies decrease steeply. As a result,
in both cases, performances correlate the most with the average FRs
(whereas, when estimating EEG, the highest correlation was with the
EEG low frequency power, see figure \ref{PFfig_spk2EEG_perf_dispersion}).
This is due to two reasons. First, the positive correlation observed
between the (mass signal-FR) synchronization and average FR (see panels
(A,B) in figures \ref{PFfig_avFR_LFP_dispersion} and \ref{PFfig_avFR_EEG_dispersion}).
Second, when reconstructing the spike train, only the position of
the peaks in FR\textsubscript{est} matters and not the whole shape
of the signal, like in the estimation of analogue (i.e., mass) signals.
\\
When evaluating mass signals, we pointed out that high amplitudes
of the low network oscillations facilitated the estimation (see section
\vref{PF_section_spk2LFP_avFR_VS_lowFreq}). This is still true, because
also in the estimated FR (prior to the spike detection), the frequencies
better estimated are the lowest (data not shown). As a results, the
correlation between the spike train estimation performances and the
product of the average firing rate for the low frequency power is
equal (LFP) or higher (EEG; but not lower) to the correlation with
the average firing rate alone (see panels (E,F) in figures \ref{PFfig_LFP2spk_perf_dispersion}
and \ref{PFfig_EEG2spk_perf_dispersion}).

\begin{figure}
\begin{centering}
\includegraphics[scale=0.5]{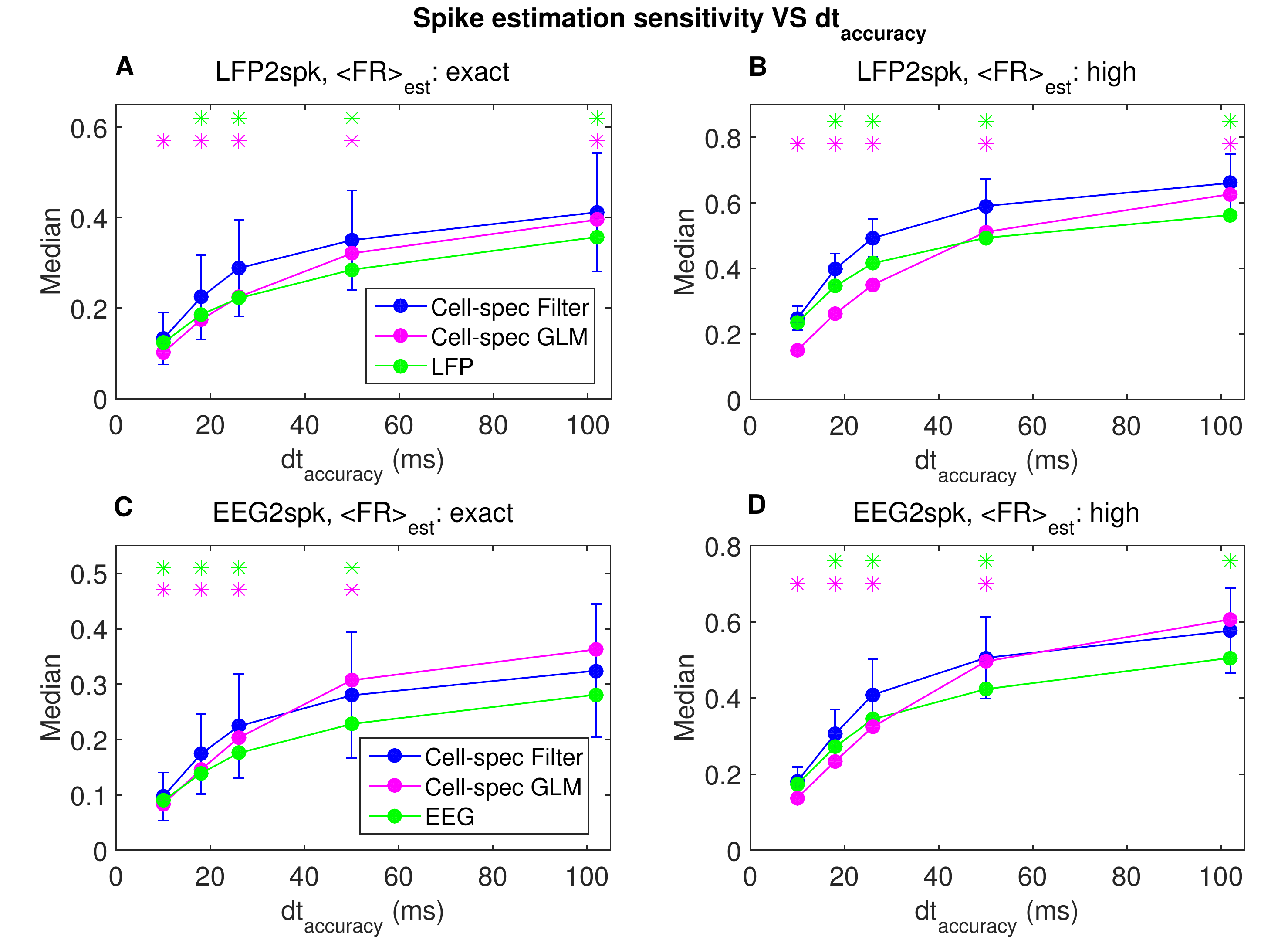}
\par\end{centering}
\begin{centering}
\caption[LFP/EEG2spk performances VS dt\textsubscript{accuracy}]{\textbf{Spike train estimation performance VS dt\protect\textsubscript{\textbf{accuracy}}.}
Median sensitivity of the estimated spike trains as a function of
the accuracy used to compare true and estimated spikes (see section
\vref{PF_section_Performance-measures}). \textbf{(A)} The spike-detection
threshold is set to obtain the same number of estimated spikes as
in the true spike train. Error bars indicates the interquartile range;
the models used to evaluate FR\protect\textsubscript{est} is specified
in the legend. {*}$p<0.03$\textbf{ }based on a one-tailed Kolmogorov-Smirnov
test comparing the estimation performances obtained from the filter
against the null hypothesis performances represented by the LFP (green
asterisks) and by the GLM (magenta asterisks).\textbf{ (B) }Same as
(A) when using a spike-detection threshold that results in an high
value of $\langle FR\rangle_{est}$ (see section\vref{PF_section_thresholds});
{*}$p<0.005$. \textbf{(C,D)} Same as respectively (A,B) when estimating
the spike trains from the EEGs; {*}$p<0.05$ in (C) and {*}$p<0.03$
in (D). The showed performances are obtained by using a cell-specific
kernel; similar results are obtained when using both trial-specific
and general kernels (data not shown).\label{PFfig_LFP/EEG2spk_perf_VS_dtACC}}
\par\end{centering}
\centering{}
\end{figure}
To conclude the analysis of the spike train estimation, we investigated
how the performances vary with the dt\textsubscript{accuracy} (in
a range between 10 and 102 ms) used to compare estimated and original
spike trains (see section \ref{PF_section_Performance-measures}).
In figure \ref{PFfig_LFP/EEG2spk_perf_VS_dtACC} we plot the median
sensitivity obtained both when using the threshold to obtain the same
number of estimated spikes as in the true spike train (panels (A,C))
and the threshold to obtained always an high average firing rate (i.e.,
9.4 Hz, see section \vref{PF_section_thresholds}; panels (B,D)).
This figure shows that for dt\textsubscript{accuracy}$\leq$50 ms,
the Wiener filter gives always performances better than the GLM. On
the other hand, with the smallest value of dt\textsubscript{accuracy}
(i.e., 10 ms), the (very low) performance obtained without filtering
the mass signals are very close to the one obtained with the filtering
procedure and in some cases not significantly lower (p>0.05, according
to one-tailed Kolmogorov-Smirnov tests).

\newpage
\subsubsection*{Firing rate estimation}

\begin{figure}
\begin{centering}
\includegraphics[scale=0.45]{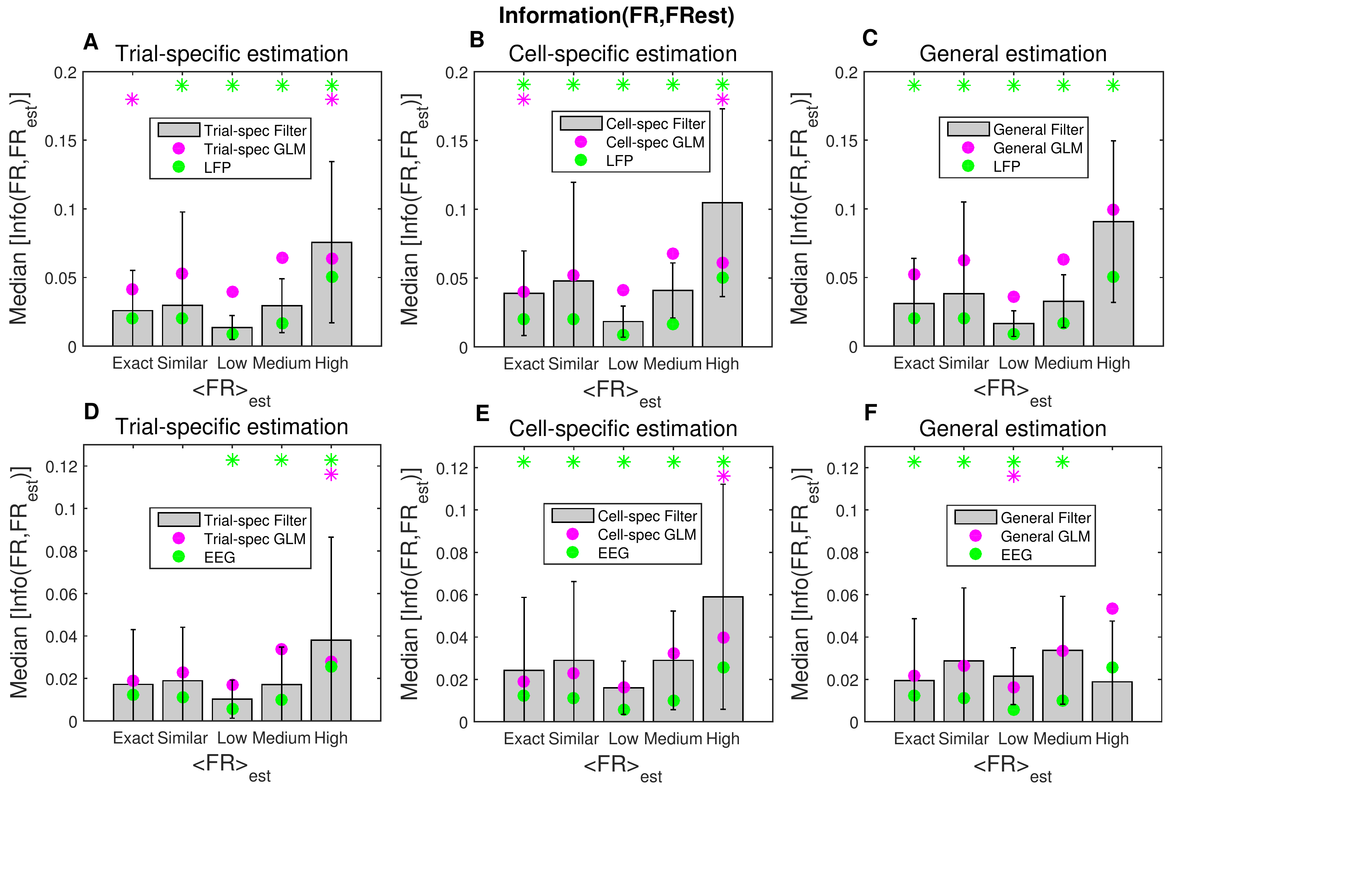}
\par\end{centering}
\centering{}\caption[Information(FR,FR\textsubscript{est})]{\textbf{Information between the true and the estimated FR}. In order
to compute information (see section \ref{section_Shannon_information}),
the original and estimated FRs are firstly smoothed with an Hann window
50ms wide and then their values are binned into two values (0 and
1). For the true FR, all the values above 0 are set to 1, while for
the FR\protect\textsubscript{est}, the values less or equal to the
spike-detection threshold are set to 0 and the ones above to 1. \textbf{(A)}
Information between FR and FR\protect\textsubscript{est} when using
a trial-specific filter as a function of the $\langle FR\rangle_{est}$
class compared with the null hypothesis given by the information obtained
from a trial-specific GLM (magenta circles) and by using directly
the LFP signal to estimate the FR (green circles). {*}$p<0.05$\textbf{
}based on one-tailed Kolmogorov-Smirnov tests.\textbf{ (B,C)} Same
as (A) in case of respectively cell-specific (B) and general (C) estimation.
\textbf{(D-F)} Same as (A-B) when the FR estimation is performed from
the EEG. We are interested in comparing the level of information across
the methods, thus, since the bias is always the same, we do not adopt
any bias correction.\label{PFfig_LFP_EEG2spk_Information-FR_FRest}}
\end{figure}

We quantified the similarity between original and estimated FRs by
means of mutual information. In particular, we computed the information
between FR and FR\textsubscript{est} after binning the signals in
two values which represent respectively a probability of firing equal
to zero or higher (for details, see caption figure \ref{PFfig_LFP_EEG2spk_Information-FR_FRest}).
Interestingly, we found (see figure \ref{PFfig_LFP_EEG2spk_Information-FR_FRest})
that by applying the filter we obtained a more precise estimation
of the time intervals when the FR is higher than zero (in addition
to the increase of similarity between the position of the peaks in
the estimated and true FRs, figures \ref{PFfig_LFP/EEG2spk_rec_spec_perf_VS_FRclass}
and \ref{PFfig_LFP/EEG2spk_general_perf_VS_FRclass}). On the other
hand, the comparison with GLM shows that the filter performances are
only in few cases better than the GLM ones. This is not surprising,
since the GLM is based on the network gamma oscillations, which, during
slow wave oscillations, are strongly locked with the firing activity
(mainly for the LFP, see panel (C) in figure \ref{PFfig_LFP/EEG2spk_Setting-GLM-parameters.})\textbf{\textcolor{red}{{}
\citep{mukovski2007detection}}}. 

\begin{figure}
\begin{centering}
\includegraphics[scale=0.45]{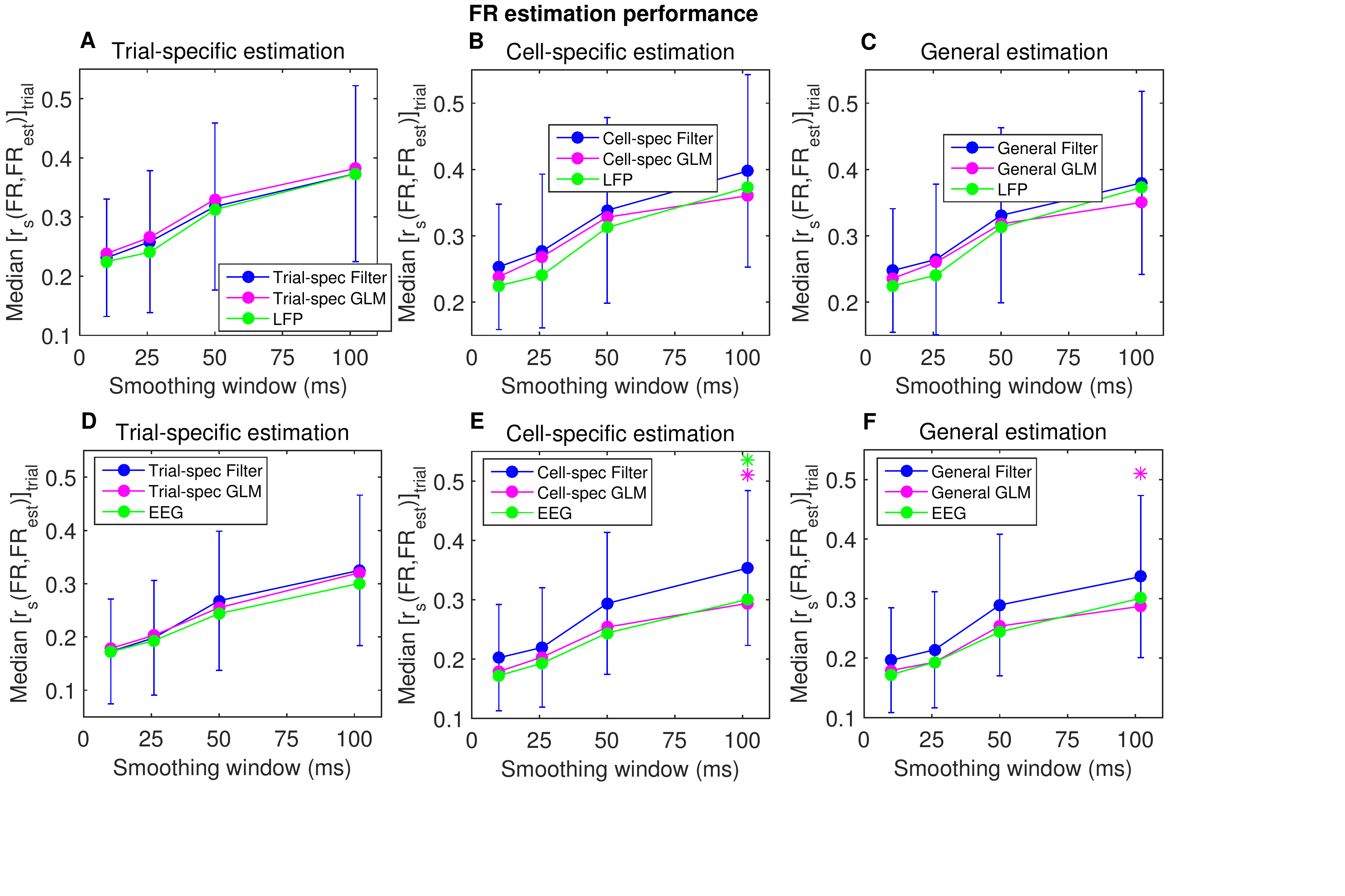}
\par\end{centering}
\centering{}\caption[LFP/EEG2FR performances VS Hann smoothing windows]{\textbf{Firing rate estimation performance}s\textbf{ as a function
of the Hann window used to smooth true and estimated FRs}. The median
value of the Spearman's correlation between the true and the estimated
FR is showed as a function of the Hann window width used to smooth
the FRs. The lowest value of smoothing window corresponds to the case
where no one smoothing was performed. \textbf{(A)} Estimation performance
of the FR obtained from the LFP by using a trial-specific Wiener kernel
compared with the null hypothesis represented by the performance obtained
from the trial-specific GLM (magenta circles) and by taking as FR\protect\textsubscript{est}
directly the LFP (green circles). The filter performance are not statistically
higher ($p>0.05$,\textbf{ }based on one-tailed Kolmogorov-Smirnov
tests).\textbf{ (B,C)} Same as (A) in case of respectively cell-specific
(B) and general (C) estimation. \textbf{(D-F)} Same as (A-B) when
the FR estimation is done from the EEG ({*}$p<0.05$). \label{PFfig_LFP_EEG2FR_perf_vs_smoothing}}
\end{figure}
We next focused on the FR estimation performances, by analyzing the
similarity between the true and the estimated FRs (prior to the use
of the spike-detection threshold). Remember that the (true) FR signal
is obtained from the spike train and depends on the spike smoothing
window chosen that, in our analysis, is 10 ms. After computing FR\textsubscript{est},
we convolved both the original and the estimated FRs with an Hann
window of a given amplitude and eventually we measured the Spearman's
correlation between the signals. In figure \ref{PFfig_LFP_EEG2FR_perf_vs_smoothing},
the median correlation is shown as a function of the amplitude of
the window used.
\begin{figure}
\begin{centering}
\includegraphics[scale=0.5]{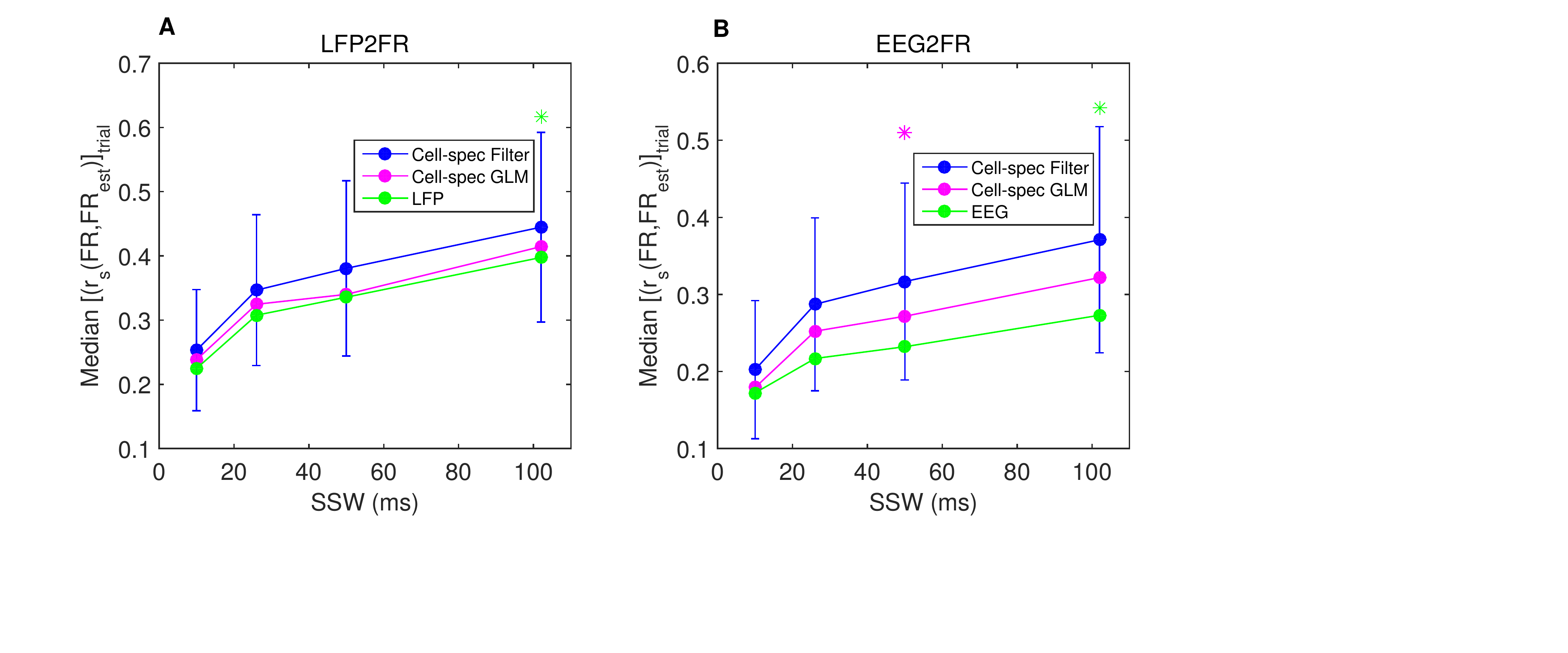}
\par\end{centering}
\centering{}\caption[LFP/EEG2FR performances VS SSWs]{\textbf{Firing rate estimation performances as a function of the
spike smoothing window used to compute the true FR}. The median values
of the Spearman's correlation between the true and the estimated FR
are showed; error bars represent the interquartile range. Note that,
with respect to the analysis showed in figure \ref{PFfig_LFP_EEG2FR_perf_vs_smoothing},
for each SSW has been computed the relative filters (and GLMs) and
no one smoothing has been applied after FR estimation. \textbf{(A)}
FR estimated from LFP. \textbf{(B)} FR estimated from EEG. ({*}$p<0.05$,\textbf{
}based on one-tailed Kolmogorov-Smirnov tests). The results are computed
from cell-specific filters (similar results are obtained when using
trial-specific and general filters; data not shown)\label{PFfig_LFP_EEG2FR_VS_binPSTH}}
\end{figure}
 In this analysis we smoothed the FRs after the estimation. However,
we obtained similar results also when using from the beginning spike
smoothing windows (to compute the original FR, see section \vpageref{PF_section_FRcomputation})
of the given amplitudes and then computing the associated filters,
without performing a final smoothing (see figure \ref{PFfig_LFP_EEG2FR_VS_binPSTH}).
Interestingly, we note that the FR estimation performances obtained
with the Wiener filter are not better than the ones obtained respectively
without filtering the mass signals and by using the GLM (i.e., $p>0.05$
according to Kolmogorov-Smirnov tests, see figures \ref{PFfig_LFP_EEG2FR_perf_vs_smoothing}
and \ref{PFfig_LFP_EEG2FR_VS_binPSTH}), whereas, by applying the
threshold to detect spikes, the filter's performances are higher than
in the other models (see panels (B,D) in figures \ref{PFfig_LFP/EEG2spk_rec_spec_perf_VS_FRclass}
and \ref{PFfig_LFP/EEG2spk_general_perf_VS_FRclass}). Thus, while
the overall shape of estimated FR is very similar over the three models
used, the positions of the peak are more close to the original ones
when using the Wiener filter to estimate the FR.

\begin{figure}
\begin{centering}
\includegraphics[scale=0.75]{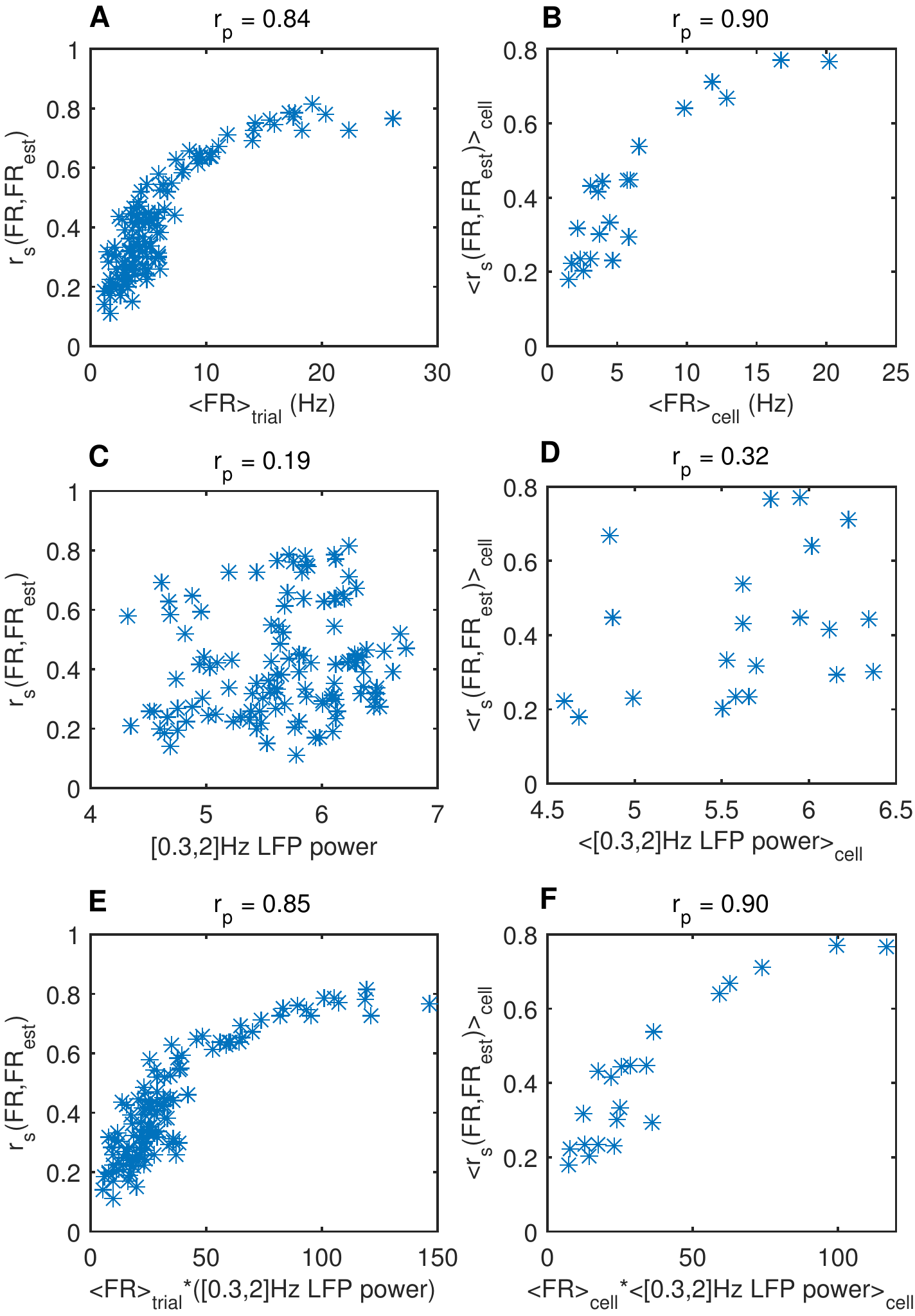}
\par\end{centering}
\centering{}\caption[LFP2FR performance scatter plots]{{\small{}}\textbf{\small{}Scatter plots of the performance of the
FRs estimated from LFPs. }{\small{}The SSW used to compute the filter
is 50 ms wide (very similar results are obtained also when SSW was
10 ms and, after estimation, we smoothed the true and estimated FR
with an Hann window of 50 ms, as done in figure \ref{PFfig_LFP_EEG2FR_perf_vs_smoothing};
data not shown).}\textbf{\small{} (A)}{\small{} Performance of the
FR estimated from the LFP (as measured by Spearman's correlation between
FR and FR\protect\textsubscript{est}) as a function of the average
FR (each point represents the values in a trial). }\textbf{\small{}(C)}{\small{}
FR estimation performance of each trial as a function of the LFP power
spectrum in the low delta band, {[}0.3 2{]}Hz. }\textbf{\small{}(E)}{\small{}
FR estimation performance of each trial as a function of the product
between the low LFP power spectrum and the average FR. }\textbf{\small{}(B,D,F)}{\small{}
same as respectively (A,C,E) when each variable is averaged over the
trials belonging to a cell. The values of the Pearson's correlation
between the plotted variables are displayed in the panel's titles.
Here we used cell-specific kernels, but similar results are obtained
when using both trial-specific and general kernels (data not shown).\label{PFfig_LFP2FR_perf_dispersion}}}
\end{figure}
\begin{figure}
\begin{centering}
\includegraphics[scale=0.75]{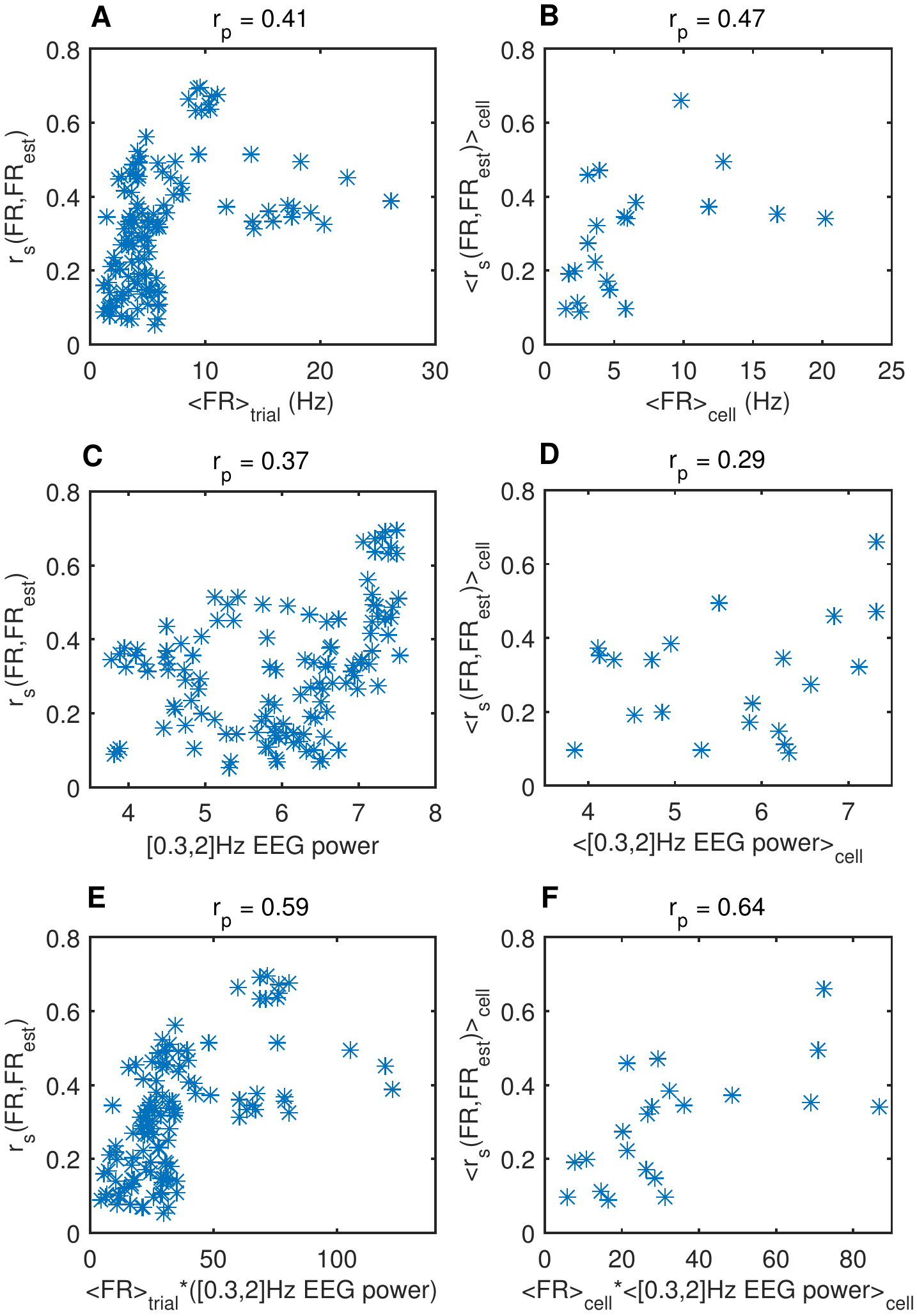}
\par\end{centering}
\centering{}\caption[EEG2FR performance scatter plots]{\textbf{Scatter plots of the performance of the FRs estimated from
EEGs.} Same analysis performed in figure \ref{PFfig_LFP2FR_perf_dispersion}
when estimating the spike trains from EEG signals.\label{PFfig_EEG2FR_perf_dispersion}}
\end{figure}
We conclude this section by showing the scatter plots of the FR estimation
performances. We already showed the same scatter plots when estimating
both the mass signals and the spike trains. In fact they are useful
to confirm our observations about the contributions of respectively
average firing rates and low frequencies of network oscillations in
shaping the relationship between single-unit firing and mass signals.
In the FR\textsubscript{est} (like in LFP\textsubscript{est} and
EEG\textsubscript{est}) the frequencies better estimated are the
lowest (as stated above). Since FRs (unlike spike trains) are analogue
signals, as a result there is an increase of the correlations between
the FR estimation performances and the low powers of the mass signals
with respect to the spike train estimation (see figures \ref{PFfig_LFP2FR_perf_dispersion}
and \ref{PFfig_EEG2FR_perf_dispersion}).%
{} However, we found that the correlation of the performances with the
product of average firing rate and low frequency power is respectively
equal (LFP) or higher (EEG; but not lower) to the correlation obtained
by considering each variable singularly (see panels (E,F) in figures
\ref{PFfig_LFP2FR_perf_dispersion} and \ref{PFfig_EEG2FR_perf_dispersion}),
like in case of spike train estimation. 

\subsection{Causality in the estimation}

As explained in section \vref{section_FILTERcausality}, the estimations
we performed above were acausal. An interesting question is whether
there is a dominant direction of causality between spiking and mass
activity, or in other words whether the spike times directly caused
the mass signal changes or instead the spike times were biased or
caused by changes in the mass signal measures. This can be investigated
by considering the temporal relationships between signals. Indeed,
in the Wiener-Granger spirit \citep{granger1980testing}, one signal
may cause the other if its changes consistently anticipate the changes
in the other signal (and thus the signal past consistently helps to
predict the other signal better than its own past alone). That this
may be the case is suggested by the fact that the peaks of the filters
are not centered on 0 time lag (and that the filters are not symmetric
with respect to the filter peak position). To investigate this issue
further, we repeated the analysis (in case of cell-specific and general
filters) using both causal and anti-causal filters. The causal filters
were obtained by setting to zero the Wiener filters for negative time
lags, whereas the anti-causal filters by setting to zero the positive
time lag values. Since the acausal filter is the optimal one, the
performances will decrease by using other filters. Thus, we quantified
the performance reductions to see if and to which extent the causal
(or anti-causal) component has a higher weight in the estimation. 

In table \ref{PFtable_spk2LFP/EEG_CAUSALITY}, it is shown the reduction
of the performances in case of LFP and EEG estimation. The performance
reduction is smaller when the filter is causal, in agreement with
the fact that the peaks of the general filters were at positive time
lags (see figure \ref{PFfig_spk2LFP/EEG-general-filters}), indeed
the performances obtained with the casual filters are statistically
higher than the ones obtained with the anti-causal filters. 
\begin{table}[H]
\begin{centering}
\begin{tabular}{|c|c|c|c|c|c|c|}
\hline 
\multirow{3}{*}{} & \multicolumn{2}{c|}{\textbf{\footnotesize{}ACAUSAL vs }} & \multicolumn{2}{c|}{\textbf{\footnotesize{}ACAUSAL vs }} & \multicolumn{2}{c|}{\textbf{\footnotesize{}CAUSAL > }}\tabularnewline
 & \multicolumn{2}{c|}{\textbf{\footnotesize{} CAUSAL}} & \multicolumn{2}{c|}{\textbf{\footnotesize{}ANTI-CAUSAL}} & \multicolumn{2}{c|}{\textbf{\footnotesize{}ANTI-CAUSAL}}\tabularnewline
\cline{2-7} 
 & \textit{\footnotesize{}LFP} & \textit{\footnotesize{}EEG} & \textit{\footnotesize{}LFP} & \textit{\footnotesize{}EEG} & \textit{\footnotesize{}LFP} & \textit{\footnotesize{}EEG}\tabularnewline
\hline 
{\footnotesize{}$r_{s}$} & {\footnotesize{}$\delta_{\%}=-9.5$ } & {\footnotesize{}$\delta_{\%}=-13.7$ } & {\footnotesize{}$\delta_{\%}=-27.2$ } & {\footnotesize{}$\delta_{\%}=-18.7$ } & {\footnotesize{}$p<10^{-10}$} & {\footnotesize{}$p=3*10^{-4}$}\tabularnewline
\hline 
{\footnotesize{}NMSD} & {\footnotesize{}$\delta_{\%}=+6.3$ } & {\footnotesize{}$\delta_{\%}=+6.2$ } & {\footnotesize{}$\delta_{\%}=+14.5$ } & {\footnotesize{}$\delta_{\%}=+9.2$ } & {\footnotesize{}$p<10^{-10}$} & {\footnotesize{}$p=6*10^{-4}$}\tabularnewline
\hline 
\end{tabular}
\par\end{centering}
\medskip{}

\centering{}\caption[Skp2LFP/EEG, causal VS anti-causal estimation]{\textbf{Comparison between the effects of performing causal and anti-causal
estimations of mass signals. }$\delta_{\%}$ is the median of the
relative performance variations observed in each trial with respect
to the acausal estimation ($\mathbf{r_{s}}$ is the Spearman's correlation,
and \textbf{NMSD} is the normalized mean squared distance, between
the true and the estimate signals). The last two columns display the
\textbf{p-values} obtained from a one-tailed Wilcoxon signed rank
test comparing the estimation performances obtained by using casual
kernels against the performances obtained with anti-causal kernels.
We found that the performance reduction is significantly smaller when
using casual kernels, both in case of LFP and EEG estimation. Results
obtained from cell-specific kernels (but similar dynamics are observed
also with a general kernel).\label{PFtable_spk2LFP/EEG_CAUSALITY}}
\end{table}

As a control test, we repeated the analysis in the opposite direction,
that is when estimating firing activity from the mass signal. We found
that, for both firing rate and spike train estimation, larger performances
are observed when using an anti-casual filter (see table \ref{PFtable_LFP/EEG2FR2spk_CAUSALITY}).
\begin{table}[h]
\begin{centering}
\begin{tabular}{|c|c|c|c|c|c|c|}
\hline 
\multirow{3}{*}{} & \multicolumn{2}{c|}{\textbf{\footnotesize{}ACAUSAL vs }} & \multicolumn{2}{c|}{\textbf{\footnotesize{}ACAUSAL vs }} & \multicolumn{2}{c|}{\textbf{\footnotesize{}CAUSAL > }}\tabularnewline
 & \multicolumn{2}{c|}{\textbf{\footnotesize{} CAUSAL}} & \multicolumn{2}{c|}{\textbf{\footnotesize{}ANTI-CAUSAL}} & \multicolumn{2}{c|}{\textbf{\footnotesize{}ANTI-CAUSAL}}\tabularnewline
\cline{2-7} 
 & \textit{\footnotesize{}LFP} & \textit{\footnotesize{}EEG} & \textit{\footnotesize{}LFP} & \textit{\footnotesize{}EEG} & \textit{\footnotesize{}LFP} & \textit{\footnotesize{}EEG}\tabularnewline
\hline 
{\footnotesize{}$r_{s}$} & {\footnotesize{}$\delta_{\%}=-17.1$ } & {\footnotesize{}$\delta_{\%}=-14.2$ } & {\footnotesize{}$\delta_{\%}=-2.7$ } & {\footnotesize{}$\delta_{\%}=-8.2$ } & {\footnotesize{}$p<10^{-10}$} & {\footnotesize{}$p=3*10^{-5}$}\tabularnewline
\hline 
{\footnotesize{}Sensitivity} & {\footnotesize{}$\delta_{\%}=-11.4$ } & {\footnotesize{}$\delta_{\%}=-6.8$ } & {\footnotesize{}$\delta_{\%}=-4.7$ } & {\footnotesize{}$\delta_{\%}=-4.6$ } & {\footnotesize{}$p=2*10^{-6}$} & {\footnotesize{}$p=0.017$}\tabularnewline
\hline 
\end{tabular}
\par\end{centering}
\medskip{}

\caption[LFP/EEG2FR, causal VS anti-causal estimation]{\textbf{Comparison between the effects of performing causal and anti-causal
estimations of the firing activity. }$\delta_{\%}$ is the median
of the relative performance variations observed in each trial with
respect to the acausal estimation. $\mathbf{r_{s}}$ is the Spearman's
correlation between $FR$ and $FR_{est}$ and it is evaluated when
the SSW was 10 ms; Sensitivity is the sensitivity of the spike times
estimation when$\langle FR\rangle_{est}$ is exact (with SSW=10ms)
and measured with an accuracy of 26 ms. The last two columns display
the p-values obtained from a one-tailed Wilcoxon signed rank test
comparing the estimation performances obtained by using anti-casual
kernels against the performances obtained from causal kernels. We
found that the performance reduction is significantly smaller when
using anti-casual kernels, both in case of estimation from LFP and
EEG. These data come from cell-specific kernels, but similar dynamics
are observed with the general kernels.\label{PFtable_LFP/EEG2FR2spk_CAUSALITY}}

\end{table}

Interestingly, the same relationships were observed when estimating
the LFP from the spiking activity of pyramidal neurons in deep layers
(see section \vref{PF_section_PYR5_LFP}), as shown in table \ref{PFtable_PYR5_spk2LFP_CAUSALITY}.
This is in agreement with the fact that, also in that case, the peak
of the general filter was at a positive time lag (see panel (B) in
figure \ref{PFfig_spk2LFP_SOM_PYR5_datasets}). 
\begin{table}[h]
\begin{centering}
\begin{tabular}{|c|c|c|c|}
\hline 
\multirow{2}{*}{\textbf{\textit{\small{}PYR.}}} & \multicolumn{1}{c|}{\textbf{\footnotesize{}ACAUSAL vs }} & \multicolumn{1}{c|}{\textbf{\footnotesize{}ACAUSAL vs }} & \textbf{\footnotesize{}CAUSAL > }\tabularnewline
 & \multicolumn{1}{c||}{\textbf{\footnotesize{} CAUSAL}} & \multicolumn{1}{c||}{\textbf{\footnotesize{}ANTI-CAUSAL}} & \textbf{\footnotesize{}ANTI-CAUSAL}\tabularnewline
\hline 
{\footnotesize{}$r_{s}$} & \textit{\scriptsize{}$\delta_{\%}=-7.7$ } & \textit{\scriptsize{}$\delta_{\%}=-42.4$ } & \textit{\scriptsize{}$p=3*10^{-5}$ }\tabularnewline
\hline 
{\footnotesize{}NMSD} & \textit{\scriptsize{}$\delta_{\%}=+4.1$ } & \textit{\scriptsize{}$\delta_{\%}=+12.4$ } & \textit{\scriptsize{}$p=2*10^{-5}$ }\tabularnewline
\hline 
\end{tabular}
\par\end{centering}
\medskip{}

\caption[Skp2LFP, causal VS anti-causal estimation]{\textbf{Comparison between the effects of performing causal and anti-causal
estimation of LFPs from SUA of excitatory neurons. }The spiking activity
comes from pyramidal neurons of deep layers (i.e., 5 or 6) and the
LFP is recorded at the same depth (3 mice, 7 cells and 23 trials).
The table shows the same analysis as in table \ref{PFtable_spk2LFP/EEG_CAUSALITY}.
We found that also with this dataset, when estimating mass signal,
the performance reduction is significantly smaller with casual kernels.\label{PFtable_PYR5_spk2LFP_CAUSALITY}}
\end{table}

Thus, we can conclude that the position of the general filter peak
indicates which signal anticipates the other. In particular, in the
spk2LFP/EEG estimation, the causal filters work better than the anti-causal,
while the opposite is true when estimating the firing activity. Therefore,
during slow wave oscillations, the spiking activity (of both inhibitory
and excitatory neurons) anticipates and causes mass signals variations,
rather than the other way around. This suggests that the types of
cells considered here play an important part in the generation of
the slow wave cycle captured by the mass signal.

\newpage{}

~

\newpage{}

\chapter{Conclusions\label{chap:Conclusions}}

\ohead{\headmark} 
\pagestyle{scrheadings}    

\lettrine[lines=2]{T}{he} motivation of this work was rooted in
the following two questions: how can we model the relationship between
the single-neuron level and the dynamics observed at the level of
population of neurons? How can we study this empirically by analyzing
joint recordings?\\
In section \ref{CONCL_frontiers} we summarized the results and discussed
their implications when the investigation was performed in a modelling
framework, while in section \ref{CONCL_Fellin}, we reported the results
obtained from the analysis of concomitant LFPs/EEGs and single-unit
activity. The implications of these findings, as well as further questions
that arise from this work, are reported in the following. 

\section[Modeling the relationship between single-neurons and population of
neurons]{Modeling the relationship between the dynamics of single neurons
and population of neurons\label{CONCL_frontiers} }

In chapter \ref{chapter_frontiers} we compared in detail the neural
population dynamics of recurrent LIF networks when adopting two different
models for the synaptic currents at the single-neuron level, namely
current-based or conductance-based models. In the former case, the
post synaptic potentials of each kind of synapses were constant (see
equation \ref{eq_5_paperFR}), while in the conductance-based models,
the PSPs depended on the membrane potential of the post synaptic neuron
(see equation \ref{eq_6_paperFR}). The comparison of network dynamics
was made on networks with all shared parameters set to an equal common
value, and with model-specific synaptic parameters set by a novel
recursive procedure that makes conductance-based networks (COBN) and
current-based networks (CUBN) directly comparable. This means that
the differences we found when analyzing the dynamics of population
of neurons in the two networks did not depend on the parameter setting,
but they were only due to the consequences of adopting a different
model at the single-neuron level. Our main result was that, although
average firing rates and peak frequency of gamma LFP oscillations
in such comparable networks were very similar over a wide range of
parameters, other aspects of neural population dynamics (such as shape
of oscillation spectra or cross-neuron correlation) were significantly
different between CUBN and COBN. In particular, oscillation spectra,
gamma synchronization and cross-neuron correlation were more markedly
modulated by the external input in COBN than in CUBN. The significance
of these findings, and their relationship with both theoretical and
experimental literature, is discussed in the following.

\subsection{Establishing comparable networks}

The first contribution of the work presented here was to provide a
new recursive algorithm to determine the COBN conductance values that
correspond to a given set of CUBN synaptic efficacies in networks
that have identical values for all the shared parameters. We found
that this procedure was able to build two networks displaying relatively
small differences, both in the average firing rates and in the gamma
frequency peak position, for an input range sufficiently large to
encompass both low- and high-conductance states \citep{Destexhe2003}.
The relationship of our new procedure with the previous work we built
on is discussed in the following. 

In a previous work addressing the issue of building equivalent CUBN
and COBN models \citep{laCamera2004}, the authors discarded the approach
of setting synaptic conductances at fixed average MP (i.e., the one
we used in this work) stating that \textquotedblleft Although this
might work for a single input, it does not work for all inputs in
a large pool (results not shown).\textquotedblright{} La Camera and
colleagues proposed instead to build equivalent networks by making
both inhibitory and excitatory connectivity free parameters, so that
the optimal equivalence was obtained when the CUBN had twice the excitatory
and half the inhibitory connectivity of the COBN. Differently from
this procedure, in our work all the common parameters of the two networks
were identical, including the connectivity matrix. This, in our view,
has the advantage that differences in network dynamics can be more
directly imputed to changes in model synaptic dynamics. \citet{Meffin2004}
determined the value of the conductances starting from a \textquotedblleft fixed
rough estimate of the average MP\textquotedblright{} set as the midpoint
between threshold and reset potential. The difference with our work
is that we used directly the actual average value of the MP of the
neurons of each population. Note that there is a discrepancy between
the two values since the true average MP was equal or slightly below
the reset potential (figure \ref{fig_5_paperFR}D). In extensive initial
simulations, we found that using the average MP, rather than the midpoint
between threshold and reset potential, made it much easier for the
comparable networks to exhibit very close firing rates and gamma spectral
peaks (results not shown). 

In summary, the comparable networks established with our procedure
exhibited average firing rate and position of the peak of the LFP
power spectrum that were both similar across network models and were
relatively robust to changes in the synaptic reversal potentials.
In our view this strengthens the value and usefulness of the setting
procedure introduced.

\subsection{Effects of synaptic models on network activity}

Previous seminal papers \citep{Meffin2004,Kuhn2004,Richardson2004}
compared the firing rate and MP of conductance- and current-based
LIF neurons. Our findings, summarized in table \ref{table_suppl_paperFR},
confirmed the main results of these previous works, and extended them
in several ways. Our main contribution was to extend the comparison
to include other aspects of neural population dynamics. In particular,
we considered the effect of the synaptic models on the spectrum of
network activity, on the cross-neuron correlations and on the stimulus
modulation of these different network features. The significance of
these advances is discussed in more detail below.

\subsection*{Correlation dynamics in the networks}

Despite the average firing rate was very similar in comparable COBNs
and CUBNs, spike trains of different neurons were more correlated
in the COBNs than in the CUBNs, with the correlation difference increasing
with the external input rate. The fact that the COBN spike train correlation
was more strongly modulated by the input rate led to the fact that
spike train correlation carried more information in the COBN.

In our networks, the neurons received inputs from the same simulated
external pool and this led to values of shared input that were likely
higher than those shared by pairs of cortical neurons recorded from
different electrodes. However, in the COBN, the dependence of correlation
on the network stimuli resembled qualitatively the one observed in
real experiments, more than in the CUBN. First, the positive correlation
between firing rate intensity and spike train correlation is often
observed in neurophysiological experiments, \citep{Kohn2005}, and
this behavior is only reproduced by the COBN. Further, MP of cortical
neurons \citep{Lampl1999} (but see also \citep{Yu2010}) are more
correlated when they receive an input triggering a stronger response
(i.e., having a higher contrast/the correct orientation). This resembles
the dynamics displayed here by the COBN, but not by the CUBN. Moreover,
in several experiments \citep{Isaacson2011} and references therein),
the correlation between AMPA and GABA synaptic inputs is stronger
the more intense is the stimulus, consistent with the COBN dynamics
shown in figure \ref{fig_8_paperFR}A. 

The high values of correlation that we found in the COBN might, at
first sight, look different from those of \citet{Renart2010} in which
a conductance-based LIF network, with a structure similar to the one
considered here, displayed a much smaller MP correlation thanks to
the decorrelation due to a precise balance between excitation and
inhibition. In other words, in that work, AMPA-GABA correlation and
cross-neuron MP correlation were described as mutually exclusive.
We think that the reason for the difference between their results
and those obtained in our work is the crucial assumption of \citet{Renart2010}
that AMPA and GABA timescales are identical. In a supplemental analysis
the authors showed indeed that, when AMPA synapses were made progressively
faster than GABA, the negative feedback was not fast enough to compensate
for excitation and hence to decorrelate the neurons; the network became
then more synchronized. When in \citet{Renart2010} the authors considered
the case in which $\tau_{r-exc}$ = 2 ms and $\tau_{r-inh}$ = 5 ms
(very close to our values, see table \ref{tab_tau_syn}), the correlation
between GABA and AMPA currents reached values above 0.5, coherent
with our results (figure \ref{fig_8_paperFR}A).

\subsection*{Frequency spectra of network activity}

We also compared the frequency spectra of the network activity (as
measured by LFP) in COBN and in CUBN. A marked difference was in the
larger amount of information and stronger stimulus modulation of the
gamma range for COBN. This, in our view, may be explained as follows.
When increasing the external input rate, we observed an increase of
the cross-neuron spike train correlation in the COBN, which was associated
with an increase of the cross-neuron correlation of the synaptic currents
(both AMPA and GABA). This caused a stronger modulation of the COBN
currents and consequently of the LFP gamma peak. The stronger modulation
of the gamma band in turn contributed to the fact that, both when
time-constant and time-varying inputs were injected, the COBN carried
more information than the CUBN in the gamma band. 

Neurophysiological recordings of LFP spectra modulation in visual
cortex during stimulation with various kinds of visual stimuli \citep{Henrie2005,Belitski2008}
reported much broader gamma peaks than the ones we found for COBNs.
The width of gamma peaks reported in cortical data was more similar
to the broad gamma peak generated by CUBN rather than to the sharp
peak generated by the COBN. We hypothesize that the sharpness of the
COBN gamma peak may be over-emphasized by the lack of neuron-to-neuron
heterogeneity in the specific network models implemented here. Introducing
a small degree of variability in neuronal parameters could decrease
the correlation in COBN while keeping it stimulus-dependent. An important
point for future research is to understand how heterogeneities in
network parameters differentially affect COBN and CUBN dynamics. 

A final point worth discussing is that the COBN, unlike the CUBN,
showed considerable amounts of information about input strength in
the LFP power in the frequency range 15\textendash{} 25 Hz. Notably,
the power of real visual cortical LFPs \citep{Belitski2008} also
did not carry information in this frequency range. Belitski and coworkers
hypothesized that the 15\textendash 25 Hz LFP frequency region related
mainly to stimulus-independent neuromodulation. The additive contribution
to the LFP of fluctuations generated by a stimulus-unrelated system
would potentially cancel out the information generated by the network
in this frequency range.

\section{Analyzing the relationship between cell-type specific single-unit
firing and mass signals\label{CONCL_Fellin}}

In chapter \ref{chapter_Fellin}, we investigated the features and
the dynamics of the empirical relationship between spiking activity
of individual neurons and mass signals. The spiking activity came
from both inhibitory and excitatory identified neurons, while the
concurrently recorded mass signals was measured as LFPs and EEGs in
mice under anesthesia. In particular, we analyzed if and to which
extent the spiking activity of single neurons can be estimated in
a general and blind way from the mass signals and vice versa. We also
characterized (i) how the estimation is significant and if it is robust
when increasing the generality of the algorithm used, (ii) which variables
mainly affect the relationship between SUAs and mass signals and finally
(iii) if there is an empirical causal direction in the relationship.

\subsection{Stability of the relationship}

We showed that, during slow wave oscillations, we can estimate in
a general way the slow oscillations of mass signals from the spiking
activity of a single neuron, both inhibitory (fast-spiking) and excitatory.
More precisely, we estimated with a strong accuracy the LFP (median
Spearman's correlation, $\langle r_{s}\rangle$, around 0.6) and,
with a good accuracy, even the EEG ($\langle r_{s}\rangle$ around
0.5) from the spiking activity of fast-spiking interneurons in layer
2. Similar results were obtained also when estimating the LFP from
the spiking activity of a pyramidal neuron ($\langle r_{s}\rangle$
around 0.5) in deep layers. On the other hand, we were able to estimate
(with a precision of 26 ms) in median the 30\% of the spike times
(mainly depending on the average firing rate) of a single fast-spiking
neuron from the mass signals recorded (i.e., LFPs and EEGs). The estimations
were performed with a simple linear model to which we added a non-linear
threshold in order to detect spike times. We found that, in both directions,
the estimations were highly significant. In particular, in case of
spike train estimation, the spikes were not simply placed at chance
level where the estimated FR was high (i.e., above thresholds), but
the positions of the peaks were actually related to the spike times.
We also verified that the filtering procedure was really useful to
increase the performances and we finally made a comparison with the
general linear model used in \citep{whittingstall2009frequency},
founding that the Wiener filter gave very similar results when estimating
FR, but higher performances when evaluating spike times. \\
Furthermore the results of all the estimations performed were remarkably
stable across cells and animals, allowing a truly general estimation
with no reduction in the performances. 

The relationship between mass signals and the underlying spiking activity
has been investigated widely using spike-triggered average and more
complex techniques during both sensory stimulation and absence of
stimulus \citep{schwartz2006spike,Rasch08,rasch2009neurons,nauhaus2009stimulus,okun2010subthreshold,zanos2012relationships,hall2014real,whittingstall2009frequency}.
In particular, \citet{rasch2009neurons} used animal-specific Wiener
filters to estimate the firing rate of MUAs from LFPs and \citet{hall2014real}
applied a similar method to estimate LFPs from the SUAs of multiple
(not specified) neurons. The same groups performed also the linear
estimation in the opposite direction, by computing the firing rate
\citep{hall2014real} and the spike times \citep{Rasch08} from the
LFP. \citet{whittingstall2009frequency} used general linear models
based on frequency decomposition of EEG (and LFP) to reconstruct the
firing rate of MUAs. However, to our knowledge, this is the first
case in which this estimation has been performed from the activity
of an individual (genetically-identified) GABAergic interneuron and
by using the same filter across all the animals to estimate mass signals.
\\
Note that LFPs integrate the postsynaptic signals coming from hundreds
to thousands of neurons \citep{Logothetis03}, and EEG integrates
signals on even a wider area than LFP. On the other hand, the activity
of a single neuron is a more localized signal than the multi-unit
activity, which has been used in the majority of the above mentioned
works. Furthermore the interneurons, due to their geometrical arrangement,
are likely to generate small dipoles (compared with pyramidal neurons)
when active \citep{murakami2006contributions}. For these reasons,
the strong and robust relationship found between mass signals and
single-unit spike trains is not trivially expected a priori. It reflects
the strong synchronization in the cortex activity observed during
slow wave oscillations, which is able to recruit also the interneuron
activity suggesting the existence of a robust control mechanism of
interneurons on network dynamics. \\
We found a very strong coupling even between the firing activity of
single pyramidal neurons and LFP, indeed, in presence of a strong
reduction in the average firing rates (median of 1.7 Hz, while it
was 4.5 Hz for fast-spiking neurons), which is a fundamental variable
when performing a linear reconstruction, the estimation performances
decrease only slightly %
. This is not surprising since the majority of the works that investigated
the relationship between LFPs/EEGs and firing activity considered
actually the firing activity of pyramidal neurons (see section \vref{PFsec:Introduction}).

\subsection{Variables shaping the relationship}

The estimation performances varied a lot from trial to trial (being
relatively constant for trials of the same cell) overall resulting
in a wide range of values. Our purpose is both to understand how the
mass signals time courses relate to the underlying spiking activity
of single neurons and to develop a general blind toolbox to estimate
the EEGs and the LFPs from SUAs and vice versa, with known estimation
accuracy. These questions are of paramount importance (i) to understand
how mass signals rely on the underlying neural computation, (ii) to
understand how the firing of single cells relates to the circuit ``context''
which led the neuron to fire (iii) and also for neuroprosthetic applications.\\
To achieve these goals, we investigated more in detail how the performances
of each trial are determined. In particular, we analyzed the correlation
properties of the performances with different features of both mass
signals and spiking activity.\\
We found that the variables mainly shaping the estimation performances
are the average firing rate and the power of the low frequencies of
the mass signals. Both of them are positively correlated with the
performances (and the highest correlation is usually observed with
the product of average firing rate and power spectrum), but their
relative weight depends on the estimation performed. \\
When estimating continuous signals (i.e., LFPs, EEGs and FRs), the
relative contributions of average firing rate and low power of the
LFP or EEG depend on the synchronization between the firing activity
and the mass signals. If the firing activity is synchronized with
the mass signals (as in the LFP case), when increasing the average
firing rate the performances strongly increase, because we have more
spikes to reconstruct the signals and in the ``right'' places\footnote{This has been found when the firing activity comes both from fast-spiking
interneurons and pyramidal neurons, data shown only for interneurons.}. On the other hand, when investigating the relationship between EEGs
and SUAs, the synchronization between FRs and mass signals is lower
and the weight of low frequencies power in determining the performances
increases. This is due to the fact that the slowest oscillations are
the strongest, therefore the ones better estimated with a linear method
(in agreement with previous results \citealp{rasch2009neurons,hall2014real}).
The more low frequencies we have, the higher the performances will
be (while a high number of spikes is not useful, since the spikes
could be no synchronized with the EEG signal).\\
When estimating spike trains, instead, the performance shows always
the largest correlation with the average firing rate (for estimation
from both LFPs and EEGs). In that case, indeed, we only take into
account the positions of the peaks in the estimated FR and (even due
to the increase of synchronization between mass signals and firing
rate when increasing the average firing rate) the more spikes there
are, the more likely the FR\textsubscript{est} peaks will be close
to the original ones. On the other hand, when the firing activity
is too sparse, a threshold applied on a linearly estimated signal
cannot efficiently detect the spike times. 

\subsection{Causality in the relationship}

An important and open issue in the comprehension of the interactions
occurring between single-cell dynamics and the dynamics of mesoscopic
and macroscopic circuits of neurons is given by the causality in the
relationship between these two levels of investigation. The LFP is
mostly generated by the totality of synaptic input and local processing
in a region \citep{Rasch08}, however, the way it is related to the
spiking output of underlying neurons is unknown. In other words: mass
signals, such as LFPs and EEGs, can be considered as the ``input''
for the network spiking dynamics, being thus responsible for the underlying
single-neuron activity, or vice versa the collective single-neuron
activities causally shape the time course of the mass signals? Probably
no one of these two extreme cases describes the truth, since the neural
circuits in the cerebral cortex are characterized by recurrent connections,
complex patterns of excitation and inhibition and inputs from multiple
structures \citep{douglas1989canonical,douglas2004neuronal}. Thus
we did not assume any a-priori causal constraint in the investigation
of the relationship between SUAs and mass signals and our estimation
were not causal (indeed the filters were not equal to zero for negative
time lags). This means that we could use spikes fired in $t>t^{*}$
to estimate mass signals in $t^{*}$and vice versa. 

To have a deeper insight about the causality issue, we tested if and
to which extent the performances are affected when imposing a causal
direction in the relation between SUAs and mass signals. We cross-validated
the results by repeated the analysis on both the directions of estimation
(i.e., spk2LFP/EEG and LFP/EEG2spk) and we found concordant results.
In particular, when assuming that the single-neuron spiking activity
causally shapes the mass signal fluctuations, the estimation performances
were less affected with respect to imposing an anti-causal relation
(analogously to what found by \citet{rasch2009neurons} when estimating
the LFPs from MUAs during visual stimulation). This was observed for
both interneuron in layer 2 and pyramidal neurons in deep layers.
\\
In conclusion, we found that spike times anticipate changes in the
time course of mass signals in a reliable way (and vice versa), thus,
from an empirical point of view, the spiking activity of single neurons
can be viewed as a ``stimulus'' for the mass signals. This result
is not obvious and requires further investigations, indeed the spiking
activity is usually considered as the output of a cortical area, whereas
the LFP as the processing of the entire subthreshold local signals
\citep{Logothetis08}.

\section{Perspectives}

In chapter \ref{chapter_frontiers}, we investigated the effects of
assuming different single-neuron models on network dynamics as a function
of the input to the network. In that network model there were two
sources of noise. The first was due to the stochastic process, which
affected the time varying rates, $\nu_{ext}(t)$, identical for all
neurons. The second source of noise, instead, was due to the fact
that each neuron received an independent realization of the Poisson
process with rate $\nu_{ext}(t)$. The fluctuations due to this second
source of noise were uncorrelated across neurons and, in particular,
they increased with the input rate. \\
However, noise in the brain is correlated across neurons, meaning
the fluctuations in the response of a neuron at fixed stimulus are
correlated with the fluctuations of other neurons \citep{Averbeck2006}.
If assuming the noise is uncorrelated, the investigation of network
dynamics is simpler and population coding is relatively well understood
\citep{Averbeck2006}. Therefore, for the computational work, it is
important to extend the theories to take into account noise correlation
and, in particular, to investigate how correlated noise affect network
dynamics. Thus, a direction of further investigation would be to disentangle
the role of changes in the mean input rate (that is a source of correlated
noise) from the ones of changes in the variance of the input across
neurons (that is a source of uncorrelated noise) in shaping network
dynamics. \\
Another very interesting direction for further research consists on
the the analysis of network dynamics when changing the topology of
the network \citep{prettejohn2011methods}. In our view, it would
be worth investigating for example, the effect of changing synaptic
dynamics and other biophysical parameters in networks composed of
clusters of strongly interconnected neurons \citep{litwin2012slow}
and compare it with the dynamics generated by the random connectivity
that we adopted. Studies like this would help to understand the relationship
between the anatomical pattern of synaptic connections and the pattern
of functional connectivity (defined as the set of statistical dependencies
between the activity of the elements of the neural network). This
is a hot topic in neuroscience \citep{Fasoli2015,deco2013resting,cabral2012modeling,eickhoff2010anatomical,ponten2010relationship,sporns2004organization}

In the work described in chapter \ref{chapter_Fellin}, we focused
on the relationship between single-neuron activity and mesoscopic
or macroscopic signals, as measured respectively by LFPs or EEGs.
We found that the relationships of SUAs with respectively LFPs and
EEGs have some similarities (for example, the shapes of the filters,
the estimation robustness when using more general filters, the empirical
causal direction of the relationships etc...). This is not surprising,
indeed, the LFPs are considered as the building blocks of EEG signals
\citep{da2013eeg} or, analogously, a more localized variant of the
EEGs \citep{whittingstall2009frequency}. Nevertheless this represents
a first approximation of the relationship occurring between LFPs and
EEGs. In order to gain a better insight about this relation, we could
take advantage of the fact that in our datasets SUA, LFP and EEG are
simultaneously recorded. Thus, we could investigate more in detail
which aspects of the relationship between SUA and LFP differs from
the one existing between SUA and EEG. In particular, we could identify
which dynamics in the relationship between EEGs and SUAs are less
reliable (with respect to the LFP-SUA case), for example, by performing
frequency decomposition of the mass signals and studying the locking
of SUAs with phase and power of network oscillations. In the end,
the next purpose will be to investigate the relationship between LFPs
and EEGs by pointing out the dynamics responsible for the performance
decrease observed when estimating single-neuron FR from EEG (and vice
versa).

\newpage{}

\bibliographystyle{agsm}
\addcontentsline{toc}{chapter}{\bibname}\bibliography{PhD_thesis}

\newpage{}

\thispagestyle{empty}

~

\newpage{}

\mbox{} 
\thispagestyle{empty}
\begin{center} 
\textsf{\textbf{\large Acknowledgements}}
\end{center}

First and foremost, my deepest gratitude goes to my Ph.D. supervisor
Professor Stefano Panzeri, which gave me the opportunity to enter
the fascinating field of neuroscience. 

I wish to thank Alberto Mazzoni and Daniel Chicharro for many valuable
suggestions and (not only scientific) discussions, for sharing thoughts
and support during these years... it was really a pleasure to share
the office with you! I cannot avoid to mention Rupan Raventos, which
has been an example of consistency for all of us.

A special thank to Pietro Salvagnini for his friendship and for being
always available to help me and to answer my questions. I am also
grateful to Cesare Magri for a first version of the code I used to
perform my work.

I am not forgetting Stefano Zucca, Tommaso Fellin and other colleagues
in the laboratory of Tommaso Fellin, which kindly provided me with
the data for the analysis performed in this thesis.

I want to thank Alessandro Maccione for his motivating words when
I really needed them and Paolo Mereghetti, which has been my last
office mate at IIT... before the ``diaspora''.

Thanks to my family which stood by me in all the good and bad moments.
I simply would not be here without their love and dedication...

Finally, thanks to my wife, Elisabetta, which was surprisingly able
to give me unexpected wonderful times in everyday life... you have
been my constant source of energy and motivation.

\vfill{}

\begin{flushleft}
\textit{\small{}PS: è stato divertente giocare in allegria,}\\
\textit{\small{}c'è un solo inconveniente che il tempo vola via...
}\\
\textit{\small{}Addio addio amici addio.}
\par\end{flushleft}{\small \par}
\end{document}